\documentclass[10pt,epsfig]{article}

\usepackage{enumerate}
\usepackage{float}
\usepackage{subfig}
\usepackage{color}
\usepackage[table]{xcolor}
\usepackage{slashed}
\usepackage{multirow}
\usepackage{cite}

\let\counterwithin\relax
\usepackage{chngcntr}
\usepackage{amssymb, amsmath,mathrsfs}

\usepackage{graphics}
\usepackage{graphicx}
\usepackage{epsf}
\usepackage{epsfig}
\usepackage{float}
\usepackage{makecell}
\usepackage{multirow}
\usepackage{color}
\usepackage{xcolor}
\usepackage{simplewick}

\usepackage[utf8]{inputenc}
\usepackage{amsmath}
\usepackage{amssymb}
\usepackage{subfig}
\usepackage[normalem]{ulem}

\usepackage{mathtools}
\usepackage{amsmath}

\newcommand\undermat[2]{
  \makebox[0.5pt][l]{$\smash{\underbrace{\phantom{%
    \begin{matrix}#2\end{matrix}}}_{ \let\scriptstyle\textstyle\text{\large $#1$}}}$}#2}
\newcommand\overmat[2]{
  \makebox[-1pt][l]{$\smash{\overbrace{\phantom{%
    \begin{matrix}#2\end{matrix}}}^{ \let\scriptstyle\textstyle\text{\large $#1$}}}$}#2}    
\usepackage{tikz}
\usepackage[vcentermath]{youngtab}
\usepackage{slashed} 
\usepackage[font=small]{caption} 
\usepackage{times} 

\usepackage{bm}       
\usepackage{bbm} 

\usepackage{relsize}  

\usepackage{autobreak}  

\usepackage{makeidx} 
\usepackage{bbding} 
\usepackage{listings} 
\usepackage{ytableau}

\usepackage[colorlinks=true,
            linkcolor=blue,
            urlcolor=red,
            citecolor=red]{hyperref}

\long\def\rpl#1!!#2!!{\textcolor{red}{#1} \textcolor{blue}{#2}}

\usepackage[top=1in, left=0.95in, bottom=1.1in, right=0.95in]{geometry}
\def\baselinestretch{1.27}
\usepackage[toc]{appendix}

\newcommand{\beq}{\begin {equation}}  
\newcommand{\eeq}{\end   {equation}} 
\newcommand{\bea}{\begin {eqnarray}} 
\newcommand{\eea}{\end   {eqnarray}}  
\newcommand{\baa}{\begin {array}   } 
\newcommand{\eaa}{\end   {array}   }     
\newcommand{\bit}{\begin {itemize} }
\newcommand{\eit}{\end   {itemize} }
\newcommand{\be }{\begin {equation}} 
\newcommand{\ee }{\end   {equation}}
\newcommand{\nn }{\nonumber        }

\newcommand{\mc}[1]{\mathcal{#1}}

\newcommand{\ket}[1]{| #1 \rangle}
\newcommand{\bra}[1]{\langle #1 |}
\newcommand{\vev}[1]{ \left\langle {#1}  \right\rangle }

\newcommand{\eq}[1]{\begin{equation}\begin{split} #1 \end{split}\end{equation}}

\newcommand{\4}{\!\!\!\!/\,}
\newcommand{\comments}[1]{}

\newcommand*\circled[1]{
	\tikz[baseline= -2.5pt]{\node[shape=circle,draw,inner sep=0.3pt] (char) {#1};}
}

\newcolumntype{M}[1]{>{\centering\arraybackslash}m{#1}}
\newcolumntype{N}{@{}m{0pt}@{}}

\allowdisplaybreaks
\begin{document}

\begin{center}

{\Large \textbf  {Complete Set of Dimension-9 Operators in the Standard Model Effective Field Theory}}\\[10mm]

Hao-Lin Li$^{a}$\footnote{lihaolin@itp.ac.cn}, Zhe Ren$^{a, b}$, Ming-Lei Xiao$^{a}$\footnote{mingleix@itp.ac.cn}, Jiang-Hao Yu$^{a, b, c}$\footnote{jhyu@itp.ac.cn}, Yu-Hui Zheng$^{a, b}$\\[10mm]

\noindent 
$^a${\em \small CAS Key Laboratory of Theoretical Physics, Institute of Theoretical Physics, Chinese Academy of Sciences,    \\ Beijing 100190, P. R. China}  \\
$^b${\em \small School of Physical Sciences, University of Chinese Academy of Sciences,   Beijing 100049, P.R. China}   \\ 
$^c${\em \small School of Fundamental Physics and Mathematical Sciences, Hangzhou Institute for Advanced Study, \\ University of Chinese Academy of Sciences, Hangzhou 310024, China} \\ [10mm]

\date{\today}   
          
\end{center}

\begin{abstract} 
	
We present a complete and independent list of the dimension 9 operator basis in the Standard Model effective field theory by an automatic algorithm based on the amplitude-operator correspondence. 
A complete basis (y-basis) is first constructed by enumerating Young tableau of an auxiliary $SU(N)$ group and the gauge groups, with the equation-of-motion and integration-by-part redundancies all removed.
In the presence of repeated fields, another basis (p-basis) with explicit flavor symmetries among them is derived from the y-basis, which further induces a basis of independent monomial operators through a systematic process called de-symmetrization.
Our form of operators have advantages over the traditional way of presenting operators constrained by flavor relations, in the simplicity of both eliminating flavor redundancies and identifying independent flavor-specified operators. 
We list the 90456 (560) operators for three (one) generations of fermions, all of which violate baryon number or lepton number conservation;  
among them we find new violation patterns as $\Delta B = 2$ and $\Delta L = 3$,  which only appear at the dimensions $d \ge 9$.

\end{abstract}

\newpage

\tableofcontents

\setcounter{footnote}{0}

\def\baselinestretch{1.5}
\counterwithin{equation}{section}

\newpage
%
%

\section{Introduction}
\label{sec:intro}

Being the most successful theory of particle physics to date, the standard model (SM) still leaves many questions about the nature of matter unanswered, which motivates direct and indirect experimental searches on new physics (NP). 
For instance, the baryon asymmetry of the universe and nonzero neutrino masses may indicate that the baryon number $\Delta B$ and the lepton number $\Delta L$ should be violated via additional new degrees of freedom. 
The absence of signals of physics beyond the SM at the Large Hadron Collider (LHC) suggests that new particles are either very weakly coupled or much heavier than the electroweak scale. 
Assuming that new particles live at high energies, $\Lambda$, well above the electroweak scale, their effects at experimental energies much below $\Lambda$ 
can be systematically described under the effective field theory (EFT) framework.

The Standard Model effective field theory (SMEFT) provides a systematic approach to describe the effects of heavy particles at low energy in a model-independent way. 
%
The SMEFT Lagrangian can be systematically organized by the dimension of effective operators in inverse powers of the heavy scale $\Lambda$, as follows
\bea
	\mathscr{L}_{\mathrm{SMEFT}}=\mathscr{L}_{\mathrm{SM}}+\frac{1}{\Lambda} \mathscr{L}_{5}+\frac{1}{\Lambda^{2}} \mathscr{L}_{6}+\frac{1}{\Lambda^{3}} \mathscr{L}_{7}+\frac{1}{\Lambda^{4}} \mathscr{L}_{8}+\frac{1}{\Lambda^{5}} \mathscr{L}_{9}+\cdots, \quad \mathscr{L}_{d}=\sum_{i} c_{i}^{(d)} \mathcal{O}_{i}^{(d)}
\eea
where each $\mathcal{O}_{i}^{(d)}$ denotes a Lorentz-and gauge-invariant operator of canonical mass dimension $d$  with SM degrees of freedom only, and its Wilson coefficient $c_{i}^{(d)}$ parameterizes size of possible deviations from SM predictions.
%
For each dimension $d$, the $\mathcal{O}_{i}^{(d)}$ construction follows that one writes all the possible Lorentz and gauge invariants using SM fields solely. 
Although it is possible to find a set of operators with Lorentz and gauge invariance for a given mass dimension $d$, these sets might be redundant due to possible relations between different operators.
%
By means of equations of motion (EOM), Fierz identities, and integration by parts (IBP), one can eliminate redundancies for each dimension
and obtain a complete and also independent operator basis.  
The operators up to dimension 7 have been listed in this way in Ref.~\cite{Weinberg:1979sa, Buchmuller:1985jz, Grzadkowski:2010es, Jenkins:2013zja, Jenkins:2013wua, Alonso:2013hga, Lehman:2014jma, Liao:2016hru}. 
At dimension 8 and higher, the number of such operators increases tremendously, which makes the task very tedious and prone to error. 
Instead, we provided a systematic and automated method~\cite{Li:2020gnx} to write a complete and independent basis directly, which has been applied to listing the complete dimension 8 operators in the SMEFT.  
At the same time, Ref.~\cite{Murphy:2020rsh} utilizing the traditional way to treat the EOM and IBP redundancies also write down the dimension 8 operators. Compared to Ref.~\cite{Murphy:2020rsh}, since we started from the operators in which the EOM is absent and the IBP is treated in the beginning, the correctness of our result is theoretically guaranteed from the first principle. It is also pointed out~\cite{Li:2020gnx} that our method provides a relatively simple way to enumerate all the independent flavor-specified operators, while the traditional method has not.

Both the origin of the matter-antimatter asymmetry and the Majorana origin of the neutrino masses are tied to the baryon number violation (BNV) or lepton number violations (LNV), which only arise at the non-renormalizable level of the SM Lagrangian. 
Therefore, the $\Delta B$- and $\Delta L$-violating processes, such as nucleon decay, neutron-antineutron oscillation, and neutrino-less double beta decay, could be parametrized systematically in the SMEFT.
The SMEFT systematically classifies effective operators with the fact that
(a) baryon and lepton number violation (BLV) has to occur with integer units and (b) $( \Delta B,  \Delta L )$ = (odd, odd) or (even, even),
and (c) the $\Delta B - \Delta L$ number violation satisfies
\bea
	|\Delta B - \Delta L|   = 
	\begin{cases} 
		0, 4, 8, \cdots, \quad {\textrm{for}\,\, d = \textrm{even}}, \\
		2, 6, \cdots, \quad {\textrm{for}\,\, d = \textrm{odd}},
	\end{cases}
\eea
based on the requirement that the operator is invariant under the weak hypercharge symmetry and the Lorentz symmetry~\cite{Alonso:2014zka, Kobach:2016ami, Helset:2019eyc}. 
Therefore, from above we learn that operators at odd dimensions must have BNV or LNV, 
and that $|\Delta B - \Delta L| = 2$ up to dimension 15~\footnote{On the other hand, for the dimension-even operators, we have $|\Delta B - \Delta L| = 0$, and  $( \Delta B,  \Delta L ) = (0, 0), (\pm 1, \pm 1)$ at the dimension 6 and 8 level. 
These operators with $( \Delta B,  \Delta L ) = (\pm 1, \pm 1)$ cause proton decay in modes such as Grand Unifications, and thus highly constrained by proton two-body decay searches. 
Starting from dimension 10, we have $|\Delta B - \Delta L| = 0, 4$ with additional BLV possibilities $( \Delta B,  \Delta L ) = (-1, 3), (0, 4)$ (dim 10), and $( \Delta B,  \Delta L ) = (-2, 2), (2, 2)$ (dim 12), etc. 
}.
At the odd dimensions, the LNV processes with $\Delta L = 2$, relevant to the leptogenesis mechanism, neutrino-less double beta decay, and the neutrino masses, exist~\cite{Babu:2001ex, deGouvea:2007qla}. 
For example, if the leptogenesis or baryogenesis occurs at temperatures above the weak scale, $B - L$ violation is required to avoid the washout effect by the  electroweak sphalerons 
and at the same time, constraints from proton two-body decays (which conserves $B-L$) are not applicable. 
At dimension 5, the only operator is the Weinberg operator~\cite{Weinberg:1979sa} with $( \Delta B,  \Delta L )$  = (0, 2), 
while at dimension 7, all the operators have BLV with possibilities $( \Delta B,  \Delta L ) = (0, \pm2), (\pm 1, \mp 1)$~\cite{Lehman:2014jma}, which either break the lepton number by 2 or induce proton two-body decay. 

Starting from dimension 9, besides the operators with $( \Delta B,  \Delta L ) = (0, \pm2), (\pm 1, \mp 1)$, there are new violation patterns in the operators with $( \Delta B,  \Delta L ) = (\pm 1, \pm 3), (\pm 2, 0)$.
First, operators relevant for $\Delta B = 2$ processes, such as neutron-antineutron oscillations, appear first at $d = 9$~\cite{Grojean:2018fus}, which are directly connected to the low-scale realization of the baryogenesis without the need for sphaleron processes. 
Second, the $\Delta L =3$ processes can only arise from dimension 9 and higher operators~\cite{Fonseca:2018ehk, Heeck:2019kgr}.
The lepton number violated only in three units implies  
the proton decay final states must be at least three-body and the new physics associated at a scale could be as low as 1 TeV, which opens the possibility of searching for such processes not only in proton decay experiments but also at the LHC~\cite{Heeck:2019kgr}. 
Finally, operators with $\Delta L = 2$ are supposed to be sub-dominate over the ones at dimension 5 and 7 levels. 
However, if the $\Delta L = 2$ operators start to appear at the dimension 9 level, new physics effect could be as low as 1 TeV and thus can be tested at the LHC in the near future. 
For example, typically the operators for the Majorana neutrino masses, such as Weinberg operators, are related to the tree-level seesaw, and thus the new physics scale is quite high. However, if the Majorana neutrino masses are generated from the tree-level mechanisms at dimension 9, the related new physics is around the TeV scale~\cite{Anamiati:2018cuq}. Thus one expects that the LHC experiment will start to explore these kinds of models in the near future.
In the neutrino-less double beta decay processes, if the dominant contributions originate from the dimension 9 operators~\cite{Graesser:2016bpz, Cirigliano:2018yza}, one expects the new physics scale should be around TeV, and thus collider experiments could also shed light on such kinds of new physics in the near future. Hence, listing a complete set of dimension 9 operators will set up the framework for these phenomenology studies.

We adopt the method in \cite{Li:2020gnx} to list the dimension 9 operators in the SMEFT. While the method is elaborated in \cite{Li:2020gnx}, we present in this paper more details about its motivation stemming from the so-called amplitude-operator correspondence. 
By establishing the one-to-one correspondence between the effective operators and the local amplitudes they generate, we first categorize them in terms of the external states in the scattering -- a certain collection of particles in the EFT. A category of operators thus found is called a type. 
For a given type of operators, we define a couple of bases for various uses as follows:
\begin{itemize}
\item y-basis: 
	Our algorithm utilizes group theory technique to enumerate an independent and complete basis as a collection of Young tableau for each factor of the operators, thus named Young tableau basis or y-basis. For the Lorentz factor, the basis is obtained as the Semi-Standard Young Tableau (SSYT) of an auxiliary $SU(N)$ group, where $N$ is the number of fields; for the gauge groups, the basis is given by Young tableau constructed from the Littlewood-Richardson (L-R) rule. 
\item m-basis:
	For practical purposes, the operator basis had better be monomials, while the y-basis operators, after transforming to the usual convention, are often long polynomials. By a systematic reduction to y-basis, we can select a set of monomial operators that have independent coordinates with respect to the y-basis. A complete basis of monomial operators selected this way is called an m-basis, which is highly non-unique. 
\item p-basis: 
	Even the y-basis is not enough when repeated fields are present, as explained in \cite{Li:2020gnx} from the operator viewpoint. In this paper, we also illustrate this extra constraint from the amplitude viewpoint, which introduces the symmetric permutation basis, or p-basis, as the symmetrized flavor-blind amplitude basis and the corresponding operators. The symmetrization procedure provides a full-rank conversion matrix from the y-basis to the p-basis, which guarantees its independence and completeness. p-basis operators when viewed as flavor tensors of a group of repeated fields, the ones that form a basis of an irrep of the symmetric group are related by certain permutations.
\item p'-basis:
	To reduce the lengths of operators in the usual notation, while keeping the flavor symmetries manifest, we develop a systematic procedure, the \emph{de-symmetrization}, to obtain a series of m-basis operators that symmetrize to independent combinations of the p-basis with the same flavor symmetry. The procedure is especially important if multiple representation spaces of the same symmetry exist. This is a new part of our method that was not developed in \cite{Li:2020gnx}.
\end{itemize}

The resulting operator basis we obtain with the above method is listed in terms of various levels of categories:
\begin{itemize}
\item Class:
	A (Lorentz) class includes types of operators with a given amount of fields under each Lorentz irreducible representation (irrep) and the same amount of covariant derivatives, such that they may share the same Lorentz structures. 
	It is different from the concept of operator ``class'' in other literature, because we distinguish the chiralities of the fields as their corresponding particles have definite helicities. In particular, fermions and gauge bosons should be written on the chiral basis in our notation. 
	The list of possible classes at a given dimension is model-independent, as we show in the tables at dimension 9, though not all of them show up in specific models.
	
\item Type:
	The definition is given previously. All the types are obtained by plugging field content of the SMEFT into the dimension 9 classes, making sure that the representations of them could form singlets for each symmetry group.
	We emphasize that our ``type'' has more rigorous definitions than those in the other literature, as we specify the way to eliminate the EOM redundancy so that the type of operators we define only corresponds to local amplitudes they can generate for a given collection of external particles. 
	
\item Term:
	The p-basis or p'-basis are operators with free flavor indices, which contract with Wilson coefficient tensors to form a (Lagrangian) term. The corresponding amplitude basis is flavor-blind. Our ``terms'' are irreducible flavor tensors with a specific flavor symmetry $\lambda$, different from the concept of ``terms'' in other literature \cite{Fonseca:2019yya,Murphy:2020rsh} where flavor tensors with different symmetries may merge into a reducible tensor. We compare the form of our terms to the traditional form of operators with flavor relations \cite{Grzadkowski:2010es,Alonso:2014zka,Liao:2016qyd} to show their equivalence, and explain the privileges of our form.
	
\item Operator:
	The number of (flavor-specified) operators per term can be understood as the independent entries in the Wilson coefficient tensor, constrained by the flavor symmetry. One can also view the independent operators as p-basis contracted with independent flavor tensor basis, labeled by flavor Young tableau. 
\end{itemize}

The paper is organized as follows. 
In section.~\ref{sec:class}, we discuss the principle to find independent and complete operators with the amplitude-operator correspondence. 
In section.~\ref{sec:eft}, we describe the general ideas of how to obtain a complete set of independent operators with free flavor indices in $y$-, $m$-, $p$-bases and how to convert them to each other. 
In section.~\ref{sec:basis}, we take a concrete example to show how to obtain a set of independent terms for a given type of operators and demonstrate the advantages of listing operators in the level of terms with definite flavor permutation symmetry. 
In sections~\ref{sec:list}, we list all the independent terms for dimension-9 in the SMEFT with different categories. 
We reach our conclusion in section.~\ref{sec:conclusion}. 
Additionally, in appendix.~\ref{app:A}, we list useful formulae transforming operators between two- and four-component spinor notations, and in appendix.~\ref{app:B} we provide a list of sub-classes up to dimension 9.

\section{On-shell Convention for Effective Operators}
\label{sec:class}

The Lagrangian of the SMEFT consists of the SM fields
\bea
	& {\rm Fermion} :& L_{\alpha i}, e_{_\mathbb{C}\alpha}, Q_{\alpha ai}, u_{_\mathbb{C}\alpha}^a, d_{_\mathbb{C}\alpha}^a, \\
	& {\rm Boson} :  & G_{\mu\nu}^A, W_{\mu\nu}^I, B_{\mu\nu}, H_i,
\eea
and their covariant derivatives $D_\mu$ along with the following group factors:
\bea
	& {\rm Lorentz} :& \ \sigma^{\mu\nu}_{\alpha\beta}, \bar\sigma^{\mu\nu}_{\dot\alpha\dot\beta}, \sigma^\mu_{\alpha\dot\alpha}, \bar\sigma^{\mu\dot\alpha\alpha}, \epsilon^{\alpha\beta}, \tilde\epsilon^{\dot\alpha\dot\beta}, \\
	& SU(2) :& \quad \epsilon^{IJK},\delta^{IJ},(\tau^I)_i^j,\epsilon_{ij},\epsilon^{ij},\\
	& SU(3) :& \quad f^{ABC},d^{ABC},\delta^{AB},(\lambda^A)_a^b,\epsilon_{abc},\epsilon^{abc},
\eea
which result in the invariant operators under the Lorentz group $SL(2,\mathbb{C}) = SU(2)_{l}\times SU(2)_{r}$ and the SM gauge group $SU(3)_C \times SU(2)_W \times U(1)_Y$.
Here the indices for the fundamental representation of the $SU(2)_l$ and $SU(2)_r$ groups are denoted by $(\alpha, \beta, \gamma, \delta)$ and $(\dot{\alpha}, \dot{\beta}, \dot{\gamma}, \dot{\delta})$. 
The indices for the fundamental and adjoint representations of the $SU(3)_C$ are $(a, b, c, d)$ and $(A, B, C, D)$ while those for the $SU(2)_W$ are $(i, j, k, l)$ and $(I, J, K, L)$, respectively. These are conventional notations for the effective operators, which we use to represent our final result -- the complete set of dim-9 operators in the SMEFT -- in section~\ref{sec:list}. 

Although operators in this notation are more familiar to phenomenologists, it is hard to systematically define an independent basis for them, given the redundancies due to the EOM and the IBP relation. The usual way to achieve this goal is to write down an over-complete basis and derive their dependencies manually \cite{Grzadkowski:2010es,Liao:2016hru,Hays:2018zze,Gripaios:2018zrz,Fonseca:2019yya,Murphy:2020rsh}. However, this has to be done model by model, and becomes extraordinarily cumbersome at higher dimensions. In Ref.~\cite{Shadmi:2018xan, Ma:2019gtx, Aoude:2019tzn, Durieux:2019eor, Falkowski:2019zdo}, it was pointed out that independent operators could be enumerated in terms of their corresponding local on-shell amplitudes, dubbed the amplitude basis. Ref.~\cite{Henning:2019enq} further proposed an algorithm to enumerate independent amplitude basis subject to momentum conservation, which is equivalent to the IBP redundancy as we will explain shortly. In Ref.~\cite{Li:2020gnx}, an integrated algorithm using the correspondence was proposed and applied to the enumeration of the dimension 8 operators in the SMEFT. In this section, we would like to elaborate the amplitude-operator correspondence and prove its applicability to the task of operator enumeration.

\subsection{Amplitude-operator correspondence}
\label{sec:amp}

The correspondence is in particular about operators as Lagrangian terms, which are Lorentz singlets, so that they directly contribute to scattering amplitudes. Among the amplitudes they contribute, the set of local amplitudes or ``amplitude basis''\footnote{%
	More specifically, the amplitude basis are featured as being ``unfactorizable'', in the sense that they do not have poles or branch cuts in the kinematic space where they should factorize due to the unitarity. This feature makes them the building blocks of any amplitudes, because the factorizable ones should ultimately factorize to them at particular kinematic configuration. } %
span a linear space isomorphic to the operator space. 
To prove the isomorphism, we first investigate the general structure of amplitude basis, which we express in terms of the spinor helicity variables $\lambda_{i\alpha},\tilde{\lambda}_i^{\dot\alpha}$, defined as
\eq{
	p_i^\mu = \lambda_i^\alpha \sigma^\mu_{\alpha\dot\alpha} \tilde\lambda_i^{\dot\alpha} \;,
} 
up to the little group transformations\footnote{%
	The definition can be extended for massive particles, whose little group is $SU(2)$ and hence the spinor variables have an extra $SU(2)$ index $I,J\dots$. In this paper we only enumerate amplitude basis for massless particles.}%
$\lambda_i \to e^{-i\varphi/2}\lambda_i$, $\tilde\lambda_i \to e^{i\varphi/2}\tilde\lambda_i$, while the spinor indices are raised and lowered by the Levi-Civita tensor $\epsilon^{12}=-\epsilon_{21}=+1$. 
The number of constituting spinors is constrained by the little group representations of the external particles, e.g. the helicities for massless particles. 
The amplitude basis $\mc{B}$ should respect the little group representations of all the external particles. For example, under the little group $U(1)_i$ for the $i$th massless particle, it should gain a phase $\mc{B} \to e^{ih_i\varphi}\mc{B}$.
Therefore in general, massless particle of helicity $h_i$ contributes a factor $\lambda_i^{r_i-h_i}\tilde\lambda_i^{r_i+h_i}$ that has the correct little group weight, where $r_i \geq |h_i|$ is a free (half-)integer parameter.
The general form of the amplitude basis reads
\eq{\label{eq:amplitude_basis}
	\mc{B}\left( \phi_1^{a_1}(p_1),\dots,\phi_N^{a_N}(p_N) \right) = T^{a_1,\dots,a_N} \mc{M}(h_1,\dots,h_N)  , \qquad \mc{M}(h_1,\dots,h_N) \sim \prod_{i=1}^{N}\lambda_i^{r_i-h_i}\tilde\lambda_i^{r_i+h_i}
}
where $\phi_i, i=1,\dots,N$ are the external particle multiplets with momenta $p_i$, and $a_i$ are the collections of group indices for them.
The mass dimension of the amplitude is determined as $ [\mc{B}] \equiv r = \sum_i r_i$. 
The \emph{kinematic factor} $\mc{M}$ is a function of the spinor variables that only depends on the helicities $h_i$ of the external particles, and characterize the energy dependency and the angular distribution of the amplitude. 
Global Lorentz invariance demands that all spinor indices are contracted, which are conventionally denoted as
\eq{
	\lambda_i^{\alpha}\lambda_{j\alpha} =: \vev{ij}, \qquad 
	\tilde\lambda_{i\dot\alpha}\tilde\lambda_j^{\dot\alpha} =:[ij],
} 
thus $\mc{M}$ must consist of $n=\frac{r-h}{2}$ number of $\vev{\cdot}$ type brackets and $\tilde{n}=\frac{r+h}{2}$ number of $[\cdot]$ type brackets, $h=\sum_ih_i$ being the total helicity. 
The \emph{group factor} $T$ is the product of tensors for each group under which the multiplets $\phi^a$ transform. 
For symmetry groups, like the gauge group or some global symmetry group, $T$ has to be invariant tensors. The index $a$ can also include the flavor degree of freedom, which not necessarily has a symmetry, while the tensor in charge does not have to be invariant tensors. 
We can define the subspace of local amplitudes with the same set of external particles $\phi_1^{a_1},\dots,\phi_N^{a_N}$ and the same mass dimension $r$ as a ``type'', in which various amplitude basis are specified by the group tensors $T$, the partition $r_i$, and the structure of spinor contractions. Furthermore, the types with the same tuple $(h_1,\dots,h_N;r)$ form a ``class'' that share the same bases of the kinematic factors $\mc{M}$.
Note that we do not have to specify the division between the initial and final states because different divisions are simply related by crossing symmetry and analytic continuation.

Here we take a simple example to illustrate: the amplitude basis for 4 left-handed fermions. Each of them contributes a factor $\lambda_i^{r_i+1/2}\tilde\lambda_i^{r_i-1/2}$ where $r_i$ is a positive half integer. The lowest mass dimension is when $r_i=1/2$ and $r=2$, and we have the following possible contractions
\eq{
	& \mc{M}_1(-1/2,-1/2,-1/2,-1/2;r=2) = \vev{12}\vev{34}, \\
	& \mc{M}_2(-1/2,-1/2,-1/2,-1/2;r=2) = \vev{13}\vev{42}, \\
	& \mc{M}_3(-1/2,-1/2,-1/2,-1/2;r=2) = \vev{14}\vev{23}.
}
Schouten identity indicates that $\mc{M}_1 + \mc{M}_2 + \mc{M}_3 = 0$, which reduces the number of independent amplitude basis by 1. This redundancy is equivalent to the Fierz identity for operators, which we will solve systematically later. For higher mass dimension $r=3$ where one of the $r_i$ takes $3/2$, there is only one $\tilde\lambda$ which cannot form Lorentz singlet. Thus the next available dimension is $r=4$, for instance $r_1=r_2=3/2$ and $r_3=r_4=1/2$, and one possible amplitude basis is
\eq{
	\mc{M}_1(-1/2,-1/2,-1/2,-1/2;r=4) = \vev{12}^2\vev{34}[12].
}
Later in section \ref{sec:lor_inv}, we will derive the full constraints on these parameters, so that we can enumerate the valid classes $(h_1,\dots,h_N;r)$ that could form Lorentz singlet.

To find the operator that generates such amplitude basis, one simply does the following translation
\eq{\label{eq:dictionary}
	\begin{array}{lcl}
		\lambda_i^{r_i\pm1}\tilde\lambda_i^{r_i\mp1}		&	\Leftrightarrow	&	D^{r_i-1}F_{{\tiny\rm L/R}\,i}			\\
		\lambda_i^{r_i\pm1/2}\tilde\lambda_i^{r_i\mp1/2}	&	\Leftrightarrow	&	D^{r_i-1/2}\psi^{(\dagger)}_i	\\
		\lambda_i^{r_i}\tilde\lambda_i^{r_i}				&	\Leftrightarrow	&	D^{r_i}\phi_i
	\end{array}
}
where $F_{\rm L/R} = \frac{1}{2}(F \mp i\tilde{F})$ are the chiral basis of the gauge bosons, and $\psi$ denotes left-handed Weyl spinors. For a unified notation, right-handed Weyl spinors are denoted as conjugates of some left-handed spinors $\psi_{\rm R}^{\dot\alpha} = \epsilon^{\dot\alpha\dot\beta}(\psi_{\rm L}^\dagger)_{\dot\beta}$. 
All the spinor indices for the operators on the right hand side are made totally symmetric, among dotted and undotted indices respectively, the same as those on the left hand side. 
These indices are contracted between such building blocks according to how the spinor variables are contracted. 
Thus the Lorentz structure corresponding to the kinematic factor in eq.~\eqref{eq:amplitude_basis} is given by
\bea
\label{eq:lorentz_format}
\mc{M}(\Phi_1,\dots,\Phi_N)_{a_1,\dots,a_N} = (\epsilon^{\alpha_i\alpha_j})^{\otimes n}(\tilde\epsilon_{\dot\alpha_i\dot\alpha_j})^{\otimes \tilde{n}} \prod_{i=1}^N (D^{r_i-|h_i|}\Phi_{i,a_i})_{\alpha_i^{r_i-h_i}}^{\dot\alpha_i^{r_i+h_i}} , 
\eea
and the operator corresponding to the full amplitude basis $\mc{B}$ is given by
\eq{
	\mc{O} = T^{a_1,\dots,a_N}\mc{M}(\Phi_1,\dots,\Phi_N)_{a_1,\dots,a_N},
}
where $T$ is the same group tensor as that in eq.~\eqref{eq:amplitude_basis}. Up to linear combinations, this is the general form of operator basis.
The notion of taking the power of spinor indices is made possible by the total symmetry among them, while the $\epsilon$'s and the $\tilde\epsilon$'s exactly correspond to the $\vev\cdot$ and $[\cdot]$ brackets in the amplitudes. The ``type'' of operators can be defined similar to that of the amplitudes, as the operators consisting of the same group of fields $\Phi_1,\dots,\Phi_N$ and at the same dimension $d=r+N$. It is easy to verify that, among the local amplitudes, these operators indeed generate, and only generate, the corresponding one. 
{One may question about the possibility of generating other local amplitudes with more gauge bosons when covariant derivatives are present, because in the Feynman rules the covariant derivatives indeed generate extra vertices with more gauge bosons. However, these vertices are not gauge invariant, and the final gauge invariant amplitudes with contributions from these operators are non-local. Consider an operator $\mc{O}^{\mu}D_{\mu}\Psi$ contributing to an amplitude with extra photon $\gamma$ from the covariant derivative of the charged field $\Psi$:
\eq{
	\acontraction{\mathcal{A}(\mathcal{O}\Psi\gamma) = \bra{\mathcal{O}}\mathcal{O}^\mu (-igA_{\mu})\Psi\ket{\Psi\gamma} + \bra{\mathcal{O}}(\mathcal{O}^\mu D}{{}_{\mu}\Psi}{)(}{J}
	\mathcal{A}(\mathcal{O}\Psi\gamma) &= \bra{\mathcal{O}}\mathcal{O}^\mu (-igA_{\mu})\Psi\ket{\Psi\gamma} + \bra{\mathcal{O}}(\mathcal{O}^\mu \partial_{\mu}\Psi)(J_{\Psi}^{\mu}A_{\mu})\ket{\Psi\gamma} \\
	& \stackrel{s=m_\Psi^2}{\longrightarrow} \bra{\mathcal{O}}\mathcal{O}^\mu \partial_{\mu}\Psi\ket{\Psi} \times \bra{\Psi}J_{\Psi}^{\mu}A_{\mu}\ket{\Psi\gamma}\;,
}
where $J_\Psi$ is the charged $\Psi$ current that minimally couple to the photon field $A$. The first term is the local but gauge dependent vertex contribution, while the sum is gauge invariant but contains a mass pole for $\Psi$, at which the amplitude factorize into an amplitude basis without the photon and an amplitude basis for the minimal coupling. 
}

Do the operators in eq.~\eqref{eq:lorentz_format} exhaust all the possible forms of gauge invariant operators? The only caveat comes from the requirement of total symmetries among the spinor indices. It turns out that if they are not totally symmetric, indicating their corresponding amplitude (of the given type) contains antisymmetric spinors from the same particle, the resulting amplitude basis must vanish due to the on-shell condition $\lambda_{i[\alpha}\lambda_{i\beta]} = \vev{ii}\epsilon_{\alpha\beta} = 0$.
It follows from the relations
\eq{
	& D_{[\alpha\dot\alpha}D_{\beta]\dot\beta} = D_{\mu}D_{\nu}\sigma^{\mu}_{[\alpha\dot\alpha}\sigma^{\nu}_{\beta]\dot\beta} = -{\color{red} D^2 }\epsilon_{\alpha\beta}\epsilon_{\dot\alpha\dot\beta} + \frac{i}{2}[D_{\mu},D_{\nu}]\epsilon_{\alpha\beta}\bar\sigma^{\mu\nu}_{\dot\alpha\dot\beta}, \\
	& D_{[\alpha\dot\alpha}\psi_{\beta]} = D_{\mu}\sigma^{\mu}_{[\alpha\dot\alpha}\psi_{\beta]} = -\epsilon_{\alpha\beta}{\color{red} (D\4\psi)_{\dot\alpha} }, \\
	& D_{[\alpha\dot\alpha}F_{{\rm L} \beta]\gamma} = D_{\mu}F_{\nu\rho} \sigma^{\mu}_{[\alpha\dot\alpha}\sigma^{\nu\rho}_{\beta]\gamma} = 2{\color{red} D^{\mu}F_{\mu\nu} } \epsilon_{\alpha\beta}\sigma^{\nu}_{\gamma\dot\alpha},
}
together with the identity $i[D_\mu,D_\nu] = \sum F_{\mu\nu}$ and the EOM that these operators would be convertible to operators of other types, where we are supposed to have obtained a complete basis.
Therefore, from the on-shell point of view, these operators of different forms from eq.~\eqref{eq:lorentz_format} belong to, or contain ingredients from, other types of operators, which are not independent. 
It is important to stick to this self-consistent definition of ``type'' for operator basis to prevent over-counting, each ``type'' forming a subspace of operators that do not overlap with each other. After this clarification, the EOM redundancy is automatically taken care of.

Up to this point, the correspondence has been set up between polynomials of spinor helicity variables\footnote{%
	Such polynomial functions are regarded as form factors of operators \cite{Caron-Huot:2016cwu} which characterize the state generated by an operator from the vacuum $F = \bra{\psi}\mc{O}\ket{0}$, which is not a physical process and does not satisfy momentum conservation. }%
and Lorentz singlet operators. 
Moving forward to on-shell amplitudes and Lagrangian terms, the extra constraint for both are the momentum conservation and the IBP redundancy, which are exactly isomorphic to each other: equal amplitudes due to the momentum conservation exactly correspond to equivalent operators related by the IBP. For example, the following equality holds for 4-point amplitudes
$$\vev{12}[23] = \bra{1}p_2|3] = -\bra{1}(p_1+p_3+p_4)|3] = -\vev{14}[43], $$
which corresponds to the operator equivalence
$$(\psi_1\sigma^\mu\bar\psi_3)D_\mu \phi_2 \phi_4 \sim -(\psi_1\sigma^\mu\bar\psi_3) \phi_2 D_\mu\phi_4.$$
Terms that convert to other types by EOM are omitted, which stems from $\vev{11}=[33]=0$. In sum, taking momentum conservation into account, the amplitude basis corresponds to an IBP non-redundant basis of operators. Inspired by this correspondence, our strategy of operator enumeration is essentially the enumeration of amplitude basis.

The physical reason for such correspondence is that the free parameters of the theory should count the same in both the Lagrangian formalism and the on-shell formalism. While it is straightforward to define them in a Lagrangian as the independent Wilson coefficients, the free parameters in the on-shell formalism should be encoded in local amplitudes because they are the ultimate outcome of the cascade of unitarity factorization of any amplitudes. Building a quantum field theory from the operator basis and their Wilson coefficients is already a textbook technique, but building a theory from the corresponding amplitude basis has not been as successful, though we show that they contain the same amount of information. The recursion relations developed in the past decade \cite{Britto:2005fq} are only applicable to certain ``on-shell constructible'' theories \cite{Cohen:2010mi}, whereas a more general on-shell formalism from amplitude basis is still waiting to be discovered.


\subsection{On-shell building blocks and Lorentz classes}
\label{sec:lor_inv}


In light of the amplitude-operator correspondence eq.~\eqref{eq:dictionary}, we adopt the chiral basis of the fields and derivatives, all with spinor indices, which are in the irreducible representations $(j_l,j_r)$ of the Lorentz group $SU(2)_l\times SU(2)_r$
\bea
&&\phi \in (0,0), \quad \psi_{\alpha }  \in (1/2,0), \quad \psi^{\dagger}_{\dot\alpha } \in (0,1/2), \\
&&F_{{\rm L}\alpha\beta} = \frac{i}{2}F_{\mu\nu}\sigma^{\mu\nu}_{\alpha\beta}\in (1,0),  \quad F_{{\rm R} \dot\alpha\dot\beta} = -\frac{i}{2}F_{\mu\nu}\bar\sigma^{\mu\nu}_{\dot\alpha\dot\beta}\in(0,1), \\
&&D_{\alpha\dot\alpha} = D_{\mu}\sigma^{\mu}_{\alpha\dot\alpha} \in (1/2,1/2),
\eea 
In this notation, we have the SMEFT field content as in the table~\ref{tab:SMEFT-field-content}, where the conjugate fields with conjugating representations and opposite helicities and charges are omitted. 
\begin{table}[t]
	\begin{center}
		\begin{tabular}{|c|cc|ccc|ccc|}
			\hline
			\text{Fields} & $SU(2)_{l}\times SU(2)_{r}$	& $h$ & $SU(3)_{C}$ & $SU(2)_{W}$ & $U(1)_{Y}$ &  Flavor & $B$ & $L$ \tabularnewline
			\hline
			$G_{\rm L\alpha\beta}^A$   & $\left(1,0\right)$  & $-1$    & $\boldsymbol{8}$ & $\boldsymbol{1}$ & 0  & $1$ & 0 & 0 \tabularnewline
			$W_{\rm L\alpha\beta}^I$   & $\left(1,0\right)$  & $-1$           & $\boldsymbol{1}$ & $\boldsymbol{3}$ & 0  & $1$ & 0 & 0 \tabularnewline
			$B_{\rm L\alpha\beta}$   & $\left(1,0\right)$    & $-1$        & $\boldsymbol{1}$ & $\boldsymbol{1}$ & 0  & $1$ & 0 & 0 \tabularnewline
			\hline
			$L_{\alpha i}$     & $\left(\frac{1}{2},0\right)$  & $-\frac12$  & $\boldsymbol{1}$ & $\boldsymbol{2}$ & $-\frac12$  & $n_f$ & 0 & $1$ \tabularnewline
			$e_{_\mathbb{C}\alpha}$ & $\left(\frac{1}{2},0\right)$ & $-\frac12$   & $\boldsymbol{1}$ & $\boldsymbol{1}$ & $1$  & $n_f$ & 0 & $-1$ \tabularnewline
			$Q_{\alpha ai}$     & $\left(\frac{1}{2},0\right)$ & $-\frac12$   & $\boldsymbol{3}$ & $\boldsymbol{2}$ & $\frac16$  & $n_f$ & $\frac13$ & 0 \tabularnewline
			$u_{_\mathbb{C}\alpha}^a$ & $\left(\frac{1}{2},0\right)$ & $-\frac12$   & $\overline{\boldsymbol{3}}$ & $\boldsymbol{1}$ & $-\frac23$  & $n_f$ & $-\frac13$ & 0 \tabularnewline
			$d_{_\mathbb{C}\alpha}^a$ & $\left(\frac{1}{2},0\right)$ & $-\frac12$   & $\overline{\boldsymbol{3}}$ & $\boldsymbol{1}$ & $\frac13$  & $n_f$ & $-\frac13$ & $0$ \tabularnewline
			\hline
			$H_i$     & $\left(0,0\right)$&  0     & $\boldsymbol{1}$ & $\boldsymbol{2}$ & $\frac12$  & $1$ & 0 & 0 \tabularnewline
			\hline
		\end{tabular}
		\caption{\label{tab:SMEFT-field-content}
			The field content of the standard model, along with their representations under the Lorentz and gauge symmetries. The representation under Lorentz group is denoted by $(j_l,j_r)$, while the helicity of the field is given by $h = j_r-j_l$ .
			The number of fermion flavors is denoted as $n_f$, which is 3 in the standard model. We also list their global charges, the baryon number $B$ and the lepton number $L$. All of the fields are accompanied with their Hermitian conjugates that are omitted, $(F_{\rm L \alpha\beta})^\dagger = F_{\rm R \dot\alpha\dot\beta}$ for gauge bosons, $(\psi_\alpha)^\dagger = (\psi^\dagger)_{\dot\alpha}$ for fermions, and $H^\dagger$ for the Higgs, which are under the conjugate representations of all the groups. }
	\end{center}
\end{table}

To enumerate the valid Lorentz classes at a given dimension $d$, denoted by $F_{\rm L}^{n_{-1}}\psi^{n_{-1/2}}\phi^{n_0}\psi^{{}_\dagger n_{1/2}}F_{\rm R}^{n_1}D^{n_D}$, corresponding to classes of amplitudes\footnote{%
	The tuple $(n_{-1},n_{-1/2},n_0,n_{1/2},n_1,n_D)$ and the tuple $(h_1,\dots,h_N,r)$ record the same information, and can be easily converted to each other.}%
$\mc{M}(h_1,\dots,h_N;r)$, one may adopt the steps described in \cite{Li:2020gnx}, where the following constraints are considered
\eq{\label{eq:lor-inv_constraint}
	& \tilde{n}+n = \sum_ir_i = r = d-N, \qquad \tilde{n}-n = \sum_ih_i \equiv h, \qquad \sum_{i=1}^N n_i = N\\
	& 2n_{-1}+n_{-1/2} = \sum_i|h_i|-h = 2n-n_D, \quad 2n_{1}+n_{1/2} = \sum_i|h_i|+h = 2\tilde{n}-n_D \\
	& \min(2n, 2\tilde{n}) \geq n_D \geq \max\begin{pmatrix} h-\sum_i|h_i|,\ \mod 2 \\ 4|\min h_i| - \sum_{h_i<0}2|h_i| \\ 4|\max h_i| - \sum_{h_i>0}2|h_i| \end{pmatrix}.
}
At dimension 9, we list all the classes in table~\ref{tab:classes9}, which is model independent. The types of operators are thus obtained by substituting the SMEFT field content from table~\ref{tab:SMEFT-field-content} into eq.~\eqref{eq:lorentz_format} with varying number of derivatives and spinor contractions, while the representations of the constituting fields under gauge groups should be able to form singlets ($U(1)$ charges should add up to zero). The classes colored in gray are those ruled out by this condition, thus they don't appear in the SMEFT. In the next section, we show the details of obtaining a complete basis for a given type of operators/amplitudes, and how to convert an arbitrary operator (basis) to our basis.
\begin{table}[htb]
	\begin{align*}
	\begin{array}{cc|llll}
	\hline
	N & (n,\tilde{n}) & & \text{Classes} & & \\
	\hline
	4 & (4,1) & \color{gray}{F^2_{\rm L}\psi^2D^2+h.c.} & \color{gray}{F^3_{\rm L}\phi D^2+h.c.} & \\ 
	& (3,2) & \psi^3\psi^{\dagger}D^3+h.c. & \psi^2\phi^2D^4+h.c. & \color{gray}{F_{\rm L}F_{\rm R}\phi^2D^2+h.c.} & \\
	& & \color{gray}{F^2_{\rm L}\psi^{\dagger 2}D^2+h.c.} & \color{gray}{F^2_{\rm L}F_{\rm R}\phi D^2+h.c.} & \color{gray}{F_{\rm L}\psi\psi^{\dagger}\phi D^3+h.c.} & \color{gray}{F_{\rm L}\phi^3D^4+h.c.} \\
	\hline
	5 & (4,0) & \color{gray}{F^3_{\rm L}\psi^2+h.c.} & \color{gray}{F^4_{\rm L}\phi+h.c.} \\ 
	& (3,1) & F_{\rm L}\psi^3\psi^{\dagger}D+h.c. & \psi^4\phi D^2+h.c. & F_{\rm L}\psi^2\phi^2D^2+h.c. & \\
	& & \color{gray}{F^3_{\rm L}\psi^{\dagger 2}+h.c.} & \color{gray}{F^2_{\rm L}\psi\psi^{\dagger}\phi D+h.c.} & \color{gray}{F^2_{\rm L}\phi^3D^2+h.c.} \\
	& (2,2) & F_{\rm R}\psi^3\psi^{\dagger}D+h.c. & \psi^2\psi^{\dagger 2}\phi D^2 & F_{\rm R}\psi^2\phi^2D^2+h.c. & \psi\psi^{\dagger}\phi^3D^3 \\
	& & \color{gray}{F_{\rm L}F_{\rm R}^2\psi^2+h.c.} & \color{gray}{F^2_{\rm L}F^2_{\rm R}\phi} & \color{gray}{F_{\rm L}F_{\rm R}\psi\psi^{\dagger}\phi D} & \color{gray}{F_{\rm L}F_{\rm R}\phi^3D^2} \\
	& & \color{gray}{\phi^5D^4} \\
	\hline
	6 & (3,0) & \psi^6+h.c. & F_{\rm L}\psi^4\phi+h.c. & F^2_{\rm L}\psi^2\phi^2+h.c. & \color{gray}{F^3_{\rm L}\phi^3+h.c.} \\
	& (2,1) & \psi^4\psi^{\dagger 2}+h.c. & F_{\rm L}\psi^2\psi^{\dagger 2}\phi+h.c. & F^2_{\rm L}\psi^{\dagger 2}\phi^2+h.c. & \psi^3\psi^{\dagger}\phi^2D+h.c. \\
	& & F_{\rm L}\psi\psi^{\dagger}\phi^3D+h.c. & \psi^2\phi^4D^2+h.c. & \color{gray}{F_{\rm L}\phi^5D^2+h.c.} & \\
	\hline
	7 & (2,0) & \psi^4\phi^3+h.c. & F_{\rm L}\psi^2\phi^4+h.c. & \color{gray}{F^2_{\rm L}\phi^5+h.c.} & \\
	& (1,1) & \psi^2\psi^{\dagger 2}\phi^3 & \psi\psi^{\dagger }\phi^5D & \color{gray}{\phi^7D^2} & \\
	\hline
	8 & (1,0) & \psi^2\phi^6+h.c. \\
	\hline
	9 & (0,0) & \color{gray}{\phi^9} \\
	\hline
	\end{array}
	\end{align*}
	\caption{All the Lorentz classes at dimension 9. Classes in gray do not appear in the SMEFT due to global symmetries, such as the odd parity for all the $SU(2)_W$ doublets that forbids quite a few of Lorentz classes with odd number of scalars. }\label{tab:classes9}
\end{table}

\section{Complete Basis for a Type of Operators}
\label{sec:eft}

\subsection{y-basis: a complete basis from Young Tableau}
\label{sec:opspace}

In this section, we briefly summarize the algorithm to obtain a complete basis for a type of local amplitudes/operators, which we elaborated in \cite{Li:2020gnx}. As explained previously, a type of local amplitudes are given by the same external particle species at certain mass dimension $r$, which consists of the kinematic factor $\mc{M}$ that describes the energy dependency and the angular distribution, and the gauge group factor $T = \prod_G T_G$ that describes the gauge group representations. Given the helicities $h_i$ and gauge representations $\mathbf{r}_i^G$ of the external particles, the two factors span linear spaces of dimension $\mc{N}_{\mc{M}}$ and $\mc{N}_G$ respectively, whose outer product is the linear space of amplitude basis. The spin statistics of identical particles will put extra constraints on this product space, which we postpone to investigate at the end of this section. According to the amplitude-operator correspondence, the space of operators with the same type should have the same structure, which has total dimension
\eq{\label{eq:dimN}
	\mc{N} = \mc{N}_{\mc{M}}\times \prod_G \mc{N}_{G}.
}
Therefore our first task is to enumerate the $\mc{N}_{\mc{M}}$ basis for a given class of $\mc{M}(h_1,\dots,h_N;r)$ and the $\mc{N}_G$ basis for group $G$ given the representations $\mathbf{r}^G_i$.

For the kinematic factor, since the EOM redundancies are removed by construction, the remaining redundancies are the momentum conservation and the Schouten identity, both mentioned in the previous section. 
We utilize an $SU(N)$ transformation introduced in \cite{Henning:2019enq}, under which the total momentum (all-out-going convention) that vanishes due to the momentum conservation is invariant. This transformation is reformulated in terms of operators and is further developed in \cite{Li:2020gnx}. 
The non-redundant amplitudes/operators thus form a particular irreducible representation space of the $SU(N)$ group, the basis of which is given by the $SU(N)$ semi-standard Young tableau (SSYT). 
Specifically, the shape of the YD for this particular irrep, called primary YD, is determined by a tuple of 3 numbers $(N,n,\tilde{n})$, where $n$, $\tilde{n}$ are the parameters introduced in the previous section, as the numbers of $\vev\cdot$ type and $[\cdot]$ type brackets in the amplitude. They can be derived from the constraints eq.~\eqref{eq:lor-inv_constraint}. 
The primary YD is given by
\begin{eqnarray}\label{eq:primary_YD}
Y_{N,n,\tilde{n}} \quad = \quad \arraycolsep=0pt\def\arraystretch{1}
\rotatebox[]{90}{\text{$N-2$}} \left\{
\begin{array}{cccccc}
\yng(1,1) &\ \ldots{}&\ \yng(1,1)& \overmat{n}{\yng(1,1)&\ \ldots{}\  &\yng(1,1)} \\
\vdotswithin{}& & \vdotswithin{}&&&\\
\undermat{\tilde{n}}{\yng(1,1)\ &\ldots{}&\ \yng(1,1)} &&&
\end{array}
\right.
\\
\nonumber 
\end{eqnarray}
which is translated to amplitudes column by column as
\eq{\label{eq:YT_translate}
	\young(i,j) \sim \vev{ij}, \qquad \begin{array}{c} \young({{k_1}},{{k_2}}) \\ \vdots \\ \young({{k_{N-3}}},{{k_{N-2}}}) \end{array} \sim \mc{E}^{k_1\dots k_{N-2}ij}[ij],
}
where the $\mc{E}$ is the Levi-Civita tensor of the $SU(N)$ group. As shown in table.~\ref{tab:classes9}, where classes are organized in terms of the tuple $(N,n,\tilde{n})$, there are typically more than one class that share the same primary YD.
It is proved in \cite{Li:2020gnx} that the classes are in one-to-one correspondence with the collection of labels to be filled in the YD.
For a given class, the number of the label $i$ in the collection is given by
\eq{
	\#i = \tilde{n} - 2h_i.
}
With the collection of labels and the YD, it is not hard to enumerate all the SSYT's and translate them into amplitudes via eq.~\eqref{eq:YT_translate}, or further into operators using the amplitude-operator correspondence. 
One can also count the number of the SSYT's $\mc{N}_{\mc{M}}$ without the label filling, as is pointed out in \cite{Li:2020gnx} that $\mc{N}_{\mc{M}}$ could be regarded as the multiplicity of the primary YD in the direct product decomposition of the one-row sub-YD for each label
\eq{
	\bigotimes_{i=1}^N  \left( {\footnotesize \underbrace{\yng(2)\ ...\ \yng(1)}_{\let\scriptstyle\textstyle\text{$\#i$}} } \right) \supset \mc{N}_{\mc{M}} Y_{N,n,\tilde{n}}.
}
The direct product decomposition is carried out by the famous Littlewood-Richardson (L-R) rule. A concrete example for the Lorentz class $F_{\rm L}\psi^3\psi^\dagger D$ is given in eq.~\eqref{eq:lsundecom}.

The gauge group sectors $T_G$ are also given by Levi-Civita tensors that contract with the fundamental indices of the fields, those of which in non-fundamental irrep (e.g. gauge field in adjoint rep) provide multiple fundamental indices with particular symmetries. As we only have adjoint and anti-fundamental representation for the SM fields their conversion is listed below:
\eq{\label{eq:example_T_gluon}
	&\epsilon_{acd}\lambda^A{}_b^d G^A = G_{abc} \sim \young(ab,c),\\
	&\epsilon_{abc}Q^{\dagger,{c}} = Q^\dagger_{ab} \sim \young(a,b),\\
	&\epsilon_{jk}\tau^I{}_i^k W^I = W_{ij}\sim \young(ij) ,\\
	&\epsilon_{ij}H^{\dagger, j} = H_i \sim \young(i) .
}
The corresponding $y$-basis group factors are obtained by constructing the singlet Young tableaux following the L-R rules with the corresponding indices filled in as discussed in Ref.~\cite{Li:2020gnx}.
The singlet Young tableaux for $SU(2)_W$ and $SU(3)_C$ constructed are in the following forms:
\begin{eqnarray}
SU(2)_W:\ \underbrace{\yng(1,1)\ ...\ \yng(1,1)}_{\let\scriptstyle\textstyle\text{\large $n_{\rm box}/2$}},\quad
SU(3)_C:\ \underbrace{\yng(1,1,1)\ ...\ \yng(1,1,1)}_{\let\scriptstyle\textstyle\text{\large $n_{\rm box}/3$}},
\end{eqnarray}
where $n_{\rm box}$ is the total number of boxes in the YD, equal to the total number of fundamental indices of the fields.
As an example, we illustrate the way to construct such singlet Young tableaux with the type $G_L d^3_{_\mc{C}} e_{_\mc{C}}^\dagger D$, which we will discuss in detail in section.~\ref{sec:basis}.
The $SU(2)_W$ group is trivial for this type of operators, we thus focus on only $SU(3)_C$ part. 
The conversion of the non-fundamental indices in this case generates correspondence:
\begin{eqnarray}
&&\young({{e_1}}{{e_2}},{{e_3}})\leftrightarrow  \epsilon_{d e_1 e_3}(\lambda^A)_{e_2}^{d} G_L^A \label{eq:gc1} \\
&&\young({{a_1}},{{a_2}})\leftrightarrow \epsilon_{a_1 a_2 a}d^a_{_\mathbb{C}}{}_p,\ \quad
\young({{b_1}},{{b_2}})\leftrightarrow \epsilon_{b_1 b_2 b}d^b_{_\mathbb{C}}{}_r,\ \quad
\young({{c_1}},{{c_2}})\leftrightarrow \epsilon_{c_1 c_2 c}d^c_{_\mathbb{C}}{}_s,\label{eq:gc2}
\end{eqnarray}  
from which we can construct the singlet Young tableaux in the following with the L-R rule in the following order:
\begin{eqnarray}
&&\Yboxdim15pt \young({{e_1}}{{e_2}},{{e_3}})\xrightarrow{\young({{a_1}},{{a_2}})}\young({{e_1}}{{e_2}}{{a_1}},{{e_3}},{{a_2}})\xrightarrow{\young({{b_1}},{{b_2}})}\young({{e_1}}{{e_2}}{{a_1}},{{e_3}}{{b_1}},{{a_2}}{{b_2}})\xrightarrow{\young({{c_1}},{{c_2}})}\young({{e_1}}{{e_2}}{{a_1}},{{e_3}}{{b_1}}{{c_1}},{{a_2}}{{b_2}}{{c_2}}) \sim \epsilon^{e_1e_3a_2}\epsilon^{e_2b_1b_2}\epsilon^{a_1c_1c_2},\\
&&\Yboxdim15pt \young({{e_1}}{{e_2}},{{e_3}})\xrightarrow{\young({{a_1}},{{a_2}})}\young({{e_1}}{{e_2}},{{e_3}}{{a_1}},{{a_2}})\xrightarrow{\young({{b_1}},{{b_2}})}\young({{e_1}}{{e_2}}{{b_1}},{{e_3}}{{a_1}},{{a_2}}{{b_2}})\xrightarrow{\young({{c_1}},{{c_2}})}\young({{e_1}}{{e_2}}{{b_1}},{{e_3}}{{a_1}}{{c_1}},{{a_2}}{{b_2}}{{c_2}}) \sim \epsilon^{e_1e_3a_2}\epsilon^{e_2a_1b_2}\epsilon^{b_1c_1c_2}.
\end{eqnarray} 
The complete basis of group factors are obtained by contracting the products of the $\epsilon$'s obtained from the Young tableau with those prefactors converting the non-fundamental indices in eq.~\eqref{eq:example_T_gluon}, which yields tensors with exactly the conjugating indices of the fields:
\begin{eqnarray}
T^y_{\rm SU3,1} =8\epsilon_{acd}(\lambda^A)_b^d,\quad
T^y_{\rm SU3,2} =4\left(\epsilon_{acd}(\lambda^A)_b^d-\epsilon_{abd}(\lambda^A)_c^d \right) ,
\label{eq:exmysu3}
\end{eqnarray}
so that they contract with the fields to form gauge singlets.
The number of complete basis can, again, be given by the direct product decomposition
\eq{
	\bigotimes_{i=1}^N \mathbf{r}^G_i \supset \mc{N}_G \mathbf{1} \;.
}
In the above example we derive through the L-R rule as
\begin{eqnarray}
\yng(2,1)\otimes \yng(1,1)\otimes \yng(1,1)\otimes \yng(1,1)\supset 2\times \yng(3,3,3) \;,
\end{eqnarray}
reproducing the number of basis we enumerated.

Since the basis obtained for both $\mc{M}$ and the gauge groups $G$ are given by Young tableau, we entitle the outer product of them as the \emph{Young tableau basis}, or \emph{y-basis}, of local amplitudes/operators. The y-basis operators are denoted as $\mc{O}^{(y)}_i$, $i=1,\dots,\mc{N}$.

\subsection{Operators reducing to y-basis}
\label{sec:basis_conv}
For operators, the y-basis defined above may not be of the most convenient form illustrated at the beginning of section~\ref{sec:class}.
However, as a complete basis, the y-basis can be used to uniquely identify any operator in the EFT, either from other literatures in some conventional form, or obtained in some particular computations like the Covariant Derivative Expansion (CDE)~\cite{Henning:2014wua}, as a coordinate in the space of operators.
To achieve this goal, it is demanded to expand an arbitrary operator in terms of the y-basis. 

For the Lorentz structure, one should first convert it into the standard form eq.~\eqref{eq:lorentz_format} with the following steps:
\begin{itemize}
	\item Decompose Dirac fermions into chiral/Weyl fermions
	\eq{
		\Psi = \left(\begin{array}{c}\psi_{\alpha} \\ \chi^{{}_\dagger \dot\alpha} \end{array}\right) ,\quad 
		\gamma^\mu = \left(\begin{array}{cc}0 & \sigma^\mu \\ \bar\sigma^{\mu} & 0 \end{array}\right) .
	}
	As we only deal with massless fields, the two Weyl components are actually independent, thus one can easily do the decomposition.
	
	\item Convert the covariant derivatives $D_\mu$ and the gauge fields $F_{\mu\nu}$ into $SU(2,\mathbb{C})$ basis, with dotted or undotted spinor indices
	\eq{
		D_{\mu} = \sigma_{\mu}^{\alpha\dot\alpha}D_{\alpha\dot\alpha}, \quad F_{\mu\nu} = F_{\rm L}{}_{\alpha\beta}\sigma_{\mu\nu}^{\alpha\beta} + F_{\rm R}{}_{\dot\alpha\dot\beta}\bar\sigma_{\mu\nu}^{\dot\alpha\dot\beta}.
	}
	All the Lorentz indices $\mu,\nu,\dots$ are on $\sigma^{\mu}$ matrices now, which contract with each other and reduce to the $\epsilon$ and $\tilde\epsilon$'s
	\eq{
		\sigma_\mu^{\alpha\dot\alpha}\sigma^{\mu \beta\dot\beta} = 2\epsilon^{\alpha\beta}\tilde\epsilon^{\dot\alpha\dot\beta}.
	}

	\item Using the $[D,D]$ identity or the EOM to convert the parts of the operator with anti-symmetric spinor indices to other types of operators, as illustrated in section~\ref{sec:class}, until the remaining part has totally symmetric spinor indices in every building block. The different types of operator shall be dealt with separately.
\end{itemize}
In the standard form, the spinor contraction structure can be translated into an $SU(N)$ Young tableau, though not necessarily SSYT. 
The group theory proves that the SSYT's are an independent and complete basis of all the Young tableau, given the Fock conditions that relate them. The Fock conditions for the primary YD eq.~\eqref{eq:primary_YD} are exactly equivalent to the momentum conservation (the IBP relation for operators) and the Schouten identities, the redundancy relations that we removed to obtain the y-basis. 
Therefore, we need a systematic replacement rule to apply these relations to an arbitrary Young tableau operator obtained above, until we get a combination of the independent y-basis
\eq{\label{eq:lor_span}
	\mc{M} = \sum_i m_i \mc{M}^{\rm (y)}.
}
We want to emphasize here that the process is not for obtaining the complete basis, but for reducing any Lorentz structure to the basis that we define. 
The replacement rule is decribed below: 
\begin{itemize}
	\item Remove all derivatives on the first field $\Phi_1$ by the IBP relation:
	\eq{
		\left(D^{r_1-|h_1|}\Phi_1\right)\dots \simeq -\Phi_1 \left(D^{r_1-|h_1|} \dots \right).
	}
	The derivatives are distributed among the rest of the building blocks by the Leibniz rule. 
	Corresponding to the convertion of spinor helicity formula is
	\eq{
		\langle i1\rangle[1j]=-\sum_{k=2}^{N}\langle ik\rangle[kj].
	}
	In the sum, the term with $k=i$ or $k=j$ would vanish, which in the corresponding operator amounts to a self-contracting building block that should be converted to other types of operators. We omit these terms and thus use the $\simeq$ for the relation, which should be understood for the following two steps as well.
	
	\item Remove derivatives on $\Phi_2$ (or $\Phi_3$) when the two spinor indices on them contract with those in building block 1 and 2, such as
	\eq{
		\Phi_{1,\alpha\dots}(D^{\alpha}_{\dot\alpha}D^{n_2-1}\Phi_2)^{\dots}_{\dots}\dots \simeq& -\Phi_{1,\alpha\dots}(D^{n_2-1}\Phi_2)^{\dots}_{\dots}\left(D^{\alpha}_{\dot\alpha}\dots\right) ,\\
		\Phi_{1,\alpha\dots}\Phi_2^{\dot\alpha\dots}(D^{\alpha}_{\dot\alpha}D^{n_2-1}\Phi_3)^{\dots}_{\dots}\dots \simeq& -\Phi_{1,\alpha\dots}\Phi_2^{\dot\alpha\dots}(D^{n_2-1}\Phi_3)^{\dots}_{\dots}\left(D^{\alpha}_{\dot\alpha}\dots\right) .
	}
	The corresponding replacement rule for amplitudes are 
	\eq{
		[1|p_2|i\rangle=-\sum_{k=3}^N[1|p_k|i\rangle,\quad &\langle 1|p_2|i]=-\sum_{k=3}^N\langle 1|p_k|i] ,\\
		[1|p_3|2\rangle=-\sum_{k=4}^N[1|p_k|2\rangle,\quad &\langle 1|p_3|2]=-\sum_{k=4}^N\langle 1|p_k|2] . 
	}
	
	\item Remove pairs of derivatives acting on $\Phi_2$ and $\Phi_3$, with indices contracting with each other, by using the following identity
	\eq{
		& \Phi_1D^2(\Phi_2\Phi_3\dots) = \Phi_1(D^2\Phi_2)\Phi_3\dots + \Phi_1\Phi_2(D^2\Phi_3)\dots \\ 
		& \quad + \Phi_1\Phi_2\Phi_3(D^2\dots) + 2\Phi_1(D\Phi_2)(D\Phi_3)\dots + 2\Phi_1(D\Phi_2)\Phi_3(D\dots) + 2\Phi_1\Phi_2(D\Phi_3)(D\dots) \;,
	}
	where the terms in the first line are all convertible to other types via the EOM. It corresponds to the following relation among Mandelstam variables
	\eq{
		p_1^2 = 2\sum_{i,j\neq1} p_i\cdot p_j = 0 .
	}
	
	\item The Schouten identity can be applied to any pair of $\epsilon$'s with all-different indices (contracting with 4 different building blocks $i<j<k<l$)
	\eq{
		\epsilon_{\alpha_i\alpha_j}\epsilon_{\alpha_k\alpha_l} + \epsilon_{\alpha_i\alpha_k}\epsilon_{\alpha_l\alpha_j} + \epsilon_{\alpha_i\alpha_l}\epsilon_{\alpha_j\alpha_k} = 0.
	}
	In spinor helicity language, it reads
	\eq{
		\langle ij\rangle\langle kl\rangle + \langle ik\rangle\langle lj\rangle + \langle il\rangle\langle jk\rangle = 0
	}
	The rule is that whenever the third term (specified by the order of the labels) shows up in the operator/amplitude, replace it by the other two terms. 
	\item Apply the Schouten identity for the $\tilde\epsilon$'s in the same manner. 
	\eq{
		\tilde{\epsilon}_{\dot\alpha_i\dot\alpha_j}\tilde{\epsilon}_{\dot\alpha_k\dot\alpha_l}+ \tilde{\epsilon}_{\dot\alpha_i\dot\alpha_k}\tilde{\epsilon}_{\dot\alpha_l\dot\alpha_j}+ \tilde{\epsilon}_{\dot\alpha_i\dot\alpha_l}\tilde{\epsilon}_{\dot\alpha_j\dot\alpha_k}=0 ,\\
		[ij][kl] + [ik][lj] + [il][jk] = 0 .
	}
\end{itemize}

For the gauge group tensor $T_G$, one can convert any bases to each other with the help of an inner product defined for the tensors:
\begin{eqnarray}
	(T_1, T_2) \equiv \sum_{a_1,a_2,...}T_1^{a_1a_2...}T_2^{a_1a_2...},
\end{eqnarray} 
as $T_G^{\rm (y)}$ are all products of Levi-Civita tensors, their contractions are easily calculated algebraically. Then using the Gram–Schmidt process, one can obtain a set of orthogonal tensors $T_G^{\rm (o)}$ span the same space of $y$-basis.
Therefore the coordinates of any group tensor $T$ in this orthogonal basis can be obtained as:
\begin{eqnarray}
	T_G &=& \sum_i t_i  T_G^{\rm (o)}{}_i ,\qquad t_i=\left(T , T_G^{\rm (o)}{}_i\right) . \label{eq:T_span}
\end{eqnarray} 

With eq.~\eqref{eq:lor_span} and eq.~\eqref{eq:T_span}, we can reduce any operator to our y-basis
\eq{
	\mc{O} = T_G \mc{M} = \sum_{i=1}^{\mc{N}_G} \sum_{j=1}^{\mc{N}_{\mc{M}}} t_i m_j T_G^{\rm (y)}{}_i \mc{M}^{\rm (y)}_j = \sum_{i,j}t_im_j \mc{O}^{\rm (y)}_{i,j}
}
In particular, we can use the reduction to build other complete bases, like a basis with conventional notation. We define such a basis of conventional monomial operators generally as m-basis. Given an over-complete set of monomial operators $\mc{O}^{\rm (m)}_i$, we can reduce them all to our y-basis and obtain a coefficient matrix
\eq{\label{eq:form_conversion}
	\mc{O}^{\rm (m)}_i = \sum_{j=1}^{\mc{N}}\bar{\mc{K}}^{\rm my}_{ij}\mc{O}^{\rm (y)}_j , \quad i=1,\dots,\mc{N},\dots
}
The m-basis is thus constructed by selecting $\mc{N}$ independent rows in the matrix $\bar{\mc{K}}^{\rm my}$ that form a full-rank square matrix $\mc{K}^{\rm my}$, which serves as the conversion matrix between the y-basis and m-basis. 
Note that the m-basis is highly non-unique, which not only depends on the notation but also depends on the selection of rows in $\bar{\mc{K}}$.

\subsection{p-basis: in the presence of repeated fields}
\label{sec:inv}

There is one more redundancy that is not yet considered for the $\mc{N}$ dimensional space of type, which is when there are repeated fields/identical particles in the operator/amplitude. While it is explained in \cite{Li:2020gnx} in terms of operators, we present here a derivation of the constraint from the amplitude point of view.

In the actual physical amplitude, identical particles should be subject to the spin-statistics, which picks out certain linear combinations of the amplitude basis (i.e. they may not be factorizable as in eq.~\eqref{eq:amplitude_basis}). These combinations are totally symmetric or totally anti-symmetric under the permutations of the bosonic or fermionic identical particles, and are thus called a p-basis:
\eq{
	\mc{B}^{\rm (p)}\big( \underbrace{\phi^{a_1}(p_1),\dots,\phi^{a_m}(p_m)}_{m},\dots \big) 
	= \mc{D}_{\phi}(\pi)\mc{B}^{\rm (p)}
	\big( \underbrace{\phi^{a_{\pi(1)}}(p_1),\dots,\phi^{a_{\pi(m)}}(p_m)}_{m},\dots \big), \\
	\mc{D}_{\phi}(\pi) = \left\{\begin{array}{ll} 1 & \text{boson } \phi \\ (-1)^\pi & \text{fermion } \phi \end{array} \right. ,\qquad \pi\in S_m ,
}
where $\mc{D}_{\phi}(\pi)$ is the representation of the permutation $\pi$ for the particle $\phi$, and $(-1)^\pi$ denotes the signature of $\pi$. 
It is reflected in the amplitude-operator correspondence by the Feynman rule that sums up all possible contractions between repeated fields and the external legs in the vertex function. 
According to the dictionary eq.~\eqref{eq:dictionary}, such amplitude basis would correspond to an operator basis, also called p-basis, with explicit permutation symmetries among the repeated fields. 

We would like to clarify that the notions of identical particles are for particle multiplets, which include the gauge group and even the flavor degrees of freedom. 
In general, the permutation symmetry of the function $\mc{B}$ stems from the inner product of the permutation symmetries of $\mc{M}$ and $T$. To explicitly show the constraint of spin-statistics on the amplitude basis for particles with flavors, we take the flavor index out of the collection $a$, and denote the flavor part of the tensor as $\kappa$, such that
\eq{
	\mc{B}\left( \phi^{f_1,a_1}(p_1),\dots,\phi^{f_m,a_m}(p_m) \right) = \kappa^{f_1,\dots,f_m}T^{a_1,\dots,a_m}\mc{M}(h_1,\dots,h_m,\dots)
}
where we omit the other possible particles and only focus on the m identical particles (multiplets) $\phi$. 
The permutation symmetries of them are denoted by the irreducible representations of the symmetric group $S_m$, which are labeled by partitions $\lambda$ of the integer m, such as $\lambda = [2,1] \vdash m=3$. They are also denoted by Young diagrams with m boxes, for instance, $[2,1]$ is denoted by $\scriptsize \yng(2,1)$. Therefore, the spin-stat requires
\eq{
	\lambda_\kappa \odot \lambda_{\rm other} \supset \left\{ 
	\begin{array}{lll}
	[m] 	& \Rightarrow \lambda_\kappa=\lambda_{\rm other} 	& \text{boson } \phi , \\ 
	{}[1^m] & \Rightarrow \lambda_\kappa=\lambda_{\rm other}^T 	& \text{fermion } \phi .
	\end{array}\right.
}
where we use the short-hand notation $[\underbrace{1,\dots,1}_{m}] = [1^m]$ for the total anti-symmetry, and the superscript ${}^T$ indicates the transpose of the Young diagram. 

In the following, $\lambda$ without subscript is short for $\lambda_\kappa$ by default, and the p-basis will be organized in terms of $\lambda$. 
First we find the $\mc{N}$ dimensional space of the flavor-blind amplitudes $T\otimes\mc{M}$ and combine the y-basis into $\lambda$ representation spaces, each being a $d_\lambda$-dimensional subspace of amplitudes. Suppose the number of representation spaces for each $\lambda$ is given by $n_\lambda$, such that 
\eq{
	\mc{N} = \sum_{\lambda\vdash m} n_\lambda d_\lambda,
}
and all the p-basis amplitudes are labeled by $\lambda$, $x=1,\dots,d_\lambda$, and $\xi=1,\dots,n_\lambda$. We will describe the derivation of the full-rank conversion matrix $\mc{K}^{\rm py}$ defined as
\eq{
	\mc{O}^{(\rm p)}_{(\lambda,x),\xi} \equiv \mc{O}^{(\rm p)}_{i} = \sum_{j=1}^{\mc{N}} \mc{K}^{\rm py}_{ij}\mc{O}^{(\rm y)}_j ,
}
in the next section.
In the meantime, the generic rank-m flavor tensor $\kappa$ with flavor number $n_f$ can be decomposed into tensor bases that form $\mc{S}(\lambda,n_f)$ number of $\lambda$ representation spaces for the group $S_m$, such that the total degrees of freedom match
\eq{
	\mc{N}_\kappa = n_f^m = \sum_{\lambda\vdash m} \mc{S}(\lambda,n_f) d_\lambda.
} 
The function $\mc{S}(\lambda,n_f)$ is known as the Hook content formula, which also counts the number of semi-standard Young tableau (SSYT). For example, with $\lambda = [3]$ and $n_f=2$, we have $\mc{S}(\lambda,n_f) = 4$, and the 4 flavor tensor bases are given by (normalization not relevant in this paper):
\eq{
	& \young(111):\qquad (\kappa^{[3]}_1)^{111} = 1 , \quad \text{otherwise } 0 , \\
	& \young(112):\qquad (\kappa^{[3]}_2)^{112} = (\kappa^{[3]}_2)^{121} = (\kappa^{[3]}_2)^{211} = 1 , \quad \text{otherwise } 0 , \\
	& \young(122):\qquad (\kappa^{[3]}_3)^{122} = (\kappa^{[3]}_3)^{212} = (\kappa^{[3]}_3)^{221} = 1 , \quad \text{otherwise } 0 , \\
	& \young(222):\qquad (\kappa^{[3]}_4)^{222} = 1 , \quad \text{otherwise } 0 .
}
Finally, from the $n_\lambda$ representation spaces for the flavor-blind amplitudes, and the $\mc{S}(\lambda,n_f)$ representation spaces for the $\kappa$ tensor, the $S_m$ inner product of an arbitrary pair of them contains a totally symmetric amplitude basis. Hence the total number of the flavor-specified amplitude basis is given by
\eq{
	\bar{\mc{N}} = \sum_\lambda n_\lambda \mc{S}(\lambda,n_f) .
}
They are labeled by $\lambda$, together with $\xi = 1,\dots,n_\lambda$ labeling each representation space of the flavor-blind amplitudes, and the flavor SSYT denoting the flavor tensor basis that also specifies the flavors of the identical multiplets in the amplitudes, all of which are explicitly derivable. The amplitude-operator correspondence translate the flavor-blind amplitude basis to operators with free flavor indices, which we define as ``term'', while the flavor tensor $\kappa$ multiplied by the Wilson coefficient becomes the Wilson coefficient flavor tensor. The symmetry among its indices, also known as the flavor relations, is exactly given by $\lambda$ as the representation of $\kappa$.

\section{Terms: Operators Organized as Irreducible Flavor Tensors}
\label{sec:basis}

\subsection{Workflow and the Master formulae}
\label{sec:motiv}
As discussed in section.~\ref{sec:inv}, when repeated fields appear in a given type of operator, the dimension of the subspace may be less then $\cal N$ calculated in eq.~\eqref{eq:dimN} due to the certain permutation symmetry among the flavor indices. Therefore, in this subsection we shall demonstrate the workflow to obtain the p-basis operator which is what we called ``\textbf{terms}" for a given type of operator, and in the next three subsections we shall illustrate the whole procedure obtaining the p-basis operator concretely with a dim-9 example: $G_L d^3_{_\mc{C}} e_{_\mc{C}}^\dagger D$.

As studied in detail in Ref.~\cite{Li:2020gnx} and discussed in section.~\ref{sec:inv}, the permutation symmetry of the flavor structure is related to that of gauge and Lorentz structure indicated in the eq.(3.10) in Ref.~\cite{Li:2020gnx}:
\begin{eqnarray}
\underbrace{\pi\circ {\cal O}^{\{f_{k},...\}}}_{\rm permute\ flavor} &=& \underbrace{\left(\pi\circ T_{{\rm SU3}}^{\{g_k,...\}}\right)\left(\pi\circ T_{{\rm SU2}}^{\{h_k,...\}}\right)}_{\rm permute\ gauge}\underbrace{\left(\pi\circ{\cal M}^{\{f_k,...\}}_{\{g_{k},...\},\{h_{k},...\}}\right)}_{\rm permute\ Lorentz},
\label{eq:tperm}
\end{eqnarray}
where $f_{k}$, $g_k$ and $h_k$ are the flavor, $SU(3)_C$ and $SU(2)_W$ gauge group indices for different sets of repeated fields respectively.
Eq.~\eqref{eq:tperm} tells that ${\cal O}^{\{f_{k},...\}}$ can be viewed as an direct product representation of the symmetric group $S_k$ permuting the repeated fields, hence it is easier to construct an operator with the definite flavor permutation symmetry from a set of symmetrized gauge group factors $T^\lambda_x$ and Lorentz structures ${\cal M}^\lambda_x$ that transform as irreps $\lambda$ of $S_k$ such that:
\begin{eqnarray}
\pi\circ T^\lambda_x = \sum_y T^\lambda_y {\cal D}^\lambda(\pi)_{yx},\quad \pi\circ {\cal M}^\lambda_x = \sum_y {\cal M}^\lambda_y {\cal D}^\lambda(\pi)_{yx}, \text{ for $\pi\in S_k$}
\end{eqnarray}
where $\lambda$ is the partition of $k$ corresponding to a certain irrep of the symmetric group, $x$ labels the basis vector of the irrep, ${\cal D}^\lambda(\pi)$ is the matrix representation of the symmetric group for this irrep.
Having introduced the concept of y-basis and m-basis in section.~\ref{sec:eft}, we shall name the  $T^\lambda_x$ and ${\cal M}^\lambda_x$ p-basis for gauge group factors and Lorentz structures. With these ingredients in hand one can construct ${\cal O}^{(p)}$ of the flavor symmetry $\lambda$ with  Clebsch-Gordon coefficients (CGCs) of the symmetric group $C_{(\lambda,x),j}^{(\lambda_{1},x_1),(\lambda_{2},x_2),(\lambda_{3},x_3)}$:
\begin{eqnarray}\label{eq:master}
{\cal O}_{(\lambda,x),j} = \sum_{x_1,x_2,x_3} C_{(\lambda,x),j}^{(\lambda_{1},x_1),(\lambda_{2},x_2),(\lambda_{3},x_3)}\mc{M}^{\lambda_{1}}_{x_1}\otimes T^{\lambda_2}_{{\rm SU3},x_2}\otimes T^{\lambda_3}_{{\rm SU2},x_3},
\end{eqnarray}
where $\lambda,\lambda_1,\lambda_2,\lambda_3$ are irreps of $S_k$ for flavor, Lorentz Structure, $SU(3)_C$ and $SU(2)_W$ group factors respectively, $x$ with and without subscripts corresponds to the labels of basis vector for each irreps of $S_k$, $j$ is the multiplicity of the resulting irreps from the decomposition.  

\begin{figure}[!ht]
\center{\includegraphics[width=0.7\textwidth]
{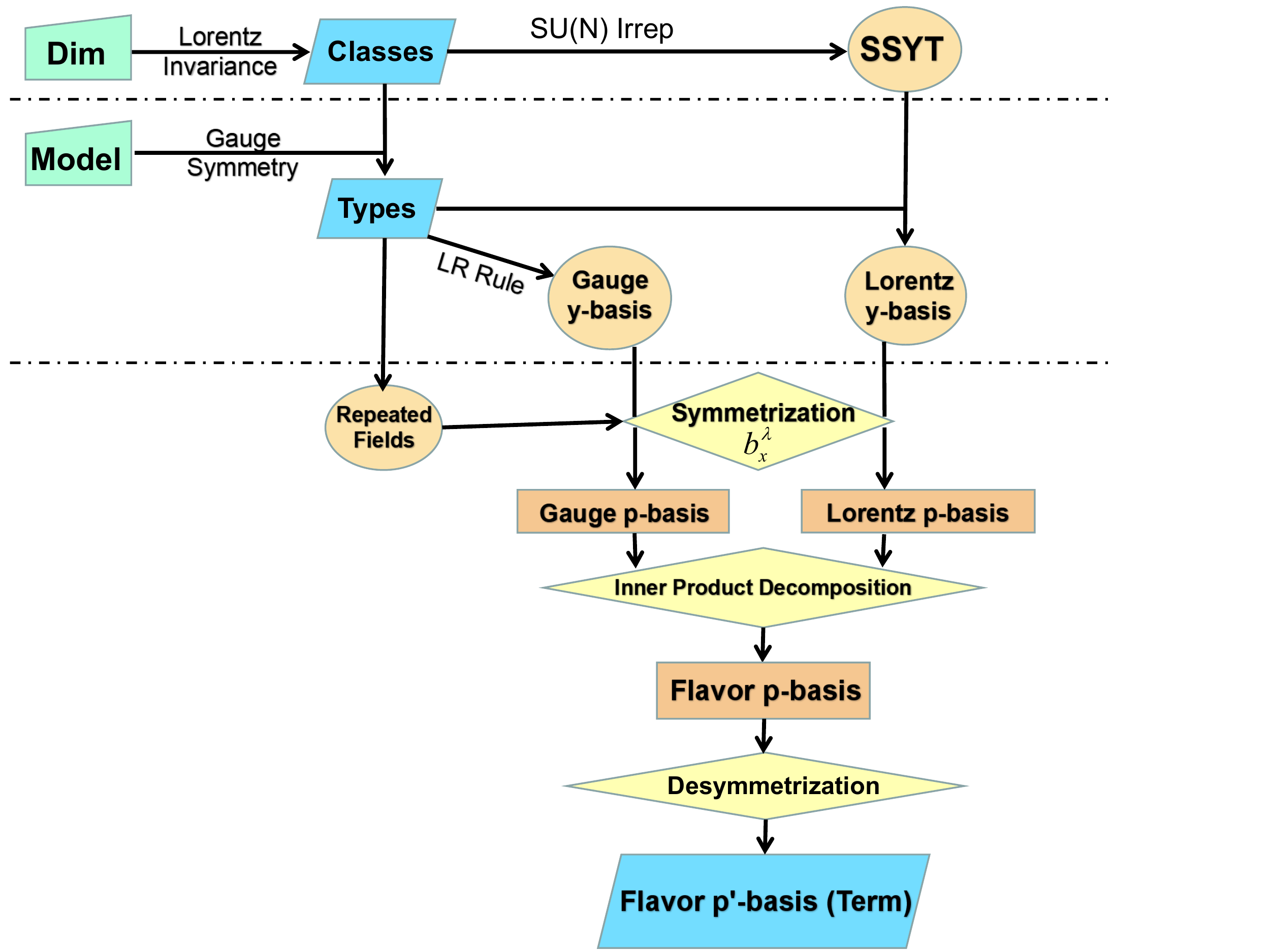}}
\caption{\label{fig:fchart}Flow chart for finding all the independent terms at a given dimension. The content above the first dash-dotted line is model independent and can be applied to any EFT. The content below the second dash-dotted line are our main contributions in this work. We automatize the whole procedure in a \textsf{Mathematica} code.}
\end{figure}

In figure.~\ref{fig:fchart}, we show our workflow obtaining the all the terms of operator for a given dimension in a flowchart and describe each step as follow:
\begin{enumerate}
\item Enumerate tuples of the numbers of fields for different helicities and the number of derivatives following the constraints in eq.~\eqref{eq:lor-inv_constraint}, and each tuple corresponds to a class of operator.
\item For each class of operators, one can filling the slots of definite helicities with concrete SM fields, and the combination of fields that can form the gauge singlet is retained as a type of operator.
\item For a given type of operators, one can enumerate the y-basis for Lorentz and gauge group structures with the corresponding SSYT.
\item for each y-basis, one can convert it to a m-basis with some group identities, the form of which is more familiar to the phenomenology community.
\item After obtaining the y-basis and m-basis, one can symmetrize them by acting on the corresponding group algebra symmetrizer $b^\lambda_x$, to obtain the p-basis for the Lorentz and gauge group structures, the appropriate irreps of the symmetric group $\lambda$ is obtained by the plethysm technique in advance.
\item With the p-basis Lorentz and gauge group structures one can construct the p-basis operators, the ``terms", with the inner product decomposition of the symmetric group related to the repeated fields. 
\item Finally, to shorten our notations for the ``terms", we perform a subtle recombination of the p-basis operators for a given type called ``de-symmetrization" to arrive at the form of terms of operator present in section.~\ref{sec:list}.
\end{enumerate}

\subsection{Lorentz and Gauge Bases}
\label{sec:group}
\subsubsection{Lorentz Structure}
As discussed in Sec.3.2 in Ref.~\cite{Li:2020gnx}, the y-basis of the Lorentz structure is enumerated by the SSYT of the corresponding primary Young Diagram of the auxiliary $SU(N)$ group determined by the tuple of three numbers $(N,n,\tilde{n})$ for the given type of the operator, where $N$ is the number of field building blocks, $n$ is the number of $\epsilon$ tensors with undotted spinor indices, while $\tilde{n}$ is that of $\epsilon$ tensors with dotted spinor indices. 

Given the operator type $G_L d^3_{_\mc{C}} e_{_\mc{C}}^\dagger D$, $N$ is obviously equal to 5, while $n=3$ and $\tilde{n}=1$ can be obtained by~\cite{Li:2020gnx}:
\begin{eqnarray}
2n_{-1}+n_{-1/2} = 2n-n_D, \quad 2n_{1}+n_{1/2} = 2\tilde{n}-n_D,
\end{eqnarray}
where $n_D=1$ is the number of derivative, $n_{-1},n_{-1/2},n_{1/2},n_{1}$ are numbers of the fields with helicities equal to $-1,-1/2,1/2,1$ respectively. 
The next step is to find the numbers of field labels $\# i$ for $i$ from 1 to 5 that need to be filled in the primary YD. 
Following the eq. (3.51) in Ref.~\cite{Li:2020gnx}:
\begin{eqnarray}
\# i = \tilde{n}-2h_i,
\end{eqnarray}
where $h_i$ is helicity of the corresponding fields.
This leads to $\# 1=3, \# 2=\# 3=\# 4 =2, \# 5=0$ where we have already arranged the fields in the order of increasing helicities. From the direct product decomposition:
\begin{eqnarray}
\yng(3)\otimes \yng(2)^{\otimes 3} \supset 4\times \yng(4,4,1)+...,
\label{eq:lsundecom}
\end{eqnarray} 
we know in advance the number of SSYT should be 4.
The corresponding SSYTs with the numbers filled in are:
\begin{eqnarray}
\young(1112,2334,4),\ \young(1112,2344,3),\ \young(1113,2234,4),\ \young(1113,2224,3),
\end{eqnarray}
which corresponds to a set of y-basis:
\begin{eqnarray}
-\tilde{\epsilon}_{\dot{\alpha}_3\dot{\alpha}_5}\epsilon^{\alpha_1\alpha_3}\epsilon^{\alpha_1\alpha_3}\epsilon^{\alpha_2\alpha_4},\ 
\tilde{\epsilon}_{\dot{\alpha}_4\dot{\alpha}_5}\epsilon^{\alpha_1\alpha_3}\epsilon^{\alpha_1\alpha_4}\epsilon^{\alpha_2\alpha_4},\ 
-\tilde{\epsilon}_{\dot{\alpha}_3\dot{\alpha}_5}\epsilon^{\alpha_1\alpha_2}\epsilon^{\alpha_1\alpha_3}\epsilon^{\alpha_3\alpha_4},\ 
\tilde{\epsilon}_{\dot{\alpha}_4\dot{\alpha}_5}\epsilon^{\alpha_1\alpha_2}\epsilon^{\alpha_1\alpha_4}\epsilon^{\alpha_3\alpha_4},
\label{eq:ybo}
\end{eqnarray} 
with operator forms from the correspondence:
\begin{eqnarray}
&&{\cal M}^{\rm y}_1=-(G^{A\alpha\beta}_L{}_{} Dd^{a\gamma}_{_\mathbb{C}}{}_{\beta\dot{\delta} p} d^{b}_{_\mathbb{C}}{}_{\alpha r} Dd^{c}_{_\mathbb{C}}{}_{\gamma\dot{\delta}s} e^{\dagger\dot{\delta}}_{_\mathbb{C}}{}_t)\label{eq:lyb1}\\
&&{\cal M}^{\rm y}_2=(G^{A\alpha\beta}_L{}_{} d^{a\gamma}_{_\mathbb{C}}{}_{ p} d^{b}_{_\mathbb{C}}{}_{\alpha r} Dd^{c}_{_\mathbb{C}}{}_{\beta\gamma\dot{\delta}s} e^{\dagger\dot{\delta}}_{_\mathbb{C}}{}_t),\\ 
&&{\cal M}^{\rm y}_3=-(G^{A\alpha\beta}_L{}_{} d^{a}_{_\mathbb{C}}{}_{\alpha p} Dd^{b}_{_\mathbb{C}}{}_{\beta\gamma\dot{\delta}r} d^{c\gamma}_{_\mathbb{C}}{}_{ s} e^{\dagger\dot{\delta}}_{_\mathbb{C}}{}_t),\\ 
&&{\cal M}^{\rm y}_4=(G^{A\alpha\beta}_L d^{a}_{_\mathbb{C}}{}_{\alpha p} d^{b\gamma}_{_\mathbb{C}}{}_{ r} Dd^{c}_{_\mathbb{C}}{}_{\gamma\beta\dot{\delta}s} e^{\dagger\dot{\delta}}_{_\mathbb{C}}{}_t),
\label{eq:lyb4}
\end{eqnarray}
In the above equations, $prst$ represent the flavor indices, $abc$ and $A$ represent color indices for the anti-fundamental representation and the adjoint representation respectively.

One can obtain the m-basis Lorentz structures by converting $G_{L\alpha\beta}$ to $G_{L\mu\nu}$ in eq.~(\ref{eq:lyb1}-\ref{eq:lyb4}) and finding independent monomials with the method discussed in section.~\ref{sec:inv}:
\begin{eqnarray}
&&{\cal M}^{\rm m}_1 = i G_L^{A\mu}{}_\nu (d^a_{p}C d^{c}_{s})(D_\mu \bar{e}_t\gamma^\nu d^b )\\
&&{\cal M}^{\rm m}_2 = i G_L^{A\mu}{}_\nu (d^a_{p} C D_\mu d^{c}_{s})(\bar{e}_t \gamma^\nu d^b_{r})\\
&&{\cal M}^{\rm m}_3 = i G_L^{A\mu}{}_\nu (\bar{e}_t\gamma^\nu d^a_{p}  )(D_\mu d^b_{r} C d^{c}_{s})\\
&&{\cal M}^{\rm m}_4 = i G_L^{A\mu}{}_\nu (\bar{e}_t \gamma^\nu  d^a_{p})(d^b_{r}C D_\mu d^{c}_{s}),
\end{eqnarray} 
where we have already converted the two component spinors to the corresponding four component ones with the formulas in appendix.~\ref{app:A}.

Now we are ready to obtain the symmetrized Lorentz factor for the repeated field $d_{_\mathbb{C}}$ in the p-basis and express them in terms of m-basis. 
As discussed in ~\ref{sec:opspace} and Ref.~\cite{Li:2020gnx}, one can view the procedure to obtain the SSYT as constructing the primary YD of $SU(N)$ group with Littlewood-Richardson rule from the outer product of totally symmetric representations of one row YD for each field labels.
Here we have $[\# 1]=\yng(3)$ and $[\# 2]=[\# 3]=[\# 4]=\footnotesize\yng(2)$ ($\# 5=0$ does not contribute), and the allowed irreps of $S_3$ for $d_{_\mathbb{C}}$ are those resulting in  primary YD after takeing plethysm with $\footnotesize\yng(2)$ . In our example we have:
\begin{eqnarray}
\nonumber \footnotesize\yng(3)\otimes \left(\footnotesize\yng(2)\ \circled{p}\  [3]\right) &=& \footnotesize\yng(9)+2\times \footnotesize\yng(6,3)+ \footnotesize\yng(8,1)+2\times \footnotesize\yng(7,2)\\ \nonumber 
&&+2\times \footnotesize\yng(5,2,2)+ \footnotesize\yng(4,2,2,1)+ \footnotesize\yng(3,2,2,2)+ \footnotesize\yng(5,4)+ \footnotesize\yng(6,2,1)+ {\color{blue}\footnotesize\yng(4,4,1)}\\ 
&&+ \footnotesize\yng(4,3,2)+ \footnotesize\yng(5,3,1)\\
\footnotesize\yng(3)\otimes \left(\footnotesize\yng(2)\ \circled{p}\  [2,1]\right) &=&  \footnotesize\yng(8,1)+2\times \footnotesize\yng(5,4)+2\times \footnotesize\yng(7,2)+2\times \footnotesize\yng(6,3)\nonumber \\
&&+ \footnotesize\yng(7,1,1)+3\times \footnotesize\yng(5,3,1)+3\times \footnotesize\yng(6,2,1)+2\times \footnotesize\yng(5,2,2)+ \footnotesize\yng(5,2,1,1)\nonumber \\
&&+2\times \footnotesize\yng(4,3,2)+ \footnotesize\yng(4,3,1,1)+ \footnotesize\yng(4,2,2,1)+ \footnotesize\yng(3,3,2,1)+{\color{blue} \footnotesize\yng(4,4,1)}\\
\footnotesize\yng(3)\otimes \left(\footnotesize\yng(2)\ \circled{p}\  [1^3]\right) &=&  \footnotesize\yng(7,1,1)+{\color{blue} \footnotesize\yng(4,4,1)}+ \footnotesize\yng(6,2,1)+2\times \footnotesize\yng(5,3,1)+ \footnotesize\yng(6,1,1,1)\nonumber \\
&&+ \footnotesize\yng(4,3,1,1)+ \footnotesize\yng(5,2,1,1)+ \footnotesize\yng(6,3)+ \footnotesize\yng(3,3,3)+ \footnotesize\yng(4,3,2)
\end{eqnarray}
after taking into account the subtleties related to the Grassmann nature of the fermion fields and the odd number of presence of $\mc E$ tensor of $SU(N)$ group converting the anti-fundametnal indices of the $\tilde{\epsilon}$ to the fundamental ones~\cite{Li:2020gnx}, the allowed symmetries remains the same (no transposition for the YDs needed). The total number of the resulting primary Young diagrams should multiply the dimension of the corresponding irrep of the symmetric group $d_\lambda$, which leads to $1+2+1=4$ distinct Lorentz structures given that $d_{[2,1]}=2$ and $d_{[3]}=d_{[1^3]}=1$ and is consistent with the result in eq.~\eqref{eq:lsundecom}.
We shall obtain the p-basis ${\cal M}^{\rm p}_{(\lambda,x)\xi}$ ($\xi$ label the multiplicity of the irrep $\lambda$ of the symmetric group) by acting corresponding group algebra projector $b^\lambda_x$ on elements in eq.~\eqref{eq:ybo} until the number of multiplicity of that $\lambda$ reaches the demanding value. Since the multiplicity equals to 1 for each $\lambda$ in our case, we simply omit the label $\xi$ in what follows.  We finally obtain the matrices $\mc{K}^{\rm pm}$ relating the p-basis and m-basis:
\begin{eqnarray}
&&\begin{pmatrix}
{\cal M}^{\rm p}_{([3],1),1}\\
{\cal M}^{\rm p}_{([2,1],1),1}\\
{\cal M}^{\rm p}_{([2,1],2),1}\\
{\cal M}^{\rm p}_{([1^3],1),1}
\end{pmatrix}
=\begin{pmatrix}
-2 & 2 & 2 & 4\\
1 & 2 & -1 & 1 \\
-2& -1 & -1 & -2\\
-2 & -2 & -2 & 0
\end{pmatrix}
\begin{pmatrix}
{\cal M}^{\rm m}_{1}\\
{\cal M}^{\rm m}_{2}\\
{\cal M}^{\rm m}_{3}\\
{\cal M}^{\rm m}_{4}
\end{pmatrix}.
\end{eqnarray}  

\subsubsection{Gauge Group}
The treatment of the gauge group is similar to that of the Lorentz group, but the usage of the Young diagram and Young Tableaux are different. 
First of all, one need to find the y-basis for each gauge group by finding the singlet Young tableaux constructed by the ordinary LR-rule with the corresponding gauge group indices of each field filled in provided that each field is expressed in terms of fundamental indices only. 
If a field is not in the fundamental representation then one can perform the conversion by contracting with the Levi-Civita tensor and the group generators. 
We have already work out the y-basis group factors in section.~\ref{sec:opspace} in eq.~\eqref{eq:exmysu3}, we list them here again 
\begin{eqnarray}
&&T^{\rm y}_{\rm SU3,1} =8\epsilon_{acd}(\lambda^A)_b^d,\nonumber \\
&&T^{\rm y}_{\rm SU3,2} =4\left(\epsilon_{acd}(\lambda^A)_b^d-\epsilon_{abd}(\lambda^A)_c^d \right).\nonumber
\end{eqnarray}
To obtain the m-basis we investigate each monomials in the above equations, and select two independent monomials as our m-basis:
\begin{eqnarray}
&&T^{\rm m}_{\rm SU3,1} = \epsilon_{acd}(\lambda^A)_b^d,\\
&&T^{\rm m}_{\rm SU3,2} = \epsilon_{abd}(\lambda^A)_c^d.
\end{eqnarray}
In practical, the independence of the monomial can be checked numerically by flattening the $T^{\rm m}$'s into a 1-D vector with components corresponding to specific tuples of $(a,b,c,A)$ in a fixed order. 
The next step is to find the proper permutation symmetry among the $SU(3)_C$ indices $abc$ of the repeated fields $d_{_\mathbb{C}}$ and to obtain the symmetrized group factors in the p-basis by acting corresponding $b^\lambda_x$ on $T_{\rm SU3}^{\rm (m)}$'s. In our example the only possible permutation symmetry is $[2,1]$, since only the plethysm of $\footnotesize \yng(1,1)$ with $[2,1]$ can generate the singlet:
\begin{eqnarray}
&&\footnotesize\yng(2,1)\otimes \left(\yng(1,1)\ \circled{p}\ [3]\right) =  \yng(5,4) +  \yng(5,3,1) +  \yng(4,3,2) + \yng(5,2,2),\\ 
&&\footnotesize\yng(2,1)\otimes \left(\yng(1,1)\ \circled{p}\ [2,1]\right) =  \yng(5,3,1) +  \yng(4,4,1) + {\color{blue}  \yng(3,3,3)} + \yng(5,2,2) + 2\times \yng(4,3,2), \\
&&\footnotesize\yng(2,1)\otimes \left(\yng(1,1)\ \circled{p}\ [3]\right) = \yng(4,3,2).  
\end{eqnarray}
Therefore one can obtain the p-basis in terms of m-basis in the following formula:
\begin{eqnarray}
\begin{pmatrix}
T_{\rm SU3,1}^{[2,1]} \\
T_{\rm SU3,2}^{[2,1]}
\end{pmatrix}
=
\begin{pmatrix}
b^{[2,1]}_1\circ T^{\rm m}_{\rm SU3,1} \\
b^{[2,1]}_2\circ T^{\rm m}_{\rm SU3,1} 
\end{pmatrix}
=
\begin{pmatrix}
\frac{4}{3} & -\frac{2}{3} \\
-\frac{2}{3} & \frac{4}{3}
\end{pmatrix}
\begin{pmatrix}
T^m_{\rm SU3,1} \\
T^m_{\rm SU3,2},
\end{pmatrix}
\end{eqnarray}
where the conversion matrix is obtained by the method  in section.~\ref{sec:basis_conv}.

\subsection{Flavor Basis from Inner Product Decomposition}
\label{sec:flavor}
As we have obtained the symmtrized gauge group factors and Lorentz structures, we shall construct p-basis through eq.~\eqref{eq:master}.
In our example, $SU(2)_W$ gauge group factor is trivial, so we only need to focus on the inner product decomposition from the Lorentz and $SU(3)_C$ parts:
\begin{eqnarray}\arraycolsep=1.4pt\def\arraystretch{2.2}
\begin{array}{ccccc}
    {\rm Lorentz}& &SU(3)_C & &{\rm Flavor} \\
     \yng(3)&\odot &\yng(2,1) &=& 1\times\yng(2,1), \\
     \yng(2,1)&\odot &\yng(2,1)  &=& 1\times\yng(3)\oplus 1\times\yng(2,1)\oplus 1\times\yng(1,1,1), \\
     \yng(1,1,1)&\odot &\yng(2,1)  &=& 1\times\yng(2,1). \\
\end{array} 
\label{eq:decompose}
\end{eqnarray}
The CGCs for each decomposition are listed in table.~\ref{tab:decomp}.

\begin{table}[ht]
\centering
\begin{tabular}{ |c| c| c| } 
\hline
Product Sym &Target Sym & Relevant CGCs  \\ 
\hline
\multirow{2}{5em}{\centering $[3]\odot [2,1]$} &\multirow{2}{5em}{\centering $[2,1]$} & $C_{([2,1],1),1}^{([3],1),([2,1],1)}=1\quad C_{([2,1],1),1}^{([3],1),([2,1],2)}=0$ \\ 
& &  $C_{([2,1],2),1}^{([3],1),([2,1],1)}=0\quad C_{([2,1],2),1}^{([3],1),([2,1],2)}=1$\\ 
\hline
\multirow{8}{5em}{\centering $[2,1]\odot [2,1]$} &\multirow{2}{5em}{\centering $[3]$} & $C_{([3],1),1}^{([2,1],1),([2,1],1)}=\frac{2}{3}\quad C_{([3],1),1}^{([3],1),([2,1],2)}=\frac{1}{3}$ \\  
& &  $C_{([3],1),1}^{([2,1],2),([2,1],1)}=\frac{1}{3}\quad C_{([3],1),1}^{([2,1],2),([2,1],2)}=\frac{2}{3}$\\
\cline{2-3}
&\multirow{4}{5em}{\centering $[2,1]$}& $C_{([2,1],1),1}^{([2,1],1),([2,1],1)}=\frac{1}{3}\quad C_{([2,1],1),1}^{([3],1),([2,1],2)}=-\frac{1}{3}$ \\  
& &  $C_{([2,1],1),1}^{([2,1],2),([2,1],1)}=-\frac{1}{3}\quad C_{([2,1],1),1}^{([2,1],2),([2,1],2)}=-\frac{2}{3}$\\
& & $C_{([2,1],2),1}^{([2,1],1),([2,1],1)}=-\frac{2}{3}\quad C_{([2,1],2),1}^{([3],1),([2,1],2)}=-\frac{1}{3}$ \\  
& &  $C_{([2,1],2),1}^{([2,1],2),([2,1],1)}=-\frac{1}{3}\quad C_{([2,1],2),1}^{([2,1],2),([2,1],2)}=\frac{1}{3}$\\
\cline{2-3}
&\multirow{2}{5em}{\centering $[1^3]$}&  $C_{([1^3],1),1}^{([2,1],1),([2,1],1)}=0\quad C_{([1^3],1),1}^{([3],1),([2,1],2)}=\frac{1}{2}$ \\  
& &  $C_{([1^3],1),1}^{([2,1],2),([2,1],1)}=-\frac{1}{2}\quad C_{([1^3],1),1}^{([2,1],2),([2,1],2)}=0$\\
\hline
\multirow{2}{5em}{\centering $[1^3]\odot [2,1]$} &\multirow{2}{5em}{\centering $[2,1]$}& $C_{([2,1],1),1}^{([1^3],1),([2,1],1)}=-\frac{1}{3}\quad C_{([2,1],1),1}^{([1^3],1),([2,1],2)}=-\frac{2}{3}$ \\ 
& &  $C_{([2,1],2),1}^{([1^3],1),([2,1],1)}=\frac{2}{3}\quad C_{([2,1],2),1}^{([1^3],1),([2,1],2)}=\frac{1}{3}$\\ 
\hline
\end{tabular}
\caption{The relevant CGCs of $S_3$ inner product decomposition}
\label{tab:decomp}
\end{table}
As an example, we explicitly show how to generate ${\cal O}_{([1^3],1),1}$ from $[2,1]\odot [2,1]$:
\begin{eqnarray}
{\cal O}_{([1^3],1),1}&=&C_{([1^3],1),1}^{([2,1],1),([2,1],1)} {\cal M}^{[2,1]}_1T^{[2,1]}_{\rm SU3,1}+C_{([1^3],1),1}^{([2,1],1),([2,1],2)} {\cal M}^{[2,1]}_1T^{[2,1]}_{\rm SU3,2}\\ \nonumber
&&+C_{([1^3],1),1}^{([2,1],2),([2,1],1)} {\cal M}^{[2,1]}_2T^{[2,1]}_{\rm SU3,1}+C_{([1^3],1),1}^{([2,1],2),([2,1],2)} {\cal M}^{[2,1]}_2T^{[2,1]}_{\rm SU3,2}\\
&=& \frac{1}{2}{\cal M}^{[2,1]}_1T^{[2,1]}_{\rm SU3,2}-\frac{1}{2}{\cal M}^{[2,1]}_2T^{[2,1]}_{\rm SU3,1}\\
& = & \frac{1}{2}\left({\cal M}^{\rm m}_1+2{\cal M}^{\rm m}_2-{\cal M}^{\rm m}_3+{\cal M}^{\rm m}_4\right)\left(\frac{4}{3}T^{\rm m}_{\rm SU3,1}-\frac{2}{3}T^{\rm m}_{\rm SU3,2}\right)\\ \nonumber
&&-\frac{1}{2}\left(-2{\cal M}^{\rm m}_1-{\cal M}^{\rm m}_2-{\cal M}^{\rm m}_3-2{\cal M}^{\rm m}_4\right)\left(-\frac{2}{3}T^{\rm m}_{\rm SU3,1}+\frac{4}{3}T^{\rm m}_{\rm SU3,2}\right)\\ \nonumber
& = &-\frac{4}{3}\left({\cal M}^{\rm m}_1 T^{m}_{\rm SU3,1}+{\cal M}^{\rm m}_2 T^{\rm m}_{\rm SU3,2}+{\cal M}^{\rm m}_3 T^{\rm m}_{\rm SU3,1}-{\cal M}^{\rm m}_3 T^{\rm m}_{\rm SU3,2}+{\cal M}^{\rm m}_4 T^{\rm m}_{\rm SU3,1}\right),\label{eq:extom}
\end{eqnarray} 
where we deliberately change the final expression into m-basis operator obtained by the direct product of m-basis of the Lorentz structures and the gauge group factors i.e. $\{{\cal O}^{\rm (m)}\}=\{{\cal M}^{\rm m}\}\otimes \{T^{\rm m}_{\rm SU3}\}$. Similarly, one can express each term in  $\{{\cal O}^{\rm (p)}_{(\lambda,x),\xi}\}$ as a linear combination of the basis vectors in $\{{\cal O}^{\rm (m)}\}$:
\begin{eqnarray}
\begin{pmatrix}
{\cal O}^{\rm (p)}_{([2,1],1),1} \\
{\cal O}^{\rm (p)}_{([2,1],2),1} \\
{\cal O}^{\rm (p)}_{([2,1],1),2} \\
{\cal O}^{\rm (p)}_{([2,1],2),2} \\
{\cal O}^{\rm (p)}_{([2,1],1),3} \\
{\cal O}^{\rm (p)}_{([2,1],2),3} \\
{\cal O}^{\rm (p)}_{([3],1),1} \\
{\cal O}^{\rm (p)}_{([1^3],1),1} 
\end{pmatrix}
=
\begin{pmatrix}
0& \frac{4}{3}& 0& \frac{4}{3}& 0& \frac{4}{3}& 0& 0\\ 
-\frac{4}{3}& 0& -\frac{4}{3}& 0& -\frac{4}{3}& 0& 0& 0\\ 
-\frac{4}{9}& -\frac{4}{9}& -\frac{8}{9}& \frac{4}{9}& \frac{4}{9}& -\frac{8}{9}& -\frac{4}{9}& -\frac{4}{9}\\ 
-\frac{4}{9}& \frac{8}{9}& \frac{4}{9}& \frac{4}{9}& -\frac{8}{9}& \frac{4}{9}& -\frac{4}{9}& \frac{8}{9}\\ 
-\frac{8}{3}& \frac{4}{3}& \frac{8}{3}& -\frac{4}{3}& \frac{8}{3}& -\frac{4}{3}& \frac{16}{3}& -\frac{8}{3}\\ 
\frac{4}{3}& -\frac{8}{3}& -\frac{4}{3}& \frac{8}{3}& -\frac{4}{3}& \frac{8}{3}& -\frac{8}{3}& \frac{16}{3}\\ 
\frac{2}{3}& -\frac{4}{3}& \frac{4}{3}& -\frac{2}{3}& -\frac{2}{3}& -\frac{2}{3}& \frac{2}{3}& -\frac{4}{3}\\ 
-\frac{4}{3}& 0& 0& -\frac{4}{3}& -\frac{4}{3}& \frac{4}{3}& -\frac{4}{3}& 0 
\end{pmatrix}
\begin{pmatrix}
{\cal M}^{\rm m}_1 T^{\rm m}_{\rm SU3,1}\\
{\cal M}^{\rm m}_1 T^{\rm m}_{\rm SU3,2}\\
{\cal M}^{\rm m}_2 T^{\rm m}_{\rm SU3,1}\\
{\cal M}^{\rm m}_2 T^{\rm m}_{\rm SU3,2}\\
{\cal M}^{\rm m}_3 T^{\rm m}_{\rm SU3,1}\\
{\cal M}^{\rm m}_3 T^{\rm m}_{\rm SU3,2}\\
{\cal M}^{\rm m}_4 T^{\rm m}_{\rm SU3,1}\\
{\cal M}^{\rm m}_4 T^{\rm m}_{\rm SU3,2}
\end{pmatrix}.
\label{eq:matrices}
\end{eqnarray}
There are two subtleties remaining in the above notation. 
First, the meaning of subscript $\xi$ in $\{{\cal O}^{\rm (p)}_{(\lambda,x),\xi}\}$  is different from that of the subscript $j$ in $\{{\cal O}_{(\lambda,x),j}\}$ as in eq.~\eqref{eq:master}. $j$ represents the label of multiplicity of the resulting irrep for a general inner product decomposition, while $\xi$ is the label of multiplicity of irrep of the flavor permutation symmetry for a certain type of operator. This multiplicity may originate from the same resulting irreps from different inner product decompositions, which is illustrated by our example: the three $[2,1]$'s come from $[3]\odot [2,1]$, $[2,1]\odot [2,1]$ and $[1^3]\odot [2,1]$ respectively.
%
The second subtlety is that the $\{{\cal O}^{\rm (p)}_{(\lambda,x),\xi}\}$ is over complete, as discussed in Ref.~\cite{Li:2020gnx} and \cite{Fonseca:2019yya} for $\lambda$ with dimension larger than one  the flavor space spaned by each basis vector is the same, so we only retain the first basis vector for these irreps. Finally we arrived at the complete set of independent terms of operator for $G_L d^3_{_\mc{C}} e_{_\mc{C}} D$:
\begin{eqnarray}
\{{\cal O}^{\rm (p)}_{([2,1],1),1},\ 
 {\cal O}^{\rm (p)}_{([2,1],1),2},\ {\cal O}^{\rm (p)}_{([2,1],1),3},\ {\cal O}^{\rm (p)}_{([3],1),1},\ {\cal O}^{\rm (p)}_{([1^3],1),1}
\}
\end{eqnarray}

\subsection{Desymmetrization: Reduction to Monomials}
\label{sec:desym}

As one can see from the above section the p-basis operators with particular flavor symmetry are often very long expressions. For instance, the ${\cal O}^{\rm (p)}_{([2,1],1),2}$ in eq.~\eqref{eq:matrices} has 8 monomial terms and cannot be simplified. 
It would be convenient though to have a simpler expression, single monomial if possible, for the operator basis, either for presenting the basis or for future applications. 
To keep track of the permutation symmetries, which is crucial as we have shown, we propose operators of the form $\mc{Y}^{[\lambda]}_x\circ\mc{O}^{\rm (m)}_i$ as our final result, called $\rm p'$-basis, in that they are nothing but some recombination of p-basis due to the symmetry imposed by the Young projection: 
\eq{\label{eq:ppbasis}
	\mc{O}^{\rm (p')}_{(\lambda,1),i} \equiv \frac{1}{d_\lambda}\mc{Y}^{[\lambda]}_1\circ\mc{O}^{\rm (m)}_i = \sum_{\xi=1}^{n_\lambda}c_{i\xi}\mc{O}^{\rm (p)}_{(\lambda,1),\xi}, \quad i=1,\dots,\mc{N},
}
where $n_\lambda$ is the number of $\lambda$ representation spaces in the operator type. 
This process is to look for certain subset of m-basis operators that need not have any permutation symmetries itself symmetrizing to independent combinations of p-basis, hence we call it \textbf{de-symmetrization}.
%
%
With the action of the Young symmetrizer, this form of operator is still intrinsically a polynomial. Another interpretation is to apply the symmetrizer to the Wilson coefficient tensor instead of the operators, so that the whole term is indeed a monomial as a singlet under the flavor group $SU(n_f)$
\begin{eqnarray}
\sum_{p_i} C_{p_1p_2...p_n}\left( {\cal Y}[\lambda]{\cal O}^{p_1p_2...p_n}\right) = \sum_{p_i}\left( {\cal Y}^{-1}[\lambda] C_{p_1p_2...p_n}\right){\cal O}^{p_1p_2...p_n}
\end{eqnarray}
where ${\cal Y}^{-1}[\lambda]$ is defined by taking inverse of each constituting permutations in ${\cal Y}[\lambda]$. The action of the symmetrizer $\left(  {\cal Y}^{-1}[\lambda] C_{p_1p_2...p_n}\right)$ will project out the $\lambda$ symmetric irreducible component of the Wilson coefficient tensor $C_{p_1p_2...p_n}$, which is exactly spanned by the $\kappa$ tensor basis we introduced in section~\ref{sec:inv}. 
Therefore by keeping the Young symmetrizer, we actually recover the necessary information of flavors we derive for the amplitude basis. 
%

We have to pick out $n_\lambda$ number of independent operators from the $\mc{N}$ projections for a given $\lambda$ in eq.~\eqref{eq:ppbasis}. It is non-trivial to guarantee the independence, unless $n_\lambda =1$ when we only need to find a non-vanishing projection. 
In general, we need to obtain the coefficient matrix $c$ in eq.~\eqref{eq:ppbasis}, from which we simply pick out $n_\lambda$ rows to form a full-rank submatrix $c_{\zeta\xi}$, where $\zeta$ takes values from an $n_\lambda$-size subset of 1 through $\mc{N}$. 
Instead of directly inspecting the matrix representation of the Young symmetrizers for the m-basis, it is easier to see how they act on the p-basis, because the p-basis already have specific symmetries. Due to the following property
\eq{
	b^\lambda_x \cdot b^{\lambda'}_y = \delta^{\lambda\lambda'}\delta_{x1}\delta_{y1}b^\lambda_1 ,\qquad b^\lambda_1 = \frac{1}{d_\lambda}\mc{Y}^{[\lambda]}_1,
}
we have
\eq{
	\frac{1}{d_\lambda}\mc{Y}^{[\lambda]}_1\circ\mc{O}^{\rm (p)}_{(\lambda',x),\xi} = \delta^{\lambda\lambda'}\delta_{1x}\mc{O}^{\rm (p)}_{(\lambda,1),\xi},\quad \xi = 1,\dots,n_\lambda.
}
Thus we obtain the matrix representation of the symmetrizer for p-basis
\eq{
	\frac{1}{d_\lambda}\mc{Y}^{[\lambda]}_1\circ\mc{O}^{\rm (p)} = \begin{pmatrix} 1_{n_\lambda\times n_\lambda} & 0 \\ 0 & 0 \end{pmatrix}_{\mc{N}\times\mc{N}} \mc{O}^{\rm (p)}
}
where we set the first $n_\lambda$ p-basis to be $\mc{O}^{\rm (p)}_{(\lambda,1),\xi}$ for convenience. Therefore we first convert the m-basis operator to p-basis using the matrix $\mc{K}^{\rm mp} = (\mc{K}^{\rm pm})^{-1}$, and obtain
\eq{
	\frac{1}{d_\lambda}\mc{Y}^{[\lambda]}_1\circ\mc{O}^{\rm (m)}_i &= \frac{1}{d_\lambda}\sum_{j=1}^{\mc{N}} \mc{K}^{\rm mp}_{ij} \mc{Y}^{[\lambda]}_1\circ\mc{O}^{\rm (p)}_j \\
	&= \sum_{j,k=1}^{\mc{N}} \begin{pmatrix} c	&	\bar{c} \end{pmatrix}_{ij} \begin{pmatrix} 1_{n_\lambda\times n_\lambda} & 0 \\ 0 & 0 \end{pmatrix}_{jk} \mc{O}^{\rm (p)}_k \\
	&= \sum_{\xi=1}^{n_\lambda} c_{i\xi} \mc{O}^{\rm (p)}_{(\lambda,1),\xi} .
}
where the matrix $c_{i\xi}$ is identified as the $n_\lambda$ columns in $\mc{K}^{\rm mp}$ that correspond to the $\mc{O}^{\rm (p)}_{(\lambda,1),\xi}$ basis. As explained above, we only need to select independent rows in $c$ that form our p'-basis for $\lambda$.

\definecolor{highlight}{rgb}{0.88,1,1}
\newcolumntype{s}{>{\columncolor{highlight}} c}

As an example, we demonstrate the desymmetrization for the $\lambda=[2,1]$ representation in eq.~\eqref{eq:matrices}, for which we find the inverse matrix
\eq{
\mc{K}^{\rm mp} = \left(
\begin{array}{scscsccc}
	{\color{red} 0} & 0 & {\color{red}-\frac{1}{2}} & -\frac{1}{4} & {\color{red}-\frac{1}{6}} &	-\frac{1}{12} & 0 & -\frac{1}{4} \\
	0 & 0 & -\frac{1}{4} & \frac{1}{4} & -\frac{1}{12} &	-\frac{1}{6} & -\frac{1}{4} & -\frac{1}{8} \\
	0 & -\frac{3}{8} & \frac{1}{4} & \frac{1}{2} &	\frac{1}{12} & \frac{1}{24} & \frac{1}{4} &	\frac{1}{8} \\
	{\color{red}\frac{3}{8}} & 0 & {\color{red}\frac{1}{2}} & \frac{1}{4} &	{\color{red}\frac{1}{24}} & \frac{1}{12} & \frac{1}{4} &	-\frac{1}{8} \\
	0 & -\frac{3}{8} & \frac{1}{4} & -\frac{1}{4} &	\frac{1}{12} & \frac{1}{24} & -\frac{1}{4} &	\frac{1}{8} \\
	{\color{red}\frac{3}{8}} & 0 & {\color{red}-\frac{1}{4}} & -\frac{1}{2} &	{\color{red}\frac{1}{24}} & \frac{1}{12} & 0 & \frac{1}{4} \\
	0 & \frac{3}{8} & -\frac{1}{2} & -\frac{1}{4} &	\frac{1}{12} & \frac{1}{24} & 0 & -\frac{1}{4} \\
	-\frac{3}{8} & 0 & -\frac{1}{4} & \frac{1}{4} &	\frac{1}{24} & \frac{1}{12} & -\frac{1}{4} &	-\frac{1}{8} \\
\end{array}
\right) ,\qquad
c_{\zeta\xi} = \begin{pmatrix} 0	&	-\frac{1}{2}	&	-\frac{1}{6}	\\
	\frac{3}{8}	&	\frac{1}{2}		&	\frac{1}{24}	\\
	\frac{3}{8}	&	-\frac{1}{4}	&	\frac{1}{24}	\end{pmatrix}
}
where the highlighted columns correspond to the p-basis $\mc{O}^{\rm (p)}_{([2,1],1),\xi}, \xi=1,2,3$, in which the rows with red color are the selected m-basis that symmetrize to the $\rm p'$-basis. The red submatrix $c_{\zeta\xi}$ sets the full-rank conversion matrix between the p-basis and the p'-basis in the $\lambda=[2,1]$ section. Therefore the final operators we get are
\begin{eqnarray}
{\cal O}^{({\rm p'})}_{([2,1],1),1} = \frac{1}{2}\mc{Y}^{[2,1]}_1 \mc{M}^{\rm m}_1T^{\rm m}_{\rm SU3,1} = {\cal Y}{\left[\footnotesize \young(pr,s)\right]} i\epsilon_{acd}\left(\lambda^A\right)_b^dG_L^{A\mu}{}_\nu (d^a_{p}C d^{c}_{s})(D_\mu \bar{e}_t\gamma^\nu d^b ),  \\
{\cal O}^{({\rm p'})}_{([2,1],1),2} = \frac{1}{2}\mc{Y}^{[2,1]}_1 \mc{M}^{\rm m}_2T^{\rm m}_{\rm SU3,2} ={\cal Y}{\left[\footnotesize \young(pr,s)\right]} i\epsilon_{abd}\left(\lambda^A\right)_c^d  G_L^{A\mu}{}_\nu (d^a_{p}C d^{c}_{s})(D_\mu \bar{e}_t\gamma^\nu d^b ),  \\ 
{\cal O}^{({\rm p'})}_{([2,1],1),3} = \frac{1}{2}\mc{Y}^{[2,1]}_1 \mc{M}^{\rm m}_3T^{\rm m}_{\rm SU3,2} = {\cal Y}{\left[\footnotesize \young(pr,s)\right]} i\epsilon_{acd}\left(\lambda^A\right)_b^d G_L^{A\mu}{}_\nu (d^a_{p} C D_\mu d^{c}_{s})(\bar{e}_t \gamma^\nu d^b_{r}),
\end{eqnarray}
where we express the Young symmetrizer by explicitly showing the Young tableau with flavor indices filled in, so that we don't have to keep track of the order of the flavor indices. Recall that ${\cal Y}^{[2,1]}_1$ acts on the flavor tensor ${\cal O}^{prst}$ as:
\begin{eqnarray}
{\cal Y}{\left[\footnotesize \young(pr,s)\right]}{\cal O}^{prst} = {\cal O}^{prst}+{\cal O}^{rpst}-{\cal O}^{srpt}-{\cal O}^{sprt}.
\end{eqnarray}

We implement the same process for all of the $\lambda$'s, like in the previous example we have for $\lambda=[3],[1^3]$:
\begin{eqnarray}
{\cal O}^{({\rm p'})}_{([3],1),1} = \mc{Y}^{[3]}_1 \mc{M}^{\rm m}_2T^{\rm m}_{\rm SU3,2} =  {\cal Y}{\left[\footnotesize \young(prs)\right]} i\epsilon_{acd}\left(\lambda^A\right)_b^d G_L^{A\mu}{}_\nu (d^a_{p}C d^{c}_{s})(D_\mu \bar{e}_t\gamma^\nu d^b ), \\
{\cal O}^{({\rm p'})}_{([1^3],1),1} = \mc{Y}^{[1^3]}_1 \mc{M}^{\rm m}_2T^{\rm m}_{\rm SU3,2} = {\cal Y}{\left[\footnotesize \young(p,r,s)\right]} i\epsilon_{acd}\left(\lambda^A\right)_b^d G_L^{A\mu}{}_\nu (d^a_{p}C d^{c}_{s})(D_\mu \bar{e}_t\gamma^\nu d^b ).
\end{eqnarray}
As one may notice, an m-basis may be selected multiple times, such as the $\mc{M}^m_2T^m_{\rm SU3,2}$ in the above case. When that happens, the final result contains terms that could merge into a single Lagrangian term in the traditional sense
\eq{
	\left(\mc{Y}^{[\lambda_1]}_1 \oplus \mc{Y}^{[\lambda_2]}_1 \oplus \dots \right) \circ\mc{O}^{\rm (m)}_i,
}
which belongs to reducible representation of the flavor group $SU(n_f)$. 
This notation is equivalent to the flavor relations in the traditional operator enumeration \cite{Grzadkowski:2010es,Liao:2016qyd,Liao:2019tep}, while the crucial difference is that in the traditional treatment, the flavor relations need to be worked out specifically for each type of operators, involving all of the operator redundancy relations like the EOM, the IBP relation and the Fierz identities for both Lorentz and gauge groups. At higher dimensions, it may even be necessary to work out flavor relations among different operators of the same type.
Suppose we have two monomial terms $\mc{O}^{(\rm (m)}_{1,2}$, which has intrinsic flavor relations that imply the following p'-basis that are independent within each merged term
\eq{
	\left(\mc{Y}^{[\lambda_1]}_1 \oplus \mc{Y}^{[\lambda_2]}_1 \right) \circ \mc{O}^{\rm (m)}_1, \quad 
	\left( \mc{Y}^{[\lambda_1]}_1 \oplus \mc{Y}^{[\lambda_2]}_1 \oplus \mc{Y}^{[\lambda_3]}_1 \right) \circ \mc{O}^{\rm (m)}_2.
}
Also suppose that in our treatment we find $n_{\lambda_1}=2$ and $n_{\lambda_2}=n_{\lambda_3}=1$, thus the two terms $\mc{Y}^{[\lambda_2]}_1\mc{O}^{\rm (m)}_{1,2}$ are actually equivalent, which is translated into a flavor relation between the two operators. 
Such flavor relations did not show up at dim-7 or lower, but are inevitable at higher dimensions when the subspaces of type become larger, and are quite tricky to work out systematically. 
Therefore, our method of operator enumeration has the privilege that we do not need to work out these relations explicitly, but rather provide an equivalent notation to represent the flavor information.
In the following, we use a dim-7 example to show the equivalence between our Young symmetrizer notation and the traditional flavor relations.

\comments{
Finally, we show that one can convert the ${\cal Y}[\lambda]$ to the Wilson coefficients tensors. We note that the following relation holds:
\begin{eqnarray}
\sum_{p_i} C_{p_1p_2...p_n}\left( \pi\circ {\cal O}^{p_1p_2...p_n}\right)  &=& \sum_{p_i} C_{p_1p_2...p_n} {\cal O}^{p_{\pi(1)}
p_{\pi(2)}...p_{\pi(n)}}\nonumber \\
& = &\sum_{p_i} C_{p_{\pi^{-1}(1)}p_{\pi^{-1}(2)}...p_{\pi^{-1}(n)}} {\cal O}^{p_1p_2...p_n}\nonumber \\
& = & \sum_{p_i} \left(\pi^{-1}\circ C_{p_1p_2...p_n}\right) {\cal O}^{p_1p_2...p_n}, 
\end{eqnarray}
}

\subsection{Flavor Tensor versus Flavor Relation}
\label{sec:relation}

To demonstrate the advantage of our notation with Young symmetrizers  we take the operator ${\cal O}^{trps}_{\bar{L}dddH}$\footnote{We have changed the order of the flavor indices in the superscripts to match our notation.} as an example and compare our result to those obtained in Ref.~\cite{Liao:2016hru} with  flavor relations:
\begin{eqnarray}
&&{\cal O}^{trps} + p{\leftrightarrow}s = 0\\
&&{\cal O}^{trps} + {\cal O}^{tpsr} +{\cal O}^{tsrp} =0.
\end{eqnarray} 
Since only the flavor indices of repeated fields $d$ are relevant for the symmetrization, we temporally neglect the index $t$ and treat the operator as a tensor of the flavor indices $p,r,s$. To make the description of the symmetric group more convenient and transparent, we change the indices $p,r,s$ to those with subscripted indices $f_1,f_2,f_3$, and the above equations are equivalent to:
\begin{eqnarray}
&&{\cal O}^{f_2f_1f_3}+{\cal O}^{f_2f_3f_1} = [(12)+(123)]\circ {\cal O}^{f_1f_2f_3}=0,
\label{eq:frelation1}\\
&&{\cal O}^{f_2f_1f_3}+{\cal O}^{f_1f_3f_2}+{\cal O}^{f_3f_2f_1} = [(12)+(23)+(13)]\circ {\cal O}^{f_1f_2f_3}=0.\label{eq:frelation2}
\end{eqnarray} 
One the other hand, the identity in the group algebra $\tilde{S}_3$ can be written as a summation of the 4 distinct primitive idempotents that are proportional to the 4 Young symmetrizers of  different SSYT:
\begin{eqnarray}
E = {\cal Y}^{[3]}_1 + \frac{1}{2}{\cal Y}^{[2,1]}_1 +\frac{1}{2}{\cal Y}^{[2,1]}_2 + {\cal Y}^{[1^3]}_1,
\label{eq:eidentity}
\end{eqnarray}
where ${\cal Y}^{[2,1]}_1=2 b_1^{[2,1]}$ is the Young symmetrizer of the Normal Young diagram $\footnotesize\young(12,3)$ we mentioned in the above subsection, and ${\cal Y}^{[2,1]}_2=2 b_2^{[2,1]}\cdot (23)$ is the Young symmetrizer of the other SSYT $\footnotesize \young(13,2)$.
Acting the identity on an arbitrary tensor yields the original tensor indicates that a 3rd rank tensor can be decomposed to 4 distinct subspace with the corresponding permutation symmetry. This is essentially the underlining reason that we have the decomposition $3\otimes 3\otimes 3 = 10\oplus 8\oplus 8\oplus 1$ for $SU(3)$.

\begin{table}
\begin{align*}
    \begin{array}{c|c}
        \multicolumn{2}{c}{\psi^2\phi^4}\\
        \hline
        \mathcal{O}_{  l^2 H{}^3 H^{\dagger}   }
        &\mathcal{Y}\left[\tiny{\young(pr)}\right]\epsilon ^{im} \epsilon ^{jn} \left(l_{pi} C l_{rj}\right) H_{m} H_{n} \left(H^{\dagger}H\right) \vspace{3ex}\\

        \multicolumn{2}{c}{\psi^2\phi^3D}\\
        \hline
        \mathcal{O}_{  e  l H{}^3 D}
        & \epsilon ^{ij} \epsilon ^{km} \left(l_{pi} C \gamma ^{\mu } e_{r}\right) H_{j} H_{m} D_{\mu } H_{k}\vspace{3ex}\\

        \multicolumn{2}{c}{\psi^2\phi^2D^2}\\
        \hline
        \mathcal{O}_{  l^2 H{}^2  }^{(1)}
        &\mathcal{Y}\left[\tiny{\young(pr)}\right]\epsilon ^{ik} \epsilon ^{jm} \left(l_{pi} C \sigma ^{\mu }{}^{\nu } l_{rj}\right) D_{\mu } H_{k} D_{\nu } H_{m}\\
        \mathcal{O}_{  l^2 H{}^2  }^{(2)}
        &\mathcal{Y}\left[\tiny{\young(pr)}\right]\epsilon ^{ik} \epsilon ^{jm} \left(l_{pi} C l_{rj}\right) D_{\mu } H_{k} D^{\mu } H_{m}\vspace{3ex}\\

        \multicolumn{2}{c}{F\psi^2\phi^2}\\
        \hline
        \mathcal{O}_{W l^2 H{}^2  }^{(1)}
        &\mathcal{Y}\left[\tiny{\young(p,r)}\right]\epsilon ^{ik} \epsilon ^{jn} \left(\tau ^I\right)_n^m W^{I}_{\mu \nu } \left(l_{pi} C \sigma ^{\mu }{}^{\nu } l_{rj}\right) H_{k} H_{m}\\
        \mathcal{O}_{W l^2 H{}^2  }^{(2)}
        &\mathcal{Y}\left[\tiny{\young(pr)}\right]\epsilon ^{ik} \epsilon ^{jn} \left(\tau ^I\right)_n^m W^{I}_{\mu \nu } \left(l_{pi} C \sigma ^{\mu }{}^{\nu } l_{rj}\right) H_{k} H_{m}\\
        \mathcal{O}_{B l^2 H{}^2  }
        &\mathcal{Y}\left[\tiny{\young(p,r)}\right]\epsilon ^{ik} \epsilon ^{jm} B{}_{\mu }{}_{\nu } \left(l_{pi} C \sigma ^{\mu }{}^{\nu } l_{rj}\right) H_{k} H_{m}\vspace{3ex}\\

        \multicolumn{2}{c}{\psi^4D}\\
        \hline
        \mathcal{O}_{  \bar{d} l^2 u    D}
        &\mathcal{Y}\left[\tiny{\young(rs)}\right]\epsilon ^{ij} \left(\overline{d}_{p}^{a} \gamma ^{\mu } u_{ta}\right) \left(l_{ri} C D_{\mu } l_{sj}\right)\\
        \mathcal{O}_{  d^2 \bar{l} q    D}(B\!\!\!\!/)
        &\mathcal{Y}\left[\tiny{\young(pr)}\right]\epsilon ^{abc} \left(\bar{l}_{s}^{i} D_{\mu} d_{rb}\right) \left(d_{pa} \gamma ^{\mu } C q_{tc i}\right)\\
        \mathcal{O}_{  d^3 \bar{e}    D}(B\!\!\!\!/)
        &\mathcal{Y}\left[\tiny{\young(prs)}\right]\epsilon ^{abc} \left(\bar{e}_{t} \gamma^{\mu} d_{pa}\right) \left(d_{rb} C D_{\mu } d_{sc}\right)
    \end{array}
    \begin{array}{c|c}
        \multicolumn{2}{c}{\psi^4\phi}\\
        \hline
        \mathcal{O}_{  \bar{d}  e l  u H   }
        & \epsilon ^{ij} \left(\bar{d}_{s}^{a} l_{tj}\right) \left(e_{p} C u_{ra}\right) H_{i}\\

       \mathcal{O}_{  \bar{d} l^2 q H  }^{(1)}
        &\mathcal{Y}\left[\tiny{\young(r,s)}\right]\epsilon ^{ik} \epsilon ^{jm} \left(\overline{d}_{p}^{a} l_{sj}\right) \left(l_{ri} C q_{ta k}\right) H_{m}\\
        \mathcal{O}_{  \bar{d} l^2 q H  }^{(2)}
        &\mathcal{Y}\left[\tiny{\young(r,s)}\right]\epsilon ^{ij} \epsilon ^{km} \left(\overline{d}_{p}^{a} l_{sj}\right) \left(l_{ri} C q_{ta k}\right) H_{m}\\
        \mathcal{O}_{  \bar{d} l^2 q H  }^{(3)}
        &\mathcal{Y}\left[\tiny{\young(rs)}\right]\epsilon ^{ik} \epsilon ^{jm} \left(\overline{d}_{p}^{a} l_{sj}\right) \left(l_{ri} C q_{ta k}\right) H_{m}\\
        \mathcal{O}_{  \bar{d} l^2 q H  }^{(4)}
        &\mathcal{Y}\left[\tiny{\young(rs)}\right]\epsilon ^{ij} \epsilon ^{km} \left(\overline{d}_{p}^{a} l_{sj}\right) \left(l_{ri} C q_{ta k}\right) H_{m}\\

        \mathcal{O}_{  \bar{e} l^3 H  }^{(1)}
        &\mathcal{Y}\left[\tiny{\young(rst)}\right]\epsilon ^{ij} \epsilon ^{km} \left(\overline{e}_{p} l_{sj}\right) \left(l_{ri} C l_{tk}\right) H_m\\
        \mathcal{O}_{  \bar{e} l^3 H  }^{(2)}
        &\mathcal{Y}\left[\tiny{\young(rs,t)}\right]\epsilon ^{ik} \epsilon ^{jm} \left(\overline{e}_{p} l_{sj}\right) \left(l_{ri} C l_{tk}\right) H_m\\
        \mathcal{O}_{  \bar{e} l^3 H  }^{(3)}
        &\mathcal{Y}\left[\tiny{\young(r,s,t)}\right]\epsilon ^{ik} \epsilon ^{jm} \left(\overline{e}_{p} l_{sj}\right) \left(l_{ri} C l_{tk}\right) H_m\\

        \mathcal{O}_{  l ^2 \bar{q} u H   }^{(1)}
        &\mathcal{Y}\left[\tiny{\young(st)}\right]\epsilon ^{jm} \left(\bar{q}_{p}^{a i} u_{ra}\right) \left(l_{si} C l_{tm}\right) H_{j}\\
        \mathcal{O}_{  l ^2 \bar{q} u H   }^{(2)}
        &\mathcal{Y}\left[\tiny{\young(s,t)}\right]\epsilon ^{jm} \left(\bar{q}_{p}^{a i} u_{ra}\right) \left(l_{si} C l_{tm}\right) H_{j}\\

         \mathcal{O}_{  d^2 \bar{l} u H^{\dagger}  }^{(1)}(B\!\!\!\!/)
        &\mathcal{Y}\left[\tiny{\young(pr)}\right]\epsilon ^{abc} \epsilon _{ij} \left(\bar{l}_{s}^{i} d_{pa}\right) \left(u_{tc} C d_{rb}\right) H^{\dagger j}\\
        \mathcal{O}_{  d^2 \bar{l} u H^{\dagger}  }^{(2)}(B\!\!\!\!/)
        &\mathcal{Y}\left[\tiny{\young(p,r)}\right]\epsilon ^{abc} \epsilon _{ij} \left(\bar{l}_{s}^{i} d_{pa}\right) \left(u_{tc} C d_{rb}\right) H^{\dagger j}\\

        \mathcal{O}_{  d^3 \bar{l} H   }(B\!\!\!\!/)
        &\mathcal{Y}\left[\tiny{\young(pr,s)}\right]\epsilon ^{abc} \left(\bar{l}_{t}^{i} d_{rb}\right) \left(d_{pa} C d_{sc}\right) H_{i}\\
        
        \mathcal{O}_{  d  \bar{l}  q^2 H^{\dagger}   }^{(1)}(B\!\!\!\!/)
        &\mathcal{Y}\left[\tiny{\young(p,r)}\right]\epsilon ^{abc} \left(\overline{l}_{t}^{j} d_{sc}\right) \left(q_{pa i} C q_{rb j}\right) H^{\dagger}{}^{i}\\
        \mathcal{O}_{  d  \bar{l}  q^2 H^{\dagger}   }^{(2)}(B\!\!\!\!/)
        &\mathcal{Y}\left[\tiny{\young(pr)}\right]\epsilon ^{abc} \left(\overline{l}_{t}^{j} d_{sc}\right) \left(q_{pa i} C q_{rb j}\right) H^{\dagger}{}^{i}\\

        \mathcal{O}_{  d ^2 \bar{e} q H^{\dagger}   }(B\!\!\!\!/)
        &\mathcal{Y}\left[\tiny{\young(s,t)}\right]\epsilon ^{abc} \left(\overline{e}_{p} q_{ra i}\right) \left(d_{sb} C d_{tc}\right) H^{\dagger}{}^{i}

    \end{array}
\end{align*}
\caption{List of dimension-7 operators in SMEFT.}
\label{tab:opdim7}
\end{table}

Therefore we can insert an identity $E$ in front of the ${\cal O}^{f_1f_2f_3}$ in eq.~\eqref{eq:frelation1} and eq.~\ref{eq:frelation2}, using the results of the group algebra multiplications:
\begin{eqnarray}
 &&[(12)+(123)]\cdot {\cal Y}^{[1^3]}_1=0,\\
 &&[(12)+(23)+(13)]\cdot {\cal Y}^{[2,1]}_{1,2}=0,
\end{eqnarray}
we convert the flavor relations into:
\begin{eqnarray}
&&[(12)+(123)]\circ [{\cal Y}^{[3]}_1 + \frac{1}{2}{\cal Y}^{[2,1]}_1 +\frac{1}{2}{\cal Y}^{[2,1]}_2 ]\circ{\cal O}^{f_1f_2f_3}=0,\label{eq:fr1}\\
&&[(12)+(23)+(13)]\circ{\cal Y}^{[3]}_1\circ{\cal O}^{f_1f_2f_3} = -[(12)+(23)+(13)]\circ{\cal Y}^{[1^3]}_1 \circ{\cal O}^{f_1f_2f_3}.
\label{eq:fr2}
\end{eqnarray}
Using the properties of the Young symmetrizer\footnote{The relation in eq.~\eqref{eq:YY} holds for the symmetric group $S_{n}$ for $n=1\sim 4$; however, starting from $n=5$ the Young symmetrizers are no longer orthogonal anymore, instead we can construct a set of orthogonal primitive idempotents by decorating the Young symmetrizers with group algebra elements.}:
\begin{eqnarray}
&&{\cal Y}^{\lambda'}_i r {\cal Y}^{\lambda}_j =0\text{ for any }r\in \tilde{S} \text{ if }\lambda'\neq\lambda\\
&&{\cal Y}^{\lambda}_i{\cal Y}^{\lambda}_i\propto {\cal Y}^{\lambda}_i,\label{eq:YY}
\end{eqnarray}
and acting ${\cal Y}^{[3]}_1$ and ${\cal Y}^{[1^3]}_1$ on the both sides of the eq.~\ref{eq:fr2} one can deduce that:
\begin{eqnarray}
{\cal Y}^{[1^3]}_1\circ {\cal O}^{f_1f_2f_3} = {\cal Y}^{[3]}_1\circ {\cal O}^{f_1f_2f_3}=0,
\end{eqnarray}  
which means that we do not have a totally symmetric or a anti-symmetric subspace for the operator ${\cal O}^{trps}_{\bar{L}dddH}$ regarding to the permutation of the flavor indices $r,p,s$ of three down quark fields.
As ${\cal Y}^{[3]}_1\circ {\cal O}^{f_1f_2f_3}=0$, eq.~\eqref{eq:fr1} becomes:
\begin{eqnarray}
[(12)+(123)]\circ {\cal Y}^{[2,1]}_1 \circ{\cal O}^{f_1f_2f_3}=-[(12)+(123)]\circ{\cal Y}^{[2,1]}_2 \circ{\cal O}^{f_1f_2f_3},
\end{eqnarray}
which indicates that the subspaces spanned by these two tensors with $[2,1]$ symmetry ${\cal Y}^{[2,1]}_1 \circ{\cal O}^{f_1f_2f_3}$ and ${\cal Y}^{[2,1]}_2 \circ{\cal O}^{f_1f_2f_3}$ are linearly dependent. So there are only one $[2,1]$ symmetric basis left, which coincides with our result:
\begin{eqnarray}
{\cal O}_{d^3\bar{l}H} = {\cal Y}\left[\footnotesize \young(pr,s)\right]{\cal O}^{trps}_{\bar{L}dddH}.
\end{eqnarray}
Moreover, our result enable ones to directly write down the independent flavor-specified components of the operator by enumerating corresponding SSYTs for $\footnotesize \young(ps,t)$: 
\begin{eqnarray}
(p,r,s)=(f_1,f_2,f_3)=(1,1,2),\ (1,1,3),\ (1,2,2),\ (1,2,3),\ (1,3,3),\ (1,3,2),\ (2,2,3),\ (2,3,3).
\end{eqnarray}
If one starts from the flavor relation, then finding out the corresponding flavor-specified operators may be difficult.

\section{Lists of the Dim-9 Operators}
\label{sec:list}

\subsection{Preview of the Result}
\label{sec:prev}
In this section, we summarize our main results for the dimension 9 operators in the SMEFT. 
%
Table.~\ref{tab:sumdim9} is a summary of the numbers of operators at different levels of categories in our result. In the second column we list all the possible classes characterized by the numbers of derivatives and fields in various Lorentz irreps. We explicitly separate the numbers of types with $\Delta L=0,1,2,3$ in the third column, and a total of 296 different types are obtained. In the fourth column we present the numbers of terms with definite flavor symmetry for each class, and get 1262 independent terms. In the fifth column we express the numbers of flavor-specified operators as functions of the number of generations of fermions $n_f$, and the total numbers of flavor-specified operators for $n_f=1$ and $3$ are 560 and 90456 respectively, again we have separated these total numbers into the sum of the numbers of operators with $\Delta L=0,1,2,3$.
We also list all the dimension 7 operators in table.~\ref{tab:opdim7} for a comparison with the result in Ref.~\cite{Lehman:2014jma, Liao:2016hru}. 

We find that there are 122 types with $(\Delta B,\Delta L)=(\pm 1,\mp 1)$ in dimension 9 that are relevant for the $B-L$ violation process needed for leptogenesis while subjecting to strong proton decay constraints. 
There are also 164 types with $(\Delta B,\Delta L)=(0,\pm 2)$ in dimension 9 that potentially contribute 
neutrino-less double beta decay, and the Majorana neutrino masse with a new physics scale possibly testable at the future LHC experiments. 
We also find that the two-fermion operators must have $(\Delta B,\Delta L)=(0,\pm2)$, while four-fermion operators can be either $(\Delta B,\Delta L)=(\pm 1,\mp 1)$ or $(\Delta B,\Delta L)=(0,\pm2)$. The new violation patterns $(\Delta B,\Delta L)=(\pm 1,\pm 3)$ and $(\Delta B,\Delta L)=(\pm 2,0)$ starting at dimension 9 only appear in the six-fermion operators. The types $u^2d^4$, $d^2q^4$, $ud^3q^2$ and their conjugates with $(\Delta B,\Delta L)=(\pm 2,0)$ will contribute to the neutron-antineutron oscillation. The types $l^3u^2q$, $el^2u^3$ and their conjugates with $(\Delta B,\Delta L)=(\pm 1,\pm 3)$ will contribute to the BLV process without stringent constraints from the proton two-body decay experiments. 

Based on the $p'$-basis, we further perform a few conversions for the convenience of phenomenologists. 
First, we have transfered the field strength tensors from the chiral basis $F_{\rm L,R}$ to the usual form $F$ and $\tilde{F}$. 
Although the chiral basis is a more natural choice from the helicity amplitudes prospect, the $F,\tilde{F}$ basis has many privileges like its hermiticity and definite CP. Moreover, a lot of mature techniques are also implemented in terms of the $F,\tilde{F}$ basis, like the program of Feynman rule calculations. 
We summarize the conversion rules between the two bases as follows\footnote{%
Note that there might be further linear combinations among the converted operators to finally obtain simple monomial operators, which has been carefully done to keep the independence of the operator basis. }:
\begin{align}
\tilde{F}^{\mu\nu}=\frac12\epsilon^{\mu\nu\rho\eta}F_{\rho\eta},\quad F_{\rm{L}/\rm{R}}=\frac12\left(F\mp i\tilde{F} \right)\;.
\label{eq:FLR1}
\end{align}
from which we can easily deduce the following useful identities
\begin{align}
\tilde{F}_{1\mu\rho}F_2{}^{\rho\nu}
=&-F_1{}^{\nu\rho}\tilde{F}_{2\rho\mu}-\frac12 (F_1\tilde{F}_2)\delta^{\nu}_{\mu}\;,\label{eq:FLR2}\\
\tilde{F}_{1\mu\rho}\tilde{F}_2{}^{\rho\nu}
=&F_1{}^{\nu\rho}F_{2\rho\mu}+\frac12(F_1F_2)\delta^{\nu}_{\mu}\;.\label{eq:FLR3}
\end{align}
After the conversion, we do not distinguish types with $F$ or $\tilde{F}$, as they are sometimes not independent of each other. Therefore the types we present in the following sections do not count the same as the numbers in the table~\ref{tab:sumdim9}.

Second, we also present the operators in the four-component form, and retain the relationship to the two-component form in the appendix~\ref{app:A}.
In the SM, the 4-component chiral fermions are related to the 2-component fermions by the following formulas:
\begin{align}
q_{\rm{L}}=\begin{pmatrix}Q\\0\end{pmatrix},\quad u_{\rm{R}}=\left(\begin{array}{c}0\\u_{_\mathbb{C}}^{\dagger}\end{array}\right),\quad d_{\rm R}=\left(\begin{array}{c}0\\d_{_\mathbb{C}}^{\dagger}\end{array}\right),\quad l_{\rm L}=\left(\begin{array}{c}L\\0\end{array}\right),\quad e_{\rm R}=\left(\begin{array}{c}0\\e_{_\mathbb{C}}^{\dagger}\end{array}\right).\\
\bar{q}_{\rm{L}}=\left(0\,,\,Q^{\dagger} \right),\quad \bar{u}_{\rm{R}}=\left(u_{_\mathbb{C}}\,,\,0 \right),\quad \bar{d}_{\rm R}=\left(d_{_\mathbb{C}}\,,\,0\right),\quad \bar{l}_{\rm L}=\left(0\,,\,L^{\dagger}\right),\quad \bar{e}_{\rm R}=\left(e_{_\mathbb{C}}\,,\,0\right).
\end{align}
The conversion rules of the fermion bilinears in the SM are obtained by substituting these fields into the relations in eq.~\eqref{eq:bilinear}, such as
\eq{
	\bar{u}\,q=u_{_\mathbb{C}}Q,&\quad
	\bar{q}\,u=Q^\dagger u^{\dagger}_{_\mathbb{C}}, \\
	\bar{u}\gamma^{\mu}d=u_{_\mathbb{C}}\sigma^{\mu}d^{\dagger}_{_\mathbb{C}},&\quad
	\bar{q}\gamma^{\mu}q=Q^\dagger\bar{\sigma}^{\mu}Q, \\
	\bar{u}\,\sigma^{\mu\nu}q=u_{_\mathbb{C}}\sigma^{\mu\nu}Q,&\quad
	\bar{q}\,\sigma^{\mu\nu}u=Q^\dagger \bar{\sigma}^{\mu\nu} u^{\dagger}_{_\mathbb{C}}, \\
	u^{\rm{T}}Cd=u^{\dagger}_{_\mathbb{C}}d^{\dagger}_{_\mathbb{C}},&\quad
	q^{\rm{T}}Cq=QQ, \\
	u^{\rm{T}}C\gamma^{\mu}q=u^{\dagger}_{_\mathbb{C}}\bar{\sigma}^{\mu}Q,&\quad
	q^{\rm{T}}C\gamma^{\mu}u=Q\sigma^{\mu}u^{\dagger}_{_\mathbb{C}}, \\
	u^{\rm{T}}C\sigma^{\mu\nu}d=u^{\dagger}_{_\mathbb{C}}\bar{\sigma}^{\mu\nu}d^{\dagger}_{_\mathbb{C}},&\quad
	q^{\rm{T}}C\sigma^{\mu\nu}q=Q\sigma^{\mu\nu}Q, \\
	\bar{u}C\bar{d}^{\rm{T}}=u_{_\mathbb{C}}d_{_\mathbb{C}},&\quad
	\bar{q}C\bar{q}^{\rm{T}}=Q^\dagger Q^\dagger, \\
	\bar{u}\gamma^{\mu}C\bar{q}^{\rm{T}}=u_{_\mathbb{C}}\sigma^{\mu}Q^\dagger,&\quad
	\bar{q}\gamma^{\mu}C\bar{u}^{\rm{T}}=Q^\dagger\bar{\sigma}^{\mu}u_{_\mathbb{C}}, \\
	\bar{u}\sigma^{\mu\nu}C\bar{d}^{\rm{T}}=u_{_\mathbb{C}}\sigma^{\mu\nu}d_{_\mathbb{C}},&\quad
	\bar{q}\sigma^{\mu\nu}C\bar{q}^{\rm{T}}=Q^\dagger\bar{\sigma}^{\mu\nu}Q^\dagger.
}
Bilinears involving leptons can be converted similarly. It should be noted that the transpose symbol $\rm{T}$ is left implicit in this section.
Derivatives, if any, do not change the spinor contraction structures.

\begin{table}
\begin{align*}
    \begin{array}{cc|c|c|c|c|c}
        \hline
        N & (n,\tilde{n}) & \text{Classes} & \mathcal{N}_{\text{type}} & \mathcal{N}_{\text{term}} & \mathcal{N}_{\text{operator}} & \text{Equations}\\
        \hline
        \hline
        4 & (3,2) & \psi^3\psi^{\dagger}D^3+h.c. & 0+4+2+0 & 10 & \frac23n^2_f(7n^2_f-1) & (\ref{cl:q3lD3})(\ref{cl:q2l2D3}) \\
         &  & \psi^2\phi^2D^4+h.c. & 0+0+2+0 & 6 & 3n_f(n_f+1) & (\ref{cl:p2h2D4}) \\
         \hline
        5 & (3,1) & F_{\rm L}\psi^3\psi^{\dagger}D+h.c. & 0+10+6+0 & 72 & 32n^4_f & (\ref{cl:fq3lD})(\ref{cl:fq2l2D})\\
         &  & \psi^4\phi D^2+h.c. & 0+4+4+0 & 100 & 40n^4_f & \multirow{1}*{(\ref{cl:q3lhD2}-\ref{cl:l4hD2})} \\
         &  & F_{\rm L}\psi^2\phi^2D^2+h.c. & 0+0+4+0 & 34 & 17n^2_f-n_f & (\ref{cl:fp2h2D21})(\ref{cl:fp2h2D22}) \\
         \cline{2-7}
         & (2,2) & F_{\rm R}\psi^3\psi^{\dagger}D+h.c. & 0+10+6+0 & 54 & 4n^3_f(6n_f+1) & (\ref{cl:fq3lD})(\ref{cl:fq2l2D})\\
         &  & \psi^2\psi^{\dagger 2}\phi D^2 & 0+4+4+0 & 84 & n^3_f(49n_f+1) & \multirow{1}*{(\ref{cl:q3lhD2}-\ref{cl:l4hD2})} \\
         &  & F_{\rm R}\psi^2\phi^2D^2+h.c. & 0+0+4+0 & 20 & 2n_f(5n_f-1) & (\ref{cl:fp2h2D21})(\ref{cl:fp2h2D22}) \\
         &  & \psi\psi^{\dagger}\phi^3D^3 & 0+0+2+0 & 6 & 6n_f^2 & (\ref{cl:p2h3D3}) \\
         \hline
        6 & (3,0) & \psi^6+h.c. & 2+4+6+0 & 116 & \frac{1}{24}n^2_f(415n^4_f+53n^3_f+59n^2_f+139n_f+6) & \multirow{1}*{(\ref{cl:p660}-\ref{cl:p606})} \\
         &  & F_{\rm L}\psi^4\phi+h.c. & 0+12+10+0 & 102 & 2n^3_f(21n_f+1) & \multirow{1}*{(\ref{cl:fq3lh}-\ref{cl:fl4h})} \\
         &  & F_{\rm L}^2\psi^2\phi^2+h.c. & 0+0+8+0 & 20 & 2n_f(5n_f+2) & (\ref{cl:f2p2h2}) \\
         \cline{2-7}
         & (2,1) & \psi^4\psi^{\dagger 2}+h.c. & 4+26+20+4 & 244 & \frac16n^3_f(382n^3_f-9n^2_f+2n_f+21) & \multirow{1}*{(\ref{cl:p660}-\ref{cl:p6242})} \\
         &  & F_{\rm L}\psi^2\psi^{\dagger 2}\phi+h.c. & 0+24+24+0 & 92 & 52n^4_f & \multirow{1}*{(\ref{cl:fq3lh}-\ref{cl:fl4h})} \\
         &  & F_{\rm L}^2\psi^{\dagger 2}\phi^2+h.c. & 0+0+8+0 & 12 & 2n_f(3n_f+2) & (\ref{cl:f2p2h2}) \\
         &  & \psi^3\psi^{\dagger}\phi^2D+h.c. & 0+12+18+0 & 186 & \frac23n^2_f(146n^2_f+1) & \multirow{1}*{(\ref{cl:q3lh2D}-\ref{cl:l4h2D})} \\
         &  & F_{\rm L}\psi\psi^{\dagger}\phi^3D+h.c. & 0+0+8+0 & 12 & 12n^2_f & (\ref{cl:fp2h3D}) \\
         &  & \psi^2\phi^4D^2+h.c. & 0+0+4+0 & 24 & 2n_f(6n_f+1) & (\ref{cl:p2h4D2}) \\
         \hline
        7 & (2,0) & \psi^4\phi^3+h.c. & 0+6+6+0 & 32 & \frac43n^2_f(10n^2_f-1) & \multirow{1}*{(\ref{cl:q3lh3}-\ref{cl:l4h3})} \\
         &  & F_{\rm L}\psi^2\phi^4+h.c. & 0+0+4+0 & 8 & 2n_f(2n_f-1) & (\ref{cl:fp2h4}) \\
         \cline{2-7}
         & (1,1) & \psi^2\psi^{\dagger 2}\phi^3 & 0+6+10+0 & 24 & 14n^4_f & \multirow{1}*{(\ref{cl:q3lh3}-\ref{cl:l4h3})} \\
         &  & \psi\psi^{\dagger}\phi^5D & 0+0+2+0 & 2 & 2n^2_f & (\ref{cl:p2h5D}) \\
         \hline
        8 & (1,0) & \psi^2\phi^6+h.c. & 0+0+2+0 & 2 & n_f^2+n_f & (\ref{cl:p2h6}) \\
        \hline\hline
        \multicolumn{2}{c|}{\multirow{2}*{\text{Total}}} & \multirow{2}*{42} & \multirow{2}*{6+122+164+4} & \multirow{2}*{1262} & 8+204+348+0\;(n_f=1) \\
         & & & & & 2862+42234+44874+486\;(n_f=3) \\
        \hline

    \end{array}
\end{align*}
\caption{We present a complete statistics of dimension 9 SMEFT operators here.
$N$ in the leftmost column shows the number of particles. $(n,\tilde{n})$ are the numbers of $\epsilon$
and $\tilde{\epsilon}$ in the Lorentz structure. $\mathcal{N}_{\rm type}$, $\mathcal{N}_{\rm term}$, and $\mathcal{N}_{\rm operator}$
show the number of types, terms and Hermitian operators respectively (independent conjugates
are counted), while the numbers under $\mathcal{N}_{\rm type}$ and the last line of $\mathcal{N}_{\rm operator}$ describe the sum of each possible $|\Delta L|$
types/operators with $\mathcal{N}=\mathcal{N}(|\Delta L|=0)+\mathcal{N}(|\Delta L|=1)+\mathcal{N}(|\Delta L|=2)+\mathcal{N}(|\Delta L|=3)$. Note that “term” in our definition is different from the other literature, so
the numbers is larger than those in, for instance, \cite{Fonseca:2019yya}. That's not surprising since they did an extra step of merging before the counting.  However the
number of operators are exactly the same as in \cite{Henning:2015alf, Fonseca:2019yya}. The links in the rightmost column refer to the list(s) of the terms in given
classes.
}
\label{tab:sumdim9}
\end{table}

Finally, unlike the dimension 8 basis in \cite{Li:2020gnx}, types are all complex here. We only present the operators without its Hermitian conjugate. 
The Hermitian conjugates of 4-component spinor bilinears can be converted using the following relations:
\eq{
	\left(\bar{\Psi}_1\Psi_2\right)^\dagger &=-\bar{\Psi}_2\Psi_1,\\
	\left(\bar{\Psi}_1\gamma^{\mu}\Psi_2\right)^\dagger &=-\bar{\Psi}_2\gamma^{\mu}\Psi_1,\\
	\left(\bar{\Psi}_1\sigma^{\mu\nu}\Psi_2\right)^\dagger &=-\bar{\Psi}_2\sigma^{\mu\nu}\Psi_1,\\
	\left(\Psi^{\rm{T}}_1C\Psi_2\right)^\dagger &=-\bar{\Psi}_2C\bar{\Psi}^{\rm{T}}_1,\\
	\left(\Psi^{\rm{T}}_1C\gamma^{\mu}\Psi_2\right)^\dagger &=-\bar{\Psi}_2\gamma^{\mu}C\bar{\Psi}_1^{\rm{T}},\\
	\left(\Psi^{\rm{T}}_1C\sigma^{\mu\nu}\Psi_2\right)^\dagger &=-\bar{\Psi}_2\sigma^{\mu\nu}C\bar{\Psi}_1^{\rm{T}}.
}



\subsection{Classes involving Two-fermions}
\label{sec:twofermion}
The classification of different types is based on the number of fermions, as there is no operator without fermion fields. We first list the operators involving two fermions, in which all operators describe $\Delta L=2$ processes, since only the leptons bilinear is allowed to appear. The type $l{}^2 H{}^4 H^{\dagger} {}^2$ can contribute to the neutrino Majorana mass. The type $W l{}^2 H{}^3 H^{\dagger}$ and $B l{}^2 H{}^3 H^{\dagger}$ may contribute to the neutrino anomalous magnetic moment. The type $W^2l^2 H^2$ contains the operators contributing to the neutrino-less double beta decay at tree-level.

\subsubsection{No gauge boson involved}
In this subsection, we deal with the classes $\psi^2 \phi^{n_D} D^{6-n_D}$. Note that for even $n_D$ we have operators with fermions of opposite helicities, or chirality conserving, while for odd $n_D$ we have operators with fermions of the same helicities or chirality violating. 

\noindent\underline{Class $\psi^2 \phi^6$}: The only Lorentz structure of this class is
\begin{align} 
	\psi_1{}{}^{\alpha } \psi _2{}_{\alpha } \phi _3 \phi _4 \phi _5 \phi _6 \phi _7 \phi_8.
\end{align}
This class involves the Weinberg operator with additional Higgses:
\begin{align}\begin{array}{c|l}

    \mathcal{O}_{  l{}^2 H{}^4 H^{\dagger} {}^2  } 
    &\mathcal{Y}\left[\tiny{\young(pr)}\right]\epsilon ^{ik} \epsilon ^{jl} \left(l_{pi} C l_{rj}\right)H_{k} H_{l} \left(H^{\dagger}H\right)^2
    
\end{array}\label{cl:p2h6}\end{align}
After taking all the Higgs vev, it can give rise to additional contributions to the Majorana neutrino masses.

\noindent\underline{Class $\psi^2 \phi^5 D$}: 
The class has to be $\psi\psi^\dagger\phi^5D$, which has the following Lorentz structures:
\begin{align}
	
    &\psi_1{}^{\alpha}\phi _2 \left(D \phi _3\right){}_{\alpha }{}_{\dot{\alpha }} \phi _4 \phi _5 \phi _6 \psi^{\dagger}_7{}^{\dot{\alpha }},\quad \psi _1{}{}^{\alpha } \phi _2 \phi _3 \left(D \phi _4\right){}_{\alpha }{}_{\dot{\alpha }}\phi _5 \phi _6 \psi^{\dagger}_7{}{}^{\dot{\alpha }} ,\\
    &\psi_1{}^{\alpha}\phi _2 \phi _3 \phi _4\left(D \phi _5\right){}_{\alpha }{}_{\dot{\alpha }} \phi _6 \psi^{\dagger}_7{}^{\dot{\alpha }} ,\quad \psi _1{}{}^{\alpha } \phi _2 \phi _3 \phi _4 \phi _5 \left(D \phi _6\right){}_{\alpha }{}_{\dot{\alpha }}\psi^{\dagger}_7{}{}^{\dot{\alpha }} .
    
\end{align}
Considering the conservation of hyper-charge, the only operator in this class is
\begin{align}\begin{array}{c|l}

    \mathcal{O}_{  e  l H{}^4 H^{\dagger}  D} 
    & \epsilon ^{ik} \epsilon ^{lm} \left(l_{pi} C \gamma _{\mu } e_{r}\right)H_{k} H_{l} D^{\mu } H_{m} \left(H^{\dagger}H\right)
    
\end{array}\label{cl:p2h5D}\end{align}

\noindent\underline{Class $\psi^2\phi^4D^2$}: 
The class $\psi^2\phi^4D^2$ contains 12 new Lorentz structures that are all absent at lower dimensions:
\begin{align}

    &\psi _1{}{}^{\alpha } \psi _2{}_{\beta } \left(D \phi _3\right){}_{\alpha }^{\dot{\alpha }} \left(D \phi _4\right){}^{\beta }_{\dot{\alpha }}\phi _5 \phi _6 ,\quad \psi _1{}{}^{\alpha } \psi _2{}_{\beta } \left(D \phi _3\right){}_{\alpha }^{\dot{\alpha }} \phi _4 \left(D \phi _5\right){}^{\beta }_{\dot{\alpha }}\phi _6 ,\quad \psi _1{}{}^{\alpha } \psi _2{}_{\beta } \left(D \phi _3\right){}_{\alpha }^{\dot{\alpha }} \phi _4 \phi _5 \left(D \phi _6\right){}^{\beta }_{\dot{\alpha }},\\
    
    &\psi _1{}{}^{\alpha } \psi _2{}_{\beta }\phi _3  \left(D \phi _4\right){}_{\alpha }^{\dot{\alpha }} \left(D \phi _5\right){}^{\beta }_{\dot{\alpha }}\phi _6 ,\quad \psi _1{}{}^{\alpha } \psi _2{}_{\beta }\phi _3  \left(D \phi _4\right){}_{\alpha }^{\dot{\alpha }} \phi _5 \left(D \phi _6\right){}^{\beta }_{\dot{\alpha }},\quad \psi _1{}{}^{\alpha } \psi _2{}_{\beta } \phi _3 \phi _4 \left(D \phi _5\right){}_{\alpha }^{\dot{\alpha }} \left(D \phi _6\right){}^{\beta }_{\dot{\alpha }},\\
    
    &\psi _1{}{}^{\alpha } \psi _2{}_{\alpha } \left(D \phi _3\right){}{}^{\beta }_{\dot{\alpha }} \left(D \phi _4\right){}_{\beta }^{\dot{\alpha }}\phi _5 \phi _6 ,\quad \psi _1{}{}^{\alpha } \psi _2{}_{\alpha } \left(D \phi _3\right){}{}^{\beta }_{\dot{\alpha }} \phi _4 \left(D \phi _5\right){}_{\beta }^{\dot{\alpha }}\phi _6 ,\quad \psi _1{}{}^{\alpha } \psi _2{}_{\alpha } \left(D \phi _3\right){}{}^{\beta }_{\dot{\alpha }} \phi _4 \phi _5 \left(D \phi _6\right){}_{\beta }^{\dot{\alpha }},\\
    
    &\psi _1{}{}^{\alpha } \psi _2{}_{\alpha } \phi _3 \left(D \phi _4\right){}{}^{\beta }_{\dot{\alpha }} \left(D \phi _5\right){}_{\beta }^{\dot{\alpha }}\phi _6 ,\quad \psi _1{}{}^{\alpha } \psi _2{}_{\alpha } \phi _3 \left(D \phi _4\right){}{}^{\beta }_{\dot{\alpha }} \phi _5 \left(D \phi _6\right){}_{\beta }^{\dot{\alpha }},\quad \psi _1{}{}^{\alpha } \psi _2{}_{\alpha }  \phi _3 \phi _4 \left(D \phi _5\right){}{}^{\beta }_{\dot{\alpha }} \left(D \phi _6\right){}_{\beta }^{\dot{\alpha }}.
    
\end{align}
The following two types are allowed in this class and the operators are listed below
\begin{align}\begin{array}{c|l}

    \multirow{6}*{$\mathcal{O}_{  l{}^2 H{}^3 H^{\dagger}  D^2}^{(1\sim 11)} $}
    
    &\mathcal{Y}\left[\tiny{\young(p,r)}\right]\epsilon ^{im} \epsilon ^{jn} \left(l_{pi} C \sigma ^{\mu }{}^{\nu } l_{rj}\right) H_{n} D_{\nu } H_{m} \left(H^{\dagger} D_{\mu } H\right),
    \quad \mathcal{Y}\left[\tiny{\young(p,r)}\right]\epsilon ^{im} \epsilon ^{jn} \left(l_{pi} C l_{rj}\right) H_{n} D^{\mu } H_{m} \left(H^{\dagger} D_{\mu } H\right),
    \\&\mathcal{Y}\left[\tiny{\young(p,r)}\right]\epsilon ^{ik} \epsilon ^{jn} \left(l_{pi} C \sigma ^{\mu }{}^{\nu } l_{rj}\right) H_{n} D_{\mu } H_{k} \left(D_{\nu } H^{\dagger}H\right),
    \quad \mathcal{Y}\left[\tiny{\young(p,r)}\right]\epsilon ^{ik} \epsilon ^{jn} \left(l_{pi} C l_{rj}\right) H_{n} D_{\mu } H_{k} \left(D^{\mu } H^{\dagger}H\right),
    \\&\mathcal{Y}\left[\tiny{\young(p,r)}\right]\epsilon ^{im} \epsilon ^{jn} \left(l_{pi} C \sigma ^{\mu }{}^{\nu } l_{rj}\right) H_{m} H_{n} \left(D_{\nu } H^{\dagger} D_{\mu } H\right),
    \quad \mathcal{Y}\left[\tiny{\young(pr)}\right]\epsilon ^{ik} \epsilon ^{jm} \left(l_{pi} C l_{rj}\right) D_{\mu } H_{k} D^{\mu } H_{m} \left(H^{\dagger}H\right),
    \\&\mathcal{Y}\left[\tiny{\young(pr)}\right]\epsilon ^{im} \epsilon ^{jn} \left(l_{pi} C \sigma ^{\mu }{}^{\nu } l_{rj}\right) H_{n} D_{\nu } H_{m} \left(H^{\dagger} D_{\mu } H\right),
    \quad \mathcal{Y}\left[\tiny{\young(pr)}\right]\epsilon ^{im} \epsilon ^{jn} \left(l_{pi} C l_{rj}\right) H_{n} D^{\mu } H_{m} \left(H^{\dagger} D_{\mu } H\right),
    \\&\mathcal{Y}\left[\tiny{\young(pr)}\right]\epsilon ^{ik} \epsilon ^{jn} \left(l_{pi} C \sigma ^{\mu }{}^{\nu } l_{rj}\right) H_{n} D_{\mu } H_{k} \left(D_{\nu } H^{\dagger}H\right),
    \quad \mathcal{Y}\left[\tiny{\young(pr)}\right]\epsilon ^{im} \epsilon ^{jn} \left(l_{pi} C l_{rj}\right) H_{m} H_{n} D^{\mu } \left(H^{\dagger} D_{\mu } H\right),
    \\&\mathcal{Y}\left[\tiny{\young(pr)}\right]\epsilon ^{ik} \epsilon ^{jn} \left(l_{pi} C l_{rj}\right) H_{n} D_{\mu } H_{k} \left(D^{\mu } H^{\dagger}H\right)
    
\vspace{2ex}\\
    
    \mathcal{O}_{  e^2 H^4 D^2} 
    
    &\mathcal{Y}\left[\tiny{\young(pr)}\right]\epsilon ^{ik} \epsilon ^{jm} \left(e_{p} C e_{r}\right) H_{k} H_{m} D_{\mu } H_{i} D^{\mu } H_{j}
    
\end{array}\label{cl:p2h4D2}\end{align}
The superscripts of the $\mc{O}$'s label the terms in particular type, in the order from left to right and from top to bottom.

\noindent\underline{Class $\psi^2\phi^3D^3$}: With three derivatives, we have 10 independent Lorentz structures as follows
\begin{align}\begin{array}{lll}

    \psi _1{}^{\alpha } \left(D \phi _2\right){}^{\beta }_{\dot{\alpha }} \left(D \phi _3\right){}_{\alpha }{}_{\dot{\beta }} \left(D \phi _4\right){}_{\beta }^{\dot{\alpha }}\psi^{\dagger}_5{}^{\dot{\beta }} ,& \psi _1{}^{\alpha } \left(D \phi _2\right){}^{\beta }_{\dot{\alpha }} \left(D \phi _3\right){}_{\alpha }{}_{\dot{\beta }} \phi _4 \left(D \psi^{\dagger}_5\right){}_{\beta }^{\dot{\alpha }\dot{\beta }},& \psi _1{}^{\alpha } \left(D \phi _2\right){}^{\beta }_{\dot{\alpha }} \phi _3 \left(D^2 \phi _4\right){}_{\dot{\beta }\alpha \beta }^{\dot{\alpha }}\psi^{\dagger}_5{}{}^{\dot{\beta }}, \\
    \psi _1{}^{\alpha } \left(D \phi _2\right){}^{\beta }_{\dot{\alpha }} \phi _3 \left(D \phi _4\right){}_{\alpha }{}_{\dot{\beta }} \left(D \psi^{\dagger}_5\right){}_{\beta }^{\dot{\alpha }\dot{\beta }},& \psi _1{}^{\alpha } \phi _2 \left(D^2 \phi _3\right){}_{\alpha\dot{\alpha }\dot{\beta }}^{\beta }\left(D \phi _4\right){}_{\beta }^{\dot{\alpha }} \psi^{\dagger}_5{}{}^{\dot{\beta }} ,& \psi _1{}^{\alpha } \phi _2 \left(D^2 \phi _3\right){}_{\alpha\dot{\alpha }\dot{\beta } }^{\beta } \phi _4 \left(D \psi^{\dagger}_5\right){}_{\beta }^{\dot{\alpha }\dot{\beta }}, \\
    \psi _1{}^{\alpha } \phi _2 \left(D \phi _3\right){}^{\beta }_{\dot{\alpha }} \left(D^2 \phi _4\right){}_{\dot{\beta }\alpha \beta }^{\dot{\alpha }}\psi^{\dagger}_5{}^{\dot{\beta }} ,& \psi _1{}^{\alpha } \phi _2 \left(D \phi _3\right){}^{\beta }_{\dot{\alpha }} \left(D \phi _4\right){}_{\alpha }{}_{\dot{\beta }} \left(D \psi^{\dagger}_5\right){}_{\beta }^{\dot{\alpha }\dot{\beta }},& \psi _1{}^{\alpha } \phi _2 \left(D \phi _3\right){}_{\alpha }{}_{\dot{\alpha }} \left(D \phi _4\right){}{}^{\beta }_{\dot{\beta }} \left(D \psi^{\dagger}_5\right){}_{\beta }^{\dot{\alpha }\dot{\beta }},\\
    \psi _1{}^{\alpha } \phi _2 \phi _3 \left(D^2 \phi _4\right){}_{\alpha\dot{\alpha }\dot{\beta } }^{\beta } \left(D \psi^{\dagger}_5\right){}_{\beta }^{\dot{\alpha }\dot{\beta }} .&&
    
\end{array}\end{align}
The Lorentz structures are also new here. There is only one possible type for these Lorentz structures: 
\begin{align}\begin{array}{c|l}

    \multirow{2}*{$\mathcal{O}_{  e  l H{}^3 D^3}^{(1\sim 3)} $}
     
    & \epsilon ^{ij} \epsilon ^{km} \left(l_{pi} C \gamma ^{\nu } D^{\mu } e_{r}\right) H_{m} D_{\mu } H_{j} D_{\nu } H_{k},
    \quad \epsilon ^{ij} \epsilon ^{km} \left(l_{pi} C \gamma ^{\nu } e_{r}\right) D_{\mu } H_{j} D_{\nu } H_{k} D^{\mu } H_{m},    
    \\& \epsilon ^{ik} \epsilon ^{jm} \left(l_{pi} C \gamma ^{\nu } D^{\mu } e_{r}\right) H_{m} D_{\mu } H_{j} D_{\nu } H_{k}
 
 \end{array}\label{cl:p2h3D3}\end{align}

\noindent\underline{Class $\psi^2\phi^2D^4$}: With four derivatives, we have 3 independent Lorentz structures as follows
\begin{align}

    &\psi _1{}{}^{\alpha } \left(D \psi _2\right){}{}^{\beta \gamma }_{\dot{\alpha }} \left(D \phi _3\right){}_{\alpha \dot{\beta }} \left(D^2 \phi _4\right){}{}_{\beta \gamma }^{\dot{\alpha }\dot{\beta }},\quad \psi _1{}{}^{\alpha } \psi _2{}{}^{\beta } \left(D^2 \phi _3\right){}{}_{\alpha\dot{\alpha }\dot{\beta } }^{\gamma } \left(D^2 \phi _4\right){}{}_{\beta \gamma }^{\dot{\alpha }\dot{\beta }},\quad \psi _1{}{}^{\alpha } \psi _2{}_{\alpha } \left(D^2 \phi _3\right){}{}_{\dot{\alpha }\dot{\beta }}^{\beta \gamma } \left(D^2 \phi _4\right){}{}_{\beta \gamma }^{\dot{\alpha }\dot{\beta }}\;.
    
\end{align}
Still there is only one possible type: 
\begin{align}\begin{array}{c|l}

    \multirow{2}*{$\mathcal{O}_{  l{}^2 H{}^2 D^4}^{(1\sim 3)} $}
    
    &\mathcal{Y}\left[\tiny{\young(pr)}\right]\epsilon ^{ik} \epsilon ^{jm} \left(l_{pi} C\sigma ^{\nu }{}^{\lambda } D_{\mu } l_{rj}\right) D_{\nu } H_{k} D_{\lambda } D^{\mu } H_{m},
    \quad \mathcal{Y}\left[\tiny{\young(pr)}\right]\epsilon ^{ik} \epsilon ^{jm} \left(l_{pi} CD_{\mu } l_{rj}\right) D_{\nu } H_{k} D^{\mu } D^{\nu } H_{m},
    \\&\mathcal{Y}\left[\tiny{\young(pr)}\right]\epsilon ^{ij} \epsilon ^{km} \left(l_{pi} C\sigma ^{\nu }{}^{\lambda } D_{\mu } l_{rj}\right) D_{\nu } H_{k} D_{\lambda } D^{\mu } H_{m}
    
\end{array}\label{cl:p2h2D4}\end{align}

\subsubsection{One gauge boson involved}

\noindent\underline{Class $F\psi^2\phi^4$}: The class has to be $F_{\rm L}\psi\psi\phi^3$, which has only one Lorentz structure 
\begin{align}
    &F_{\rm{L}1}{}^{\alpha\beta}\psi_2{}_{\alpha}\psi _3{}_{\beta } \phi _4 \phi _5 \phi _6 \phi _7 
\end{align}
There are two possible types, the anti-symmetric flavor representations of which contribute to neutrino anomalous magnetic moment:
\begin{align}\begin{array}{c|l}

    \multirow{3}*{$\mathcal{O}_{W l{}^2 H{}^3 H^{\dagger}   }^{(1\sim 3)} $}
    
    &\mathcal{Y}\left[\tiny{\young(p,r)}\right] \epsilon ^{im} \epsilon ^{kl} \left(\tau ^I\right)_l^n W^{I}_{\mu \nu } \left(l_{pi} C \sigma ^{\mu }{}^{\nu } l_{rj}\right) H_{k} H_{m} H_{n} H^{\dagger}{}^{j},
    \\&\mathcal{Y}\left[\tiny{\young(p,r)}\right] \epsilon ^{im} \epsilon ^{jl} \left(\tau ^I\right)_l^n W^{I}_{\mu \nu } \left(l_{pi} C \sigma ^{\mu }{}^{\nu } l_{rj}\right) H_{m} H_{n} \left(H^{\dagger}H\right),
    \\&\mathcal{Y}\left[\tiny{\young(pr)}\right] \epsilon ^{im} \epsilon ^{kl} \left(\tau ^I\right)_l^n W^{I}_{\mu \nu } \left(l_{pi} C \sigma ^{\mu }{}^{\nu } l_{rj}\right) H_{k} H_{m} H_{n} H^{\dagger}{}^{j}
    
\vspace{2ex}\\
    
    \mathcal{O}_{B l{}^2 H{}^3 H^{\dagger}   } 
    
    &\mathcal{Y}\left[\tiny{\young(p,r)}\right]\epsilon ^{im} \epsilon ^{jn} B_{\mu }{}_{\nu } \left(l_{pi} C \sigma ^{\mu }{}^{\nu } l_{rj}\right) H_{m} H_{n} \left(H^{\dagger}H\right)
    
\end{array}\label{cl:fp2h4}\end{align}
The fermion bilinear terms here are always chirality-violating.

\noindent\underline{Class $F\psi^2\phi^3D$}: In this class, gauge boson contracts with the fermion current and the Higgs current. There are  3 independent Lorentz structures as follows
\begin{align}

    &F_{\rm{L}1}{}{}^{\alpha }{}{}^{\beta } \psi _2{}_{\alpha } \left(D \phi _3\right){}_{\beta }{}_{\dot{\alpha }} \phi _4 \phi _5 \psi^{\dagger}_6{}{}^{\dot{\alpha }} ,\quad F_{\rm{L}1}{}{}^{\alpha }{}{}^{\beta } \psi _2{}_{\alpha } \phi _3 \left(D \phi _4\right){}_{\beta }{}_{\dot{\alpha }} \phi _5 \psi^{\dagger}_6{}{}^{\dot{\alpha }} ,\quad F_{\rm{L}1}{}{}^{\alpha }{}{}^{\beta } \psi _2{}_{\alpha } \phi _3 \phi _4 \left(D \phi _5\right){}_{\beta }{}_{\dot{\alpha }} \psi^{\dagger}_6{}{}^{\dot{\alpha }} .
    
\end{align}
Two types are written as follows:
\begin{align}\begin{array}{c|l}

    \multirow{2}*{$\mathcal{O}_{W e  l H{}^3 D}^{(1\sim 4)} $}
    
    & \epsilon ^{ik} \epsilon ^{jn} \left(\tau ^I\right)_n^m W^{I}_{\mu \nu } \left(l_{pi} C \gamma ^{\nu } e_{r}\right) H_{k} H_{m} D^{\mu } H_{j},
    \quad \epsilon ^{ik} \epsilon ^{jn} \left(\tau ^I\right)_n^m \tilde{W}^{I}_{\mu \nu } \left(l_{pi} C \gamma ^{\nu } e_{r}\right) H_{k} H_{m} D^{\mu } H_{j},
    \\& \epsilon ^{in} \epsilon ^{jk} \left(\tau ^I\right)_n^m W^{I}_{\mu \nu } \left(l_{pi} C \gamma ^{\nu } e_{r}\right) H_{k} H_{m} D^{\mu } H_{j},
    \quad \epsilon ^{in} \epsilon ^{jk} \left(\tau ^I\right)_n^m \tilde{W}^{I}_{\mu \nu } \left(l_{pi} C \gamma ^{\nu } e_{r}\right) H_{k} H_{m} D^{\mu } H_{j}
\vspace{2ex}\\
    
    \mathcal{O}_{B e  l H{}^3 D} ^{(1,2)}
    
    & \epsilon ^{ik} \epsilon ^{jm}  B{}^{\mu }{}_{\nu } \left(l_{pi} C \gamma ^{\nu } e_{r}\right) H_{k} H_{m}D^{\mu } H_{j},
    \quad \epsilon ^{ik} \epsilon ^{jm}  \tilde{B}{}^{\mu }{}_{\nu } \left(l_{pi} C \gamma ^{\nu } e_{r}\right) H_{k} H_{m}D^{\mu } H_{j}
\end{array}\label{cl:fp2h3D}\end{align}

\noindent\underline{Class $F\psi^2\phi^2D^2$}: There  are  2  classes  of  this  form.   One  is $F_{\rm L}\psi^2\phi^2D^2$,  a  dimension  7  class
$F_{\rm L}\psi^2\phi^2$
with  two additional derivatives, which has 7 independent Lorentz structures:
\begin{align}\begin{array}{lll}

    F_{\rm{L}1}{}{}^{\alpha }{}{}^{\beta } \psi _2{}{}^{\gamma } \left(D \psi _3\right){}_{\alpha }{}_{\beta }{}_{\dot{\alpha }} \left(D \phi _4\right){}_{\gamma }^{\dot{\alpha }}\phi _5 ,& F_{\rm{L}1}{}{}^{\alpha }{}{}^{\beta } \psi _2{}{}^{\gamma } \left(D \psi _3\right){}_{\alpha }{}_{\beta }{}_{\dot{\alpha }} \phi _4 \left(D \phi _5\right){}_{\gamma }^{\dot{\alpha }},& F_{\rm{L}1}{}{}^{\alpha }{}{}^{\beta } \psi _2{}{}^{\gamma } \psi _3{}_{\alpha } \left(D \phi _4\right){}_{\beta }{}_{\dot{\alpha }} \left(D \phi _5\right){}_{\gamma }^{\dot{\alpha }},\\
    F_{\rm{L}1}{}{}^{\alpha }{}{}^{\beta } \psi _2{}_{\alpha } \left(D \psi _3\right){}_{\beta }^{\gamma }{}{}_{\dot{\alpha }} \left(D \phi _4\right){}_{\gamma }^{\dot{\alpha }} \phi _5,& F_{\rm{L}1}{}{}^{\alpha }{}{}^{\beta } \psi _2{}_{\alpha } \left(D \psi _3\right){}_{\beta }^{\gamma }{}{}_{\dot{\alpha }}\phi _4 \left(D \phi _5\right){}_{\gamma }^{\dot{\alpha }} ,& F_{\rm{L}1}{}{}^{\alpha }{}{}^{\beta } \psi _2{}_{\alpha } \psi _3{}{}^{\gamma } \left(D \phi _4\right){}_{\beta }{}_{\dot{\alpha }} \left(D \phi _5\right){}_{\gamma }^{\dot{\alpha }}\\
    F_{\rm{L}1}{}{}^{\alpha }{}{}^{\beta } \psi _2{}_{\alpha } \psi _3{}_{\beta } \left(D \phi _4\right){}{}^{\gamma }_{\dot{\alpha }} \left(D \phi _5\right){}_{\gamma }^{\dot{\alpha }}.&
    
\end{array}\end{align}
The other class is $F_{\rm R}\psi^2\phi^2D^2$, where the flip of helicity for the gauge boson is made possible by the presence of the
two additional derivatives. The Lorentz structures of this class are
\begin{align}
\begin{array}{ll}
    \psi _1{}{}^{\alpha } \psi _2{}{}^{\beta } \left(D^2 \phi _3\right)_{\alpha\beta\dot{\alpha }\dot{\beta }}\phi _4 F_{\rm{R}5}{}{}^{\dot{\alpha }}{}{}^{\dot{\beta }} ,
    &\psi _1{}{}^{\alpha } \psi _2{}{}^{\beta } \left(D \phi _3\right){}_{\alpha }{}_{\dot{\alpha }} \left(D \phi _4\right){}_{\beta }{}_{\dot{\beta }} F_{\rm{R}5}{}{}^{\dot{\alpha }}{}{}^{\dot{\beta }}\\
    
    \psi _1{}{}^{\alpha } \psi _2{}{}^{\beta } \phi _3 \left(D^2 \phi _4\right)_{\alpha\beta\dot{\alpha }\dot{\beta } }F_{\rm{R}5}{}{}^{\dot{\alpha }}{}{}^{\dot{\beta }} ,
    &\psi _1{}{}^{\alpha } \psi _2{}_{\alpha } \left(D \phi _3\right)^{\beta }_{\dot{\alpha }} \left(D \phi _4\right){}_{\beta }{}_{\dot{\beta }} F_{\rm{R}5}{}{}^{\dot{\alpha }}{}{}^{\dot{\beta }} .
\end{array}   
\end{align}
After converting to the $F,\tilde{F}$ basis, these two classes mix together.  
\begin{align}\begin{array}{c|l}

    \multirow{16}*{$\mathcal{O}_{W l{}^2 H{}^2 D^2}^{(1\sim 16)} $}
    
    &\mathcal{Y}\left[\tiny{\young(p,r)}\right] \epsilon ^{jn} \epsilon ^{km} \left(\tau ^I\right)_n^i W^{I}{}_{\lambda }{}^{\mu } \left(l_{pi} C\sigma ^{\lambda }{}^{\nu } D_{\mu } l_{rj}\right) H_{m} D_{\nu } H_{k},
    \\& \mathcal{Y}\left[\tiny{\young(p,r)}\right] \epsilon ^{jn} \epsilon ^{km} \left(\tau ^I\right)_n^i \tilde{W}^{I}{}_{\lambda }{}^{\mu } \left(l_{pi} C\sigma ^{\lambda }{}^{\nu } D_{\mu } l_{rj}\right) H_{m} D_{\nu } H_{k},
    \\&\mathcal{Y}\left[\tiny{\young(p,r)}\right] \epsilon ^{ik} \epsilon ^{jn} \left(\tau ^I\right)_n^m W^{I}{}_{\lambda }{}^{\mu } \left(l_{pi} C\sigma ^{\lambda }{}^{\nu } D_{\mu } l_{rj}\right) H_{m} D_{\nu } H_{k}, 
    \\& \mathcal{Y}\left[\tiny{\young(p,r)}\right] \epsilon ^{ik} \epsilon ^{jn} \left(\tau ^I\right)_n^m \tilde{W}^{I}{}_{\lambda }{}^{\mu } \left(l_{pi} C\sigma ^{\lambda }{}^{\nu } D_{\mu } l_{rj}\right) H_{m} D_{\nu } H_{k},
    \\&\mathcal{Y}\left[\tiny{\young(p,r)}\right] \epsilon ^{in} \epsilon ^{jk} \left(\tau ^I\right)_n^m W^{I}{}_{\lambda }{}^{\mu } \left(l_{pi} C\sigma ^{\lambda }{}^{\nu } D_{\mu } l_{rj}\right) H_{m} D_{\nu } H_{k}, 
    \\& \mathcal{Y}\left[\tiny{\young(p,r)}\right] \epsilon ^{in} \epsilon ^{jk} \left(\tau ^I\right)_n^m \tilde{W}^{I}{}_{\lambda }{}^{\mu } \left(l_{pi} C\sigma ^{\lambda }{}^{\nu } D_{\mu } l_{rj}\right) H_{m} D_{\nu } H_{k},
    \\&\mathcal{Y}\left[\tiny{\young(p,r)}\right] \epsilon ^{jn} \epsilon ^{km} \left(\tau ^I\right)_n^i W^{I}_{\nu \lambda } \left(l_{pi} C\sigma ^{\nu }{}^{\lambda } D_{\mu } l_{rj}\right) H_{m} D^{\mu } H_{k},
    \\& \mathcal{Y}\left[\tiny{\young(p,r)}\right] \epsilon ^{ik} \epsilon ^{jn} \left(\tau ^I\right)_n^m W^{I}_{\nu \lambda } \left(l_{pi} C\sigma ^{\nu }{}^{\lambda } D_{\mu } l_{rj}\right) H_{m} D^{\mu } H_{k},
    \\&\mathcal{Y}\left[\tiny{\young(pr)}\right] \epsilon ^{jn} \epsilon ^{km} \left(\tau ^I\right)_n^i W^{I}{}_{\lambda }{}^{\mu } \left(l_{pi} C\sigma ^{\lambda }{}^{\nu } D_{\mu } l_{rj}\right) H_{m} D_{\nu } H_{k},
    \\& \mathcal{Y}\left[\tiny{\young(pr)}\right] \epsilon ^{jn} \epsilon ^{km} \left(\tau ^I\right)_n^i \tilde{W}^{I}{}_{\lambda }{}^{\mu } \left(l_{pi} C\sigma ^{\lambda }{}^{\nu } D_{\mu } l_{rj}\right) H_{m} D_{\nu } H_{k},
    \\&\mathcal{Y}\left[\tiny{\young(pr)}\right] \epsilon ^{ik} \epsilon ^{jn} \left(\tau ^I\right)_n^m W^{I}{}_{\lambda }{}^{\mu } \left(l_{pi} C\sigma ^{\lambda }{}^{\nu } D_{\mu } l_{rj}\right) H_{m} D_{\nu } H_{k},
    \\& \mathcal{Y}\left[\tiny{\young(pr)}\right] \epsilon ^{ik} \epsilon ^{jn} \left(\tau ^I\right)_n^m \tilde{W}^{I}{}_{\lambda }{}^{\mu } \left(l_{pi} C\sigma ^{\lambda }{}^{\nu } D_{\mu } l_{rj}\right) H_{m} D_{\nu } H_{k}
    \\&\mathcal{Y}\left[\tiny{\young(pr)}\right] \epsilon ^{in} \epsilon ^{jk} \left(\tau ^I\right)_n^m W^{I}{}_{\lambda }{}^{\mu } \left(l_{pi} C\sigma ^{\lambda }{}^{\nu } D_{\mu } l_{rj}\right) H_{m} D_{\nu } H_{k},
    \\& \mathcal{Y}\left[\tiny{\young(pr)}\right] \epsilon ^{in} \epsilon ^{jk} \left(\tau ^I\right)_n^m \tilde{W}^{I}{}_{\lambda }{}^{\mu } \left(l_{pi} C\sigma ^{\lambda }{}^{\nu } D_{\mu } l_{rj}\right) H_{m} D_{\nu } H_{k},
    \\&\mathcal{Y}\left[\tiny{\young(pr)}\right] \epsilon ^{jn} \epsilon ^{km} \left(\tau ^I\right)_n^i W^{I}_{\nu \lambda } \left(l_{pi} C\sigma ^{\nu }{}^{\lambda } D_{\mu } l_{rj}\right) H_{m} D^{\mu } H_{k},
    \\& \mathcal{Y}\left[\tiny{\young(pr)}\right] \epsilon ^{ik} \epsilon ^{jn} \left(\tau ^I\right)_n^m W^{I}_{\nu \lambda } \left(l_{pi} C\sigma ^{\nu }{}^{\lambda } D_{\mu } l_{rj}\right) H_{m} D^{\mu } H_{k}
    
\end{array}\label{cl:fp2h2D21}\end{align}
\begin{align}\begin{array}{c|l}
    
    \multirow{6}*{$\mathcal{O}_{B l{}^2 H{}^2 D^2}^{(1\sim 11)} $}
    
    &\mathcal{Y}\left[\tiny{\young(p,r)}\right]\epsilon ^{ik} \epsilon ^{jm} B{}_{\lambda }{}{}^{\mu } H_{m} D_{\nu } H_{k} \left(l_{pi} C\sigma ^{\lambda }{}^{\nu } D_{\mu } l_{rj}\right),
    \quad \mathcal{Y}\left[\tiny{\young(p,r)}\right]\epsilon ^{ik} \epsilon ^{jm} \tilde{B}{}_{\lambda }{}{}^{\mu } H_{m} D_{\nu } H_{k} \left(l_{pi} C\sigma ^{\lambda }{}^{\nu } D_{\mu } l_{rj}\right),
    \\&\mathcal{Y}\left[\tiny{\young(p,r)}\right]\epsilon ^{ij} \epsilon ^{km} B{}_{\lambda }{}{}^{\mu } H_{m} D_{\nu } H_{k} \left(l_{pi} C\sigma ^{\lambda }{}^{\nu } D_{\mu } l_{rj}\right),
    \quad \mathcal{Y}\left[\tiny{\young(p,r)}\right]\epsilon ^{ij} \epsilon ^{km} \tilde{B}{}_{\lambda }{}{}^{\mu } H_{m} D_{\nu } H_{k} \left(l_{pi} C\sigma ^{\lambda }{}^{\nu } D_{\mu } l_{rj}\right),
    \\&\mathcal{Y}\left[\tiny{\young(p,r)}\right]\epsilon ^{ik} \epsilon ^{jm} B{}_{\nu }{}_{\lambda } H_{m} D^{\mu } H_{k} \left(l_{pi} C\sigma ^{\nu }{}^{\lambda } D_{\mu } l_{rj}\right),
    \quad \mathcal{Y}\left[\tiny{\young(p,r)}\right]\epsilon ^{ij} \epsilon ^{km} B{}_{\nu }{}_{\lambda } H_{m} D^{\mu } H_{k} \left(l_{pi} C\sigma ^{\nu }{}^{\lambda } D_{\mu } l_{rj}\right),
    \\&\mathcal{Y}\left[\tiny{\young(pr)}\right]\epsilon ^{ik} \epsilon ^{jm} B{}_{\lambda }{}{}^{\mu } H_{m} D_{\nu } H_{k} \left(l_{pi} C\sigma ^{\lambda }{}^{\nu } D_{\mu } l_{rj}\right),
    \quad \mathcal{Y}\left[\tiny{\young(pr)}\right]\epsilon ^{ik} \epsilon ^{jm} \tilde{B}{}_{\lambda }{}{}^{\mu } H_{m} D_{\nu } H_{k} \left(l_{pi} C\sigma ^{\lambda }{}^{\nu } D_{\mu } l_{rj}\right),
    \\&\mathcal{Y}\left[\tiny{\young(pr)}\right]\epsilon ^{ij} \epsilon ^{km} B{}_{\lambda }{}{}^{\mu } H_{m} D_{\nu } H_{k} \left(l_{pi} C\sigma ^{\lambda }{}^{\nu } D_{\mu } l_{rj}\right),
    \quad \mathcal{Y}\left[\tiny{\young(pr)}\right]\epsilon ^{ij} \epsilon ^{km} \tilde{B}{}_{\lambda }{}{}^{\mu } H_{m} D_{\nu } H_{k} \left(l_{pi} C\sigma ^{\lambda }{}^{\nu } D_{\mu } l_{rj}\right),
    \\&\mathcal{Y}\left[\tiny{\young(pr)}\right]\epsilon ^{ik} \epsilon ^{jm} B{}_{\nu }{}_{\lambda } H_{m} D^{\mu } H_{k} \left(l_{pi} C\sigma ^{\nu }{}^{\lambda } D_{\mu } l_{rj}\right)
    
\end{array}\label{cl:fp2h2D22}\end{align}

\subsubsection{Two gauge boson involved}

\noindent\underline{Cass $F^2\psi^2\phi^2$}: Two classes are involved, with the same and opposite helicities for the gauge bosons and fermions.  For
the class $F_{\rm L}^2\psi^2\phi^2$, we obtained 2 independent Lorentz structures:
\begin{align}
    F_{\rm{L}1}{}{}^{\alpha }{}{}^{\beta } F_{\rm{L}2}{}_{\alpha \gamma }\psi _3{}_{\beta } \psi _4{}^{\gamma }\phi _5 \phi_6 ,\quad F_{\rm{L}1}{}{}^{\alpha }{}{}^{\beta }F_{\rm{L}2}{}_{\alpha }{}_{\beta }\psi _3{}{}^{\gamma }  \psi _4{}_{\gamma } \phi _5\phi_6. \label{LB:ffpp1}
\end{align}
while for $F_{\rm R}^2\psi^2\phi$ we have only 1 independent Lorentz structure
\begin{align}
    \psi _1{}{}^{\alpha }  \psi _2{}_{\alpha } \phi _3\phi_4F_{\rm{R}5}{}{}^{\dot\alpha }{}{}^{\dot\beta }F_{\rm{R}6}{}_{\dot\alpha }{}_{\dot\beta }. \label{LB:ffpp2}
\end{align}
After converting to the $F,\tilde{F}$ basis, the terms with the second Lorentz structure in eq.~(\ref{LB:ffpp1}) and those with the Lorentz structure in eq.~(\ref{LB:ffpp2}) combine to the form as the Weinberg operator with an extra $F^2$ or $F\tilde{F}$. The terms with the first Lorentz structure in eq.~(\ref{LB:ffpp1}) are left as they are.
\begin{align}\begin{array}{c|l}

    \mathcal{O}_{G^2 l^2 H{}^2  }
    
    &\mathcal{Y}\left[\tiny{\young(pr)}\right]\epsilon ^{ik} \epsilon ^{jm} G^{A}_{\mu \nu } G^{A}{}^{\mu }{}^{\nu } \left(l_{pi} C l_{rj}\right) H_{k} H_{m}
    \quad \mathcal{Y}\left[\tiny{\young(pr)}\right]\epsilon ^{ik} \epsilon ^{jm} G^{A}_{\mu \nu } \tilde{G}^{A}{}^{\mu }{}^{\nu } \left(l_{pi} C l_{rj}\right) H_{k} H_{m}

\vspace{2ex}\\
    
    \multirow{5}*{$\mathcal{O}_{W^2 l^2 H{}^2  }^{(1\sim 6)}$}
    &\mathcal{Y}\left[\tiny{\young(pr)}\right]\epsilon ^{ik} \epsilon ^{jm} W^{I}_{\mu \nu } W^{I}{}^{\mu }{}^{\nu } \left(l_{pi} C l_{rj}\right) H_{k} H_{m},
    \quad \mathcal{Y}\left[\tiny{\young(pr)}\right]\epsilon ^{ik} \epsilon ^{jm} W^{I}_{\mu \nu } \tilde{W}^{I}{}^{\mu }{}^{\nu } \left(l_{pi} C l_{rj}\right) H_{k} H_{m},
    \\&\mathcal{Y}\left[\tiny{\young(p,r)}\right]\epsilon ^{jm} \epsilon ^{kn} \epsilon ^{IJK} \left(\tau ^K\right)_n^i W^{I}_{\mu \nu } W^{J}{}^{\mu }{}{}_{\lambda } \left(l_{pi} C \sigma ^{\nu }{}^{\lambda } l_{rj}\right) H_{k} H_{m},
    \\&\mathcal{Y}\left[\tiny{\young(pr)}\right]\epsilon ^{kn} \epsilon ^{mo} \left(\tau ^I\right)_o^i \left(\tau ^J\right)_n^j W^{I}_{\mu \nu } W^{J}{}^{\mu }{}{}_{\lambda } \left(l_{pi} C \sigma ^{\nu }{}^{\lambda } l_{rj}\right) H_{k} H_{m},
    \\&\mathcal{Y}\left[\tiny{\young(pr)}\right]\epsilon ^{kn} \epsilon ^{mo} \left(\tau ^I\right)_o^i \left(\tau ^J\right)_n^j W^{I}_{\mu \nu } W^{J}{}^{\mu }{}^{\nu } \left(l_{pi} C l_{rj}\right) H_{k} H_{m},
    \\&\mathcal{Y}\left[\tiny{\young(pr)}\right]\epsilon ^{kn} \epsilon ^{mo} \left(\tau ^I\right)_o^i \left(\tau ^J\right)_n^j W^{I}_{\mu \nu } \tilde{W}^{J}{}^{\mu }{}^{\nu } \left(l_{pi} C l_{rj}\right) H_{k} H_{m}
    
\vspace{2ex}\\
    
    \multirow{4}*{$\mathcal{O}_{B W l^2 H{}^2  }^{(1\sim 6)}$}
    
    &\mathcal{Y}\left[\tiny{\young(p,r)}\right]\epsilon ^{ik} \epsilon ^{jn} \left(\tau ^I\right)_n^m B{}_{\mu }{}_{\nu } W^{I}{}^{\mu }{}{}_{\lambda } \left(l_{pi} C \sigma ^{\nu }{}^{\lambda } l_{rj}\right) H_{k} H_{m},
    \\&\mathcal{Y}\left[\tiny{\young(p,r)}\right]\epsilon ^{ik} \epsilon ^{jn} \left(\tau ^I\right)_n^m B{}_{\mu }{}_{\nu } W^{I}{}^{\mu }{}^{\nu } \left(l_{pi} C l_{rj}\right) H_{k} H_{m},
    \quad \mathcal{Y}\left[\tiny{\young(p,r)}\right]\epsilon ^{ik} \epsilon ^{jn} \left(\tau ^I\right)_n^m B{}_{\mu }{}_{\nu } \tilde{W}^{I}{}^{\mu }{}^{\nu } \left(l_{pi} C l_{rj}\right) H_{k} H_{m},
    \\&\mathcal{Y}\left[\tiny{\young(pr)}\right]\epsilon ^{ik} \epsilon ^{jn} \left(\tau ^I\right)_n^m B{}_{\mu }{}_{\nu } W^{I}{}^{\mu }{}{}_{\lambda } \left(l_{pi} C \sigma ^{\nu }{}^{\lambda } l_{rj}\right) H_{k} H_{m},
    \\&\mathcal{Y}\left[\tiny{\young(pr)}\right]\epsilon ^{ik} \epsilon ^{jn} \left(\tau ^I\right)_n^m B{}_{\mu }{}_{\nu } W^{I}{}^{\mu }{}^{\nu } \left(l_{pi} C l_{rj}\right) H_{k} H_{m},
    \quad \mathcal{Y}\left[\tiny{\young(pr)}\right]\epsilon ^{ik} \epsilon ^{jn} \left(\tau ^I\right)_n^m B{}_{\mu }{}_{\nu } \tilde{W}^{I}{}^{\mu }{}^{\nu } \left(l_{pi} C l_{rj}\right) H_{k} H_{m}    
\vspace{2ex}\\
    
    \mathcal{O}_{B^2 l^2 H{}^2  }
    
    &\mathcal{Y}\left[\tiny{\young(pr)}\right]\epsilon ^{ik} \epsilon ^{jm} B{}_{\mu }{}_{\nu } B{}^{\mu }{}^{\nu } \left(l_{pi} C l_{rj}\right) H_{k} H_{m},
    \quad \mathcal{Y}\left[\tiny{\young(pr)}\right]\epsilon ^{ik} \epsilon ^{jm} B{}_{\mu }{}_{\nu } \tilde{B}{}^{\mu }{}^{\nu } \left(l_{pi} C l_{rj}\right) H_{k} H_{m}

\end{array}\label{cl:f2p2h2}\end{align}

\subsection{Classes involving Four-fermions}
\label{sec:fourfermion}
In this sub-section, quarks begin to appear in operators, and $|B-L|$ is always equal to 2, such that only $(\Delta B,\Delta L)=(\pm 1,\mp 1)$ and $(\Delta B,\Delta L)=(0,2)$ are allowed. The classes involve three quarks and one lepton, or two quarks and two leptons, or four leptons. The operators with $(\Delta B,\Delta L)=(\pm 1,\mp 1)$ usually contribute to the proton two-body decay processes, while the $\Delta L=2$ operators could give rise to contribution to the neutrino-less double beta decay processes, such as the operator type $W u\bar{d}l^2D$ at tree-level.  
We're going to present the operators in terms of the number of quarks. Operators with $\Delta B=-1$ or $\Delta L=-2$ are taken conjugate to make them look a bit neater. 


\subsubsection{No guage boson involved}
\noindent\underline{Class $ \psi^4 \phi^3 $}: There are two classes in this form: $\psi^2\psi^{\dagger 2}\phi^3$ and $\psi^4\phi^3$, and the independent Lorentz structures are
\bea
&\psi _1^{\alpha } \psi _2{}_{\alpha } \phi _3 \phi _4 \phi _5 \psi^{\dagger}_6{}_{\dot{\alpha }} \psi^{\dagger}_7{}{}^{\dot{\alpha }},\\
&\psi _1^{\alpha } \psi _2^{\beta } \psi _3{}_{\alpha } \psi _4{}_{\beta } \phi _5 \phi _6 \phi _7,\quad
&\psi _1^{\alpha } \psi _2{}_{\alpha } \psi _3^{\beta } \psi _4{}_{\beta } \phi _5 \phi _6 \phi _7.
\eea
Operators of this class contribute to the four-fermion interactions if the Higgs fields take their vev, and operators involving two or three $l$'s are relevant to the neutrino non-standard interactions. 

\noindent 1. Operators involving three quarks with $\Delta B=1$ and $\Delta L=-1$:
\begin{align}\begin{array}{c|l}

\mathcal{O}_{ q{}^3 \overline{e}  H^{\dagger} {}^3  }

&\mathcal{Y}\left[\tiny{\young(rs,t)}\right]\epsilon ^{abc} \left(\overline{e}_{p} q_{sb j}\right) \left(q_{ra i} C q_{tc k}\right) H^{\dagger}{}^{i} H^{\dagger}{}^{j} H^{\dagger}{}^{k}

\vspace{2ex}\\

\mathcal{O}_{ q{}^2 u \overline{l}  H^{\dagger} {}^3  }

&\mathcal{Y}\left[\tiny{\young(p,r)}\right]\epsilon ^{abc} \epsilon _{no} \left(\overline{l}_{s}^{o} u_{tc}\right) \left(q_{pa i} C q_{rb j}\right) H^{\dagger}{}^{i} H^{\dagger}{}^{j} H^{\dagger}{}^{n}

\vspace{2ex}\\

\multirow{2}*{$\mathcal{O}_{ q{}^2 d  \overline{l}  H H^{\dagger} {}^2  }^{(1 \sim 3)}$}

&\mathcal{Y}\left[\tiny{\young(p,r)}\right]\epsilon ^{abc} \left(\overline{l}_{t}^{k} d_{sc}\right) \left(q_{pa i} C q_{rb j}\right) H_{k} H^{\dagger}{}^{i} H^{\dagger}{}^{j}

,\quad\mathcal{Y}\left[\tiny{\young(p,r)}\right]\epsilon ^{abc} \left(\overline{l}_{t}^{i} d_{sc}\right) \left(q_{pa i} C q_{rb j}\right) H_{k} H^{\dagger}{}^{j} H^{\dagger}{}^{k},

\\&\mathcal{Y}\left[\tiny{\young(pr)}\right]\epsilon ^{abc} \left(\overline{l}_{t}^{i} d_{sc}\right) \left(q_{pa i} C q_{rb j}\right) H_{k} H^{\dagger}{}^{j} H^{\dagger}{}^{k}

\vspace{2ex}\\

\mathcal{O}_{ q d {}^2 \overline{e}  H H^{\dagger} {}^2  }

&\mathcal{Y}\left[\tiny{\young(s,t)}\right]\epsilon ^{abc} \left(\overline{e}_{p} q_{ra i}\right) \left(d_{sb} C d_{tc}\right) H_{j} H^{\dagger}{}^{i} H^{\dagger}{}^{j}

\vspace{2ex}\\

\multirow{1}*{$\mathcal{O}_{ u d{}^2 \overline{l}  H H^{\dagger}{}^2   }^{(1 \sim 2)}$}

&\mathcal{Y}\left[\tiny{\young(pr)}\right]\epsilon ^{abc} \epsilon _{ik}\left(\overline{l}_{s}^{i} d_{pa} \right) \left(u_{tc} C d_{rb}\right) H_{j} H^{\dagger}{}^{j} H^{\dagger}{}^{k} 

,\quad\mathcal{Y}\left[\tiny{\young(p,r)}\right]\epsilon ^{abc} \epsilon _{ik}\left(\overline{l}_{s}^{i} d_{pa} \right) \left(u_{tc} C d_{rb}\right) H_{j} H^{\dagger}{}^{j} H^{\dagger}{}^{k} 

\vspace{2ex}\\

\mathcal{O}_{  d{}^3 \overline{l} H {}^2 H^{\dagger}  }

&\mathcal{Y}\left[\tiny{\young(pr,s)}\right]\epsilon ^{abc} \left(\overline{l}_{t}^{i} d_{rb} \right) \left(d_{sc} C d_{pa}\right) H_{i} H_{j} H^{\dagger}{}^{j}

\end{array}\label{cl:q3lh3}\end{align}

\noindent 2. Operators involving two leptons and two quarks with $\Delta L=2$:
\begin{align}\begin{array}{c|l}

\mathcal{O}_{ q \overline{q} l e H{}^3  } 

& \epsilon ^{im} \epsilon ^{jn} \left(\overline{q}_{t}^{a k} e_{s}\right) \left(l_{pi} C q_{ra j}\right) H_{k} H_{m} H_{n}

\vspace{2ex}\\

\multirow{1}*{$\mathcal{O}_{ q \overline{u} l{}^2 H{}^3  }^{(1,2)} $}

&\mathcal{Y}\left[\tiny{\young(p,r)}\right]\epsilon ^{im} \epsilon ^{jn} \epsilon ^{ko} \left(\overline{u}_{t}^{a} l_{rj}\right) \left(l_{pi} C q_{sa k}\right) H_{m} H_{n} H_{o}

,\quad\mathcal{Y}\left[\tiny{\young(pr)}\right]\epsilon ^{im} \epsilon ^{jn} \epsilon ^{ko} \left(\overline{u}_{t}^{a} l_{rj}\right) \left(l_{pi} C q_{sa k}\right) H_{m} H_{n} H_{o}

\vspace{2ex}\\

\multirow{2}*{$\mathcal{O}_{ \overline{q} u l {}^2  H {}^2 H^{\dagger}  }^{(1 \sim 3)} $}

&\mathcal{Y}\left[\tiny{\young(st)}\right]\epsilon ^{ko} \left(\overline{q}_{p}^{a i} u_{ra} \right) \left(l_{to} C l_{sj}\right) H_{i} H_{k} H^{\dagger}{}^{j}

,\quad\mathcal{Y}\left[\tiny{\young(st)}\right]\epsilon ^{mo} \left(\overline{q}_{p}^{a i} u_{ra} \right) \left(l_{to} C l_{si}\right) H_{j} H_{m} H^{\dagger}{}^{j},

\\&\mathcal{Y}\left[\tiny{\young(s,t)}\right]\epsilon ^{ko} \left(\overline{q}_{p}^{a i} u_{ra} \right) \left(l_{to} C l_{sj}\right) H_{i} H_{k} H^{\dagger}{}^{j}

\vspace{2ex}\\

\multirow{3}*{$\mathcal{O}_{ q \overline{d} l{}^2  H{}^2 H^{\dagger}   }^{(1 \sim 6)} $}

&\mathcal{Y}\left[\tiny{\young(r,s)}\right]\epsilon ^{im} \epsilon ^{jn} \left(\overline{d}_{p}^{a} l_{sj}\right) \left(l_{ri} C q_{ta k}\right) H_{m} H_{n} H^{\dagger}{}^{k}

,\quad\mathcal{Y}\left[\tiny{\young(r,s)}\right]\epsilon ^{ik} \epsilon ^{jn} \left(\overline{d}_{p}^{a} l_{sj}\right) \left(l_{ri} C q_{ta k}\right) H_{m} H_{n} H^{\dagger}{}^{m},

\\&\mathcal{Y}\left[\tiny{\young(r,s)}\right]\epsilon ^{ij} \epsilon ^{kn} \left(\overline{d}_{p}^{a} l_{sj}\right) \left(l_{ri} C q_{ta k}\right) H_{m} H_{n} H^{\dagger}{}^{m}

,\quad\mathcal{Y}\left[\tiny{\young(rs)}\right]\epsilon ^{im} \epsilon ^{jn} \left(\overline{d}_{p}^{a} l_{sj}\right) \left(l_{ri} C q_{ta k}\right) H_{m} H_{n} H^{\dagger}{}^{k},

\\&\mathcal{Y}\left[\tiny{\young(rs)}\right]\epsilon ^{ik} \epsilon ^{jn} \left(\overline{d}_{p}^{a} l_{sj}\right) \left(l_{ri} C q_{ta k}\right) H_{m} H_{n} H^{\dagger}{}^{m}

,\quad\mathcal{Y}\left[\tiny{\young(rs)}\right]\epsilon ^{ij} \epsilon ^{kn} \left(\overline{d}_{p}^{a} l_{sj}\right) \left(l_{ri} C q_{ta k}\right) H_{m} H_{n} H^{\dagger}{}^{m}

\vspace{2ex}\\

\mathcal{O}_{ \overline{q} d l {}^2  H {}^3  } 

&\mathcal{Y}\left[\tiny{\young(st)}\right]\epsilon ^{ko} \epsilon ^{mn} \left(\overline{q}_{r}^{a i} d_{pa} \right) \left(l_{to} C l_{sn}\right) H_{i} H_{k} H_{m}

\vspace{2ex}\\

\mathcal{O}_{ u \overline{d} l e  H {}^2 H^{\dagger}  } 

& \epsilon ^{km} \left(\overline{d}_{s}^{a} l_{tm} \right) \left(u_{ra} C e_{p}\right) H_{i} H_{k} H^{\dagger}{}^{i}

\end{array}\label{cl:q2l2h3}\end{align}

\noindent 3. Operators involving only leptons with $\Delta L=2$:
\begin{align}\begin{array}{c|l}

\multirow{2}*{$\mathcal{O}_{ l{}^3 \overline{e}  H{}^2 H^{\dagger}   }^{(1 \sim 4)} $}

&\mathcal{Y}\left[\tiny{\young(rs,t)}\right]\epsilon ^{im} \epsilon ^{jn} \left(\overline{e}_{p} l_{sj}\right) \left(l_{ri} C l_{tk}\right) H_{m} H_{n} H^{\dagger}{}^{k}

,\quad\mathcal{Y}\left[\tiny{\young(rs,t)}\right]\epsilon ^{im} \epsilon ^{jn} \left(\overline{e}_{p} l_{ri}\right) \left(l_{sj} C l_{tk}\right) H_{m} H_{n} H^{\dagger}{}^{k},

\\&\mathcal{Y}\left[\tiny{\young(rst)}\right]\epsilon ^{im} \epsilon ^{jn} \left(\overline{e}_{p} l_{sj}\right) \left(l_{ri} C l_{tk}\right) H_{m} H_{n} H^{\dagger}{}^{k}

,\quad\mathcal{Y}\left[\tiny{\young(r,s,t)}\right]\epsilon ^{im} \epsilon ^{jn} \left(\overline{e}_{p} l_{sj}\right) \left(l_{ri} C l_{tk}\right) H_{m} H_{n} H^{\dagger}{}^{k}

\vspace{2ex}\\

\mathcal{O}_{ l{}^2 \overline{l} e  H{}^3  } 

&\mathcal{Y}\left[\tiny{\young(pr)}\right]\epsilon ^{im} \epsilon ^{jn} \left(\overline{l}_{t}^{k} e_{s}\right) \left(l_{pi} C l_{rj}\right) H_{k} H_{m} H_{n}

\end{array}\label{cl:l4h3}\end{align}

\noindent\underline{Class $ \psi^4 \phi^2 D $}: The class of this form must contain 3 spinors of the same helicities and 1 spinor of the opposite helicity, namely $\psi^3\psi^\dagger \phi^2 D$. A total of 5 independent Lorentz structures exist in this class
\bea
&\psi _1^{\alpha } \psi _2^{\beta } \left(D \psi _3\right)_{\alpha\beta\dot{\alpha }} \phi _4 \phi _5 \psi^{\dagger}_6{}{}^{\dot{\alpha }},&\quad
\psi _1^{\alpha } \psi _2^{\beta } \psi _3{}_{\alpha } \left(D \phi _4\right)_{\beta \dot{\alpha }} \phi _5 \psi^{\dagger}_6{}{}^{\dot{\alpha }},\nonumber \\
&\psi _1^{\alpha } \psi _2^{\beta } \psi _3{}_{\alpha } \phi _4 \left(D \phi _5\right)_{\beta \dot{\alpha }} \psi^{\dagger}_6{}{}^{\dot{\alpha }},&\quad
\psi _1^{\alpha } \psi _2{}_{\alpha } \psi _3^{\beta } \left(D \phi _4\right)_{\beta \dot{\alpha }} \phi _5 \psi^{\dagger}_6{}{}^{\dot{\alpha }},\quad
\psi _1^{\alpha } \psi _2{}_{\alpha } \psi _3^{\beta } \phi _4 \left(D \phi _5\right)_{\beta \dot{\alpha }} \psi^{\dagger}_6{}{}^{\dot{\alpha }}.
\eea

\noindent 1. Operators involving three quarks with $\Delta B=1$ and $\Delta L=-1$:
\begin{align}\begin{array}{c|l}

\multirow{5}*{$\mathcal{O}_{ q{}^3 \overline{l}  H^{\dagger} {}^2 D}^{(1 \sim 9)}$}

&\mathcal{Y}\left[\tiny{\young(prs)}\right] \epsilon ^{abc} \left(\overline{l}_{t}^{k} \gamma _{\mu } q_{pa i}\right) \left(q_{rb j} C D^{\mu } q_{sc k}\right) H^{\dagger}{}^{i} H^{\dagger}{}^{j}

,\quad\mathcal{Y}\left[\tiny{\young(prs)}\right] \epsilon ^{abc} \left(\overline{l}_{t}^{i} \gamma _{\mu } q_{pa i}\right) \left(q_{rb j} C D^{\mu } q_{sc k}\right) H^{\dagger}{}^{j} H^{\dagger}{}^{k},

\\&\mathcal{Y}\left[\tiny{\young(prs)}\right] \epsilon ^{abc} \left(\overline{l}_{t}^{k} \gamma ^{\mu } q_{rb j}\right) \left(q_{pa i} C q_{sc k}\right) H^{\dagger}{}^{j} D_{\mu } H^{\dagger}{}^{i}

,\quad\mathcal{Y}\left[\tiny{\young(pr,s)}\right] \epsilon ^{abc} \left(\overline{l}_{t}^{k} \gamma _{\mu } q_{pa i}\right) \left(q_{rb j} C D^{\mu } q_{sc k}\right) H^{\dagger}{}^{i} H^{\dagger}{}^{j},

\\&\mathcal{Y}\left[\tiny{\young(pr,s)}\right] \epsilon ^{abc} \left(\overline{l}_{t}^{i} \gamma _{\mu } q_{pa i}\right) \left(q_{rb j} C D^{\mu } q_{sc k}\right) H^{\dagger}{}^{j} H^{\dagger}{}^{k}

,\quad\mathcal{Y}\left[\tiny{\young(pr,s)}\right] \epsilon ^{abc} \left(\overline{l}_{t}^{j} \gamma _{\mu } q_{pa i}\right) \left(q_{rb j} C D^{\mu } q_{sc k}\right) H^{\dagger}{}^{i} H^{\dagger}{}^{k},

\\&\mathcal{Y}\left[\tiny{\young(pr,s)}\right] \epsilon ^{abc} \left(\overline{l}_{t}^{k} \gamma ^{\mu } q_{rb j}\right) \left(q_{pa i} C q_{sc k}\right) H^{\dagger}{}^{j} D_{\mu } H^{\dagger}{}^{i}

,\quad\mathcal{Y}\left[\tiny{\young(p,r,s)}\right] \epsilon ^{abc} \left(\overline{l}_{t}^{i} \gamma _{\mu } q_{pa i}\right) \left(q_{rb j} C D^{\mu } q_{sc k}\right) H^{\dagger}{}^{j} H^{\dagger}{}^{k},

\\&\mathcal{Y}\left[\tiny{\young(p,r,s)}\right] \epsilon ^{abc} \left(\overline{l}_{t}^{k} \gamma ^{\mu } q_{rb j}\right) \left(q_{pa i} C q_{sc k}\right) H^{\dagger}{}^{j} D_{\mu } H^{\dagger}{}^{i}

\vspace{2ex}\\

\multirow{3}*{$\mathcal{O}_{ q{}^2 d  \overline{e}  H^{\dagger} {}^2 D}^{(1 \sim 5)}$}

&\mathcal{Y}\left[\tiny{\young(rs)}\right] \epsilon ^{abc} \left(\overline{e}_{p} \gamma _{\mu } d_{tc}\right) \left(q_{ra i} C D^{\mu } q_{sb j}\right) H^{\dagger}{}^{i} H^{\dagger}{}^{j}

,\quad\mathcal{Y}\left[\tiny{\young(rs)}\right] \epsilon ^{abc} \left(\overline{e}_{p} q_{sb j}\right) \left(q_{ra i} C \gamma ^{\mu } d_{tc}\right) H^{\dagger}{}^{j} D_{\mu } H^{\dagger}{}^{i},

\\&\mathcal{Y}\left[\tiny{\young(rs)}\right] \epsilon ^{abc} \left(\overline{e}_{p} q_{sb j}\right) \left(q_{ra i} C \gamma ^{\mu } d_{tc}\right) H^{\dagger}{}^{i} D_{\mu } H^{\dagger}{}^{j}

,\quad\mathcal{Y}\left[\tiny{\young(r,s)}\right] \epsilon ^{abc} \left(\overline{e}_{p} \gamma _{\mu } d_{tc}\right) \left(q_{ra i} C D^{\mu } q_{sb j}\right) H^{\dagger}{}^{i} H^{\dagger}{}^{j},

\\&\mathcal{Y}\left[\tiny{\young(r,s)}\right] \epsilon ^{abc} \left(\overline{e}_{p} q_{sb j}\right) \left(q_{ra i} C \gamma ^{\mu } d_{tc}\right) H^{\dagger}{}^{j} D_{\mu } H^{\dagger}{}^{i}

\vspace{2ex}\\

\multirow{3}*{$\mathcal{O}_{ q  u d \overline{l}  H^{\dagger}{}^2 D}^{(1 \sim 5)}$}

&  \epsilon ^{abc} \epsilon _{ik} \left(\overline{l}_{r}^{i} D^{\mu } u_{sb} \right) \left(q_{tc j} C \gamma _{\mu } d_{pa}\right) H^{\dagger}{}^{j} H^{\dagger}{}^{k}

,\quad  \epsilon ^{abc} \epsilon _{ik} \left(\overline{l}_{r}^{i} \gamma ^{\mu } q_{tc j}\right) \left(u_{sb} C d_{pa}\right) H^{\dagger}{}^{k} D_{\mu } H^{\dagger}{}^{j},

\\&  \epsilon ^{abc} \epsilon _{ij} \left(\overline{l}_{r}^{i} \gamma ^{\mu } q_{tc k}\right) \left(u_{sb} C d_{pa}\right) H^{\dagger}{}^{k} D_{\mu } H^{\dagger}{}^{j}

,\quad  \epsilon ^{abc} \epsilon _{ik} \left(\overline{l}_{r}^{i} d_{pa}\right) \left(q_{tc j} C \gamma _{\mu } u_{sb}\right) H^{\dagger}{}^{k} D_{\mu } H^{\dagger}{}^{j},

\\&  \epsilon ^{abc} \epsilon _{ij} \left(\overline{l}_{r}^{i} d_{pa}\right) \left(q_{tc j} C \gamma _{\mu } u_{sb}\right) H^{\dagger}{}^{k} D_{\mu } H^{\dagger}{}^{k}

\vspace{2ex}\\

\multirow{5}*{$\mathcal{O}_{ q d{}^2 \overline{l}  H H^{\dagger}  D}^{(1 \sim 10)}$}

&\mathcal{Y}\left[\tiny{\young(pr)}\right] \epsilon ^{abc} \left(D^{\mu } \overline{l}_{s}^{i} d_{rb} \right) \left(q_{tc j} C \gamma _{\mu } d_{pa}\right) H_{i} H^{\dagger}{}^{j}

,\quad\mathcal{Y}\left[\tiny{\young(pr)}\right] \epsilon ^{abc} \left(D^{\mu } \overline{l}_{s}^{i} d_{rb} \right) \left(q_{tc i} C \gamma _{\mu } d_{pa}\right) H_{j} H^{\dagger}{}^{j},

\\&\mathcal{Y}\left[\tiny{\young(pr)}\right] \epsilon ^{abc} \left(\overline{l}_{s}^{i} d_{pa} \right) \left(q_{tc j} C \gamma ^{\mu } d_{rb}\right) H_{i} D_{\mu } H^{\dagger}{}^{j}

,\quad\mathcal{Y}\left[\tiny{\young(pr)}\right] \epsilon ^{abc} \left(\overline{l}_{s}^{i} d_{pa} \right) \left(q_{tc i} C \gamma ^{\mu } d_{rb}\right) H_{j} D_{\mu } H^{\dagger}{}^{j},

\\&\mathcal{Y}\left[\tiny{\young(pr)}\right] \epsilon ^{abc} \left(\overline{l}_{s}^{i} d_{pa} \right) \left(q_{tc j} C \gamma ^{\mu } d_{rb}\right) H^{\dagger}{}^{j} D_{\mu } H_{i}

,\quad\mathcal{Y}\left[\tiny{\young(pr)}\right] \epsilon ^{abc} \left(\overline{l}_{s}^{i} d_{pa} \right) \left(q_{tc i} C \gamma ^{\mu } d_{rb}\right) H^{\dagger}{}^{j} D_{\mu } H_{j},

\\&\mathcal{Y}\left[\tiny{\young(p,r)}\right] \epsilon ^{abc} \left(\overline{l}_{s}^{i} d_{pa} \right) \left(q_{tc j} C \gamma ^{\mu } d_{rb}\right) H_{i} D_{\mu } H^{\dagger}{}^{j}

,\quad\mathcal{Y}\left[\tiny{\young(p,r)}\right] \epsilon ^{abc} \left(\overline{l}_{s}^{i} d_{pa} \right) \left(q_{tc i} C \gamma ^{\mu } d_{rb}\right) H_{j} D_{\mu } H^{\dagger}{}^{j},

\\&\mathcal{Y}\left[\tiny{\young(p,r)}\right] \epsilon ^{abc} \left(\overline{l}_{s}^{i} d_{pa} \right) \left(q_{tc j} C \gamma ^{\mu } d_{rb}\right) H^{\dagger}{}^{j} D_{\mu } H_{i}

,\quad\mathcal{Y}\left[\tiny{\young(p,r)}\right] \epsilon ^{abc} \left(\overline{l}_{s}^{i} d_{pa} \right) \left(q_{tc i} C \gamma ^{\mu } d_{rb}\right) H^{\dagger}{}^{j} D_{\mu } H_{j}

\vspace{2ex}\\

\multirow{1}*{$\mathcal{O}_{ u d{}^2 \overline{e} H^{\dagger}{}^2 D}^{(1,2)}$}

&\mathcal{Y}\left[\tiny{\young(pr)}\right] \epsilon ^{abc} \epsilon _{ij} \left(\overline{e}_{t} \gamma ^{\mu } d_{rb}\right) \left(u_{sc} C d_{pa}\right) H^{\dagger}{}^{j} D_{\mu } H^{\dagger}{}^{i}

,\quad\mathcal{Y}\left[\tiny{\young(p,r)}\right] \epsilon ^{abc} \epsilon _{ij} \left(\overline{e}_{t} \gamma ^{\mu } d_{rb}\right) \left(u_{sc} C d_{pa}\right) H^{\dagger}{}^{j} D_{\mu } H^{\dagger}{}^{i}

\vspace{2ex}\\

\multirow{2}*{$\mathcal{O}_{  d{}^3 \overline{e}  H H^{\dagger}  D}^{(1 \sim 3)}$}

&\mathcal{Y}\left[\tiny{\young(prs)}\right] \epsilon ^{abc} \left(\overline{e}_{t} \gamma _{\mu } d_{pa}\right) \left(D^{\mu } d_{sc} C d_{rb}\right) H_{i} H^{\dagger}{}^{i}

,\quad\mathcal{Y}\left[\tiny{\young(pr,s)}\right] \epsilon ^{abc} \left(\overline{e}_{t} \gamma _{\mu } d_{pa}\right) \left(D^{\mu } d_{sc} C d_{rb}\right) H_{i} H^{\dagger}{}^{i},

\\&\mathcal{Y}\left[\tiny{\young(pr,s)}\right] \epsilon ^{abc} \left(\overline{e}_{t} \gamma ^{\mu } d_{rb}\right) \left(d_{sc} C d_{pa}\right) H_{i} D_{\mu } H^{\dagger}{}^{i}

\end{array}\label{cl:q3lh2D}\end{align}

\noindent 2. Operators involving two leptons and two quarks with $\Delta B=0$ and $\Delta L=2$:
\begin{align}\begin{array}{c|l}

\multirow{7}*{$\mathcal{O}_{ q \overline{q} l{}^2  H{}^2 D}^{(1 \sim 13)} $}

&\mathcal{Y}\left[\tiny{\young(p,r)}\right] \epsilon ^{im} \epsilon ^{jn} \left(\overline{q}_{t}^{a k} \gamma _{\mu } l_{pi}\right) \left(l_{rj} C D^{\mu } q_{sa k}\right) H_{m} H_{n}

,\quad\mathcal{Y}\left[\tiny{\young(p,r)}\right] \epsilon ^{ik} \epsilon ^{jn} \left(\overline{q}_{t}^{a m} \gamma _{\mu } l_{pi}\right) \left(l_{rj} C D^{\mu } q_{sa k}\right) H_{m} H_{n},

\\&\mathcal{Y}\left[\tiny{\young(p,r)}\right] \epsilon ^{im} \epsilon ^{jn} \left(\overline{q}_{t}^{a k} \gamma ^{\mu } l_{rj}\right) \left(l_{pi} C q_{sa k}\right) H_{n} D_{\mu } H_{m}

,\quad\mathcal{Y}\left[\tiny{\young(p,r)}\right] \epsilon ^{ik} \epsilon ^{jn} \left(\overline{q}_{t}^{a m} \gamma ^{\mu } l_{rj}\right) \left(l_{pi} C q_{sa k}\right) H_{n} D_{\mu } H_{m},

\\&\mathcal{Y}\left[\tiny{\young(p,r)}\right] \epsilon ^{ik} \epsilon ^{jm} \left(\overline{q}_{t}^{a n} \gamma ^{\mu } l_{rj}\right) \left(l_{pi} C q_{sa k}\right) H_{n} D_{\mu } H_{m}

,\quad\mathcal{Y}\left[\tiny{\young(p,r)}\right] \epsilon ^{ij} \epsilon ^{kn} \left(\overline{q}_{t}^{a m} \gamma ^{\mu } l_{rj}\right) \left(l_{pi} C q_{sa k}\right) H_{n} D_{\mu } H_{m},

\\&\mathcal{Y}\left[\tiny{\young(p,r)}\right] \epsilon ^{ij} \epsilon ^{km} \left(\overline{q}_{t}^{a n} \gamma ^{\mu } l_{rj}\right) \left(l_{pi} C q_{sa k}\right) H_{n} D_{\mu } H_{m}

,\quad\mathcal{Y}\left[\tiny{\young(pr)}\right] \epsilon ^{ik} \epsilon ^{jn} \left(\overline{q}_{t}^{a m} \gamma _{\mu } l_{pi}\right) \left(l_{rj} C D^{\mu } q_{sa k}\right) H_{m} H_{n},

\\&\mathcal{Y}\left[\tiny{\young(pr)}\right] \epsilon ^{im} \epsilon ^{jn} \left(\overline{q}_{t}^{a k} \gamma ^{\mu } l_{rj}\right) \left(l_{pi} C q_{sa k}\right) H_{n} D_{\mu } H_{m}

,\quad\mathcal{Y}\left[\tiny{\young(pr)}\right] \epsilon ^{ik} \epsilon ^{jn} \left(\overline{q}_{t}^{a m} \gamma ^{\mu } l_{rj}\right) \left(l_{pi} C q_{sa k}\right) H_{n} D_{\mu } H_{m},

\\&\mathcal{Y}\left[\tiny{\young(pr)}\right] \epsilon ^{ik} \epsilon ^{jm} \left(\overline{q}_{t}^{a n} \gamma ^{\mu } l_{rj}\right) \left(l_{pi} C q_{sa k}\right) H_{n} D_{\mu } H_{m}

,\quad\mathcal{Y}\left[\tiny{\young(pr)}\right] \epsilon ^{ij} \epsilon ^{kn} \left(\overline{q}_{t}^{a m} \gamma ^{\mu } l_{rj}\right) \left(l_{pi} C q_{sa k}\right) H_{n} D_{\mu } H_{m},

\\&\mathcal{Y}\left[\tiny{\young(pr)}\right] \epsilon ^{ij} \epsilon ^{km} \left(\overline{q}_{t}^{a n} \gamma ^{\mu } l_{rj}\right) \left(l_{pi} C q_{sa k}\right) H_{n} D_{\mu } H_{m}

\end{array}\label{cl:q2l2h2D2}\end{align}
\begin{align}\begin{array}{c|l}

\multirow{3}*{$\mathcal{O}_{ \overline{q} u l e  H {}^2 D}^{(1 \sim 5)} $}

&  \epsilon ^{km} \left(\overline{q}_{r}^{a i} D^{\mu } u_{sa} \right) \left(l_{tm} C \gamma _{\mu } e_{p}\right) H_{i} H_{k}

,\quad  \epsilon ^{km} \left(\overline{q}_{r}^{a i} \gamma ^{\mu } l_{tm}\right) \left(u_{sa} C e_{p}\right) H_{k} D_{\mu } H_{i},

\\&  \epsilon ^{jk} \left(\overline{q}_{r}^{a i} \gamma ^{\mu } l_{ti}\right) \left(u_{sa} C e_{p}\right) H_{k} D_{\mu } H_{j}

,\quad  \epsilon ^{km} \left(\overline{q}_{r}^{a i} e_{p} \right) \left(l_{tm} C \gamma _{\mu } u_{sa}\right) H_{k} D^{\mu } H_{i},

\\&  \epsilon ^{jk} \left(\overline{q}_{r}^{a i} e_{p} \right) \left(l_{ti} C \gamma _{\mu } u_{sa}\right) H_{k} D^{\mu } H_{j}

\vspace{2ex}\\

\multirow{3}*{$\mathcal{O}_{ q \overline{d} l e  H{}^2 D}^{(1 \sim 5)} $}

&  \epsilon ^{ik} \epsilon ^{jm} \left(\overline{d}_{p}^{a} \gamma _{\mu } e_{t}\right) \left(l_{ri} C D^{\mu } q_{sa j}\right) H_{k} H_{m}

,\quad  \epsilon ^{ik} \epsilon ^{jm} \left(\overline{d}_{p}^{a} q_{sa j}\right) \left(l_{ri} C \gamma ^{\mu } e_{t}\right) H_{m} D_{\mu } H_{k},

\\&  \epsilon ^{ij} \epsilon ^{km} \left(\overline{d}_{p}^{a} q_{sa j}\right) \left(l_{ri} C \gamma ^{\mu } e_{t}\right) H_{m} D_{\mu } H_{k}

,\quad  \epsilon ^{ik} \epsilon ^{jm} \left(\overline{d}_{p}^{a} l_{ri}\right) \left(q_{sa j} C \gamma _{\mu } e_{t}\right) H_{m} D^{\mu } H_{k},

\\&  \epsilon ^{ij} \epsilon ^{km} \left(\overline{d}_{p}^{a} l_{ri}\right) \left(q_{sa j} C \gamma _{\mu } e_{t}\right) H_{m} D^{\mu } H_{k}

\vspace{2ex}\\

\multirow{3}*{$\mathcal{O}_{ u \overline{u} l{}^2  H{}^2 D}^{(1 \sim 5)} $}

&\mathcal{Y}\left[\tiny{\young(p,r)}\right] \epsilon ^{ik} \epsilon ^{jm} \left(D^{\mu } \overline{u}_{s}^{a} l_{rj}\right) \left(l_{pi} C \gamma _{\mu } u_{ta}\right) H_{k} H_{m}

,\quad\mathcal{Y}\left[\tiny{\young(p,r)}\right] \epsilon ^{ik} \epsilon ^{jm} \left(\overline{u}_{s}^{a} l_{pi}\right) \left(l_{rj} C \gamma ^{\mu } u_{ta}\right) H_{m} D_{\mu } H_{k},

\\&\mathcal{Y}\left[\tiny{\young(p,r)}\right] \epsilon ^{ij} \epsilon ^{km} \left(\overline{u}_{s}^{a} l_{pi}\right) \left(l_{rj} C \gamma ^{\mu } u_{ta}\right) H_{m} D_{\mu } H_{k}

,\quad\mathcal{Y}\left[\tiny{\young(pr)}\right] \epsilon ^{ik} \epsilon ^{jm} \left(\overline{u}_{s}^{a} l_{pi}\right) \left(l_{rj} C \gamma ^{\mu } u_{ta}\right) H_{m} D_{\mu } H_{k},

\\&\mathcal{Y}\left[\tiny{\young(pr)}\right] \epsilon ^{ij} \epsilon ^{km} \left(\overline{u}_{s}^{a} l_{pi}\right) \left(l_{rj} C \gamma ^{\mu } u_{ta}\right) H_{m} D_{\mu } H_{k}

\vspace{2ex}\\

\multirow{5}*{$\mathcal{O}_{ u \overline{d} l{}^2  H H^{\dagger}  D}^{(1 \sim 10)} $}

&\mathcal{Y}\left[\tiny{\young(r,s)}\right] \epsilon ^{ik} \left(\overline{d}_{p}^{a} \gamma _{\mu } u_{ta}\right) \left(l_{ri} C D^{\mu } l_{sj}\right) H_{k} H^{\dagger}{}^{j}

,\quad\mathcal{Y}\left[\tiny{\young(r,s)}\right] \epsilon ^{ij} \left(\overline{d}_{p}^{a} \gamma _{\mu } u_{ta}\right) \left(l_{ri} C D^{\mu } l_{sj}\right) H_{k} H^{\dagger}{}^{k},

\\&\mathcal{Y}\left[\tiny{\young(r,s)}\right] \epsilon ^{ik} \left(\overline{d}_{p}^{a} l_{sj}\right) \left(l_{ri} C \gamma ^{\mu } u_{ta}\right) H^{\dagger}{}^{j} D_{\mu } H_{k}

,\quad\mathcal{Y}\left[\tiny{\young(r,s)}\right] \epsilon ^{ij} \left(\overline{d}_{p}^{a} l_{sj}\right) \left(l_{ri} C \gamma ^{\mu } u_{ta}\right) H^{\dagger}{}^{k} D_{\mu } H_{k},

\\&\mathcal{Y}\left[\tiny{\young(r,s)}\right] \epsilon ^{ik} \left(\overline{d}_{p}^{a} l_{sj}\right) \left(l_{ri} C \gamma ^{\mu } u_{ta}\right) H_{k} D_{\mu } H^{\dagger}{}^{j}

,\quad\mathcal{Y}\left[\tiny{\young(rs)}\right] \epsilon ^{ik} \left(\overline{d}_{p}^{a} \gamma _{\mu } u_{ta}\right) \left(l_{ri} C D^{\mu } l_{sj}\right) H_{k} H^{\dagger}{}^{j},

\\&\mathcal{Y}\left[\tiny{\young(rs)}\right] \epsilon ^{ij} \left(\overline{d}_{p}^{a} \gamma _{\mu } u_{ta}\right) \left(l_{ri} C D^{\mu } l_{sj}\right) H_{k} H^{\dagger}{}^{k}

,\quad\mathcal{Y}\left[\tiny{\young(rs)}\right] \epsilon ^{ik} \left(\overline{d}_{p}^{a} l_{sj}\right) \left(l_{ri} C \gamma ^{\mu } u_{ta}\right) H^{\dagger}{}^{j} D_{\mu } H_{k},

\\&\mathcal{Y}\left[\tiny{\young(rs)}\right] \epsilon ^{ij} \left(\overline{d}_{p}^{a} l_{sj}\right) \left(l_{ri} C \gamma ^{\mu } u_{ta}\right) H^{\dagger}{}^{k} D_{\mu } H_{k}

,\quad\mathcal{Y}\left[\tiny{\young(rs)}\right] \epsilon ^{ik} \left(\overline{d}_{p}^{a} l_{sj}\right) \left(l_{ri} C \gamma ^{\mu } u_{ta}\right) H_{k} D_{\mu } H^{\dagger}{}^{j}

\vspace{2ex}\\

\multirow{1}*{$\mathcal{O}_{ u \overline{d}  e{}^2  H {}^2 D}^{(1,2)} $}

&\mathcal{Y}\left[\tiny{\young(p,r)}\right] \epsilon ^{ij} \left(\overline{d}_{t}^{a} \gamma ^{\mu } e_{r}\right) \left(u_{sa} C e_{p}\right) H_{j} D_{\mu } H_{i}

,\quad\mathcal{Y}\left[\tiny{\young(pr)}\right] \epsilon ^{ij} \left(\overline{d}_{t}^{a} \gamma ^{\mu } e_{r}\right) \left(u_{sa} C e_{p}\right) H_{j} D_{\mu } H_{i}
\vspace{2ex}\\

\multirow{3}*{$\mathcal{O}_{  d \overline{d} l{}^2 H{}^2 D}^{(1 \sim 5)} $}

&\mathcal{Y}\left[\tiny{\young(r,s)}\right] \epsilon ^{ik} \epsilon ^{jm} \left(\overline{d}_{p}^{a} \gamma _{\mu } d_{ta}\right) \left(l_{ri} C D^{\mu } l_{sj}\right) H_{k} H_{m}

,\quad\mathcal{Y}\left[\tiny{\young(r,s)}\right] \epsilon ^{ik} \epsilon ^{jm} \left(\overline{d}_{p}^{a} l_{sj}\right) \left(l_{ri} C \gamma ^{\mu } d_{ta}\right) H_{m} D_{\mu } H_{k},

\\&\mathcal{Y}\left[\tiny{\young(r,s)}\right] \epsilon ^{ij} \epsilon ^{km} \left(\overline{d}_{p}^{a} l_{sj}\right) \left(l_{ri} C \gamma ^{\mu } d_{ta}\right) H_{m} D_{\mu } H_{k}

,\quad\mathcal{Y}\left[\tiny{\young(rs)}\right] \epsilon ^{ik} \epsilon ^{jm} \left(\overline{d}_{p}^{a} \gamma _{\mu } d_{ta}\right) \left(l_{ri} C D^{\mu } l_{sj}\right) H_{k} H_{m},

\\&\mathcal{Y}\left[\tiny{\young(rs)}\right] \epsilon ^{ik} \epsilon ^{jm} \left(\overline{d}_{p}^{a} l_{sj}\right) \left(l_{ri} C \gamma ^{\mu } d_{ta}\right) H_{m} D_{\mu } H_{k}

\end{array}\label{cl:q2l2h2D2}\end{align}

\noindent 3. Operators involving only leptons with $\Delta L=2$:
\begin{align}\begin{array}{c|l}

\multirow{5}*{$\mathcal{O}_{  l{}^3 \overline{l}  H{}^2 D}^{(1 \sim 9)} $}

&\mathcal{Y}\left[\tiny{\young(p,r,s)}\right] \epsilon ^{im} \epsilon ^{jn} \left(\overline{l}_{t}^{k} \gamma _{\mu } l_{pi}\right) \left(l_{rj} C D^{\mu } l_{sk}\right) H_{m} H_{n}

,\quad\mathcal{Y}\left[\tiny{\young(p,r,s)}\right] \epsilon ^{ik} \epsilon ^{jn} \left(\overline{l}_{t}^{m} \gamma _{\mu } l_{pi}\right) \left(l_{rj} C D^{\mu } l_{sk}\right) H_{m} H_{n},

\vspace{0.5ex}

\\&\mathcal{Y}\left[\tiny{\young(p,r,s)}\right] \epsilon ^{im} \epsilon ^{jn} \left(\overline{l}_{t}^{k} \gamma ^{\mu } l_{rj}\right) \left(l_{pi} C l_{sk}\right) H_{n} D_{\mu } H_{m}

,\quad\mathcal{Y}\left[\tiny{\young(pr,s)}\right] \epsilon ^{im} \epsilon ^{jn} \left(\overline{l}_{t}^{k} \gamma _{\mu } l_{pi}\right) \left(l_{rj} C D^{\mu } l_{sk}\right) H_{m} H_{n},

\\&\mathcal{Y}\left[\tiny{\young(pr,s)}\right] \epsilon ^{ik} \epsilon ^{jn} \left(\overline{l}_{t}^{m} \gamma _{\mu } l_{pi}\right) \left(l_{rj} C D^{\mu } l_{sk}\right) H_{m} H_{n}

,\quad\mathcal{Y}\left[\tiny{\young(pr,s)}\right] \epsilon ^{ij} \epsilon ^{kn} \left(\overline{l}_{t}^{m} \gamma _{\mu } l_{pi}\right) \left(l_{rj} C D^{\mu } l_{sk}\right) H_{m} H_{n},

\\&\mathcal{Y}\left[\tiny{\young(pr,s)}\right] \epsilon ^{im} \epsilon ^{jn} \left(\overline{l}_{t}^{k} \gamma ^{\mu } l_{rj}\right) \left(l_{pi} C l_{sk}\right) H_{n} D_{\mu } H_{m}

,\quad\mathcal{Y}\left[\tiny{\young(prs)}\right] \epsilon ^{ik} \epsilon ^{jn} \left(\overline{l}_{t}^{m} \gamma _{\mu } l_{pi}\right) \left(l_{rj} C D^{\mu } l_{sk}\right) H_{m} H_{n},

\\&\mathcal{Y}\left[\tiny{\young(prs)}\right] \epsilon ^{im} \epsilon ^{jn} \left(\overline{l}_{t}^{k} \gamma ^{\mu } l_{rj}\right) \left(l_{pi} C l_{sk}\right) H_{n} D_{\mu } H_{m}

\vspace{2ex}\\

\multirow{3}*{$\mathcal{O}_{ l{}^2 e \overline{e}  H{}^2 D}^{(1 \sim 5)} $}

&\mathcal{Y}\left[\tiny{\young(r,s)}\right] \epsilon ^{ik} \epsilon ^{jm} \left(\overline{e}_{p} \gamma _{\mu } e_{t}\right) \left(l_{ri} C D^{\mu } l_{sj}\right) H_{k} H_{m}

,\quad\mathcal{Y}\left[\tiny{\young(r,s)}\right] \epsilon ^{ik} \epsilon ^{jm} \left(\overline{e}_{p} l_{sj}\right) \left(l_{ri} C \gamma ^{\mu } e_{t}\right) H_{m} D_{\mu } H_{k},

\\&\mathcal{Y}\left[\tiny{\young(r,s)}\right] \epsilon ^{ij} \epsilon ^{km} \left(\overline{e}_{p} l_{sj}\right) \left(l_{ri} C \gamma ^{\mu } e_{t}\right) H_{m} D_{\mu } H_{k}

,\quad\mathcal{Y}\left[\tiny{\young(rs)}\right] \epsilon ^{ik} \epsilon ^{jm} \left(\overline{e}_{p} \gamma _{\mu } e_{t}\right) \left(l_{ri} C D^{\mu } l_{sj}\right) H_{k} H_{m},

\\&\mathcal{Y}\left[\tiny{\young(rs)}\right] \epsilon ^{ik} \epsilon ^{jm} \left(\overline{e}_{p} l_{sj}\right) \left(l_{ri} C \gamma ^{\mu } e_{t}\right) H_{m} D_{\mu } H_{k}

\end{array}\label{cl:l4h2D}\end{align}

\noindent\underline{Class $\psi^4 \phi D^2$}: This form involves two classes: $\psi ^4 \phi D^2$ with 10 Lorentz structure
\bea
&\psi _1^{\alpha } \left(D \psi _2\right)^{\beta \gamma }_{\dot{\alpha }} \psi _3{}_{\alpha } \left(D \psi _4\right)_{\beta \gamma }^{\dot{\alpha }} \phi _5,&\quad
\psi _1^{\alpha } \left(D \psi _2\right)^{\beta \gamma }_{\dot{\alpha }} \psi _3{}_{\alpha } \psi _4{}_{\beta } \left(D \phi _5\right)_{\gamma }^{\dot{\alpha }},\quad
\psi _1^{\alpha } \psi _2^{\beta } \left(D \psi _3\right)_{\alpha \dot{\alpha }}^{\gamma } \left(D \psi _4\right)_{\beta \gamma }^{\dot{\alpha }} \phi _5, \nn \\
&\psi _1^{\alpha } \psi _2^{\beta } \left(D \psi _3\right)_{\alpha \dot{\alpha }}^{\gamma } \psi _4{}_{\beta } \left(D \phi _5\right)_{\gamma }^{\dot{\alpha }},&\quad
\psi _1^{\alpha } \psi _2^{\beta } \psi _3^{\gamma } \left(D \psi _4\right)_{\alpha \beta \dot{\alpha }} \left(D \phi _5\right)_{\gamma }^{\dot{\alpha }},\quad
\psi _1^{\alpha } \psi _2^{\beta } \left(D \psi _3\right)_{\alpha {\beta \dot{\alpha }}} \psi _4^{\gamma } \left(D \phi _5\right)_{\gamma }^{\dot{\alpha }}, \nn \\
&\psi _1^{\alpha } \psi _2^{\beta } \psi _3{}_{\alpha } \left(D \psi _4\right)_{\beta \dot{\alpha }}^{\gamma } \left(D \phi _5\right)_{\gamma }^{\dot{\alpha }},&\quad
\psi _1^{\alpha } \psi _2{}_{\alpha } \left(D \psi _3\right)^{\beta \gamma }_{\dot{\alpha }} \left(D \psi _4\right)_{\beta \gamma }^{\dot{\alpha }} \phi _5,\quad
\psi _1^{\alpha } \psi _2{}_{\alpha } \left(D \psi _3\right)^{\beta \gamma }_{\dot{\alpha }} \psi _4{}_{\beta } \left(D \phi _5\right)_{\gamma }^{\dot{\alpha }}, \nn \\
&\psi _1^{\alpha } \psi _2{}_{\alpha } \psi _3^{\beta } \left(D \psi _4\right)_{\beta \dot{\alpha }}^{\gamma } \left(D \phi _5\right)_{\gamma }^{\dot{\alpha }},&
\eea
and $\psi ^2 \psi ^{\dagger 2} \phi D^2$ with 7 independent Lorentz structures
\bea
&\psi _1^{\alpha } \psi _2^{\beta } (D^2 \phi _3)_{\dot{\alpha } \dot{\beta } \alpha \beta } \psi^{\dagger}_4{}{}^{\dot{\alpha }} \psi^{\dagger}_5{}{}^{\dot{\beta }},&\quad
\psi _1^{\alpha } \psi _2^{\beta } \left(D \phi _3\right)_{\alpha \dot{\alpha }} (D \psi^{\dagger}_4)_{\beta \dot{\beta }}^{\dot{\alpha }} \psi^{\dagger}_5{}{}^{\dot{\beta }},\quad 
\psi _1^{\alpha } \psi _2^{\beta } \left(D \phi _3\right)_{\alpha \dot{\alpha }} \psi^{\dagger}_4{}_{\dot{\beta }} (D \psi^{\dagger}_5)_{\beta }^{\dot{\alpha } \dot{\beta }}\nn \\
&\psi _1^{\alpha } \psi _2^{\beta } \phi _3 (D \psi^{\dagger}_4)_{\alpha \dot{\alpha } \dot{\beta }} (D \psi^{\dagger}_5)_{\beta }^{\dot{\alpha } \dot{\beta }},&\quad
\psi _1^{\alpha } \psi _2{}_{\alpha } \left(D \phi _3\right)^{\beta \dot{\alpha }} (D \psi^{\dagger}_4)_{\beta \dot{\beta }}^{\dot{\alpha }} \psi^{\dagger}_5{}{}^{\dot{\beta }},\quad
\psi _1^{\alpha } \psi _2{}_{\alpha } \left(D \phi _3\right)^{\beta }_{\dot{\alpha }} \psi^{\dagger}_4{}_{\dot{\beta }} (D \psi^{\dagger}_5)_{\beta }^{\dot{\alpha } \dot{\beta }},\nn \\
&\psi _1^{\alpha } \psi _2{}_{\alpha } \phi _3 (D \psi^{\dagger}_4)^{\beta }_{\dot{\alpha } \dot{\beta }} (D \psi^{\dagger}_5)_{\beta }^{\dot{\alpha } \dot{\beta }}.&
\eea

\noindent 1. Operators involving three quarks with $\Delta B=1$ and $\Delta L=-1$:
\begin{align}\begin{array}{c|l}

\multirow{7}*{$\mathcal{O}_{ q{}^2 d  \overline{l} H^{\dagger} D^2  }^{(1 \sim 14)}$}

&\mathcal{Y}\left[\tiny{\young(pr)}\right]\epsilon ^{abc} \left(\overline{l}_{t}^{j} \gamma ^{\nu } q_{rb j}\right) \left(q_{pa i} C \gamma ^{\mu } d_{sc}\right) D_{\mu } D_{\nu } H^{\dagger}{}^{i}

,\quad\mathcal{Y}\left[\tiny{\young(pr)}\right]\epsilon ^{abc} \left(\overline{l}_{t}^{j} \gamma ^{\nu } q_{rb j}\right) \left(q_{pa i} C \gamma ^{\mu } D_{\nu } d_{sc}\right) D_{\mu } H^{\dagger}{}^{i},

\\&\mathcal{Y}\left[\tiny{\young(pr)}\right]\epsilon ^{abc} \left(\overline{l}_{t}^{i} \gamma ^{\nu } q_{rb j}\right) \left(q_{pa i} C \gamma ^{\mu } D_{\nu } d_{sc}\right) D_{\mu } H^{\dagger}{}^{j}

,\quad\mathcal{Y}\left[\tiny{\young(pr)}\right] \epsilon ^{abc} \left(D_{\nu } \overline{l}_{t}^{j} d_{sc}\right) \left(q_{pa i} C \sigma ^{\mu }{}^{\nu } q_{rb j}\right) D_{\mu } H^{\dagger}{}^{i},

\\&\mathcal{Y}\left[\tiny{\young(pr)}\right]\epsilon ^{abc} \left(D^{\mu } \overline{l}_{t}^{j} d_{sc}\right) \left(q_{pa i} C q_{rb j}\right) D_{\mu } H^{\dagger}{}^{i}

,\quad\mathcal{Y}\left[\tiny{\young(pr)}\right] \epsilon ^{abc} \left(D_{\nu } \overline{l}_{t}^{j} D_{\mu } d_{sc}\right) \left(q_{pa i} C \sigma ^{\mu }{}^{\nu } q_{rb j}\right) H^{\dagger}{}^{i},

\\&\mathcal{Y}\left[\tiny{\young(pr)}\right]\epsilon ^{abc} \left(D^{\mu } \overline{l}_{t}^{j} D_{\mu } d_{sc}\right) \left(q_{pa i} C q_{rb j}\right) H^{\dagger}{}^{i}

,\quad\mathcal{Y}\left[\tiny{\young(p,r)}\right]\epsilon ^{abc} \left(\overline{l}_{t}^{j} \gamma ^{\nu } q_{rb j}\right) \left(q_{pa i} C \gamma ^{\mu } d_{sc}\right) D_{\mu } D_{\nu } H^{\dagger}{}^{i},

\\&\mathcal{Y}\left[\tiny{\young(p,r)}\right]\epsilon ^{abc} \left(\overline{l}_{t}^{j} \gamma ^{\nu } q_{rb j}\right) \left(q_{pa i} C \gamma ^{\mu } D_{\nu } d_{sc}\right) D_{\mu } H^{\dagger}{}^{i}

,\quad\mathcal{Y}\left[\tiny{\young(p,r)}\right]\epsilon ^{abc} \left(\overline{l}_{t}^{i} \gamma ^{\nu } q_{rb j}\right) \left(q_{pa i} C \gamma ^{\mu } D_{\nu } d_{sc}\right) D_{\mu } H^{\dagger}{}^{j},

\\&\mathcal{Y}\left[\tiny{\young(p,r)}\right] \epsilon ^{abc} \left(D_{\nu } \overline{l}_{t}^{j} d_{sc}\right) \left(q_{pa i} C \sigma ^{\mu }{}^{\nu } q_{rb j}\right) D_{\mu } H^{\dagger}{}^{i}

,\quad\mathcal{Y}\left[\tiny{\young(p,r)}\right]\epsilon ^{abc} \left(D^{\mu } \overline{l}_{t}^{j} d_{sc}\right) \left(q_{pa i} C q_{rb j}\right) D_{\mu } H^{\dagger}{}^{i},

\\&\mathcal{Y}\left[\tiny{\young(p,r)}\right] \epsilon ^{abc} \left(D_{\nu } \overline{l}_{t}^{j} D_{\mu } d_{sc}\right) \left(q_{pa i} C \sigma ^{\mu }{}^{\nu } q_{rb j}\right) H^{\dagger}{}^{i}

,\quad\mathcal{Y}\left[\tiny{\young(p,r)}\right]\epsilon ^{abc} \left(D^{\mu } \overline{l}_{t}^{j} D_{\mu } d_{sc}\right) \left(q_{pa i} C q_{rb j}\right) H^{\dagger}{}^{i}

\vspace{2ex}\\

\multirow{4}*{$\mathcal{O}_{ q d {}^2 \overline{e} H^{\dagger} D^2  }^{(1 \sim 7)}$}

&\mathcal{Y}\left[\tiny{\young(st)}\right]\epsilon ^{abc} \left(\overline{e}_{p} \gamma ^{\mu } d_{sb}\right) \left(q_{ra i} C \gamma ^{\nu } d_{tc}\right) D_{\mu } D_{\nu } H^{\dagger}{}^{i}

,\quad\mathcal{Y}\left[\tiny{\young(st)}\right]\epsilon ^{abc} \left(\overline{e}_{p} \gamma ^{\mu } D_{\nu } d_{sb}\right) \left(q_{ra i} C \gamma ^{\nu } d_{tc}\right) D_{\mu } H^{\dagger}{}^{i},

\\&\mathcal{Y}\left[\tiny{\young(st)}\right] \epsilon ^{abc} \left(\overline{e}_{p} \sigma ^{\mu }{}^{\nu } q_{ra i}\right) \left(d_{sb} C D_{\nu } d_{tc}\right) D_{\mu } H^{\dagger}{}^{i}

,\quad\mathcal{Y}\left[\tiny{\young(st)}\right] \epsilon ^{abc} \left(\overline{e}_{p} \sigma ^{\mu }{}^{\nu } q_{ra i}\right) \left(D_{\mu } d_{sb} C D_{\nu } d_{tc}\right) H^{\dagger}{}^{i},

\\&\mathcal{Y}\left[\tiny{\young(s,t)}\right]\epsilon ^{abc} \left(\overline{e}_{p} \gamma ^{\mu } D_{\nu } d_{sb}\right) \left(q_{ra i} C \gamma ^{\nu } d_{tc}\right) D_{\mu } H^{\dagger}{}^{i}

,\quad\mathcal{Y}\left[\tiny{\young(s,t)}\right] \epsilon ^{abc} \left(\overline{e}_{p} \sigma ^{\mu }{}^{\nu } q_{ra i}\right) \left(d_{sb} C D_{\nu } d_{tc}\right) D_{\mu } H^{\dagger}{}^{i},

\\&\mathcal{Y}\left[\tiny{\young(s,t)}\right]\epsilon ^{abc} \left(\overline{e}_{p} q_{ra i}\right) \left(D_{\mu } d_{sb} C D^{\mu } d_{tc}\right) H^{\dagger}{}^{i}

\vspace{2ex}\\

\multirow{5}*{$\mathcal{O}_{ u d{}^2 \overline{l} H^{\dagger} D^2 }^{(1 \sim 10)}$}

&\mathcal{Y}\left[\tiny{\young(pr)}\right]\epsilon ^{abc} \epsilon _{ij} \left(\overline{l}_{s}^{i} d_{pa}\right) \left(D^{\mu } u_{tc} C D_{\mu } d_{rb}\right) H^{\dagger}{}^{j}

,\quad\mathcal{Y}\left[\tiny{\young(pr)}\right]\epsilon ^{abc} \epsilon _{ij} \left(\overline{l}_{s}^{i} d_{pa}\right) \left(u_{tc} C D_{\mu } d_{rb}\right) D^{\mu }  H^{\dagger}{}^{j},

\\&\mathcal{Y}\left[\tiny{\young(pr)}\right] \epsilon ^{abc} \epsilon _{ij} \left(D_{\mu } \overline{l}_{s}^{i} D_{\nu } u_{tc} \right) \left(d_{rb} C \sigma ^{\mu }{}^{\nu } d_{pa}\right) H^{\dagger}{}^{j}

,\quad\mathcal{Y}\left[\tiny{\young(pr)}\right]\epsilon ^{abc} \epsilon _{ij} \left(D_{\mu } \overline{l}_{s}^{i} d_{pa}\right) \left(u_{tc} C d_{rb}\right) D^{\mu } H^{\dagger}{}^{j},

\\&\mathcal{Y}\left[\tiny{\young(pr)}\right] \epsilon ^{abc} \epsilon _{ij} \left(\overline{l}_{s}^{i} \sigma ^{\mu }{}^{\nu } d_{pa}\right) \left(D_{\mu } u_{tc} C d_{rb}\right) D_{\nu } H^{\dagger}{}^{j}

,\quad\mathcal{Y}\left[\tiny{\young(p,r)}\right]\epsilon ^{abc} \epsilon _{ij} \left(\overline{l}_{s}^{i} d_{pa}\right) \left(D^{\mu } u_{tc} C D_{\mu } d_{rb}\right) H^{\dagger}{}^{j},

\\&\mathcal{Y}\left[\tiny{\young(p,r)}\right]\epsilon ^{abc} \epsilon _{ij} \left(\overline{l}_{s}^{i} d_{pa}\right) \left(u_{tc} C D_{\mu } d_{rb}\right) D^{\mu }  H^{\dagger}{}^{j}

,\quad\mathcal{Y}\left[\tiny{\young(p,r)}\right]\epsilon ^{abc} \epsilon _{ij} \left(D_{\mu } \overline{l}_{s}^{i} D^{\nu } u_{tc} \right) \left(d_{rb} C d_{pa}\right) H^{\dagger}{}^{j},

\\&\mathcal{Y}\left[\tiny{\young(p,r)}\right]\epsilon ^{abc} \epsilon _{ij} \left(D_{\mu } \overline{l}_{s}^{i} d_{pa}\right) \left(u_{tc} C d_{rb}\right) D^{\mu } H^{\dagger}{}^{j}

,\quad\mathcal{Y}\left[\tiny{\young(p,r)}\right] \epsilon ^{abc} \epsilon _{ij} \left(\overline{l}_{s}^{i} \sigma ^{\mu }{}^{\nu } d_{pa}\right) \left(D_{\mu } u_{tc} C d_{rb}\right) D_{\nu } H^{\dagger}{}^{j}

\vspace{2ex}\\

\multirow{4}*{$\mathcal{O}_{ d{}^3 \overline{l} H  D^2 }^{(1 \sim 7)}$}

&\mathcal{Y}\left[\tiny{\young(prs)}\right] \epsilon ^{abc} \left(D_{\nu } \overline{l}_{t}^{i} D_{\mu } d_{sc} \right) \left(d_{rb} C \sigma ^{\mu }{}^{\nu } d_{pa}\right) H_{i}

,\quad\mathcal{Y}\left[\tiny{\young(prs)}\right]\epsilon ^{abc} \left(\overline{l}_{t}^{i} d_{rb}\right) \left(D_{\mu } d_{sc} C d_{pa}\right) D^{\mu } H_{i},

\\&\mathcal{Y}\left[\tiny{\young(pr,s)}\right]\epsilon ^{abc} \left(D^{\mu } \overline{l}_{t}^{i} D_{\mu } d_{rb} \right) \left(d_{sc} C d_{pa}\right) H_{i}

,\quad\mathcal{Y}\left[\tiny{\young(pr,s)}\right]\epsilon ^{abc} \left(\overline{l}_{t}^{i} D_{\mu } d_{rb} \right) \left(d_{sc} C d_{pa}\right) D^{\mu } H_{i},

\\&\mathcal{Y}\left[\tiny{\young(pr,s)}\right] \epsilon ^{abc} \left(D_{\nu } \overline{l}_{t}^{i} D_{\mu } d_{sc} \right) \left(d_{rb} C \sigma ^{\mu }{}^{\nu } d_{pa}\right) H_{i}

,\quad\mathcal{Y}\left[\tiny{\young(p,r,s)}\right]\epsilon ^{abc} \left(D^{\mu } \overline{l}_{t}^{i} D_{\mu } d_{rb} \right) \left(d_{sc} C d_{pa}\right) H_{i},

\\&\mathcal{Y}\left[\tiny{\young(p,r,s)}\right]\epsilon ^{abc} \left(\overline{l}_{t}^{i} D_{\mu } d_{rb} \right) \left(d_{sc} C d_{pa}\right) D^{\mu } H_{i}

\end{array}\label{cl:q3lhD2}\end{align}

\noindent 2. Operators involving two leptons and two quarks with $\Delta L=2$:
\begin{align}\begin{array}{c|l}

\multirow{7}*{$\mathcal{O}_{ \overline{q} u l {}^2 H D^2  }^{(1 \sim 14)} $}

&\mathcal{Y}\left[\tiny{\young(s,t)}\right]\epsilon ^{jm} \left(\overline{q}_{p}^{a i} \gamma ^{\mu } l_{si}\right) \left(l_{tm} C \gamma ^{\nu } u_{ra}\right) D_{\mu } D_{\nu } H_{j}

,\quad\mathcal{Y}\left[\tiny{\young(s,t)}\right]\epsilon ^{jm} \left(\overline{q}_{p}^{a i} \gamma ^{\mu } D_{\nu } l_{si}\right) \left(l_{tm} C \gamma ^{\nu } u_{ra}\right) D_{\mu } H_{j},

\\&\mathcal{Y}\left[\tiny{\young(s,t)}\right]\epsilon ^{km} \left(\overline{q}_{p}^{a i} \gamma ^{\mu } D_{\nu } l_{sk}\right) \left(l_{tm} C \gamma ^{\nu } u_{ra}\right) D_{\mu } H_{i}

,\quad\mathcal{Y}\left[\tiny{\young(s,t)}\right] \epsilon ^{jm} \left(\overline{q}_{p}^{a i} \sigma ^{\mu }{}^{\nu } u_{ra}\right) \left(D_{\nu } l_{tm} C l_{si}\right) D_{\mu } H_{j},

\\&\mathcal{Y}\left[\tiny{\young(s,t)}\right] \epsilon ^{km} \left(\overline{q}_{p}^{a i} \sigma ^{\mu }{}^{\nu } u_{ra}\right) \left(D_{\nu } l_{tm} C l_{sk}\right) D_{\mu } H_{i}

,\quad\mathcal{Y}\left[\tiny{\young(s,t)}\right] \epsilon ^{jm} \left(\overline{q}_{p}^{a i} \sigma ^{\mu }{}^{\nu } u_{ra}\right) \left(D_{\nu } l_{tm} C D_{\mu } l_{si}\right) H_{j},

\\&\mathcal{Y}\left[\tiny{\young(s,t)}\right]\epsilon ^{jm} \left(\overline{q}_{p}^{a i} u_{ra}\right) \left(D^{\mu } l_{tm} C D_{\mu } l_{si}\right) H_{j}

,\quad\mathcal{Y}\left[\tiny{\young(st)}\right]\epsilon ^{jm} \left(\overline{q}_{p}^{a i} \gamma ^{\mu } l_{si}\right) \left(l_{tm} C \gamma ^{\nu } u_{ra}\right) D_{\mu } D_{\nu } H_{j},

\\&\mathcal{Y}\left[\tiny{\young(st)}\right]\epsilon ^{jm} \left(\overline{q}_{p}^{a i} \gamma ^{\mu } D_{\nu } l_{si}\right) \left(l_{tm} C \gamma ^{\nu } u_{ra}\right) D_{\mu } H_{j}

,\quad\mathcal{Y}\left[\tiny{\young(st)}\right]\epsilon ^{km} \left(\overline{q}_{p}^{a i} \gamma ^{\mu } D_{\nu } l_{sk}\right) \left(l_{tm} C \gamma ^{\nu } u_{ra}\right) D_{\mu } H_{i},

\\&\mathcal{Y}\left[\tiny{\young(st)}\right] \epsilon ^{jm} \left(\overline{q}_{p}^{a i} \sigma ^{\mu }{}^{\nu } u_{ra}\right) \left(D_{\nu } l_{tm} C l_{si}\right) D_{\mu } H_{j}

,\quad\mathcal{Y}\left[\tiny{\young(st)}\right] \epsilon ^{km} \left(\overline{q}_{p}^{a i} \sigma ^{\mu }{}^{\nu } u_{ra}\right) \left(D_{\nu } l_{tm} C l_{sk}\right) D_{\mu } H_{i},

\\&\mathcal{Y}\left[\tiny{\young(st)}\right] \epsilon ^{jm} \left(\overline{q}_{p}^{a i} \sigma ^{\mu }{}^{\nu } u_{ra}\right) \left(D_{\nu } l_{tm} C D_{\mu } l_{si}\right) H_{j}

,\quad\mathcal{Y}\left[\tiny{\young(st)}\right]\epsilon ^{jm} \left(\overline{q}_{p}^{a i} u_{ra}\right) \left(D^{\mu } l_{tm} C D_{\mu } l_{si}\right) H_{j}

\end{array}\label{cl:q2l2hD21}\end{align}
\begin{align}\begin{array}{c|l}

\multirow{10}*{$\mathcal{O}_{ q \overline{d} l{}^2 H D^2 }^{(1 \sim 20)} $}

&\mathcal{Y}\left[\tiny{\young(r,s)}\right]\epsilon ^{ik} \epsilon ^{jm} \left(\overline{d}_{p}^{a} l_{sj}\right) \left(D_{\mu } l_{ri} C D^{\mu } q_{ta k}\right) H_{m}

,\quad\mathcal{Y}\left[\tiny{\young(r,s)}\right]\epsilon ^{ij} \epsilon ^{km} \left(\overline{d}_{p}^{a} l_{sj}\right) \left(D_{\mu } l_{ri} C D^{\mu } q_{ta k}\right) H_{m},

\\&\mathcal{Y}\left[\tiny{\young(r,s)}\right]\epsilon ^{ik} \epsilon ^{jm} \left(\overline{d}_{p}^{a} l_{sj}\right) \left(D_{\mu } l_{ri} C q_{ta k}\right) D^{\mu } H_{m}

,\quad\mathcal{Y}\left[\tiny{\young(r,s)}\right]\epsilon ^{ij} \epsilon ^{km} \left(\overline{d}_{p}^{a} l_{sj}\right) \left(D_{\mu } l_{ri} C q_{ta k}\right) D^{\mu } H_{m},

\\&\mathcal{Y}\left[\tiny{\young(r,s)}\right] \epsilon ^{ik} \epsilon ^{jm} \left(\overline{d}_{p}^{a} \sigma ^{\mu }{}^{\nu } l_{ri}\right) \left(D_{\mu } l_{sj} C D_{\nu } q_{ta k}\right) H_{m}

,\quad\mathcal{Y}\left[\tiny{\young(r,s)}\right] \epsilon ^{ij} \epsilon ^{km} \left(\overline{d}_{p}^{a} \sigma ^{\mu }{}^{\nu } l_{ri}\right) \left(D_{\mu } l_{sj} C D_{\nu } q_{ta k}\right) H_{m},

\\&\mathcal{Y}\left[\tiny{\young(r,s)}\right]\epsilon ^{ik} \epsilon ^{jm} \left(\overline{d}_{p}^{a} D_{\mu } l_{sj}\right) \left(l_{ri} C q_{ta k}\right) D^{\mu } H_{m}

,\quad\mathcal{Y}\left[\tiny{\young(r,s)}\right]\epsilon ^{ij} \epsilon ^{km} \left(\overline{d}_{p}^{a} D_{\mu } l_{sj}\right) \left(l_{ri} C q_{ta k}\right) D^{\mu } H_{m},

\\&\mathcal{Y}\left[\tiny{\young(r,s)}\right] \epsilon ^{ik} \epsilon ^{jm} \left(\overline{d}_{p}^{a} \sigma ^{\mu }{}^{\nu } l_{sj}\right) \left(l_{ri} C D_{\mu } q_{ta k}\right) D_{\nu } H_{m}

,\quad\mathcal{Y}\left[\tiny{\young(r,s)}\right] \epsilon ^{ij} \epsilon ^{km} \left(\overline{d}_{p}^{a} \sigma ^{\mu }{}^{\nu } l_{sj}\right) \left(l_{ri} C D_{\mu } q_{ta k}\right) D_{\nu } H_{m},

\\&\mathcal{Y}\left[\tiny{\young(rs)}\right]\epsilon ^{ik} \epsilon ^{jm} \left(\overline{d}_{p}^{a} l_{sj}\right) \left(D_{\mu } l_{ri} C D^{\mu } q_{ta k}\right) H_{m}

,\quad\mathcal{Y}\left[\tiny{\young(rs)}\right]\epsilon ^{ij} \epsilon ^{km} \left(\overline{d}_{p}^{a} l_{sj}\right) \left(D_{\mu } l_{ri} C D^{\mu } q_{ta k}\right) H_{m},

\\&\mathcal{Y}\left[\tiny{\young(rs)}\right]\epsilon ^{ik} \epsilon ^{jm} \left(\overline{d}_{p}^{a} l_{sj}\right) \left(D_{\mu } l_{ri} C q_{ta k}\right) D^{\mu } H_{m}

,\quad\mathcal{Y}\left[\tiny{\young(rs)}\right]\epsilon ^{ij} \epsilon ^{km} \left(\overline{d}_{p}^{a} l_{sj}\right) \left(D_{\mu } l_{ri} C q_{ta k}\right) D^{\mu } H_{m},

\\&\mathcal{Y}\left[\tiny{\young(rs)}\right] \epsilon ^{ik} \epsilon ^{jm} \left(\overline{d}_{p}^{a} \sigma ^{\mu }{}^{\nu } l_{ri}\right) \left(D_{\mu } l_{sj} C D_{\nu } q_{ta k}\right) H_{m}

,\quad\mathcal{Y}\left[\tiny{\young(rs)}\right] \epsilon ^{ij} \epsilon ^{km} \left(\overline{d}_{p}^{a} \sigma ^{\mu }{}^{\nu } l_{ri}\right) \left(D_{\mu } l_{sj} C D_{\nu } q_{ta k}\right) H_{m},

\\&\mathcal{Y}\left[\tiny{\young(rs)}\right]\epsilon ^{ik} \epsilon ^{jm} \left(\overline{d}_{p}^{a} D_{\mu } l_{sj}\right) \left(l_{ri} C q_{ta k}\right) D^{\mu } H_{m}

,\quad\mathcal{Y}\left[\tiny{\young(rs)}\right]\epsilon ^{ij} \epsilon ^{km} \left(\overline{d}_{p}^{a} D_{\mu } l_{sj}\right) \left(l_{ri} C q_{ta k}\right) D^{\mu } H_{m},

\\&\mathcal{Y}\left[\tiny{\young(rs)}\right] \epsilon ^{ik} \epsilon ^{jm} \left(\overline{d}_{p}^{a} \sigma ^{\mu }{}^{\nu } l_{sj}\right) \left(l_{ri} C D_{\mu } q_{ta k}\right) D_{\nu } H_{m}

,\quad\mathcal{Y}\left[\tiny{\young(rs)}\right] \epsilon ^{ij} \epsilon ^{km} \left(\overline{d}_{p}^{a} \sigma ^{\mu }{}^{\nu } l_{sj}\right) \left(l_{ri} C D_{\mu } q_{ta k}\right) D_{\nu } H_{m}

\vspace{2ex}\\

\multirow{4}*{$\mathcal{O}_{ u \overline{d} l e  H D^2  }^{(1 \sim 7)} $}

& \epsilon ^{ij} \left(\overline{d}_{s}^{a} \gamma ^{\mu } e_{p}\right) \left(l_{tj} C \gamma ^{\nu } u_{ra}\right) D_{\mu } D_{\nu } H_{i}

,\quad \epsilon ^{ij} \left(D_{\nu } \overline{d}_{s}^{a} \gamma ^{\mu } e_{p}\right) \left(l_{tj} C \gamma ^{\nu } u_{ra}\right) D_{\mu } H_{i},

\\& \epsilon ^{ij} \left(\overline{d}_{s}^{a} D_{\nu } l_{tj} \right) \left(u_{ra} C \sigma ^{\mu }{}^{\nu } e_{p}\right) D_{\mu } H_{i}

,\quad \epsilon ^{ij} \left(\overline{d}_{s}^{a} D^{\mu } l_{tj} \right) \left(u_{ra} C e_{p}\right) D_{\mu } H_{i},

\\& \epsilon ^{ij} \left(D_{\mu } \overline{d}_{s}^{a} D_{\nu } l_{tj} \right) \left(u_{ra} C \sigma ^{\mu }{}^{\nu } e_{p}\right) H_{i}

,\quad \epsilon ^{ij} \left(D_{\mu } \overline{d}_{s}^{a} D^{\mu } l_{tj} \right) \left(u_{ra} C e_{p}\right) H_{i},

\\& \epsilon ^{ij} \left(D^{\mu } \overline{d}_{s}^{a} l_{tj} \right) \left(u_{ra} C e_{p}\right) D_{\mu } H_{i}

\end{array}\label{cl:q2l2hD22}\end{align}

\noindent 3. Operators involving only leptons with $\Delta L=2$:
\begin{align}\begin{array}{c|l}

\multirow{7}*{$\mathcal{O}_{  l{}^3 \overline{e} H D^2 }^{(1 \sim 13)} $}

&\mathcal{Y}\left[\tiny{\young(rs,t)}\right]\epsilon ^{ik} \epsilon ^{jm} \left(\overline{e}_{p} l_{sj}\right) \left(D_{\mu } l_{ri} C D^{\mu } l_{tk}\right) H_{m}

,\quad\mathcal{Y}\left[\tiny{\young(rs,t)}\right]\epsilon ^{ik} \epsilon ^{jm} \left(\overline{e}_{p} l_{sj}\right) \left(D_{\mu } l_{ri} C l_{tk}\right) D^{\mu } H_{m},

\\&\mathcal{Y}\left[\tiny{\young(rs,t)}\right]\epsilon ^{ij} \epsilon ^{km} \left(\overline{e}_{p} l_{sj}\right) \left(D_{\mu } l_{ri} C l_{tk}\right) D^{\mu } H_{m}

,\quad\mathcal{Y}\left[\tiny{\young(rs,t)}\right]\epsilon ^{ik} \epsilon ^{jm} \left(\overline{e}_{p} \sigma ^{\mu }{}^{\nu } l_{ri}\right) \left(D_{\mu } l_{sj} C D_{\nu } l_{tk}\right) H_{m},

\\&\mathcal{Y}\left[\tiny{\young(rs,t)}\right]\epsilon ^{ij} \epsilon ^{km} \left(\overline{e}_{p} \sigma ^{\mu }{}^{\nu } l_{ri}\right) \left(D_{\mu } l_{sj} C D_{\nu } l_{tk}\right) H_{m}

,\quad\mathcal{Y}\left[\tiny{\young(rs,t)}\right]\epsilon ^{ik} \epsilon ^{jm} \left(\overline{e}_{p} l_{ri}\right) \left(D_{\mu } l_{sj} C D^{\mu } l_{tk}\right) H_{m},

\\&\mathcal{Y}\left[\tiny{\young(rs,t)}\right]\epsilon ^{ij} \epsilon ^{km} \left(\overline{e}_{p} \sigma ^{\mu }{}^{\nu } l_{tk}\right) \left(l_{ri} C D_{\mu } l_{sj}\right) D_{\nu } H_{m}

,\quad\mathcal{Y}\left[\tiny{\young(rst)}\right]\epsilon ^{ij} \epsilon ^{km} \left(\overline{e}_{p} l_{sj}\right) \left(D_{\mu } l_{ri} C D^{\mu } l_{tk}\right) H_{m},

\\&\mathcal{Y}\left[\tiny{\young(rst)}\right]\epsilon ^{ik} \epsilon ^{jm} \left(\overline{e}_{p} l_{sj}\right) \left(D_{\mu } l_{ri} C l_{tk}\right) D^{\mu } H_{m}

,\quad\mathcal{Y}\left[\tiny{\young(rst)}\right]\epsilon ^{ij} \epsilon ^{km} \left(\overline{e}_{p} l_{sj}\right) \left(D_{\mu } l_{ri} C l_{tk}\right) D^{\mu } H_{m},

\\&\mathcal{Y}\left[\tiny{\young(r,s,t)}\right]\epsilon ^{ik} \epsilon ^{jm} \left(\overline{e}_{p} l_{sj}\right) \left(D_{\mu } l_{ri} C D^{\mu } l_{tk}\right) H_{m}

,\quad\mathcal{Y}\left[\tiny{\young(r,s,t)}\right]\epsilon ^{ik} \epsilon ^{jm} \left(\overline{e}_{p} l_{sj}\right) \left(D_{\mu } l_{ri} C l_{tk}\right) D^{\mu } H_{m},

\vspace{0.5ex}

\\&\mathcal{Y}\left[\tiny{\young(r,s,t)}\right]\epsilon ^{ij} \epsilon ^{km} \left(\overline{e}_{p} l_{sj}\right) \left(D_{\mu } l_{ri} C l_{tk}\right) D^{\mu } H_{m}

\end{array}\label{cl:l4hD2}\end{align}

\noindent\underline{Class $\psi^4 D^3$}: There is only one class here: $\psi ^3 \psi ^{\dagger } D^3$, with independent Lorentz structures
\bea
\psi _1^{\alpha } \left(D \psi _2\right)^{\beta \gamma \dot{\alpha }} \left(D \psi _3\right)_{\alpha \beta \dot{\beta }} (D \psi^{\dagger}_4)_{\gamma }^{\dot{\alpha } \dot{\beta }},\quad
\psi _1^{\alpha } \psi _2^{\beta } (D^2 \psi _3)_{\dot{\alpha } \dot{\beta } \alpha \beta }^{\gamma } (D \psi^{\dagger}_4)_{\gamma }^{\dot{\alpha } \dot{\beta }}.
\eea
\noindent 1. Operators involving three quarks with $\Delta B=1$ and $\Delta L=-1$:
\begin{align}\begin{array}{c|l}
\multirow{1}*{$\mathcal{O}_{ q d{}^2 \overline{l} D^3    }^{(1,2)}$}

&\mathcal{Y}\left[\tiny{\young(pr)}\right]\epsilon ^{abc} \left(D_{\nu } \overline{l}_{s}^{i} D_{\mu } d_{rb} \right) \left(D^{\mu } q_{tc i} C \gamma ^{\nu } d_{pa}\right)

,\quad\mathcal{Y}\left[\tiny{\young(p,r)}\right]\epsilon ^{abc} \left(D_{\nu } \overline{l}_{s}^{i} D_{\mu } d_{rb} \right) \left(D^{\mu } q_{tc i} C \gamma ^{\nu } d_{pa}\right)

\vspace{2ex}\\

\mathcal{O}_{  d{}^3 \overline{e} D^3    }

&\mathcal{Y}\left[\tiny{\young(pr,s)}\right]\epsilon ^{abc} \left(D^{\mu } \overline{e}_{t} \gamma ^{\nu } d_{rb}\right) \left(D_{\mu } D_{\nu } d_{sc} C d_{pa}\right)

\end{array}\label{cl:q3lD3}\end{align}

\noindent 2. Operators involving two leptons and two quarks with $\Delta L=2$:
\begin{align}\begin{array}{c|l}

\multirow{1}*{$\mathcal{O}_{ u \overline{d} l{}^2 D^3    }^{(1,2)} $}

&\mathcal{Y}\left[\tiny{\young(rs)}\right]\epsilon ^{ij} \left(\overline{d}_{p}^{a} \gamma ^{\nu } D^{\mu } u_{ta}\right) \left(D_{\mu } l_{ri} C D_{\nu } l_{sj}\right)

,\quad\mathcal{Y}\left[\tiny{\young(r,s)}\right]\epsilon ^{ij} \left(\overline{d}_{p}^{a} \gamma ^{\nu } D^{\mu } u_{ta}\right) \left(D_{\mu } l_{ri} C D_{\nu } l_{sj}\right)

\end{array}\label{cl:q2l2D3}\end{align}

\subsubsection{One guage boson involved}
\noindent\underline{Class $F \psi^4 \phi $}: There are two class involved: $F_L \psi ^4 \phi$ and $F_L \psi ^2 \psi ^{\dagger 2} \phi$, and the independent Lorentz structures are
\begin{align}
&F_{\rm{L}1}{}{}^{\alpha \beta } \psi _2{}{}^{\gamma } \psi _3{}_{\alpha } \psi _4{}_{\beta } \psi _5{}_{\gamma } \phi _6,\quad
F_{\rm{L}1}{}{}^{\alpha \beta } \psi _2{}_{\alpha } \psi _3{}{}^{\gamma } \psi _4{}_{\beta } \psi _5{}_{\gamma } \phi _6, \quad 
F_{\rm{L}1}{}{}^{\alpha \beta } \psi _2{}_{\alpha } \psi _3{}_{\beta } \psi _4{}{}^{\gamma } \psi _5{}_{\gamma } \phi _6,\\
&F_{\rm{L}1}{}{}^{\alpha \beta }\psi _2{}_{\alpha } \psi _3{}_{\beta } \phi _4  \psi^{\dagger}_5{}_{\dot{\alpha }} \psi^{\dagger}_6{}{}^{\dot{\alpha }}.
\end{align}
Via a simple relation
$
    F_{\rm{R}\mu\nu}\sigma^{\mu\nu}_{\alpha\beta}=0,\; F_{\mu\nu}\sigma^{\mu\nu}_{\alpha\beta}=F_{\rm{L}\mu\nu}\sigma^{\mu\nu}_{\alpha\beta},
$
we replace all $F_{\rm L}$ with $F$. All types follow this replacing rule.

\noindent 1. Operators involving three quarks with $\Delta B=1$ and $\Delta L=-1$:
\begin{align}\begin{array}{c|l}

\multirow{4}*{$\mathcal{O}_{G q{}^2 d  \overline{l}  H^{\dagger}   }^{(1\sim 8)}$}

&\mathcal{Y}\left[\tiny{\young(pr)}\right]\epsilon ^{ace} \left(\lambda ^A\right)_e^b   G^{A}_{\mu\nu } \left(\overline{l}_{t}^{j} d_{sc}\right) \left(q_{pa i} C \sigma ^{\mu }{}^{\nu } q_{rb j}\right) H^{\dagger}{}^{i}

,\quad\mathcal{Y}\left[\tiny{\young(pr)}\right]\epsilon ^{ace} \left(\lambda ^A\right)_e^b   G^{A}_{\mu\nu } \left(\overline{l}_{t}^{i} d_{sc}\right) \left(q_{pa i} C \sigma ^{\mu }{}^{\nu } q_{rb j}\right) H^{\dagger}{}^{j},

\\&\mathcal{Y}\left[\tiny{\young(pr)}\right]\epsilon ^{ace} \left(\lambda ^A\right)_e^b  G^{A}_{\mu\nu } \left(\overline{l}_{t}^{i} \sigma ^{\mu }{}^{\nu } d_{sa}\right) \left(q_{rc j} C q_{pb i}\right) H^{\dagger}{}^{j}

,\quad\mathcal{Y}\left[\tiny{\young(pr)}\right]\epsilon ^{ace} \left(\lambda ^A\right)_e^b  G^{A}_{\mu\nu } \left(\overline{l}_{t}^{i} \sigma ^{\mu }{}^{\nu } d_{sa}\right) \left(q_{rc i} C q_{pb j}\right) H^{\dagger}{}^{j},

\\&\mathcal{Y}\left[\tiny{\young(p,r)}\right]\epsilon ^{ace} \left(\lambda ^A\right)_e^b   G^{A}_{\mu\nu } \left(\overline{l}_{t}^{j} d_{sc}\right) \left(q_{pa i} C \sigma ^{\mu }{}^{\nu } q_{rb j}\right) H^{\dagger}{}^{i}

,\quad\mathcal{Y}\left[\tiny{\young(p,r)}\right]\epsilon ^{ace} \left(\lambda ^A\right)_e^b   G^{A}_{\mu\nu } \left(\overline{l}_{t}^{i} d_{sc}\right) \left(q_{pa i} C \sigma ^{\mu }{}^{\nu } q_{rb j}\right) H^{\dagger}{}^{j},

\\&\mathcal{Y}\left[\tiny{\young(p,r)}\right]\epsilon ^{ace} \left(\lambda ^A\right)_e^b  G^{A}_{\mu\nu } \left(\overline{l}_{t}^{i} \sigma ^{\mu }{}^{\nu } d_{sa}\right) \left(q_{rc j} C q_{pb i}\right) H^{\dagger}{}^{j}

,\quad\mathcal{Y}\left[\tiny{\young(p,r)}\right]\epsilon ^{ace} \left(\lambda ^A\right)_e^b  G^{A}_{\mu\nu } \left(\overline{l}_{t}^{i} \sigma ^{\mu }{}^{\nu } d_{sa}\right) \left(q_{rc i} C q_{pb j}\right) H^{\dagger}{}^{j}

\vspace{2ex}\\

\multirow{3}*{$\mathcal{O}_{W q{}^2 d \overline{l} H^{\dagger}   }^{(1\sim 6)}$}

&\mathcal{Y}\left[\tiny{\young(pr)}\right]\epsilon ^{abc}  \left(\tau ^I\right)_k^j W^{I}_{\mu\nu } \left(\overline{l}_{t}^{i} d_{sc}\right) \left(q_{pa i} C \sigma ^{\mu }{}^{\nu } q_{rb j}\right) H^{\dagger}{}^{k}

,\quad\mathcal{Y}\left[\tiny{\young(pr)}\right]\epsilon ^{abc}  \left(\tau ^I\right)_m^j W^{I}_{\mu\nu } \left(\overline{l}_{t}^{m} d_{sc}\right) \left(q_{pa i} C \sigma ^{\mu }{}^{\nu } q_{rb j}\right) H^{\dagger}{}^{i},

\\&\mathcal{Y}\left[\tiny{\young(pr)}\right]\epsilon ^{abc} \left(\tau ^I\right)_j^m  W^{I}_{\mu\nu } \left(\overline{l}_{t}^{i} \sigma ^{\mu }{}^{\nu } d_{sa}\right) \left(q_{rc m} C q_{pb i}\right) H^{\dagger}{}^{j}

,\quad\mathcal{Y}\left[\tiny{\young(p,r)}\right]\epsilon ^{abc}  \left(\tau ^I\right)_k^j W^{I}_{\mu\nu } \left(\overline{l}_{t}^{i} d_{sc}\right) \left(q_{pa i} C \sigma ^{\mu }{}^{\nu } q_{rb j}\right) H^{\dagger}{}^{k},

\\&\mathcal{Y}\left[\tiny{\young(s,t)}\right]\epsilon ^{abc} \left(\tau ^I\right)_j^m  W^{I}_{\mu\nu } \left(\overline{l}_{t}^{i} \sigma ^{\mu }{}^{\nu } d_{sa}\right) \left(q_{rc m} C q_{pb i}\right) H^{\dagger}{}^{j}

,\quad\mathcal{Y}\left[\tiny{\young(s,t)}\right]\epsilon ^{abc} \left(\tau ^I\right)_i^m  W^{I}_{\mu\nu } \left(\overline{l}_{t}^{i} \sigma ^{\mu }{}^{\nu } d_{sa}\right) \left(q_{rc m} C q_{pb j}\right) H^{\dagger}{}^{j}

\vspace{2ex}\\

\multirow{2}*{$\mathcal{O}_{B q{}^2 d \overline{l} H^{\dagger}   }^{(1 \sim 4)}$}

&\mathcal{Y}\left[\tiny{\young(pr)}\right]\epsilon ^{abc} B_{\mu\nu }  \left(\overline{l}_{t}^{j} d_{sc}\right) \left(q_{pa i} C \sigma ^{\mu }{}^{\nu } q_{rb j}\right) H^{\dagger}{}^{i}

,\quad\mathcal{Y}\left[\tiny{\young(pr)}\right]\epsilon ^{abc}  B_{\mu\nu } \left(\overline{l}_{t}^{i} \sigma ^{\mu }{}^{\nu } d_{sa}\right) \left(q_{rc j} C q_{pb i}\right) H^{\dagger}{}^{j},

\\&\mathcal{Y}\left[\tiny{\young(p,r)}\right]\epsilon ^{abc} B_{\mu\nu }  \left(\overline{l}_{t}^{j} d_{sc}\right) \left(q_{pa i} C \sigma ^{\mu }{}^{\nu } q_{rb j}\right) H^{\dagger}{}^{i}

,\quad\mathcal{Y}\left[\tiny{\young(s,t)}\right]\epsilon ^{abc}  B_{\mu\nu } \left(\overline{l}_{t}^{i} \sigma ^{\mu }{}^{\nu } d_{sa}\right) \left(q_{rc j} C q_{pb i}\right) H^{\dagger}{}^{j}

\vspace{2ex}\\

\multirow{2}*{$\mathcal{O}_{G q d {}^2 \overline{e} H^{\dagger}   }^{(1 \sim 4)}$}

&\mathcal{Y}\left[\tiny{\young(st)}\right]\epsilon ^{ace} \left(\lambda ^A\right)_e^b  G^{A}_{\mu\nu } \left(\overline{e}_{p} \sigma ^{\mu }{}^{\nu } q_{ra i}\right) \left(d_{sb} C d_{tc}\right) H^{\dagger}{}^{i}

,\quad\mathcal{Y}\left[\tiny{\young(st)}\right]\epsilon ^{ace} \left(\lambda ^A\right)_e^b G^{A}_{\mu\nu } \left(\overline{e}_{p} q_{rc i} \right) \left(d_{tb} C \sigma ^{\mu }{}^{\nu } d_{sa}\right) H^{\dagger}{}^{i},

\\&\mathcal{Y}\left[\tiny{\young(s,t)}\right]\epsilon ^{ace} \left(\lambda ^A\right)_e^b G^{A}_{\mu\nu } \left(\overline{e}_{p} \sigma ^{\mu }{}^{\nu } q_{ra i}\right) \left(d_{sb} C d_{tc}\right) H^{\dagger}{}^{i}

,\quad\mathcal{Y}\left[\tiny{\young(s,t)}\right]\epsilon ^{ace} \left(\lambda ^A\right)_e^b G^{A}_{\mu\nu } \left(\overline{e}_{p} q_{rc i} \right) \left(d_{tb} C \sigma ^{\mu }{}^{\nu } d_{sa}\right) H^{\dagger}{}^{i}

\vspace{2ex}\\

\multirow{1}*{$\mathcal{O}_{W q d {}^2 \overline{e}  H^{\dagger}   }^{(1,2)}$}

&\mathcal{Y}\left[\tiny{\young(s,t)}\right]\epsilon ^{abc} \left(\tau ^I\right)_j^i  W^{I}_{\mu\nu } \left(\overline{e}_{p} \sigma ^{\mu }{}^{\nu } q_{ra i}\right) \left(d_{sb} C d_{tc}\right) H^{\dagger}{}^{j}

,\quad\mathcal{Y}\left[\tiny{\young(st)}\right]\epsilon ^{abc} \left(\tau ^I\right)_i^j W^{I}_{\mu\nu } \left(\overline{e}_{p} q_{rc j} \right) \left(d_{tb} C \sigma ^{\mu }{}^{\nu } d_{sa}\right) H^{\dagger}{}^{i}

\vspace{2ex}\\

\multirow{1}*{$\mathcal{O}_{B q d {}^2 \overline{e}  H^{\dagger}   }^{(1,2)}$}

&\mathcal{Y}\left[\tiny{\young(s,t)}\right]\epsilon ^{abc} B_{\mu\nu } \left(\overline{e}_{p} \sigma ^{\mu }{}^{\nu } q_{ra i}\right) \left(d_{sb} C d_{tc}\right) H^{\dagger}{}^{i}

,\quad\mathcal{Y}\left[\tiny{\young(st)}\right]\epsilon ^{abc} B_{\mu\nu } \left(\overline{e}_{p} q_{rc i} \right) \left(d_{tb} C \sigma ^{\mu }{}^{\nu } d_{sa}\right) H^{\dagger}{}^{i}

\vspace{2ex}\\

\multirow{3}*{$\mathcal{O}_{G u d{}^2 \overline{l}  H^{\dagger}  }^{(1 \sim 6)}$}

&\mathcal{Y}\left[\tiny{\young(p,r)}\right]\epsilon ^{ace} \epsilon _{ij} \left(\lambda ^A\right)_e^b G^{A}_{\mu\nu } \left(\overline{l}_{s}^{i} \sigma ^{\mu }{}^{\nu } d_{rb}\right) \left(u_{tc} C d_{pa}\right) H^{\dagger}{}^{j}

,\quad\mathcal{Y}\left[\tiny{\young(p,r)}\right]\epsilon ^{abe} \epsilon _{ij} \left(\lambda ^A\right)_e^c G^{A}_{\mu\nu } \left(\overline{l}_{s}^{i} \sigma ^{\mu }{}^{\nu } d_{rb}\right) \left(u_{tc} C d_{pa}\right) H^{\dagger}{}^{j},

\\&\mathcal{Y}\left[\tiny{\young(p,r)}\right]\epsilon ^{ace} \epsilon _{ij} \left(\lambda ^A\right)_e^b G^{A}_{\mu\nu } \left(\overline{l}_{s}^{i} u_{tc} \right) \left(d_{rb} C \sigma ^{\mu }{}^{\nu } d_{pa}\right) H^{\dagger}{}^{j}

,\quad\mathcal{Y}\left[\tiny{\young(pr)}\right]\epsilon ^{ace} \epsilon _{ij} \left(\lambda ^A\right)_e^b G^{A}_{\mu\nu } \left(\overline{l}_{s}^{i} \sigma ^{\mu }{}^{\nu } d_{rb}\right) \left(u_{tc} C d_{pa}\right) H^{\dagger}{}^{j},

\\&\mathcal{Y}\left[\tiny{\young(pr)}\right]\epsilon ^{abe} \epsilon _{ij} \left(\lambda ^A\right)_e^c G^{A}_{\mu\nu } \left(\overline{l}_{s}^{i} \sigma ^{\mu }{}^{\nu } d_{rb}\right) \left(u_{tc} C d_{pa}\right) H^{\dagger}{}^{j}

,\quad\mathcal{Y}\left[\tiny{\young(pr)}\right]\epsilon ^{ace} \epsilon _{ij} \left(\lambda ^A\right)_e^b G^{A}_{\mu\nu } \left(\overline{l}_{s}^{i} u_{tc} \right) \left(d_{rb} C \sigma ^{\mu }{}^{\nu } d_{pa}\right) H^{\dagger}{}^{j}

\vspace{2ex}\\

\multirow{2}*{$\mathcal{O}_{W u d{}^2 \overline{l}  H^{\dagger}  }^{(1 \sim 3)}$}

&\mathcal{Y}\left[\tiny{\young(pr)}\right]\epsilon ^{abc} \epsilon _{jk} \left(\tau ^I\right)_i^k W^{I}_{\mu\nu } \left(\overline{l}_{s}^{i} \sigma ^{\mu }{}^{\nu } d_{rb}\right) \left(u_{tc} C d_{pa}\right) H^{\dagger}{}^{j}

,\quad\mathcal{Y}\left[\tiny{\young(pr)}\right]\epsilon ^{abc} \epsilon _{jk} \left(\tau ^I\right)_i^k W^{I}_{\mu\nu } \left(\overline{l}_{s}^{i} u_{tc} \right) \left(d_{rb} C \sigma ^{\mu }{}^{\nu } d_{pa}\right) H^{\dagger}{}^{j},

\\&\mathcal{Y}\left[\tiny{\young(p,r)}\right]\epsilon ^{abc} \epsilon _{jk} \left(\tau ^I\right)_i^k W^{I}_{\mu\nu } \left(\overline{l}_{s}^{i} \sigma ^{\mu }{}^{\nu } d_{rb}\right) \left(u_{tc} C d_{pa}\right) H^{\dagger}{}^{j}

\vspace{2ex}\\

\multirow{2}*{$\mathcal{O}_{B u d{}^2 \overline{l}  H^{\dagger}  }^{(1 \sim 3)}$}

&\mathcal{Y}\left[\tiny{\young(pr)}\right]\epsilon ^{abc} \epsilon _{ij} B_{\mu\nu } \left(\overline{l}_{s}^{i} \sigma ^{\mu }{}^{\nu } d_{rb}\right) \left(u_{tc} C d_{pa}\right) H^{\dagger}{}^{j}

,\quad\mathcal{Y}\left[\tiny{\young(pr)}\right]\epsilon ^{abc} \epsilon _{ij} B_{\mu\nu } \left(\overline{l}_{s}^{i} u_{tc} \right) \left(d_{rb} C \sigma ^{\mu }{}^{\nu } d_{pa}\right) H^{\dagger}{}^{j},

\\&\mathcal{Y}\left[\tiny{\young(p,r)}\right]\epsilon ^{abc} \epsilon _{ij} B_{\mu\nu } \left(\overline{l}_{s}^{i} \sigma ^{\mu }{}^{\nu } d_{rb}\right) \left(u_{tc} C d_{pa}\right) H^{\dagger}{}^{j}

\vspace{2ex}\\

\multirow{2}*{$\mathcal{O}_{G d{}^3 \overline{l} H   }^{(1 \sim 4)}$}

&\mathcal{Y}\left[\tiny{\young(pr,s)}\right]\epsilon ^{ace} \left(\lambda ^A\right)_e^b  G^{A}_{\mu\nu } \left(\overline{l}_{t}^{i} d_{pa} \right) \left(d_{sc} C \sigma ^{\mu }{}^{\nu } d_{rb}\right) H_{i}

,\quad\mathcal{Y}\left[\tiny{\young(pr,s)}\right]\epsilon ^{abe} \left(\lambda ^A\right)_e^c  G^{A}_{\mu\nu } \left(\overline{l}_{t}^{i} d_{pa} \right) \left(d_{sc} C \sigma ^{\mu }{}^{\nu } d_{rb}\right) H_{i},

\\&\mathcal{Y}\left[\tiny{\young(prs)}\right]\epsilon ^{ace} \left(\lambda ^A\right)_e^b  G^{A}_{\mu\nu } \left(\overline{l}_{t}^{i} d_{pa} \right) \left(d_{sc} C \sigma ^{\mu }{}^{\nu } d_{rb}\right) H_{i}

,\quad\mathcal{Y}\left[\tiny{\young(p,r,s)}\right]\epsilon ^{ace} \left(\lambda ^A\right)_e^b  G^{A}_{\mu\nu } \left(\overline{l}_{t}^{i} d_{pa} \right) \left(d_{sc} C \sigma ^{\mu }{}^{\nu } d_{rb}\right) H_{i}

\vspace{2ex}\\

\multirow{1}*{$\mathcal{O}_{W d{}^3 \overline{l} H   }^{(1,2)}$}

&\mathcal{Y}\left[\tiny{\young(prs)}\right]\epsilon ^{abc} \left(\tau ^I\right)_i^j  W^{I}_{\mu\nu } \left(\overline{l}_{t}^{i} d_{pa} \right) \left(d_{sc} C \sigma ^{\mu }{}^{\nu } d_{rb}\right) H_{j}

,\quad\mathcal{Y}\left[\tiny{\young(pr,s)}\right]\epsilon ^{abc} \left(\tau ^I\right)_i^j  W^{I}_{\mu\nu } \left(\overline{l}_{t}^{i} d_{pa} \right) \left(d_{sc} C \sigma ^{\mu }{}^{\nu } d_{rb}\right) H_{j}

\vspace{2ex}\\

\multirow{1}*{$\mathcal{O}_{B d{}^3 \overline{l} H   }^{(1,2)}$}

&\mathcal{Y}\left[\tiny{\young(prs)}\right]\epsilon ^{abc} B_{\mu\nu } \left(\overline{l}_{t}^{i} d_{pa} \right) \left(d_{sc} C \sigma ^{\mu }{}^{\nu } d_{rb}\right) H_{i}

,\quad\mathcal{Y}\left[\tiny{\young(pr,s)}\right]\epsilon ^{abc} B_{\mu\nu } \left(\overline{l}_{t}^{i} d_{pa} \right) \left(d_{sc} C \sigma ^{\mu }{}^{\nu } d_{rb}\right) H_{i}

\end{array}\label{cl:fq3lh}\end{align}

\noindent 2. Operators involving two leptons and two quarks with $\Delta L=2$:
\begin{align}\begin{array}{c|l}

\multirow{2}*{$\mathcal{O}_{G \overline{q} u l {}^2  H   }^{(1 \sim 4)} $}

&\mathcal{Y}\left[\tiny{\young(st)}\right]\epsilon ^{jm} \left(\lambda ^A\right)_a^b  G^{A}_{\mu\nu } \left(\overline{q}_{p}^{a i} \sigma ^{\mu }{}^{\nu } u_{rb}\right) \left(l_{tm} C l_{si}\right) H_{j}

,\quad\mathcal{Y}\left[\tiny{\young(st)}\right]  \epsilon ^{ik} \left(\lambda ^A\right)_a^b G^{A}_{\mu\nu } \left(\overline{q}_{p}^{a j} u_{rb} \right) \left(l_{si} C \sigma ^{\mu }{}^{\nu } l_{tj}\right) H_{k},

\\&\mathcal{Y}\left[\tiny{\young(s,t)}\right]\epsilon ^{jm} \left(\lambda ^A\right)_a^b  G^{A}_{\mu\nu } \left(\overline{q}_{p}^{a i} \sigma ^{\mu }{}^{\nu } u_{rb}\right) \left(l_{tm} C l_{si}\right) H_{j}

,\quad\mathcal{Y}\left[\tiny{\young(s,t)}\right]  \epsilon ^{ik} \left(\lambda ^A\right)_a^b G^{A}_{\mu\nu } \left(\overline{q}_{p}^{a j} u_{rb} \right) \left(l_{si} C \sigma ^{\mu }{}^{\nu } l_{tj}\right) H_{k}
\vspace{2ex}\\

\multirow{3}*{$\mathcal{O}_{W \overline{q} u l {}^2  H   }^{(1\sim 6)} $}

&\mathcal{Y}\left[\tiny{\young(st)}\right]\epsilon ^{kn} \left(\tau ^I\right)_n^j  W^{I}_{\mu\nu } \left(\overline{q}_{p}^{a i} \sigma ^{\mu }{}^{\nu } u_{ra}\right) \left(l_{ti} C l_{sk}\right) H_{j}

,\quad\mathcal{Y}\left[\tiny{\young(st)}\right]\epsilon ^{kn} \left(\tau ^I\right)_n^m  W^{I}_{\mu\nu } \left(\overline{q}_{p}^{a i} \sigma ^{\mu }{}^{\nu } u_{ra}\right) \left(l_{tm} C l_{sk}\right) H_{i},

\\&\mathcal{Y}\left[\tiny{\young(st)}\right]\epsilon ^{kn} \left(\tau ^I\right)_n^j W^{I}_{\mu\nu } \left(\overline{q}_{p}^{a i} u_{ra} \right) \left(l_{si} C \sigma ^{\mu }{}^{\nu } l_{tj}\right) H_{k}

,\quad\mathcal{Y}\left[\tiny{\young(s,t)}\right] \epsilon ^{kn} \left(\tau ^I\right)_n^j  W^{I}_{\mu\nu } \left(\overline{q}_{p}^{a i} \sigma ^{\mu }{}^{\nu } u_{ra}\right) \left(l_{ti} C l_{sk}\right) H_{j},

\\&\mathcal{Y}\left[\tiny{\young(s,t)}\right]\epsilon ^{kn} \left(\tau ^I\right)_n^j W^{I}_{\mu\nu } \left(\overline{q}_{p}^{a i} u_{ra} \right) \left(l_{si} C \sigma ^{\mu }{}^{\nu } l_{tj}\right) H_{k}

,\quad\mathcal{Y}\left[\tiny{\young(s,t)}\right]\epsilon ^{jn} \left(\tau ^I\right)_n^i W^{I}_{\mu\nu } \left(\overline{q}_{p}^{a k} u_{ra} \right) \left(l_{si} C \sigma ^{\mu }{}^{\nu } l_{tj}\right) H_{k}

\vspace{2ex}\\

\multirow{2}*{$\mathcal{O}_{B \overline{q} u l {}^2  H   }^{(1 \sim 4)} $}

&\mathcal{Y}\left[\tiny{\young(st)}\right]\epsilon ^{jm} B_{\mu\nu } \left(\overline{q}_{p}^{a i} \sigma ^{\mu }{}^{\nu } u_{ra}\right) \left(l_{tm} C l_{si}\right) H_{j}

,\quad\mathcal{Y}\left[\tiny{\young(st)}\right]\epsilon ^{ik} B_{\mu\nu } \left(\overline{q}_{p}^{a j} u_{ra} \right) \left(l_{si} C \sigma ^{\mu }{}^{\nu } l_{tj}\right) H_{k},

\\&\mathcal{Y}\left[\tiny{\young(st)}\right]\epsilon ^{jm} B_{\mu\nu } \left(\overline{q}_{p}^{a i} \sigma ^{\mu }{}^{\nu } u_{ra}\right) \left(l_{tm} C l_{si}\right) H_{j}

,\quad\mathcal{Y}\left[\tiny{\young(st)}\right]\epsilon ^{ik} B_{\mu\nu } \left(\overline{q}_{p}^{a j} u_{ra} \right) \left(l_{si} C \sigma ^{\mu }{}^{\nu } l_{tj}\right) H_{k}

\vspace{2ex}\\

\multirow{3}*{$\mathcal{O}_{G q \overline{d} l{}^2  H  }^{(1 \sim 6)} $}

&\mathcal{Y}\left[\tiny{\young(r,s)}\right]\epsilon ^{ik} \epsilon ^{jm} \left(\lambda ^A\right)_a^b G^{A}_{\mu\nu } \left(\overline{d}_{p}^{a} q_{tb k}\right) \left(l_{ri} C \sigma ^{\mu }{}^{\nu } l_{sj}\right) H_{m}

,\quad\mathcal{Y}\left[\tiny{\young(r,s)}\right]\epsilon ^{ik} \epsilon ^{jm} \left(\lambda ^A\right)_a^b G^{A}_{\mu\nu } \left(\overline{d}_{p}^{a} \sigma _{\mu }{}_{\nu } l_{sj}\right) \left(l_{ri} C q_{tb k}\right) H_{m},

\\&\mathcal{Y}\left[\tiny{\young(r,s)}\right]\epsilon ^{ij} \epsilon ^{km} \left(\lambda ^A\right)_a^b G^{A}_{\mu\nu } \left(\overline{d}_{p}^{a} \sigma _{\mu }{}_{\nu } l_{sj}\right) \left(l_{ri} C q_{tb k}\right) H_{m}

,\quad\mathcal{Y}\left[\tiny{\young(rs)}\right]\epsilon ^{ik} \epsilon ^{jm} \left(\lambda ^A\right)_a^b G^{A}_{\mu\nu } \left(\overline{d}_{p}^{a} q_{tb k}\right) \left(l_{ri} C \sigma ^{\mu }{}^{\nu } l_{sj}\right) H_{m},

\\&\mathcal{Y}\left[\tiny{\young(rs)}\right]\epsilon ^{ik} \epsilon ^{jm} \left(\lambda ^A\right)_a^b G^{A}_{\mu\nu } \left(\overline{d}_{p}^{a} \sigma _{\mu }{}_{\nu } l_{sj}\right) \left(l_{ri} C q_{tb k}\right) H_{m}

,\quad\mathcal{Y}\left[\tiny{\young(rs)}\right]\epsilon ^{ij} \epsilon ^{km} \left(\lambda ^A\right)_a^b G^{A}_{\mu\nu } \left(\overline{d}_{p}^{a} \sigma _{\mu }{}_{\nu } l_{sj}\right) \left(l_{ri} C q_{tb k}\right) H_{m}

\vspace{2ex}\\

\multirow{5}*{$\mathcal{O}_{W q \overline{d} l{}^2  H  }^{(1 \sim 9)} $}

&\mathcal{Y}\left[\tiny{\young(r,s)}\right]\epsilon ^{jn} \epsilon ^{km} \left(\tau ^I\right)_n^i W^{I}_{\mu\nu } \left(\overline{d}_{p}^{a} q_{ta k}\right) \left(l_{ri} C \sigma ^{\mu }{}^{\nu } l_{sj}\right) H_{m}

,\quad\mathcal{Y}\left[\tiny{\young(r,s)}\right]\epsilon ^{ik} \epsilon ^{jn} \left(\tau ^I\right)_n^m W^{I}_{\mu\nu } \left(\overline{d}_{p}^{a} q_{ta k}\right) \left(l_{ri} C \sigma ^{\mu }{}^{\nu } l_{sj}\right) H_{m},

\\&\mathcal{Y}\left[\tiny{\young(r,s)}\right]\epsilon ^{jn} \epsilon ^{km} \left(\tau ^I\right)_n^i W^{I}_{\mu\nu } \left(\overline{d}_{p}^{a} \sigma _{\mu }{}_{\nu } l_{sj}\right) \left(l_{ri} C q_{ta k}\right) H_{m}

,\quad\mathcal{Y}\left[\tiny{\young(r,s)}\right]\epsilon ^{ik} \epsilon ^{jn} \left(\tau ^I\right)_n^m W^{I}_{\mu\nu } \left(\overline{d}_{p}^{a} \sigma _{\mu }{}_{\nu } l_{sj}\right) \left(l_{ri} C q_{ta k}\right) H_{m},

\\&\mathcal{Y}\left[\tiny{\young(r,s)}\right]\epsilon ^{in} \epsilon ^{jk} \left(\tau ^I\right)_n^m W^{I}_{\mu\nu } \left(\overline{d}_{p}^{a} \sigma _{\mu }{}_{\nu } l_{sj}\right) \left(l_{ri} C q_{ta k}\right) H_{m}

,\quad\mathcal{Y}\left[\tiny{\young(rs)}\right]\epsilon ^{ik} \epsilon ^{jn} \left(\tau ^I\right)_n^m W^{I}_{\mu\nu } \left(\overline{d}_{p}^{a} q_{ta k}\right) \left(l_{ri} C \sigma ^{\mu }{}^{\nu } l_{sj}\right) H_{m},

\\&\mathcal{Y}\left[\tiny{\young(rs)}\right]\epsilon ^{jn} \epsilon ^{km} \left(\tau ^I\right)_n^i W^{I}_{\mu\nu } \left(\overline{d}_{p}^{a} \sigma _{\mu }{}_{\nu } l_{sj}\right) \left(l_{ri} C q_{ta k}\right) H_{m}

,\quad\mathcal{Y}\left[\tiny{\young(rs)}\right]\epsilon ^{ik} \epsilon ^{jn} \left(\tau ^I\right)_n^m W^{I}_{\mu\nu } \left(\overline{d}_{p}^{a} \sigma _{\mu }{}_{\nu } l_{sj}\right) \left(l_{ri} C q_{ta k}\right) H_{m},

\\&\mathcal{Y}\left[\tiny{\young(rs)}\right]\epsilon ^{in} \epsilon ^{jk} \left(\tau ^I\right)_n^m W^{I}_{\mu\nu } \left(\overline{d}_{p}^{a} \sigma _{\mu }{}_{\nu } l_{sj}\right) \left(l_{ri} C q_{ta k}\right) H_{m}

\vspace{2ex}\\

\multirow{3}*{$\mathcal{O}_{B q \overline{d} l{}^2  H  }^{(1 \sim 6)} $}

&\mathcal{Y}\left[\tiny{\young(r,s)}\right]\epsilon ^{ik} \epsilon ^{jm} B_{\mu\nu } \left(\overline{d}_{p}^{a} q_{ta k}\right) \left(l_{ri} C \sigma ^{\mu }{}^{\nu } l_{sj}\right) H_{m}

,\quad\mathcal{Y}\left[\tiny{\young(r,s)}\right]\epsilon ^{ik} \epsilon ^{jm} B_{\mu\nu } \left(\overline{d}_{p}^{a} \sigma _{\mu }{}_{\nu } l_{sj}\right) \left(l_{ri} C q_{ta k}\right) H_{m},

\\&\mathcal{Y}\left[\tiny{\young(r,s)}\right]\epsilon ^{ij} \epsilon ^{km} B_{\mu\nu } \left(\overline{d}_{p}^{a} \sigma _{\mu }{}_{\nu } l_{sj}\right) \left(l_{ri} C q_{ta k}\right) H_{m}

,\quad\mathcal{Y}\left[\tiny{\young(rs)}\right]\epsilon ^{ik} \epsilon ^{jm} B_{\mu\nu } \left(\overline{d}_{p}^{a} q_{ta k}\right) \left(l_{ri} C \sigma ^{\mu }{}^{\nu } l_{sj}\right) H_{m},

\\&\mathcal{Y}\left[\tiny{\young(rs)}\right]\epsilon ^{ik} \epsilon ^{jm} B_{\mu\nu } \left(\overline{d}_{p}^{a} \sigma _{\mu }{}_{\nu } l_{sj}\right) \left(l_{ri} C q_{ta k}\right) H_{m}

,\quad\mathcal{Y}\left[\tiny{\young(rs)}\right]\epsilon ^{ij} \epsilon ^{km} B_{\mu\nu } \left(\overline{d}_{p}^{a} \sigma _{\mu }{}_{\nu } l_{sj}\right) \left(l_{ri} C q_{ta k}\right) H_{m}

\vspace{2ex}\\

\multirow{1}*{$\mathcal{O}_{G u \overline{d} l e  H   }^{(1,2)}$} 

& \epsilon ^{ij} \left(\lambda ^A\right)_b^a G^{A}_{\mu\nu } \left(\overline{d}_{s}^{b} l_{tj} \right) \left(u_{ra} C \sigma ^{\mu }{}^{\nu } e_{p}\right) H_{i}

,\quad \epsilon ^{ij} \left(\lambda ^A\right)_a^b G^{A}_{\mu\nu } \left(\overline{d}_{s}^{a} \sigma ^{\mu }{}^{\nu } l_{ti}\right) \left(e_{p} C u_{rb}\right) H_{j}

\vspace{2ex}\\

\multirow{1}*{$\mathcal{O}_{W u \overline{d} l e  H   }^{(1,2)}$}

& \epsilon ^{jk} \left(\tau ^I\right)_k^i W^{I}_{\mu\nu } \left(\overline{d}_{s}^{a} l_{tj} \right) \left(u_{ra} C \sigma ^{\mu }{}^{\nu } e_{p}\right) H_{i}

,\quad \epsilon ^{jk} \left(\tau ^I\right)_k^i W^{I}_{\mu\nu } \left(\overline{d}_{s}^{a} \sigma ^{\mu }{}^{\nu } l_{ti}\right) \left(e_{p} C u_{ra}\right) H_{j}

\vspace{2ex}\\

\multirow{1}*{$\mathcal{O}_{B u \overline{d} l e  H   }^{(1,2)}$} 

& \epsilon ^{ij} B_{\mu\nu } \left(\overline{d}_{s}^{a} l_{tj} \right) \left(u_{ra} C \sigma ^{\mu }{}^{\nu } e_{p}\right) H_{i}

,\quad \epsilon ^{ij} B_{\mu\nu } \left(\overline{d}_{s}^{a} \sigma ^{\mu }{}^{\nu } l_{ti}\right) \left(e_{p} C u_{ra}\right) H_{j}

\end{array}\label{cl:fq2l2h}\end{align}

\noindent 3. Operators involving only leptons with $\Delta L=2$:
\begin{align}\begin{array}{c|l}

\multirow{3}*{$\mathcal{O}_{W l{}^3 \overline{e}  H  }^{(1 \sim 6)} $}

&\mathcal{Y}\left[\tiny{\young(r,s,t)}\right]\epsilon ^{jn} \epsilon ^{km} \left(\tau ^I\right)_n^i W^{I}_{\mu\nu } \left(\overline{e}_{p} l_{tk}\right) \left(l_{ri} C \sigma ^{\mu }{}^{\nu } l_{sj}\right) H_{m}

,\quad\mathcal{Y}\left[\tiny{\young(r,s,t)}\right]\epsilon ^{ik} \epsilon ^{jn} \left(\tau ^I\right)_n^m W^{I}_{\mu\nu } \left(\overline{e}_{p} l_{tk}\right) \left(l_{ri} C \sigma ^{\mu }{}^{\nu } l_{sj}\right) H_{m},

\vspace{0.5ex}

\\&\mathcal{Y}\left[\tiny{\young(rs,t)}\right]\epsilon ^{jn} \epsilon ^{km} \left(\tau ^I\right)_n^i W^{I}_{\mu\nu } \left(\overline{e}_{p} l_{tk}\right) \left(l_{ri} C \sigma ^{\mu }{}^{\nu } l_{sj}\right) H_{m}

,\quad\mathcal{Y}\left[\tiny{\young(rs,t)}\right]\epsilon ^{ik} \epsilon ^{jn} \left(\tau ^I\right)_n^m W^{I}_{\mu\nu } \left(\overline{e}_{p} l_{tk}\right) \left(l_{ri} C \sigma ^{\mu }{}^{\nu } l_{sj}\right) H_{m},

\\&\mathcal{Y}\left[\tiny{\young(rs,t)}\right]\epsilon ^{in} \epsilon ^{jk} \left(\tau ^I\right)_n^m W^{I}_{\mu\nu } \left(\overline{e}_{p} l_{tk}\right) \left(l_{ri} C \sigma ^{\mu }{}^{\nu } l_{sj}\right) H_{m}

,\quad\mathcal{Y}\left[\tiny{\young(rst)}\right]\epsilon ^{ik} \epsilon ^{jn} \left(\tau ^I\right)_n^m W^{I}_{\mu\nu } \left(\overline{e}_{p} l_{tk}\right) \left(l_{ri} C \sigma ^{\mu }{}^{\nu } l_{sj}\right) H_{m}

\vspace{2ex}\\

\multirow{2}*{$\mathcal{O}_{B l{}^3 \overline{e}  H  }^{(1 \sim 4)} $}

&\mathcal{Y}\left[\tiny{\young(rs,t)}\right]\epsilon ^{ik} \epsilon ^{jm} B_{\mu\nu } \left(\overline{e}_{p} l_{tk}\right) \left(l_{ri} C \sigma ^{\mu }{}^{\nu } l_{sj}\right) H_{m}

,\quad\mathcal{Y}\left[\tiny{\young(rs,t)}\right]\epsilon ^{ij} \epsilon ^{km} B_{\mu\nu } \left(\overline{e}_{p} l_{tk}\right) \left(l_{ri} C \sigma ^{\mu }{}^{\nu } l_{sj}\right) H_{m},

\vspace{0.5ex}

\\&\mathcal{Y}\left[\tiny{\young(rst)}\right]\epsilon ^{ik} \epsilon ^{jm} B_{\mu\nu } \left(\overline{e}_{p} l_{tk}\right) \left(l_{ri} C \sigma ^{\mu }{}^{\nu } l_{sj}\right) H_{m}

,\quad\mathcal{Y}\left[\tiny{\young(r,s,t)}\right]\epsilon ^{ik} \epsilon ^{jm} B_{\mu\nu } \left(\overline{e}_{p} l_{tk}\right) \left(l_{ri} C \sigma ^{\mu }{}^{\nu } l_{sj}\right) H_{m}

\end{array}\label{cl:fl4h}\end{align}

\noindent\underline{Class $F \psi^4 D$}: The classes have to be either $F_{\rm R}\psi ^3 \psi ^{\dagger } D$ with 3 Lorentz structure,
\bea
\psi _1^{\alpha } \psi _2^{\beta } \left(D \psi _3\right)_{\alpha \beta \dot{\alpha }} \psi^{\dagger}_4{}_{\dot{\beta }} F_{\rm{R}5}{}{}^{\dot{\alpha } \dot{\beta }},\quad
\psi _1^{\alpha } \psi _2^{\beta } \psi _3{}_{\alpha } (D \psi^{\dagger}_4)_{\beta \dot{\alpha } \dot{\beta }} F_{\rm{R}5}{}{}^{\dot{\alpha } \dot{\beta }},\quad
\psi _1^{\alpha } \psi _2{}_{\alpha } \psi _3^{\beta } (D \psi^{\dagger}_4)_{\beta \dot{\alpha } \dot{\beta }} F_{\rm{R}5}{}{}^{\dot{\alpha } \dot{\beta }},
\eea
or $F_{\rm L} \psi ^3 \psi ^{\dagger } D$ with 4 independent Lorentz structures
\bea
&F_{\rm{L}1}{}{}^{\alpha \beta } \psi _2^{\gamma } \left(D \psi _3\right)_{\alpha \beta \dot{\alpha }} \psi _4{}_{\gamma } \psi^{\dagger}_5{}{}^{\dot{\alpha }},\quad
&F_{\rm{L}1}{}{}^{\alpha \beta } \psi _2^{\gamma } \psi _3{}_{\alpha } \left(D \psi _4\right)_{\beta \gamma \dot{\alpha }} \psi^{\dagger}_5{}{}^{\dot{\alpha }}, \nn \\
&F_{\rm{L}1}{}{}^{\alpha \beta } \psi _2{}_{\alpha } \left(D \psi _3\right)_{\beta \dot{\alpha }}^{\gamma } \psi _4{}_{\gamma } \psi^{\dagger}_5{}{}^{\dot{\alpha }},\quad
&F_{\rm{L}1}{}{}^{\alpha \beta } \psi _2{}_{\alpha } \psi _3^{\gamma } \left(D \psi _4\right)_{\beta \gamma \dot{\alpha }} \psi^{\dagger}_5{}{}^{\dot{\alpha }}.
\eea
After converting to the $F,\tilde{F}$ basis,  these two classes mix together.

\noindent 1. Operators involving three quarks with $\Delta B=1$ and $\Delta L=-1$:
\begin{align}\begin{array}{c|l}

\multirow{7}*{$\mathcal{O}_{G q d{}^2 \overline{l}  D}^{(1\sim 14)}$}

&\mathcal{Y}\left[\tiny{\young(p,r)}\right]\epsilon ^{ace} \left(\lambda ^A\right)_e^b G^{A}{}^{\mu }{}{}_{\nu } \left(\overline{l}_{s}^{i} d_{pa} \right) \left(q_{tc i} C \gamma ^{\nu } D_{\mu } d_{rb}\right),

\quad\mathcal{Y}\left[\tiny{\young(p,r)}\right]\epsilon ^{abe} \left(\lambda ^A\right)_e^c G^{A}{}^{\mu }{}{}_{\nu } \left(\overline{l}_{s}^{i} d_{pa} \right) \left(q_{tc i} C \gamma ^{\nu } D_{\mu } d_{rb}\right),

\\&\mathcal{Y}\left[\tiny{\young(p,r)}\right]\epsilon ^{ace} \left(\lambda ^A\right)_e^b G^{A}{}^{\mu }{}{}_{\nu } \left(D_{\mu } \overline{l}_{s}^{i} d_{pa} \right) \left(q_{tc i} C \gamma ^{\nu } d_{rb}\right),

\quad\mathcal{Y}\left[\tiny{\young(p,r)}\right]\epsilon ^{abe} \left(\lambda ^A\right)_e^c G^{A}{}^{\mu }{}{}_{\nu } \left(D_{\mu } \overline{l}_{s}^{i} d_{pa} \right) \left(q_{tc i} C \gamma ^{\nu } d_{rb}\right),

\\&\mathcal{Y}\left[\tiny{\young(p,r)}\right]\epsilon ^{ace} \left(\lambda ^A\right)_e^b \tilde{G}^{A}{}^{\mu }{}{}_{\nu } \left(D_{\mu } \overline{l}_{s}^{i} d_{rb} \right) \left(q_{tc i} C \gamma ^{\nu } d_{pa}\right),

\quad\mathcal{Y}\left[\tiny{\young(p,r)}\right]\epsilon ^{ace} \left(\lambda ^A\right)_e^b \tilde{G}^{A}{}^{\mu }{}{}_{\nu } \left(\overline{l}_{s}^{i} d_{pa}\right) \left(D_{\mu } q_{tc i} C \gamma ^{\nu } d_{rb}\right),

\\&\mathcal{Y}\left[\tiny{\young(p,r)}\right]\epsilon ^{bce} \left(\lambda ^A\right)_e^a \tilde{G}^{A}{}^{\mu }{}{}_{\nu } \left(\overline{l}_{s}^{i} d_{pa}\right) \left(D_{\mu } q_{tc i} C \gamma ^{\nu } d_{rb}\right),

\quad\mathcal{Y}\left[\tiny{\young(pr)}\right]\epsilon ^{ace} \left(\lambda ^A\right)_e^b G^{A}{}^{\mu }{}{}_{\nu } \left(\overline{l}_{s}^{i} d_{pa} \right) \left(q_{tc i} C \gamma ^{\nu } D_{\mu } d_{rb}\right),

\\&\mathcal{Y}\left[\tiny{\young(pr)}\right]\epsilon ^{abe} \left(\lambda ^A\right)_e^c G^{A}{}^{\mu }{}{}_{\nu } \left(\overline{l}_{s}^{i} d_{pa} \right) \left(q_{tc i} C \gamma ^{\nu } D_{\mu } d_{rb}\right),

\quad\mathcal{Y}\left[\tiny{\young(pr)}\right]\epsilon ^{ace} \left(\lambda ^A\right)_e^b G^{A}{}^{\mu }{}{}_{\nu } \left(D_{\mu } \overline{l}_{s}^{i} d_{pa} \right) \left(q_{tc i} C \gamma ^{\nu } d_{rb}\right),

\\&\mathcal{Y}\left[\tiny{\young(pr)}\right]\epsilon ^{abe} \left(\lambda ^A\right)_e^c G^{A}{}^{\mu }{}{}_{\nu } \left(D_{\mu } \overline{l}_{s}^{i} d_{pa} \right) \left(q_{tc i} C \gamma ^{\nu } d_{rb}\right),

\quad\mathcal{Y}\left[\tiny{\young(pr)}\right]\epsilon ^{ace} \left(\lambda ^A\right)_e^b \tilde{G}^{A}{}^{\mu }{}{}_{\nu } \left(D_{\mu } \overline{l}_{s}^{i} d_{rb} \right) \left(q_{tc i} C \gamma ^{\nu } d_{pa}\right),

\\&\mathcal{Y}\left[\tiny{\young(pr)}\right]\epsilon ^{ace} \left(\lambda ^A\right)_e^b \tilde{G}^{A}{}^{\mu }{}{}_{\nu } \left(\overline{l}_{s}^{i} d_{pa}\right) \left(D_{\mu } q_{tc i} C \gamma ^{\nu } d_{rb}\right),

\quad\mathcal{Y}\left[\tiny{\young(pr)}\right]\epsilon ^{bce} \left(\lambda ^A\right)_e^a \tilde{G}^{A}{}^{\mu }{}{}_{\nu } \left(\overline{l}_{s}^{i} d_{pa}\right) \left(D_{\mu } q_{tc i} C \gamma ^{\nu } d_{rb}\right)

\vspace{2ex}\\

\multirow{4}*{$\mathcal{O}_{W q d{}^2 \overline{l}  D}^{(1\sim 7)}$}

&\mathcal{Y}\left[\tiny{\young(pr)}\right]\epsilon ^{abc} \left(\tau ^I\right)_i^j W^{I}{}^{\mu }{}{}_{\nu } \left(\overline{l}_{s}^{i} d_{pa} \right) \left(q_{tc j} C \gamma ^{\nu } D_{\mu } d_{rb}\right)

,\quad\mathcal{Y}\left[\tiny{\young(pr)}\right]\epsilon ^{abc} \left(\tau ^I\right)_i^j W^{I}{}^{\mu }{}{}_{\nu } \left(D_{\mu } \overline{l}_{s}^{i} d_{pa} \right) \left(q_{tc j} C \gamma ^{\nu } d_{rb}\right),

\\&\mathcal{Y}\left[\tiny{\young(pr)}\right]\epsilon ^{abc} \left(\tau ^I\right)_i^j \tilde{W}^{I}{}^{\mu }{}{}_{\nu } \left(D_{\mu } \overline{l}_{s}^{i} d_{rb}\right) \left(q_{tc j} C \gamma ^{\nu } d_{pa}\right)

,\quad\mathcal{Y}\left[\tiny{\young(pr)}\right]\epsilon ^{abc} \left(\tau ^I\right)_i^j \tilde{W}^{I}{}^{\mu }{}{}_{\nu } \left(\overline{l}_{s}^{i} d_{pa}\right) \left(D_{\mu } q_{tc j} C \gamma ^{\nu } d_{rb}\right),

\\&\mathcal{Y}\left[\tiny{\young(p,r)}\right]\epsilon ^{abc} \left(\tau ^I\right)_i^j W^{I}{}^{\mu }{}{}_{\nu } \left(\overline{l}_{s}^{i} d_{pa} \right) \left(q_{tc j} C \gamma ^{\nu } D_{\mu } d_{rb}\right)

,\quad\mathcal{Y}\left[\tiny{\young(p,r)}\right]\epsilon ^{abc} \left(\tau ^I\right)_i^j W^{I}{}^{\mu }{}{}_{\nu } \left(D_{\mu } \overline{l}_{s}^{i} d_{pa} \right) \left(q_{tc j} C \gamma ^{\nu } d_{rb}\right),

\\&\mathcal{Y}\left[\tiny{\young(p,r)}\right]\epsilon ^{abc} \left(\tau ^I\right)_i^j \tilde{W}^{I}{}^{\mu }{}{}_{\nu } \left(\overline{l}_{s}^{i} d_{pa}\right) \left(D_{\mu } q_{tc j} C \gamma ^{\nu } d_{rb}\right)

\vspace{2ex}\\

\multirow{4}*{$\mathcal{O}_{B q d{}^2 \overline{l}  D}^{(1\sim 7)}$}

&\mathcal{Y}\left[\tiny{\young(pr)}\right]\epsilon ^{abc} B{}^{\mu }{}{}_{\nu } \left(\overline{l}_{s}^{i} d_{pa} \right) \left(q_{tc i} C \gamma ^{\nu } D_{\mu } d_{rb}\right)

,\quad\mathcal{Y}\left[\tiny{\young(pr)}\right]\epsilon ^{abc} B{}^{\mu }{}{}_{\nu } \left(D_{\mu } \overline{l}_{s}^{i} d_{pa} \right) \left(q_{tc i} C \gamma ^{\nu } d_{rb}\right),

\\&\mathcal{Y}\left[\tiny{\young(pr)}\right]\epsilon ^{abc} \tilde{B}{}^{\mu }{}{}_{\nu } \left(D_{\mu } \overline{l}_{s}^{i} d_{rb}\right) \left(q_{tc i} C \gamma ^{\nu } d_{pa}\right)

,\quad\mathcal{Y}\left[\tiny{\young(pr)}\right]\epsilon ^{abc} \tilde{B}{}^{\mu }{}{}_{\nu } \left(\overline{l}_{s}^{i} d_{pa}\right) \left(D_{\mu } q_{tc i} C \gamma ^{\nu } d_{rb}\right),

\\&\mathcal{Y}\left[\tiny{\young(p,r)}\right]\epsilon ^{abc} B{}^{\mu }{}{}_{\nu } \left(\overline{l}_{s}^{i} d_{pa} \right) \left(q_{tc i} C \gamma ^{\nu } D_{\mu } d_{rb}\right)

,\quad\mathcal{Y}\left[\tiny{\young(p,r)}\right]\epsilon ^{abc} B{}^{\mu }{}{}_{\nu } \left(D_{\mu } \overline{l}_{s}^{i} d_{pa} \right) \left(q_{tc i} C \gamma ^{\nu } d_{rb}\right),

\\&\mathcal{Y}\left[\tiny{\young(p,r)}\right]\epsilon ^{abc} \tilde{B}{}^{\mu }{}{}_{\nu } \left(\overline{l}_{s}^{i} d_{pa}\right) \left(D_{\mu } q_{tc i} C \gamma ^{\nu } d_{rb}\right)

\vspace{2ex}\\

\multirow{5}*{$\mathcal{O}_{G d{}^3 \overline{e}    D}^{(1\sim 9)}$}

&\mathcal{Y}\left[\tiny{\young(pr,s)}\right]\epsilon ^{ace} \left(\lambda ^A\right)_e^b G^{A}{}^{\mu }{}{}_{\nu } \left(\overline{e}_{t} \gamma ^{\nu } D_{\mu } d_{rb}\right) \left(d_{sc} C d_{pa}\right)

,\quad\mathcal{Y}\left[\tiny{\young(pr,s)}\right]\epsilon ^{abe} \left(\lambda ^A\right)_e^c G^{A}{}^{\mu }{}{}_{\nu } \left(\overline{e}_{t} \gamma ^{\nu } d_{rb}\right) \left(D_{\mu } d_{sc} C d_{pa}\right),

\\&\mathcal{Y}\left[\tiny{\young(pr,s)}\right]\epsilon ^{abe} \left(\lambda ^A\right)_e^c G^{A}{}^{\mu }{}{}_{\nu } \left(\overline{e}_{t} \gamma ^{\nu } d_{pa}\right) \left(d_{sc} C D_{\mu } d_{rb}\right)

,\quad\mathcal{Y}\left[\tiny{\young(pr,s)}\right]\epsilon ^{ace} \left(\lambda ^A\right)_e^b \tilde{G}^{A}{}^{\mu }{}{}_{\nu } \left(\overline{e}_{t} \gamma ^{\nu } d_{pa}\right) \left(D_{\mu } d_{sc} C d_{rb}\right),

\\&\mathcal{Y}\left[\tiny{\young(pr,s)}\right]\epsilon ^{abe} \left(\lambda ^A\right)_e^c \tilde{G}^{A}{}^{\mu }{}{}_{\nu } \left(\overline{e}_{t} \gamma ^{\nu } d_{pa}\right) \left(D_{\mu } d_{sc} C d_{rb}\right)

,\quad\mathcal{Y}\left[\tiny{\young(prs)}\right]\epsilon ^{abe} \left(\lambda ^A\right)_e^c G^{A}{}^{\mu }{}{}_{\nu } \left(\overline{e}_{t} \gamma ^{\nu } D_{\mu } d_{rb}\right) \left(d_{sc} C d_{pa}\right),

\\&\mathcal{Y}\left[\tiny{\young(prs)}\right]\epsilon ^{ace} \left(\lambda ^A\right)_e^b \tilde{G}^{A}{}^{\mu }{}{}_{\nu } \left(\overline{e}_{t} \gamma ^{\nu } d_{pa}\right) \left(D_{\mu } d_{sc} C d_{rb}\right)

,\quad\mathcal{Y}\left[\tiny{\young(p,r,s)}\right]\epsilon ^{ace} \left(\lambda ^A\right)_e^b G^{A}{}^{\mu }{}{}_{\nu } \left(\overline{e}_{t} \gamma ^{\nu } D_{\mu } d_{rb}\right) \left(d_{sc} C d_{pa}\right),

\\&\mathcal{Y}\left[\tiny{\young(p,r,s)}\right]\epsilon ^{ace} \left(\lambda ^A\right)_e^b \tilde{G}^{A}{}^{\mu }{}{}_{\nu } \left(\overline{e}_{t} \gamma ^{\nu } d_{pa}\right) \left(D_{\mu } d_{sc} C d_{rb}\right)

\vspace{2ex}\\

\multirow{3}*{$\mathcal{O}_{B d{}^3 \overline{e}    D}^{(1\sim 5)}$}

&\mathcal{Y}\left[\tiny{\young(prs)}\right]\epsilon ^{abc} B{}^{\mu }{}{}_{\nu } \left(\overline{e}_{t} \gamma ^{\nu } d_{rb}\right) \left(D_{\mu } d_{sc} C d_{pa}\right)

,\quad\mathcal{Y}\left[\tiny{\young(prs)}\right]\epsilon ^{abc} \tilde{B}{}^{\mu }{}{}_{\nu } \left(\overline{e}_{t} \gamma ^{\nu } d_{pa}\right) \left(D_{\mu } d_{sc} C d_{rb}\right),

\\&\mathcal{Y}\left[\tiny{\young(pr,s)}\right]\epsilon ^{abc} B{}^{\mu }{}{}_{\nu } \left(\overline{e}_{t} \gamma ^{\nu } D_{\mu } d_{rb}\right) \left(d_{sc} C d_{pa}\right)

,\quad\mathcal{Y}\left[\tiny{\young(pr,s)}\right]\epsilon ^{abc} \tilde{B}{}^{\mu }{}{}_{\nu } \left(\overline{e}_{t} \gamma ^{\nu } d_{pa}\right) \left(D_{\mu } d_{sc} C d_{rb}\right),

\vspace{0.5ex}

\\&\mathcal{Y}\left[\tiny{\young(p,r,s)}\right]\epsilon ^{abc} B{}^{\mu }{}{}_{\nu } \left(\overline{e}_{t} \gamma ^{\nu } D_{\mu } d_{rb}\right) \left(d_{sc} C d_{pa}\right)

\end{array}\label{cl:fq3lD}\end{align}

\noindent 2. Operators involving two leptons and two quarks with $\Delta L=2$, the type $W u \overline{d} l{}^2    D$ contain the operators contributing the neutrinoless double beta decay at tree-level:
\begin{align}\begin{array}{c|l}

\multirow{4}*{$\mathcal{O}_{G u \overline{d} l{}^2    D}^{(1\sim 7)} $}

&\mathcal{Y}\left[\tiny{\young(rs)}\right]\epsilon ^{ij} \left(\lambda ^A\right)_a^b G^{A}{}^{\mu }{}{}_{\nu } \left(\overline{d}_{p}^{a} l_{sj}\right) \left(D_{\mu } l_{ri} C \gamma ^{\nu } u_{tb}\right)

,\quad\mathcal{Y}\left[\tiny{\young(rs)}\right]\epsilon ^{ij} \left(\lambda ^A\right)_a^b G^{A}{}^{\mu }{}{}_{\nu } \left(\overline{d}_{p}^{a} D_{\mu } l_{sj}\right) \left(l_{ri} C \gamma ^{\nu } u_{tb}\right),

\\&\mathcal{Y}\left[\tiny{\young(rs)}\right]\epsilon ^{ij} \left(\lambda ^A\right)_a^b \tilde{G}^{A}{}^{\mu }{}{}_{\nu } \left(\overline{d}_{p}^{a} \gamma ^{\nu } u_{tb}\right) \left(l_{ri} C D_{\mu } l_{sj}\right)

,\quad\mathcal{Y}\left[\tiny{\young(rs)}\right]\epsilon ^{ij} \left(\lambda ^A\right)_a^b \tilde{G}^{A}{}^{\mu }{}{}_{\nu } \left(\overline{d}_{p}^{a} l_{sj}\right) \left(l_{ri} C \gamma ^{\nu } D_{\mu } u_{tb}\right),

\\&\mathcal{Y}\left[\tiny{\young(r,s)}\right]\epsilon ^{ij} \left(\lambda ^A\right)_a^b G^{A}{}^{\mu }{}{}_{\nu } \left(\overline{d}_{p}^{a} l_{sj}\right) \left(D_{\mu } l_{ri} C \gamma ^{\nu } u_{tb}\right)

,\quad\mathcal{Y}\left[\tiny{\young(r,s)}\right]\epsilon ^{ij} \left(\lambda ^A\right)_a^b G^{A}{}^{\mu }{}{}_{\nu } \left(\overline{d}_{p}^{a} D_{\mu } l_{sj}\right) \left(l_{ri} C \gamma ^{\nu } u_{tb}\right),

\\&\mathcal{Y}\left[\tiny{\young(r,s)}\right]\epsilon ^{ij} \left(\lambda ^A\right)_a^b \tilde{G}^{A}{}^{\mu }{}{}_{\nu } \left(\overline{d}_{p}^{a} \gamma ^{\nu } u_{tb}\right) \left(l_{ri} C D_{\mu } l_{sj}\right)

\vspace{2ex}\\

\multirow{4}*{$\mathcal{O}_{W u \overline{d} l{}^2    D}^{(1\sim 7)} $}

&\mathcal{Y}\left[\tiny{\young(r,s)}\right]\epsilon ^{jk} \left(\tau ^I\right)_k^i W^{I}{}^{\mu }{}{}_{\nu } \left(\overline{d}_{p}^{a} l_{sj}\right) \left(D_{\mu } l_{ri} C \gamma ^{\nu } u_{ta}\right)

,\quad\mathcal{Y}\left[\tiny{\young(r,s)}\right]\epsilon ^{jk} \left(\tau ^I\right)_k^i W^{I}{}^{\mu }{}{}_{\nu } \left(\overline{d}_{p}^{a} D_{\mu } l_{sj}\right) \left(l_{ri} C \gamma ^{\nu } u_{ta}\right),

\\&\mathcal{Y}\left[\tiny{\young(r,s)}\right]\epsilon ^{ik} \left(\tau ^I\right)_k^j \tilde{W}^{I}{}^{\mu }{}{}_{\nu } \left(\overline{d}_{p}^{a} \gamma ^{\nu } u_{ta}\right) \left(l_{ri} C D_{\mu } l_{sj}\right)

,\quad\mathcal{Y}\left[\tiny{\young(r,s)}\right]\epsilon ^{ik} \left(\tau ^I\right)_k^j \tilde{W}^{I}{}^{\mu }{}{}_{\nu } \left(\overline{d}_{p}^{a} l_{sj}\right) \left(l_{ri} C \gamma ^{\nu } D_{\mu } u_{ta}\right),

\\&\mathcal{Y}\left[\tiny{\young(rs)}\right]\epsilon ^{jk} \left(\tau ^I\right)_k^i W^{I}{}^{\mu }{}{}_{\nu } \left(\overline{d}_{p}^{a} l_{sj}\right) \left(D_{\mu } l_{ri} C \gamma ^{\nu } u_{ta}\right)

,\quad\mathcal{Y}\left[\tiny{\young(rs)}\right]\epsilon ^{jk} \left(\tau ^I\right)_k^i W^{I}{}^{\mu }{}{}_{\nu } \left(\overline{d}_{p}^{a} D_{\mu } l_{sj}\right) \left(l_{ri} C \gamma ^{\nu } u_{ta}\right),

\\&\mathcal{Y}\left[\tiny{\young(rs)}\right]\epsilon ^{ik} \left(\tau ^I\right)_k^j \tilde{W}^{I}{}^{\mu }{}{}_{\nu } \left(\overline{d}_{p}^{a} \gamma ^{\nu } u_{ta}\right) \left(l_{ri} C D_{\mu } l_{sj}\right)

\vspace{2ex}\\

\multirow{4}*{$\mathcal{O}_{B u \overline{d} l{}^2    D}^{(1\sim 7)} $}

&\mathcal{Y}\left[\tiny{\young(rs)}\right]\epsilon ^{ij} B{}^{\mu }{}{}_{\nu } \left(\overline{d}_{p}^{a} l_{sj}\right) \left(D_{\mu } l_{ri} C \gamma ^{\nu } u_{ta}\right)

,\quad\mathcal{Y}\left[\tiny{\young(rs)}\right]\epsilon ^{ij} B{}^{\mu }{}{}_{\nu } \left(\overline{d}_{p}^{a} D_{\mu } l_{sj}\right) \left(l_{ri} C \gamma ^{\nu } u_{ta}\right),

\\&\mathcal{Y}\left[\tiny{\young(rs)}\right]\epsilon ^{ij} \tilde{B}{}^{\mu }{}{}_{\nu } \left(\overline{d}_{p}^{a} \gamma ^{\nu } u_{ta}\right) \left(l_{ri} C D_{\mu } l_{sj}\right)

,\quad\mathcal{Y}\left[\tiny{\young(rs)}\right]\epsilon ^{ij} \tilde{B}{}^{\mu }{}{}_{\nu } \left(\overline{d}_{p}^{a} l_{sj}\right) \left(l_{ri} C \gamma ^{\nu } D_{\mu } u_{ta}\right),

\\&\mathcal{Y}\left[\tiny{\young(r,s)}\right]\epsilon ^{ij} B{}^{\mu }{}{}_{\nu } \left(\overline{d}_{p}^{a} l_{sj}\right) \left(D_{\mu } l_{ri} C \gamma ^{\nu } u_{ta}\right)

,\quad\mathcal{Y}\left[\tiny{\young(r,s)}\right]\epsilon ^{ij} B{}^{\mu }{}{}_{\nu } \left(\overline{d}_{p}^{a} D_{\mu } l_{sj}\right) \left(l_{ri} C \gamma ^{\nu } u_{ta}\right),

\\&\mathcal{Y}\left[\tiny{\young(r,s)}\right]\epsilon ^{ij} \tilde{B}{}^{\mu }{}{}_{\nu } \left(\overline{d}_{p}^{a} \gamma ^{\nu } u_{ta}\right) \left(l_{ri} C D_{\mu } l_{sj}\right)

\end{array}\label{cl:fq2l2D}\end{align}

\subsection{Classes involving Six-fermions}
\label{sec:sixfermion}
All the Lorentz structures in this section are new. 
This class of Lorentz structures contains both processes with $(\Delta B,\Delta L)=(\pm 1,\mp 1)$ and $(\Delta B,\Delta L)=(0,2)$ that are already present at lower dimension, and  with $\Delta B=2$ or $\Delta L=3$  that are absent at lower dimensions and relevant for the neutron-antineutron oscillation or the proton 3-body decay processes. 
Meanwhile, $|B-L|$ is still equal to 2. Only two classes involve six fermions. In the first class, 6 fermion with the same helicities contract with each other, $\psi^6$, producing 5 independent Lorentz structures
\begin{align}\begin{array}{lll}

    \psi _1{}{}^{\alpha } \psi _2{}{}^{\beta } \psi _3{}{}^{\gamma } \psi _4{}_{\alpha } \psi _5{}_{\beta } \psi _6{}_{\gamma },
    &\psi _1{}{}^{\alpha } \psi _2{}{}^{\beta } \psi _3{}_{\alpha } \psi _4{}{}^{\gamma } \psi _5{}_{\beta } \psi _6{}_{\gamma },    
    &\psi _1{}{}^{\alpha } \psi _2{}{}^{\beta } \psi _3{}_{\alpha } \psi _4{}_{\beta } \psi _5{}{}^{\gamma } \psi _6{}_{\gamma }\\
    \psi _1{}{}^{\alpha } \psi _2{}_{\alpha } \psi _3{}{}^{\beta } \psi _4{}{}^{\gamma } \psi _5{}_{\beta } \psi _6{}_{\gamma },
    &\psi _1{}{}^{\alpha } \psi _2{}_{\alpha } \psi _3{}{}^{\beta } \psi _4{}_{\beta } \psi _5{}{}^{\gamma } \psi _6{}_{\gamma }.
    
\end{array}\end{align}
In the second class, two of them have opposite helicities, $\psi^4\psi^{\dagger 2}$, giving two other Lorentz structures.
\begin{align}

    &\psi _1{}{}^{\alpha } \psi _2{}{}^{\beta } \psi _3{}_{\alpha } \psi _4{}_{\beta } \psi^{\dagger}_5{}_{\dot{\alpha }} \psi^{\dagger}_6{}{}^{\dot{\alpha }},\quad \psi _1{}{}^{\alpha } \psi _2{}_{\alpha } \psi _3{}{}^{\beta } \psi _4{}_{\beta } \psi^{\dagger}_5{}_{\dot{\alpha }} \psi^{\dagger}_6{}{}^{\dot{\alpha }}.
    
\end{align}
Similarly, we present the operators in terms of the number of quarks, and conjugate those operators with $\Delta B< 0$ or $\Delta L=2$.
The only type relevant for the proton three body decay is $l^3qu^2$, while for the neutron-antineutron oscillation, two types $d^2q^4$ and $d^3q^2u$ are relevant. \\

1. Operators involving only quarks with $\Delta B=2$, only two types $d^2q^4$ and $d^3q^2u$ contain operators involving the neutron-antineutron oscillation:
\begin{align}\begin{array}{c|l}

    \multirow{10}*{$\mathcal{O}_{  u^2 d^4    }^{(1\sim 10)}$}
    
    &\mathcal{Y}\left[\tiny{\young(prs,t)},\tiny{\young(u,v)}\right]\epsilon ^{ace} \epsilon ^{bdf} \left(d_{pa} C d_{td}\right) \left(u_{ue} C d_{rb}\right) \left(u_{vf} C d_{sc}\right),
    \\& \mathcal{Y}\left[\tiny{\young(prs,t)},\tiny{\young(u,v)}\right]\epsilon ^{adf} \epsilon ^{bce} \left(d_{pa} C d_{td}\right) \left(u_{ue} C d_{rb}\right) \left(u_{vf} C d_{sc}\right),
    \\&\mathcal{Y}\left[\tiny{\young(pr,s,t)},\tiny{\young(u,v)}\right]\epsilon ^{ace} \epsilon ^{bdf} \left(d_{pa} C d_{td}\right) \left(u_{ue} C d_{rb}\right) \left(u_{vf} C d_{sc}\right),
    \\& \mathcal{Y}\left[\tiny{\young(pr,s,t)},\tiny{\young(u,v)}\right]\epsilon ^{adf} \epsilon ^{bce} \left(d_{pa} C d_{td}\right) \left(u_{ue} C d_{rb}\right) \left(u_{vf} C d_{sc}\right),
    \\&\mathcal{Y}\left[\tiny{\young(prst)},\tiny{\young(uv)}\right]\epsilon ^{ace} \epsilon ^{bdf} \left(d_{pa} C d_{td}\right) \left(u_{ue} C d_{rb}\right) \left(u_{vf} C d_{sc}\right),
    \\& \mathcal{Y}\left[\tiny{\young(prst)},\tiny{\young(uv)}\right]\epsilon ^{abe} \epsilon ^{cdf} \left(d_{pa} C d_{td}\right) \left(u_{ue} C d_{rb}\right) \left(u_{vf} C d_{sc}\right),
    \\&\mathcal{Y}\left[\tiny{\young(prs,t)},\tiny{\young(uv)}\right]\epsilon ^{adf} \epsilon ^{bce} \left(d_{pa} C d_{td}\right) \left(u_{ue} C d_{rb}\right) \left(u_{vf} C d_{sc}\right),
    \\& \mathcal{Y}\left[\tiny{\young(pr,st)},\tiny{\young(uv)}\right]\epsilon ^{ace} \epsilon ^{bdf} \left(d_{pa} C d_{td}\right) \left(u_{ue} C d_{rb}\right) \left(u_{vf} C d_{sc}\right),
    \\&\mathcal{Y}\left[\tiny{\young(pr,st)},\tiny{\young(uv)}\right]\epsilon ^{abf} \epsilon ^{cde} \left(d_{pa} C d_{td}\right) \left(u_{ue} C d_{rb}\right) \left(u_{vf} C d_{sc}\right),
    \\& \mathcal{Y}\left[\tiny{\young(pr,s,t)},\tiny{\young(uv)}\right]\epsilon ^{ace} \epsilon ^{bdf} \left(d_{pa} C d_{td}\right) \left(u_{ue} C d_{rb}\right) \left(u_{vf} C d_{sc}\right)
    
\vspace{2ex}\\

    \multirow{8}*{$\mathcal{O}_{  d ^2 q^4    }^{(1\sim 8)}$}
    
    &\mathcal{Y}\left[\tiny{\young(prst)},\tiny{\young(uv)}\right]\epsilon ^{ace} \epsilon ^{bdf} \epsilon ^{ik} \epsilon ^{jm} \left(d_{ue} C d_{vf}\right) \left(q_{pa i} C q_{sc k}\right) \left(q_{rb j} C q_{td m}\right),
    \\&\mathcal{Y}\left[\tiny{\young(pr,st)},\tiny{\young(uv)}\right]\epsilon ^{ace} \epsilon ^{bdf} \epsilon ^{ij} \epsilon ^{km} \left(d_{ue} C d_{vf}\right) \left(q_{pa i} C q_{sc k}\right) \left(q_{rb j} C q_{td m}\right),
    \\&\mathcal{Y}\left[\tiny{\young(pr,st)},\tiny{\young(uv)}\right]\epsilon ^{abe} \epsilon ^{cdf} \epsilon ^{ik} \epsilon ^{jm} \left(d_{ue} C d_{vf}\right) \left(q_{pa i} C q_{sc k}\right) \left(q_{rb j} C q_{td m}\right),
    \\&\mathcal{Y}\left[\tiny{\young(pr,st)},\tiny{\young(uv)}\right]\epsilon ^{ace} \epsilon ^{bdf} \epsilon ^{ik} \epsilon ^{jm} \left(d_{ue} C d_{vf}\right) \left(q_{pa i} C q_{rb j}\right) \left(q_{sc k} C q_{td m}\right),
    \\&\mathcal{Y}\left[\tiny{\young(prs,t)},\tiny{\young(u,v)}\right]\epsilon ^{ace} \epsilon ^{bdf} \epsilon ^{ik} \epsilon ^{jm} \left(d_{ue} C d_{vf}\right) \left(q_{pa i} C q_{sc k}\right) \left(q_{rb j} C q_{td m}\right),
    \\&\mathcal{Y}\left[\tiny{\young(prs,t)},\tiny{\young(u,v)}\right]\epsilon ^{acd} \epsilon ^{bef} \epsilon ^{ik} \epsilon ^{jm} \left(d_{ue} C d_{vf}\right) \left(q_{pa i} C q_{sc k}\right) \left(q_{rb j} C q_{td m}\right),
    \\&\mathcal{Y}\left[\tiny{\young(pr,s,t)},\tiny{\young(u,v)}\right]\epsilon ^{ace} \epsilon ^{bdf} \epsilon ^{ij} \epsilon ^{km} \left(d_{ue} C d_{vf}\right) \left(q_{pa i} C q_{sc k}\right) \left(q_{rb j} C q_{td m}\right),
    \\&\mathcal{Y}\left[\tiny{\young(pr,s,t)},\tiny{\young(u,v)}\right]\epsilon ^{acd} \epsilon ^{bef} \epsilon ^{ij} \epsilon ^{km} \left(d_{ue} C d_{vf}\right) \left(q_{pa i} C q_{sc k}\right) \left(q_{rb j} C q_{td m}\right)
    
\vspace{2ex}\\

    \multirow{7}*{$\mathcal{O}_{  d^3 q ^2 u    }^{(1\sim 7)}$}
    
    &\mathcal{Y}\left[\tiny{\young(prs)},\tiny{\young(u,v)}\right]\epsilon ^{ade} \epsilon ^{bcf} \epsilon ^{ij} \left(d_{pa} C d_{sc}\right) \left(d_{rb} C u_{td}\right) \left(q_{ue i} C q_{vf j}\right),
    \\& \mathcal{Y}\left[\tiny{\young(pr,s)},\tiny{\young(u,v)}\right]\epsilon ^{acf} \epsilon ^{bde} \epsilon ^{ij} \left(d_{pa} C d_{sc}\right) \left(d_{rb} C u_{td}\right) \left(q_{ue i} C q_{vf j}\right),
    \\&\mathcal{Y}\left[\tiny{\young(p,r,s)},\tiny{\young(u,v)}\right]\epsilon ^{acf} \epsilon ^{bde} \epsilon ^{ij} \left(d_{pa} C d_{sc}\right) \left(d_{rb} C u_{td}\right) \left(q_{ue i} C q_{vf j}\right),
    \\& \mathcal{Y}\left[\tiny{\young(prs)},\tiny{\young(uv)}\right]\epsilon ^{aef} \epsilon ^{bcd} \epsilon ^{ij} \left(d_{pa} C d_{sc}\right) \left(d_{rb} C u_{td}\right) \left(q_{ue i} C q_{vf j}\right),
    \\&\mathcal{Y}\left[\tiny{\young(pr,s)},\tiny{\young(uv)}\right]\epsilon ^{aef} \epsilon ^{bcd} \epsilon ^{ij} \left(d_{pa} C d_{sc}\right) \left(d_{rb} C u_{td}\right) \left(q_{ue i} C q_{vf j}\right),
    \\& \mathcal{Y}\left[\tiny{\young(pr,s)},\tiny{\young(uv)}\right]\epsilon ^{ade} \epsilon ^{bcf} \epsilon ^{ij} \left(d_{pa} C d_{sc}\right) \left(d_{rb} C u_{td}\right) \left(q_{ue i} C q_{vf j}\right),
    \\&\mathcal{Y}\left[\tiny{\young(p,r,s)},\tiny{\young(uv)}\right]\epsilon ^{aef} \epsilon ^{bcd} \epsilon ^{ij} \left(d_{pa} C d_{sc}\right) \left(d_{rb} C u_{td}\right) \left(q_{ue i} C q_{vf j}\right)
 
\end{array}\label{cl:p660}\end{align}

2. Operators involving 5 quarks and 1 lepton with $\Delta B=1$ and $\Delta L=-1$:
\begin{align}\begin{array}{c|l}

    \multirow{5}*{$\mathcal{O}_{  d^3 \bar{l} \bar{q} u    }^{(1\sim 10)}$}
    
    &\mathcal{Y}\left[\tiny{\young(pr,s)}\right]\epsilon ^{ace} \epsilon _{ij} \left(\bar{l}_{t}^{i} d_{pa}\right) \left(\bar{q}_{u}^{b j} d_{rb}\right) \left(u_{ve} C d_{sc}\right),
    \quad \mathcal{Y}\left[\tiny{\young(pr,s)}\right]\epsilon ^{bce} \epsilon _{ij} \left(\bar{l}_{t}^{i} d_{pa}\right) \left(\bar{q}_{u}^{a j} d_{rb}\right) \left(u_{ve} C d_{sc}\right),
    \\&\mathcal{Y}\left[\tiny{\young(pr,s)}\right]\epsilon ^{abe} \epsilon _{ij} \left(\bar{l}_{t}^{i} d_{pa}\right) \left(\bar{q}_{u}^{c j} d_{rb}\right) \left(u_{ve} C d_{sc}\right),
    \quad \mathcal{Y}\left[\tiny{\young(pr,s)}\right]\epsilon ^{ace} \epsilon _{ij} \left(\bar{l}_{t}^{i} d_{rb}\right) \left(\bar{q}_{u}^{b j} u_{ve}\right) \left(d_{pa} C d_{sc}\right),
    \\&\mathcal{Y}\left[\tiny{\young(pr,s)}\right]\epsilon ^{bce} \epsilon _{ij} \left(\bar{l}_{t}^{i} d_{rb}\right) \left(\bar{q}_{u}^{a j} u_{ve}\right) \left(d_{pa} C d_{sc}\right),
    \quad \mathcal{Y}\left[\tiny{\young(prs)}\right]\epsilon ^{ace} \epsilon _{ij} \left(\bar{l}_{t}^{i} d_{pa}\right) \left(\bar{q}_{u}^{b j} d_{rb}\right) \left(u_{ve} C d_{sc}\right),
    \\&\mathcal{Y}\left[\tiny{\young(prs)}\right]\epsilon ^{bce} \epsilon _{ij} \left(\bar{l}_{t}^{i} d_{pa}\right) \left(\bar{q}_{u}^{a j} d_{rb}\right) \left(u_{ve} C d_{sc}\right),
    \quad \mathcal{Y}\left[\tiny{\young(prs)}\right]\epsilon ^{abe} \epsilon _{ij} \left(\bar{l}_{t}^{i} d_{pa}\right) \left(\bar{q}_{u}^{c j} d_{rb}\right) \left(u_{ve} C d_{sc}\right),
    \\&\mathcal{Y}\left[\tiny{\young(p,r,s)}\right]\epsilon ^{ace} \epsilon _{ij} \left(\bar{l}_{t}^{i} d_{pa}\right) \left(\bar{q}_{u}^{b j} d_{rb}\right) \left(u_{ve} C d_{sc}\right),
    \quad \mathcal{Y}\left[\tiny{\young(p,r,s)}\right]\epsilon ^{bce} \epsilon _{ij} \left(\bar{l}_{t}^{i} d_{pa}\right) \left(\bar{q}_{u}^{a j} d_{rb}\right) \left(u_{ve} C d_{sc}\right)
    
\vspace{2ex}\\

    \multirow{4}*{$\mathcal{O}_{  d  \bar{l}  q^3 \bar{u}    }^{(1\sim 8)}$}
    
    &\mathcal{Y}\left[\tiny{\young(prs)}\right]\epsilon ^{bce} \epsilon ^{ik} \left(\overline{u}_{t}^{a} q_{rb j}\right) \left(\overline{l}_{v}^{j} d_{ue}\right) \left(q_{pa i} C q_{sc k}\right),
    \quad \mathcal{Y}\left[\tiny{\young(prs)}\right]\epsilon ^{bce} \epsilon ^{ij} \left(\overline{u}_{t}^{a} q_{rb j}\right) \left(\overline{l}_{v}^{k} d_{ue}\right) \left(q_{pa i} C q_{sc k}\right),
    \\&\mathcal{Y}\left[\tiny{\young(pr,s)}\right]\epsilon ^{bce} \epsilon ^{ik} \left(\overline{u}_{t}^{a} q_{rb j}\right) \left(\overline{l}_{v}^{j} d_{ue}\right) \left(q_{pa i} C q_{sc k}\right),
    \quad \mathcal{Y}\left[\tiny{\young(pr,s)}\right]\epsilon ^{bce} \epsilon ^{ij} \left(\overline{u}_{t}^{a} q_{rb j}\right) \left(\overline{l}_{v}^{k} d_{ue}\right) \left(q_{pa i} C q_{sc k}\right),
    \\&\mathcal{Y}\left[\tiny{\young(pr,s)}\right]\epsilon ^{ace} \epsilon ^{ij} \left(\overline{u}_{t}^{b} q_{rb j}\right) \left(\overline{l}_{v}^{k} d_{ue}\right) \left(q_{pa i} C q_{sc k}\right),
    \quad \mathcal{Y}\left[\tiny{\young(pr,s)}\right]\epsilon ^{bce} \epsilon ^{ik} \left(\overline{u}_{t}^{a} q_{sc k}\right) \left(\overline{l}_{v}^{j} d_{ue}\right) \left(q_{pa i} C q_{rb j}\right),
    \\&\mathcal{Y}\left[\tiny{\young(p,r,s)}\right]\epsilon ^{bce} \epsilon ^{ik} \left(\overline{u}_{t}^{a} q_{rb j}\right) \left(\overline{l}_{v}^{j} d_{ue}\right) \left(q_{pa i} C q_{sc k}\right),
    \quad \mathcal{Y}\left[\tiny{\young(p,r,s)}\right]\epsilon ^{bce} \epsilon ^{ij} \left(\overline{u}_{t}^{a} q_{rb j}\right) \left(\overline{l}_{v}^{k} d_{ue}\right) \left(q_{pa i} C q_{sc k}\right)
    
\vspace{2ex}\\

    \multirow{3}*{$\mathcal{O}_{  d ^2 \bar{e} q^2 \bar{u}    }^{(1\sim 6)}$}
    
    &\mathcal{Y}\left[\tiny{\young(rs)},\tiny{\young(u,v)}\right]\epsilon ^{bde} \epsilon ^{ij} \left(\overline{u}_{t}^{a} q_{ra i}\right) \left(\overline{e}_{p} q_{sb j}\right) \left(d_{ud} C d_{ve}\right),
    \quad \mathcal{Y}\left[\tiny{\young(rs)},\tiny{\young(u,v)}\right]\epsilon ^{ade} \epsilon ^{ij} \left(\overline{u}_{t}^{b} q_{ra i}\right) \left(\overline{e}_{p} q_{sb j}\right) \left(d_{ud} C d_{ve}\right),
    \\&\mathcal{Y}\left[\tiny{\young(r,s)},\tiny{\young(uv)}\right]\epsilon ^{abe} \epsilon ^{ij} \left(\overline{u}_{t}^{c} q_{ra i}\right) \left(\overline{e}_{p} q_{sb j}\right) \left(d_{uc} C d_{ve}\right),
    \quad \mathcal{Y}\left[\tiny{\young(r,s)},\tiny{\young(u,v)}\right]\epsilon ^{bde} \epsilon ^{ij} \left(\overline{u}_{t}^{a} q_{ra i}\right) \left(\overline{e}_{p} q_{sb j}\right) \left(d_{ud} C d_{ve}\right),
    \\&\mathcal{Y}\left[\tiny{\young(r,s)},\tiny{\young(u,v)}\right]\epsilon ^{ade} \epsilon ^{ij} \left(\overline{u}_{t}^{b} q_{ra i}\right) \left(\overline{e}_{p} q_{sb j}\right) \left(d_{ud} C d_{ve}\right),
    \quad \mathcal{Y}\left[\tiny{\young(rs)},\tiny{\young(uv)}\right]\epsilon ^{abe} \epsilon ^{ij} \left(\overline{u}_{t}^{c} q_{ra i}\right) \left(\overline{e}_{p} q_{sb j}\right) \left(d_{uc} C d_{ve}\right)
    
\vspace{2ex}\\

    \multirow{2}*{$\mathcal{O}_{  d^3 \bar{e}  q \bar{q}     }^{(1\sim 4)}$}
    
    &\mathcal{Y}\left[\tiny{\young(prs)}\right]\epsilon ^{bce} \left(\bar{q}_{t}^{a i} d_{rb}\right) \left(\bar{e}_{u} q_{ve i}\right) \left(d_{pa} C d_{sc}\right),
    \quad \mathcal{Y}\left[\tiny{\young(pr,s)}\right]\epsilon ^{ace} \left(\bar{q}_{t}^{b i} d_{rb}\right) \left(\bar{e}_{u} q_{ve i}\right) \left(d_{pa} C d_{sc}\right),
    \\&\mathcal{Y}\left[\tiny{\young(pr,s)}\right]\epsilon ^{bce} \left(\bar{q}_{t}^{a i} d_{rb}\right) \left(\bar{e}_{u} q_{ve i}\right) \left(d_{pa} C d_{sc}\right),
    \quad \mathcal{Y}\left[\tiny{\young(p,r,s)}\right]\epsilon ^{ace} \left(\bar{q}_{t}^{b i} d_{rb}\right) \left(\bar{e}_{u} q_{ve i}\right) \left(d_{pa} C d_{sc}\right)
    
\vspace{2ex}\\

    \multirow{6}*{$\mathcal{O}_{  d^2 \bar{l} q^2 \bar{q}    }^{(1\sim 12)}$}
    
    &\mathcal{Y}\left[\tiny{\young(p,r)},\tiny{\young(u,v)}\right]\epsilon ^{bde} \left(\bar{q}_{t}^{a j} d_{rb}\right) \left(\bar{l}_{s}^{i} d_{pa}\right) \left(q_{ud i} C q_{ve j}\right),
    \quad \mathcal{Y}\left[\tiny{\young(p,r)},\tiny{\young(u,v)}\right]\epsilon ^{ade} \left(\bar{q}_{t}^{b j} d_{rb}\right) \left(\bar{l}_{s}^{i} d_{pa}\right) \left(q_{ud i} C q_{ve j}\right),
    \\&\mathcal{Y}\left[\tiny{\young(p,r)},\tiny{\young(u,v)}\right]\epsilon ^{abe} \left(\bar{q}_{t}^{c j} d_{rb}\right) \left(\bar{l}_{s}^{i} d_{pa}\right) \left(q_{uc i} C q_{ve j}\right),
    \quad \mathcal{Y}\left[\tiny{\young(p,r)},\tiny{\young(uv)}\right]\epsilon ^{bde} \left(\bar{q}_{t}^{a j} d_{rb}\right) \left(\bar{l}_{s}^{i} d_{pa}\right) \left(q_{ud i} C q_{ve j}\right),
    \\&\mathcal{Y}\left[\tiny{\young(p,r)},\tiny{\young(uv)}\right]\epsilon ^{ade} \left(\bar{q}_{t}^{b j} d_{rb}\right) \left(\bar{l}_{s}^{i} d_{pa}\right) \left(q_{ud i} C q_{ve j}\right),
    \quad \mathcal{Y}\left[\tiny{\young(p,r)},\tiny{\young(uv)}\right]\epsilon ^{abe} \left(\bar{q}_{t}^{c j} d_{rb}\right) \left(\bar{l}_{s}^{i} d_{pa}\right) \left(q_{uc i} C q_{ve j}\right),
    \\&\mathcal{Y}\left[\tiny{\young(pr)},\tiny{\young(uv)}\right]\epsilon ^{bde} \left(\bar{q}_{t}^{a j} d_{rb}\right) \left(\bar{l}_{s}^{i} d_{pa}\right) \left(q_{ud i} C q_{ve j}\right),
    \quad \mathcal{Y}\left[\tiny{\young(pr)},\tiny{\young(uv)}\right]\epsilon ^{ade} \left(\bar{q}_{t}^{b j} d_{rb}\right) \left(\bar{l}_{s}^{i} d_{pa}\right) \left(q_{ud i} C q_{ve j}\right),
    \\&\mathcal{Y}\left[\tiny{\young(pr)},\tiny{\young(uv)}\right]\epsilon ^{abe} \left(\bar{q}_{t}^{c j} d_{rb}\right) \left(\bar{l}_{s}^{i} d_{pa}\right) \left(q_{uc i} C q_{ve j}\right),
    \quad \mathcal{Y}\left[\tiny{\young(pr)},\tiny{\young(u,v)}\right]\epsilon ^{bde} \left(\bar{q}_{t}^{a j} d_{rb}\right) \left(\bar{l}_{s}^{i} d_{pa}\right) \left(q_{ud i} C q_{ve j}\right),
    \\&\mathcal{Y}\left[\tiny{\young(pr)},\tiny{\young(u,v)}\right]\epsilon ^{ade} \left(\bar{q}_{t}^{b j} d_{rb}\right) \left(\bar{l}_{s}^{i} d_{pa}\right) \left(q_{ud i} C q_{ve j}\right),
    \quad \mathcal{Y}\left[\tiny{\young(pr)},\tiny{\young(u,v)}\right]\epsilon ^{abe} \left(\bar{q}_{t}^{c j} d_{rb}\right) \left(\bar{l}_{s}^{i} d_{pa}\right) \left(q_{uc i} C q_{ve j}\right)
    
\vspace{2ex}\\

    \multirow{2}*{$\mathcal{O}_{  d^3 \bar{e}  u \bar{u}     }^{(1\sim 4)}$}
    
    &\mathcal{Y}\left[\tiny{\young(prs)}\right]\epsilon ^{bcd} \left(\bar{e}_{u} C \bar{u}_{va}\right) \left(d_{pa} C d_{sc}\right) \left(d_{rb} C u_{td}\right),
    \quad \mathcal{Y}\left[\tiny{\young(pr,s)}\right]\epsilon ^{bcd} \left(\bar{e}_{u} C \bar{u}_{va}\right) \left(d_{pa} C d_{sc}\right) \left(d_{rb} C u_{td}\right),
    \\&\mathcal{Y}\left[\tiny{\young(pr,s)}\right]\epsilon ^{acd} \left(\bar{e}_{u} C \bar{u}_{vb}\right) \left(d_{pa} C d_{sc}\right) \left(d_{rb} C u_{td}\right),
    \quad \mathcal{Y}\left[\tiny{\young(p,r,s)}\right]\epsilon ^{bcd} \left(\bar{e}_{u} C \bar{u}_{va}\right) \left(d_{pa} C d_{sc}\right) \left(d_{rb} C u_{td}\right)
    
\vspace{2ex}\\

    \multirow{3}*{$\mathcal{O}_{  d^2 \bar{l} q  u \bar{u}     }^{(1\sim 6)}$}
    
    &\mathcal{Y}\left[\tiny{\young(p,r)}\right]\epsilon ^{acd} \left(\bar{u}_{v}^{b} q_{ud i}\right) \left(\bar{l}_{s}^{i} d_{pa}\right) \left(d_{rb} C u_{tc}\right),
    \quad \mathcal{Y}\left[\tiny{\young(p,r)}\right]\epsilon ^{abc} \left(\bar{u}_{v}^{d} q_{ud i}\right) \left(\bar{l}_{s}^{i} d_{pa}\right) \left(d_{rb} C u_{tc}\right),
    \\&\mathcal{Y}\left[\tiny{\young(p,r)}\right]\epsilon ^{bcd} \left(\bar{u}_{v}^{a} q_{ud i}\right) \left(\bar{l}_{s}^{i} d_{pa}\right) \left(d_{rb} C u_{tc}\right),
    \quad \mathcal{Y}\left[\tiny{\young(pr)}\right]\epsilon ^{acd} \left(\bar{u}_{v}^{b} q_{ud i}\right) \left(\bar{l}_{s}^{i} d_{pa}\right) \left(d_{rb} C u_{tc}\right),
    \\&\mathcal{Y}\left[\tiny{\young(pr)}\right]\epsilon ^{abc} \left(\bar{u}_{v}^{d} q_{ud i}\right) \left(\bar{l}_{s}^{i} d_{pa}\right) \left(d_{rb} C u_{tc}\right),
    \quad \mathcal{Y}\left[\tiny{\young(pr)}\right]\epsilon ^{bcd} \left(\bar{u}_{v}^{a} q_{ud i}\right) \left(\bar{l}_{s}^{i} d_{pa}\right) \left(d_{rb} C u_{tc}\right)
    
\vspace{2ex}\\

    \mathcal{O}_{  d^4 \bar{d}  \bar{e}     }^{(1,2)}
    
    &\mathcal{Y}\left[\tiny{\young(prs,t)}\right]\epsilon ^{bcd} \left(\bar{e}_{v} C \bar{d}_u^a\right) \left(d_{pa} C d_{sc}\right) \left(d_{rb} C d_{td}\right),
    \quad \mathcal{Y}\left[\tiny{\young(pr,s,t)}\right]\epsilon ^{bcd} \left(\bar{e}_{v} C \bar{d}_u^a\right) \left(d_{pa} C d_{sc}\right) \left(d_{rb} C d_{td}\right)
    
\vspace{2ex}\\

    \multirow{2}*{$\mathcal{O}_{  d^3 \bar{d}  \bar{l} q     }^{(1\sim 4)}$}
    
    &\mathcal{Y}\left[\tiny{\young(prs)}\right]\epsilon ^{bce} \left(\bar{l}_{t}^{i} d_{rb}\right) \left(\bar{d}_{u}^{a} q_{ve i}\right) \left(d_{pa} C d_{sc}\right),
    \quad \mathcal{Y}\left[\tiny{\young(pr,s)}\right]\epsilon ^{ace} \left(\bar{l}_{t}^{i} d_{rb}\right) \left(\bar{d}_{u}^{b} q_{ve i}\right) \left(d_{pa} C d_{sc}\right),
    \\&\mathcal{Y}\left[\tiny{\young(pr,s)}\right]\epsilon ^{bce} \left(\bar{l}_{t}^{i} d_{rb}\right) \left(\bar{d}_{u}^{a} q_{ve i}\right) \left(d_{pa} C d_{sc}\right),
    \quad \mathcal{Y}\left[\tiny{\young(p,r,s)}\right]\epsilon ^{ace} \left(\bar{l}_{t}^{i} d_{rb}\right) \left(\bar{d}_{u}^{b} q_{ve i}\right) \left(d_{pa} C d_{sc}\right)
    
\end{array}\label{cl:p651}\end{align}
3. Operators involving 4 quarks and 2 leptons with $\Delta B=0$ and $\Delta L=-2$. Except the last two types $d\bar{l}^2 u\bar{u}^2$ and $d^2\bar{d} \bar{l}^2 \bar{u}$, all the types contain operators contributing the neutrinoless double beta decay at tree-level:
\begin{align}\begin{array}{c|l}

    \multirow{10}*{$\mathcal{O}_{  d^2 \bar{l}^2 \bar{q}^2    }^{(1\sim 20)} $}
    
    &\mathcal{Y}\left[\tiny{\young(p,r)},\tiny{\young(s,t)},\tiny{\young(u,v)}\right]\epsilon _{ik} \epsilon _{jm} \left(\bar{l}_{t}^{j} d_{pa}\right) \left(\bar{q}_{u}^{a k} d_{rb}\right) \left(\bar{l}_{s}^{i} C \bar{q}_{v}^{b m}\right),
    \quad \mathcal{Y}\left[\tiny{\young(p,r)},\tiny{\young(st)},\tiny{\young(uv)}\right]\epsilon _{ik} \epsilon _{jm} \left(\bar{l}_{t}^{j} d_{pa}\right) \left(\bar{q}_{u}^{a k} d_{rb}\right) \left(\bar{l}_{s}^{i} C \bar{q}_{v}^{b m}\right),
    \\&\mathcal{Y}\left[\tiny{\young(pr)},\tiny{\young(s,t)},\tiny{\young(uv)}\right]\epsilon _{ik} \epsilon _{jm} \left(\bar{l}_{t}^{j} d_{pa}\right) \left(\bar{q}_{u}^{a k} d_{rb}\right) \left(\bar{l}_{s}^{i} C \bar{q}_{v}^{b m}\right),
    \quad \mathcal{Y}\left[\tiny{\young(pr)},\tiny{\young(st)},\tiny{\young(u,v)}\right]\epsilon _{ik} \epsilon _{jm} \left(\bar{l}_{t}^{j} d_{pa}\right) \left(\bar{q}_{u}^{a k} d_{rb}\right) \left(\bar{l}_{s}^{i} C \bar{q}_{v}^{b m}\right),
    \\&\mathcal{Y}\left[\tiny{\young(p,r)},\tiny{\young(s,t)},\tiny{\young(uv)}\right]\epsilon _{ik} \epsilon _{jm} \left(\bar{l}_{t}^{j} d_{pa}\right) \left(\bar{q}_{u}^{a k} d_{rb}\right) \left(\bar{l}_{s}^{i} C \bar{q}_{v}^{b m}\right),
    \quad \mathcal{Y}\left[\tiny{\young(p,r)},\tiny{\young(s,t)},\tiny{\young(uv)}\right]\epsilon _{ij} \epsilon _{km} \left(\bar{l}_{t}^{j} d_{pa}\right) \left(\bar{q}_{u}^{a k} d_{rb}\right) \left(\bar{l}_{s}^{i} C \bar{q}_{v}^{b m}\right),
    \\&\mathcal{Y}\left[\tiny{\young(p,r)},\tiny{\young(s,t)},\tiny{\young(uv)}\right]\epsilon _{ik} \epsilon _{jm} \left(\bar{l}_{t}^{j} d_{pa}\right) \left(\bar{q}_{u}^{b k} d_{rb}\right) \left(\bar{l}_{s}^{i} C \bar{q}_{v}^{a m}\right),
    \quad \mathcal{Y}\left[\tiny{\young(p,r)},\tiny{\young(s,t)},\tiny{\young(uv)}\right]\epsilon _{ij} \epsilon _{km} \left(\bar{l}_{t}^{j} d_{pa}\right) \left(\bar{q}_{u}^{b k} d_{rb}\right) \left(\bar{l}_{s}^{i} C \bar{q}_{v}^{a m}\right),
    \\&\mathcal{Y}\left[\tiny{\young(p,r)},\tiny{\young(st)},\tiny{\young(u,v)}\right]\epsilon _{ik} \epsilon _{jm} \left(\bar{l}_{t}^{j} d_{pa}\right) \left(\bar{q}_{u}^{a k} d_{rb}\right) \left(\bar{l}_{s}^{i} C \bar{q}_{v}^{b m}\right),
    \quad \mathcal{Y}\left[\tiny{\young(p,r)},\tiny{\young(st)},\tiny{\young(u,v)}\right]\epsilon _{ij} \epsilon _{km} \left(\bar{l}_{t}^{j} d_{pa}\right) \left(\bar{q}_{u}^{a k} d_{rb}\right) \left(\bar{l}_{s}^{i} C \bar{q}_{v}^{b m}\right),
    \\&\mathcal{Y}\left[\tiny{\young(p,r)},\tiny{\young(st)},\tiny{\young(u,v)}\right]\epsilon _{ik} \epsilon _{jm} \left(\bar{l}_{t}^{j} d_{pa}\right) \left(\bar{q}_{u}^{b k} d_{rb}\right) \left(\bar{l}_{s}^{i} C \bar{q}_{v}^{a m}\right),
    \quad \mathcal{Y}\left[\tiny{\young(p,r)},\tiny{\young(st)},\tiny{\young(u,v)}\right]\epsilon _{ij} \epsilon _{km} \left(\bar{l}_{t}^{j} d_{pa}\right) \left(\bar{q}_{u}^{b k} d_{rb}\right) \left(\bar{l}_{s}^{i} C \bar{q}_{v}^{a m}\right),
    \\&\mathcal{Y}\left[\tiny{\young(pr)},\tiny{\young(s,t)},\tiny{\young(u,v)}\right]\epsilon _{ik} \epsilon _{jm} \left(\bar{l}_{t}^{j} d_{pa}\right) \left(\bar{q}_{u}^{a k} d_{rb}\right) \left(\bar{l}_{s}^{i} C \bar{q}_{v}^{b m}\right),
    \quad \mathcal{Y}\left[\tiny{\young(pr)},\tiny{\young(s,t)},\tiny{\young(u,v)}\right]\epsilon _{ij} \epsilon _{km} \left(\bar{l}_{t}^{j} d_{pa}\right) \left(\bar{q}_{u}^{a k} d_{rb}\right) \left(\bar{l}_{s}^{i} C \bar{q}_{v}^{b m}\right),
    \\&\mathcal{Y}\left[\tiny{\young(pr)},\tiny{\young(s,t)},\tiny{\young(u,v)}\right]\epsilon _{ik} \epsilon _{jm} \left(\bar{l}_{t}^{j} d_{pa}\right) \left(\bar{q}_{u}^{b k} d_{rb}\right) \left(\bar{l}_{s}^{i} C \bar{q}_{v}^{a m}\right),
    \quad \mathcal{Y}\left[\tiny{\young(pr)},\tiny{\young(s,t)},\tiny{\young(u,v)}\right]\epsilon _{ij} \epsilon _{km} \left(\bar{l}_{t}^{j} d_{pa}\right) \left(\bar{q}_{u}^{b k} d_{rb}\right) \left(\bar{l}_{s}^{i} C \bar{q}_{v}^{a m}\right),
    \\&\mathcal{Y}\left[\tiny{\young(pr)},\tiny{\young(st)},\tiny{\young(uv)}\right]\epsilon _{ik} \epsilon _{jm} \left(\bar{l}_{t}^{j} d_{pa}\right) \left(\bar{q}_{u}^{a k} d_{rb}\right) \left(\bar{l}_{s}^{i} C \bar{q}_{v}^{b m}\right),
    \quad \mathcal{Y}\left[\tiny{\young(pr)},\tiny{\young(st)},\tiny{\young(uv)}\right]\epsilon _{ij} \epsilon _{km} \left(\bar{l}_{t}^{j} d_{pa}\right) \left(\bar{q}_{u}^{a k} d_{rb}\right) \left(\bar{l}_{s}^{i} C \bar{q}_{v}^{b m}\right),
    \\&\mathcal{Y}\left[\tiny{\young(pr)},\tiny{\young(st)},\tiny{\young(uv)}\right]\epsilon _{ik} \epsilon _{jm} \left(\bar{l}_{t}^{j} d_{pa}\right) \left(\bar{q}_{u}^{b k} d_{rb}\right) \left(\bar{l}_{s}^{i} C \bar{q}_{v}^{a m}\right),
    \quad \mathcal{Y}\left[\tiny{\young(pr)},\tiny{\young(st)},\tiny{\young(uv)}\right]\epsilon _{ij} \epsilon _{km} \left(\bar{l}_{t}^{j} d_{pa}\right) \left(\bar{q}_{u}^{b k} d_{rb}\right) \left(\bar{l}_{s}^{i} C \bar{q}_{v}^{a m}\right)
    
\vspace{2ex}\\
    
    \multirow{4}*{$\mathcal{O}_{  \bar{l} ^2 q^2 \bar{u}^2    }^{(1\sim 8)}$}
    
    &\mathcal{Y}\left[\tiny{\young(p,r)},\tiny{\young(s,t)},\tiny{\young(uv)}\right]\left(\bar{u}_{s}^{a} q_{pa i}\right) \left(\bar{u}_{t}^{b} q_{rb j}\right) \left(\bar{l}_{u}^{i} C \bar{l}_{v}^{j}\right),
    \quad \mathcal{Y}\left[\tiny{\young(p,r)},\tiny{\young(s,t)},\tiny{\young(uv)}\right]\left(\bar{u}_{s}^{b} q_{pa i}\right) \left(\bar{u}_{t}^{a} q_{rb j}\right) \left(\bar{l}_{u}^{i} C \bar{l}_{v}^{j}\right),
    \\&\mathcal{Y}\left[\tiny{\young(pr)},\tiny{\young(s,t)},\tiny{\young(u,v)}\right]\left(\bar{u}_{s}^{a} q_{pa i}\right) \left(\bar{u}_{t}^{b} q_{rb j}\right) \left(\bar{l}_{u}^{i} C \bar{l}_{v}^{j}\right),
    \quad \mathcal{Y}\left[\tiny{\young(pr)},\tiny{\young(s,t)},\tiny{\young(u,v)}\right]\left(\bar{u}_{s}^{b} q_{pa i}\right) \left(\bar{u}_{t}^{a} q_{rb j}\right) \left(\bar{l}_{u}^{i} C \bar{l}_{v}^{j}\right),
    \\&\mathcal{Y}\left[\tiny{\young(pr)},\tiny{\young(st)},\tiny{\young(uv)}\right]\left(\bar{u}_{s}^{a} q_{pa i}\right) \left(\bar{u}_{t}^{b} q_{rb j}\right) \left(\bar{l}_{u}^{i} C \bar{l}_{v}^{j}\right),
    \quad \mathcal{Y}\left[\tiny{\young(pr)},\tiny{\young(st)},\tiny{\young(uv)}\right]\left(\bar{u}_{s}^{b} q_{pa i}\right) \left(\bar{u}_{t}^{a} q_{rb j}\right) \left(\bar{l}_{u}^{i} C \bar{l}_{v}^{j}\right),
    \\&\mathcal{Y}\left[\tiny{\young(p,r)},\tiny{\young(st)},\tiny{\young(u,v)}\right]\left(\bar{u}_{s}^{a} q_{pa i}\right) \left(\bar{u}_{t}^{b} q_{rb j}\right) \left(\bar{l}_{u}^{i} C \bar{l}_{v}^{j}\right),
    \quad \mathcal{Y}\left[\tiny{\young(p,r)},\tiny{\young(st)},\tiny{\young(u,v)}\right]\left(\bar{u}_{s}^{b} q_{pa i}\right) \left(\bar{u}_{t}^{a} q_{rb j}\right) \left(\bar{l}_{u}^{i} C \bar{l}_{v}^{j}\right)
    
\vspace{2ex}\\

    \multirow{2}*{$\mathcal{O}_{  d  \bar{e} \bar{l} q \bar{u}^2    }^{(1\sim 4)}$}
    
    &\mathcal{Y}\left[\tiny{\young(s,t)}\right]\left(\bar{u}_{t}^{a} q_{ra i}\right) \left(\bar{l}_{v}^{i} d_{ub}\right) \left(\bar{e}_{p} C \bar{u}_{s}^{b}\right),
    \quad \mathcal{Y}\left[\tiny{\young(s,t)}\right]\left(\bar{u}_{t}^{b} q_{ra i}\right) \left(\bar{l}_{v}^{i} d_{ub}\right) \left(\bar{e}_{p} C \bar{u}_{s}^{a}\right),
    \\&\mathcal{Y}\left[\tiny{\young(st)}\right]\left(\bar{u}_{t}^{a} q_{ra i}\right) \left(\bar{l}_{v}^{i} d_{ub}\right) \left(\bar{e}_{p} C \bar{u}_{s}^{b}\right),
    \quad \mathcal{Y}\left[\tiny{\young(st)}\right]\left(\bar{u}_{t}^{b} q_{ra i}\right) \left(\bar{l}_{v}^{i} d_{ub}\right) \left(\bar{e}_{p} C \bar{u}_{s}^{a}\right)
    
\vspace{2ex}\\

    \multirow{2}*{$\mathcal{O}_{  d^2 \bar{e}  \bar{l} \bar{q} \bar{u}     }^{(1\sim 4)}$}
    
    &\mathcal{Y}\left[\tiny{\young(p,r)}\right]\epsilon _{ij} \left(\bar{q}_{t}^{a j} d_{rb}\right) \left(\bar{l}_{s}^{i} d_{pa}\right) \left(\bar{e}_{u} C \bar{u}_{v}^{b}\right),
    \quad \mathcal{Y}\left[\tiny{\young(p,r)}\right]\epsilon _{ij} \left(\bar{q}_{t}^{b j} d_{rb}\right) \left(\bar{l}_{s}^{i} d_{pa}\right) \left(\bar{e}_{u} C \bar{u}_{v}^{a}\right),
    \\&\mathcal{Y}\left[\tiny{\young(pr)}\right]\epsilon _{ij} \left(\bar{q}_{t}^{a j} d_{rb}\right) \left(\bar{l}_{s}^{i} d_{pa}\right) \left(\bar{e}_{u} C \bar{u}_{v}^{b}\right),
    \quad \mathcal{Y}\left[\tiny{\young(pr)}\right]\epsilon _{ij} \left(\bar{q}_{t}^{b j} d_{rb}\right) \left(\bar{l}_{s}^{i} d_{pa}\right) \left(\bar{e}_{u} C \bar{u}_{v}^{a}\right)
    
\vspace{2ex}\\

    \multirow{4}*{$\mathcal{O}_{  d \bar{l}^2 q \bar{q}  \bar{u}     }^{(1\sim 8)}$}
    
    &\mathcal{Y}\left[\tiny{\young(r,s)}\right]\epsilon _{ik} \left(\bar{u}_{v}^{a} q_{ub j}\right) \left(\bar{l}_{s}^{j} d_{pa}\right) \left(\bar{l}_{r}^{i} C \bar{q}_{t}^{b k}\right),
    \quad \mathcal{Y}\left[\tiny{\young(r,s)}\right]\epsilon _{ij} \left(\bar{u}_{v}^{a} q_{ub k}\right) \left(\bar{l}_{s}^{j} d_{pa}\right) \left(\bar{l}_{r}^{i} C \bar{q}_{t}^{b k}\right),
    \\&\mathcal{Y}\left[\tiny{\young(r,s)}\right]\epsilon _{ik} \left(\bar{u}_{v}^{c} q_{uc j}\right) \left(\bar{l}_{s}^{j} d_{pa}\right) \left(\bar{l}_{r}^{i} C \bar{q}_{t}^{a k}\right),
    \quad \mathcal{Y}\left[\tiny{\young(r,s)}\right]\epsilon _{ij} \left(\bar{u}_{v}^{c} q_{uc k}\right) \left(\bar{l}_{s}^{j} d_{pa}\right) \left(\bar{l}_{r}^{i} C \bar{q}_{t}^{a k}\right),
    \\&\mathcal{Y}\left[\tiny{\young(rs)}\right]\epsilon _{ik} \left(\bar{u}_{v}^{a} q_{ub j}\right) \left(\bar{l}_{s}^{j} d_{pa}\right) \left(\bar{l}_{r}^{i} C \bar{q}_{t}^{b k}\right),
    \quad \mathcal{Y}\left[\tiny{\young(rs)}\right]\epsilon _{ij} \left(\bar{u}_{v}^{a} q_{ub k}\right) \left(\bar{l}_{s}^{j} d_{pa}\right) \left(\bar{l}_{r}^{i} C \bar{q}_{t}^{b k}\right),
    \\&\mathcal{Y}\left[\tiny{\young(rs)}\right]\epsilon _{ik} \left(\bar{u}_{v}^{c} q_{uc j}\right) \left(\bar{l}_{s}^{j} d_{pa}\right) \left(\bar{l}_{r}^{i} C \bar{q}_{t}^{a k}\right),
    \quad \mathcal{Y}\left[\tiny{\young(rs)}\right]\epsilon _{ij} \left(\bar{u}_{v}^{c} q_{uc k}\right) \left(\bar{l}_{s}^{j} d_{pa}\right) \left(\bar{l}_{r}^{i} C \bar{q}_{t}^{a k}\right)
        
\vspace{2ex}\\

    \multirow{2}*{$\mathcal{O}_{  d ^2 \bar{e}^2 \bar{u}^2    }^{(1\sim 4)}$}
    
    &\mathcal{Y}\left[\tiny{\young(p,r)},\tiny{\young(s,t)},\tiny{\young(uv)}\right]\left(d_{ua} C d_{vb}\right) \left(\bar{e}_{p} C \bar{u}_{s}^{a}\right) \left(\bar{e}_{r} C \bar{u}_{t}^{b}\right),
    \quad \mathcal{Y}\left[\tiny{\young(p,r)},\tiny{\young(st)},\tiny{\young(u,v)}\right]\left(d_{ua} C d_{vb}\right) \left(\bar{e}_{p} C \bar{u}_{s}^{a}\right) \left(\bar{e}_{r} C \bar{u}_{t}^{b}\right),
    \\&\mathcal{Y}\left[\tiny{\young(pr)},\tiny{\young(st)},\tiny{\young(uv)}\right]\left(d_{ua} C d_{vb}\right) \left(\bar{e}_{p} C \bar{u}_{s}^{a}\right) \left(\bar{e}_{r} C \bar{u}_{t}^{b}\right),
    \quad \mathcal{Y}\left[\tiny{\young(pr)},\tiny{\young(s,t)},\tiny{\young(u,v)}\right]\left(d_{ua} C d_{vb}\right) \left(\bar{e}_{p} C \bar{u}_{s}^{a}\right) \left(\bar{e}_{r} C \bar{u}_{t}^{b}\right)
    
\vspace{2ex}\\

    \multirow{2}*{$\mathcal{O}_{  d \bar{l}^2 u \bar{u} ^2    }^{(1\sim 4)}$}
    
    &\mathcal{Y}\left[\tiny{\young(rs)},\tiny{\young(uv)}\right]\epsilon _{ij} \left(\bar{l}_{s}^{j} d_{pa}\right) \left(\bar{l}_{r}^{i} u_{tb}\right) \left(\bar{u}_{u}^{a} C \bar{u}_{v}^{b}\right),
    \quad \mathcal{Y}\left[\tiny{\young(rs)},\tiny{\young(u,v)}\right]\epsilon _{ij} \left(\bar{l}_{s}^{j} d_{pa}\right) \left(\bar{l}_{r}^{i} u_{tb}\right) \left(\bar{u}_{u}^{a} C \bar{u}_{v}^{b}\right),
    \\&\mathcal{Y}\left[\tiny{\young(r,s)},\tiny{\young(uv)}\right]\epsilon _{ij} \left(\bar{l}_{s}^{j} d_{pa}\right) \left(\bar{l}_{r}^{i} u_{tb}\right) \left(\bar{u}_{u}^{a} C \bar{u}_{v}^{b}\right),
    \quad \mathcal{Y}\left[\tiny{\young(r,s)},\tiny{\young(u,v)}\right]\epsilon _{ij} \left(\bar{l}_{s}^{j} d_{pa}\right) \left(\bar{l}_{r}^{i} u_{tb}\right) \left(\bar{u}_{u}^{a} C \bar{u}_{v}^{b}\right)
    
\vspace{2ex}\\

    \multirow{2}*{$\mathcal{O}_{  d^2 \bar{d}  \bar{l}^2 \bar{u}     }^{(1\sim 4)}$}
    
    &\mathcal{Y}\left[\tiny{\young(p,r)},\tiny{\young(st)}\right]\epsilon _{ij} \left(\bar{l}_{s}^{i} d_{pa}\right) \left(\bar{l}_{t}^{j} d_{rb}\right) \left(\bar{d}_{u}^{a} C \bar{u}_{v}^{b}\right),
    \quad \mathcal{Y}\left[\tiny{\young(pr)},\tiny{\young(st)}\right]\epsilon _{ij} \left(\bar{l}_{s}^{i} d_{pa}\right) \left(\bar{l}_{t}^{j} d_{rb}\right) \left(\bar{d}_{u}^{a} C \bar{u}_{v}^{b}\right),
    \\&\mathcal{Y}\left[\tiny{\young(pr)},\tiny{\young(s,t)}\right]\epsilon _{ij} \left(\bar{l}_{s}^{i} d_{pa}\right) \left(\bar{l}_{t}^{j} d_{rb}\right) \left(\bar{d}_{u}^{a} C \bar{u}_{v}^{b}\right),
    \quad \mathcal{Y}\left[\tiny{\young(p,r)},\tiny{\young(s,t)}\right]\epsilon _{ij} \left(\bar{l}_{s}^{i} d_{pa}\right) \left(\bar{l}_{t}^{j} d_{rb}\right) \left(\bar{d}_{u}^{a} C \bar{u}_{v}^{b}\right)
    
\end{array}\label{cl:p642}\end{align}
Compared with \cite{Graesser:2016bpz}, after taking flavor symmetries of fermion into account, operators we listed are more complete than theirs. 
Some basis here could change into their form using the Fierz Identities $\left(\lambda^A\right)^a_b\left(\lambda^A\right)^c_d=\delta^a_d\delta^c_b-\frac13\delta^a_b\delta^c_d$ , $\epsilon^{\alpha\beta}\delta^{\gamma_{\kappa}}+\epsilon^{\beta\gamma}\delta^{\alpha}_{\kappa}+\epsilon^{\gamma\alpha}\delta^{\beta}_{\gamma}=0$. For example,
\begin{align}
    \left\{\begin{array}{c}
        \mathcal{Y}\left[\tiny{\young(pr)},\tiny{\young(s,t)},\tiny{\young(u,v)}\right]\epsilon _{ik} \epsilon _{jm} \left(\bar{l}_{t}^{j} d_{pa}\right) \left(\bar{q}_{u}^{a k} d_{rb}\right) \left(\bar{l}_{s}^{i} C \bar{q}_{v}^{b m}\right)\\
        \mathcal{Y}\left[\tiny{\young(pr)},\tiny{\young(s,t)},\tiny{\young(u,v)}\right]\epsilon _{ij} \epsilon _{km} \left(\bar{l}_{t}^{j} d_{pa}\right) \left(\bar{q}_{u}^{a k} d_{rb}\right) \left(\bar{l}_{s}^{i} C \bar{q}_{v}^{b m}\right)\\
        \mathcal{Y}\left[\tiny{\young(pr)},\tiny{\young(s,t)},\tiny{\young(u,v)}\right]\epsilon _{ik} \epsilon _{jm} \left(\bar{l}_{t}^{j} d_{pa}\right) \left(\bar{q}_{u}^{b k} d_{rb}\right) \left(\bar{l}_{s}^{i} C \bar{q}_{v}^{a m}\right)\\
        \mathcal{Y}\left[\tiny{\young(pr)},\tiny{\young(s,t)},\tiny{\young(u,v)}\right]\epsilon _{ij} \epsilon _{km} \left(\bar{l}_{t}^{j} d_{pa}\right) \left(\bar{q}_{u}^{b k} d_{rb}\right) \left(\bar{l}_{s}^{i} C \bar{q}_{v}^{a m}\right)
    \end{array}\right.\Longrightarrow\left\{\begin{array}{c}
        \mathcal{Y}\left[\tiny{\young(pr)},\tiny{\young(s,t)},\tiny{\young(u,v)}\right]\epsilon _{ik} \epsilon _{jm} \left(\bar{q}_{v}^{m} d_p\right) \left(\bar{q}_{u}^{k} d_r\right) \left(\bar{l}_{s}^{i} C \bar{l}_{t}^{j}\right)\\
        \mathcal{Y}\left[\tiny{\young(pr)},\tiny{\young(s,t)},\tiny{\young(u,v)}\right]\epsilon _{ij} \epsilon _{km} \left(\bar{q}_{v}^{m} d_p\right) \left(\bar{q}_{u}^{k} d_r\right) \left(\bar{l}_{s}^{i} C \bar{l}_{t}^{j}\right)\\
        \mathcal{Y}\left[\tiny{\young(pr)},\tiny{\young(s,t)},\tiny{\young(u,v)}\right]\epsilon _{ik} \epsilon _{jm} \left(\bar{q}_{v}^{m} \lambda^A d_p\right) \left(\bar{q}_{u}^{k} \lambda^A d_r\right) \left(\bar{l}_{s}^{i} C \bar{l}_{t}^{j}\right)\\
        \mathcal{Y}\left[\tiny{\young(pr)},\tiny{\young(s,t)},\tiny{\young(u,v)}\right]\epsilon _{ij} \epsilon _{km} \left(\bar{q}_{v}^{m} \lambda^A d_p\right) \left(\bar{q}_{u}^{k} \lambda^A d_r\right) \left(\bar{l}_{s}^{i} C \bar{l}_{t}^{j}\right)
       
    \end{array}\right.
\end{align}
In most cases, the Fierz Identities won't help us simplify the operators any further or put them in more familiar forms. 
\\4. Operators involving 3 quarks and 3 lepton with $\Delta B=1$ and $\Delta L=-1$ or 3, the type $l^3qu^2$ is relevant for the proton three body decay:
\begin{align}\begin{array}{c|l}

    \multirow{3}*{$\mathcal{O}_{  d^2 e \bar{l}^2 u   }^{(1\sim 5)}$}
    
    &\mathcal{Y}\left[\tiny{\young(pr)},\tiny{\young(tu)}\right]\epsilon ^{abc} \epsilon _{ij} \left(\bar{l}_{t}^{i} d_{pa}\right) \left(\bar{l}_{u}^{j} d_{rb}\right) \left(e_{s} C u_{vc}\right),
    \quad \mathcal{Y}\left[\tiny{\young(pr)},\tiny{\young(tu)}\right]\epsilon ^{abc} \epsilon _{ij} \left(\bar{l}_{u}^{j} d_{rb}\right) \left(\bar{l}_{t}^{i} u_{vc}\right) \left(e_{s} C d_{pa}\right),
    \\&\mathcal{Y}\left[\tiny{\young(pr)},\tiny{\young(t,u)}\right]\epsilon ^{abc} \epsilon _{ij} \left(\bar{l}_{u}^{j} d_{rb}\right) \left(\bar{l}_{t}^{i} u_{vc}\right) \left(e_{s} C d_{pa}\right),
    \quad \mathcal{Y}\left[\tiny{\young(p,r)},\tiny{\young(tu)}\right]\epsilon ^{abc} \epsilon _{ij} \left(\bar{l}_{u}^{j} d_{rb}\right) \left(\bar{l}_{t}^{i} u_{vc}\right) \left(e_{s} C d_{pa}\right),
    \\&\mathcal{Y}\left[\tiny{\young(p,r)},\tiny{\young(t,u)}\right]\epsilon ^{abc} \epsilon _{ij} \left(\bar{l}_{t}^{i} d_{pa}\right) \left(\bar{l}_{u}^{j} d_{rb}\right) \left(e_{s} C u_{vc}\right)
    
\vspace{2ex}\\

    \multirow{3}*{$\mathcal{O}_{  l^3 q u ^2    }^{(1\sim 3)}$}
    
    &\mathcal{Y}\left[\tiny{\young(prs)},\tiny{\young(u,v)}\right]\epsilon ^{abc} \epsilon ^{ij} \epsilon ^{km} \left(u_{ub} C u_{vc}\right) \left(l_{pi} C l_{sk}\right) \left(l_{rj} C q_{ta m}\right),
    \\&\mathcal{Y}\left[\tiny{\young(pr,s)},\tiny{\young(u,v)}\right]\epsilon ^{abc} \epsilon ^{ik} \epsilon ^{jm} \left(u_{ub} C u_{vc}\right) \left(l_{pi} C l_{sk}\right) \left(l_{rj} C q_{ta m}\right),
    \\& \mathcal{Y}\left[\tiny{\young(p,r,s)},\tiny{\young(u,v)}\right]\epsilon ^{abc} \epsilon ^{ik} \epsilon ^{jm} \left(u_{ub} C u_{vc}\right) \left(l_{pi} C l_{sk}\right) \left(l_{rj} C q_{ta m}\right)
    
\vspace{2ex}\\

    \mathcal{O}_{  e l ^2 u^3    }
    
    &\mathcal{Y}\left[\tiny{\young(rs,t)},\tiny{\young(u,v)}\right]\epsilon ^{abc} \epsilon ^{ij} \left(u_{ra} C u_{tc}\right) \left(e_{p} C u_{sb}\right) \left(l_{ui} C l_{vj}\right)
    
\vspace{2ex}\\

    \mathcal{O}_{  d^3 \bar{e}  \bar{l} l     }
    
    &\mathcal{Y}\left[\tiny{\young(pr,s)}\right]\epsilon ^{abc} \left(\bar{e}_{u} l_{vi}\right) \left(\bar{l}_{t}^{i} d_{rb}\right) \left(d_{pa} C d_{sc}\right)
    
\vspace{2ex}\\

    \mathcal{O}_{  d^3 e \bar{e} ^2    }
    
    &\mathcal{Y}\left[\tiny{\young(pr,s)},\tiny{\young(uv)}\right]\epsilon ^{abc} \left(\bar{e}_{u} C \bar{e}_{v}\right) \left(d_{pa} C d_{sc}\right) \left(d_{rb} C e_{t}\right)
    
\vspace{2ex}\\

    \multirow{2}*{$\mathcal{O}_{  d^2 l \bar{l}^2  q     }^{(1\sim 4)}$}
    
    &\mathcal{Y}\left[\tiny{\young(pr)},\tiny{\young(s,t)}\right]\epsilon ^{abc} \left(\bar{l}_{s}^{i} d_{pa}\right) \left(\bar{l}_{t}^{j} d_{rb}\right) \left(l_{ui} C q_{vc j}\right),
    \quad \mathcal{Y}\left[\tiny{\young(pr)},\tiny{\young(st)}\right]\epsilon ^{abc} \left(\bar{l}_{s}^{i} d_{pa}\right) \left(\bar{l}_{t}^{j} d_{rb}\right) \left(l_{ui} C q_{vc j}\right),
    \\&\mathcal{Y}\left[\tiny{\young(p,r)},\tiny{\young(st)}\right]\epsilon ^{abc} \left(\bar{l}_{s}^{i} d_{pa}\right) \left(\bar{l}_{t}^{j} d_{rb}\right) \left(l_{ui} C q_{vc j}\right),
    \quad \mathcal{Y}\left[\tiny{\young(p,r)},\tiny{\young(s,t)}\right]\epsilon ^{abc} \left(\bar{l}_{s}^{i} d_{pa}\right) \left(\bar{l}_{t}^{j} d_{rb}\right) \left(l_{ui} C q_{vc j}\right)
    
\vspace{2ex}\\

    \mathcal{O}_{  d^2 e \bar{e}  \bar{l} q     }^{(1,2)}
    
    &\mathcal{Y}\left[\tiny{\young(pr)}\right]\epsilon ^{abc} \left(\bar{l}_{t}^{i} d_{rb}\right) \left(\bar{e}_{u} q_{vc i}\right) \left(e_{s} C d_{pa}\right),
    \quad \mathcal{Y}\left[\tiny{\young(p,r)}\right]\epsilon ^{abc} \left(\bar{l}_{t}^{i} d_{rb}\right) \left(\bar{e}_{u} q_{vc i}\right) \left(e_{s} C d_{pa}\right)
    
\vspace{2ex}\\

    \multirow{2}*{$\mathcal{O}_{  d e \bar{l}^2 q ^2    }^{(1\sim 4)}$}
    
    &\mathcal{Y}\left[\tiny{\young(s,t)},\tiny{\young(u,v)}\right]\epsilon ^{abc} \left(\bar{l}_{t}^{j} e_r\right) \left(\bar{l}_{s}^{i} d_{pa}\right) \left(q_{ub i} C q_{vc j}\right),
    \quad \mathcal{Y}\left[\tiny{\young(st)},\tiny{\young(uv)}\right]\epsilon ^{abc} \left(\bar{l}_{t}^{j} e_r\right) \left(\bar{l}_{s}^{i} d_{pa}\right) \left(q_{ub i} C q_{vc j}\right),
    \\&\mathcal{Y}\left[\tiny{\young(st)},\tiny{\young(u,v)}\right]\epsilon ^{abc} \left(\bar{l}_{t}^{j} e_r\right) \left(\bar{l}_{s}^{i} d_{pa}\right) \left(q_{ub i} C q_{vc j}\right),
    \quad \mathcal{Y}\left[\tiny{\young(s,t)},\tiny{\young(uv)}\right]\epsilon ^{abc} \left(\bar{l}_{t}^{j} e_r\right) \left(\bar{l}_{s}^{i} d_{pa}\right) \left(q_{ub i} C q_{vc j}\right)
    
\end{array}\label{cl:p633}\end{align}
5. Operators involving 2 quarks and 4 leptons with $\Delta B=0$ and $\Delta L=-2$:
\begin{align}\begin{array}{c|l}

    \multirow{4}*{$\mathcal{O}_{  d e \bar{l}^3 \bar{q}    }^{(1\sim 7)} $}
    
    &\mathcal{Y}\left[\tiny{\young(st,u)}\right]\epsilon _{ik} \epsilon _{jm} \left(\bar{l}_{u}^{k} e_r\right) \left(\bar{l}_{t}^{j} d_{pa}\right) \left(\bar{l}_{s}^{i} C \bar{q}_{v}^{a m}\right),
    \quad \mathcal{Y}\left[\tiny{\young(st,u)}\right]\epsilon _{ij} \epsilon _{km} \left(\bar{l}_{u}^{k} e_r\right) \left(\bar{l}_{t}^{j} d_{pa}\right) \left(\bar{l}_{s}^{i} C \bar{q}_{v}^{a m}\right),
    \\&\mathcal{Y}\left[\tiny{\young(st,u)}\right]\epsilon _{ik} \epsilon _{jm} \left(\bar{l}_{u}^{k} e_r\right) \left(\bar{l}_{s}^{i} d_{pa}\right) \left(\bar{l}_{t}^{j} C \bar{q}_{v}^{a m}\right),
    \quad \mathcal{Y}\left[\tiny{\young(stu)}\right]\epsilon _{ik} \epsilon _{jm} \left(\bar{l}_{u}^{k} e_r\right) \left(\bar{l}_{t}^{j} d_{pa}\right) \left(\bar{l}_{s}^{i} C \bar{q}_{v}^{a m}\right),
    \\&\mathcal{Y}\left[\tiny{\young(stu)}\right]\epsilon _{ij} \epsilon _{km} \left(\bar{l}_{u}^{k} e_r\right) \left(\bar{l}_{t}^{j} d_{pa}\right) \left(\bar{l}_{s}^{i} C \bar{q}_{v}^{a m}\right),
    \quad \mathcal{Y}\left[\tiny{\young(s,t,u)}\right]\epsilon _{ik} \epsilon _{jm} \left(\bar{l}_{u}^{k} e_r\right) \left(\bar{l}_{t}^{j} d_{pa}\right) \left(\bar{l}_{s}^{i} C \bar{q}_{v}^{a m}\right),
    \\&\mathcal{Y}\left[\tiny{\young(s,t,u)}\right]\epsilon _{ij} \epsilon _{km} \left(\bar{l}_{u}^{k} e_r\right) \left(\bar{l}_{t}^{j} d_{pa}\right) \left(\bar{l}_{s}^{i} C \bar{q}_{v}^{a m}\right)
    
\end{array}\label{cl:p6241}\end{align}    
\begin{align}\begin{array}{c|l}
    
    \multirow{2}*{$\mathcal{O}_{  d l\bar{l}^3  \bar{u}     }^{(1\sim 3)}$}
    
    &\mathcal{Y}\left[\tiny{\young(rst)}\right]\epsilon _{ij} \left(\bar{u}_{v}^{a} l_{uk}\right) \left(\bar{l}_{s}^{j} d_{pa}\right) \left(\bar{l}_{r}^{i} C \bar{l}_{t}^{k}\right),
    \\&\mathcal{Y}\left[\tiny{\young(rs,t)}\right]\epsilon _{ik} \left(\bar{u}_{v}^{a} l_{uj}\right) \left(\bar{l}_{s}^{j} d_{pa}\right) \left(\bar{l}_{r}^{i} C \bar{l}_{t}^{k}\right),
    \quad \mathcal{Y}\left[\tiny{\young(r,s,t)}\right]\epsilon _{ik} \left(\bar{u}_{v}^{a} l_{uj}\right) \left(\bar{l}_{s}^{j} d_{pa}\right) \left(\bar{l}_{r}^{i} C \bar{l}_{t}^{k}\right)
    
\vspace{2ex}\\

    \mathcal{O}_{  d e \bar{e}  \bar{l}^2 \bar{u}     }^{(1,2)}
    
    &\mathcal{Y}\left[\tiny{\young(st)}\right]\epsilon _{ij} \left(\bar{l}_{t}^{j} e_r\right) \left(\bar{l}_{s}^{i} d_{pa}\right) \left(\bar{e}_{u} C \bar{u}_{v}^{a}\right),
    \quad \mathcal{Y}\left[\tiny{\young(s,t)}\right]\epsilon _{ij} \left(\bar{l}_{t}^{j} e_r\right) \left(\bar{l}_{s}^{i} d_{pa}\right) \left(\bar{e}_{u} C \bar{u}_{v}^{a}\right)
    
\vspace{2ex}\\
    
    \multirow{2}*{$\mathcal{O}_{  e \bar{l}^3 q  \bar{u}     }^{(1\sim 3)}$}
    
    &\mathcal{Y}\left[\tiny{\young(rst)}\right]\epsilon _{ij} \left(\bar{u}_{v}^{a} q_{ua k}\right) \left(\bar{l}_{s}^{j} e_p\right) \left(\bar{l}_{r}^{i} C \bar{l}_{t}^{k}\right),
    \\&\mathcal{Y}\left[\tiny{\young(rs,t)}\right]\epsilon _{ik} \left(\bar{u}_{v}^{a} q_{ua i}\right) \left(\bar{l}_{s}^{j} e_p\right) \left(\bar{l}_{r}^{i} C \bar{l}_{t}^{k}\right),
    \quad \mathcal{Y}\left[\tiny{\young(r,s,t)}\right]\epsilon _{ik} \left(\bar{u}_{v}^{a} q_{ua i}\right) \left(\bar{l}_{s}^{j} e_p\right) \left(\bar{l}_{r}^{i} C \bar{l}_{t}^{k}\right)
   
\end{array}\label{cl:p6242}\end{align}
6. Operators involving 6 leptons with $\Delta L=-2$:
\begin{align}\begin{array}{c|l}

    \multirow{2}*{$\mathcal{O}_{  e^2 \bar{l}^4    }^{(1\sim 4)} $}
    
    &\mathcal{Y}\left[\tiny{\young(p,r)},\tiny{\young(stu,v)}\right]\epsilon _{ik} \epsilon _{jm} \left(\bar{l}_{t}^{j} e_p\right) \left(\bar{l}_{u}^{k} e_r\right) \left(\bar{l}_{s}^{i} C \bar{l}_{v}^{m}\right),
    \quad \mathcal{Y}\left[\tiny{\young(p,r)},\tiny{\young(st,u,v)}\right]\epsilon _{ik} \epsilon _{jm} \left(\bar{l}_{t}^{j} e_p\right) \left(\bar{l}_{u}^{k} e_r\right) \left(\bar{l}_{s}^{i} C \bar{l}_{v}^{m}\right),
    \\&\mathcal{Y}\left[\tiny{\young(pr)},\tiny{\young(stuv)}\right]\epsilon _{ik} \epsilon _{jm} \left(\bar{l}_{t}^{j} e_p\right) \left(\bar{l}_{u}^{k} e_r\right) \left(\bar{l}_{s}^{i} C \bar{l}_{v}^{m}\right),
    \quad \mathcal{Y}\left[\tiny{\young(pr)},\tiny{\young(st,uv)}\right]\epsilon _{ik} \epsilon _{jm} \left(\bar{l}_{t}^{j} e_p\right) \left(\bar{l}_{u}^{k} e_r\right) \left(\bar{l}_{s}^{i} C \bar{l}_{v}^{m}\right)
    
\end{array}\label{cl:p606}\end{align}

\section{Conclusion}
\label{sec:conclusion}

In this paper, we provided the full result of the independent dimension 9 operator basis in the SMEFT. The numbers of operators at different levels are summarized in the table~\ref{tab:sumdim9}. According to our study, we find a total of 1262 terms, in which the total number of flavor-specified operators for 3 fermion generations is 90456, while that for 1 fermion generation is 560, agreeing with the previous counting\footnote{The number of terms does not match that in \cite{Fonseca:2019yya} because we have slightly different definitions of the concept of ``term''. } \cite{Henning:2015alf,Fonseca:2019yya}. 
A complete list of dim-9 SMEFT operators is very meaningful,
because new operators with $(\Delta B, \Delta L) = (\pm 1,\pm 3),(\pm 2, 0)$ start to appear at dimension 9, which signal the phenomenologies of the neutron-antineutron oscillations and the proton three-body decays with new physics scale reachable for the future LHC experiments. 

Our operator enumeration method starts from the amplitude-operator correspondence, where we one-to-one map effective operators to local amplitudes. The correspondence provides a natural way to unambiguously divide the operator space into subspaces that we call types, each consisting of operators at a given dimension that only generates local amplitudes for a given set of external particles. Our categorization of types makes use of the EOM, and is more rigorous than the old definition that only counts the apparent numbers of fields and derivatives in the operator.
Moreover, the repeated field issue for operators that becomes important at higher dimensions also has exact correspondence with the spin-statistic constraint for amplitudes. 
The correspondence thus provides a novel path towards operator enumeration -- by enumerating the amplitude basis. 
We claim that the correspondence has deep physical reason, as both sides are a complete basis of input for effective theories -- on the amplitude side, it is interesting to see how the same amount of input could be used to construct the whole theory in an on-shell way.

To enumerate independent flavor-specified operators for a given type, 
we introduce the concepts of the y-basis, m-basis and p-basis operators. 
We develop the algorithm to enumerate the y-basis with the help of the auxiliary $SU(N)$ group for the Lorentz structures and the novel L-R procedure technique for the gauge groups, whose completeness and independence are guaranteed by group theory. 
Based on the y-basis, the m-basis is obtained by converting y-basis operators into a set of independent monomials that are familiar to the phenomenology community.
%
%
%
%
The p-basis of Lorentz structures and gauge group tensors of an operator, are obtained from either y-basis or m-basis by acting on them with a set of symmetrizers, which are the basis of the left ideal in the symmetric group algebra for the repeated fields.
Combining the factors by the inner product decomposition, we build the ``terms'' as irreducible flavor tensors of operators.


The p-basis operators are usually combinations of multiple monomials as a result of symmetrization, which makes them lengthy to express. In this work we provided a systematical way that we call ``de-symmetrization'' to solve the problem by expressing our final result in the form of a Young symmetrizer acting on a monomial operator. The algorithm was not discussed in detail in our previous paper. 
The subtlety was to guarantee the independence among the operators with the same symmetrizer acting on different monomials, which is necessary only when $n_\lambda$, the number of certain representation space in a given type, is greater than one. 
The de-symmetrization procedure results in independent combinations of the $p$-basis, thus named $p'$-basis, which have quite concise expressions with the Young symmetrizer denoting the flavor symmetry. 

One may also make the Young symmetrizer to act on the Wilson coefficient tensor with which the monomial operator contracts, so that the operator becomes a monomial genuinely while the Young symmetrizer only serves as a reminder of the symmetry of the Wilson coefficients. It is equivalent to the flavor relations that the traditional method of operator enumeration applies to solve the repeated field issue.
We summarize the advantages of our method and final notation over the traditional method and notation:
\begin{itemize}
\item The completeness and independence are guaranteed by the underlining mathematical principle. The flavor symmetries among the Wilson coefficients are given systematically, unlike in the traditional treatment where flavor relations should be found manually.
\item It enables one to directly write down the flavor-specified operators by enumerating the flavor SSYT's of the corresponding flavor symmetry. This is the most important reason that we insist in expressing our final result as the irreducible flavor tensors, as it is tricky to list the independent operators from the flavor relations accompanied by the traditional form of the operators.
\item We provide a systematic way to convert any basis into our $y$-basis without any ambiguity, or, by using the conversion matrices that we also obtained, into any other basis that we provide here. 
Therefore, our basis could serve as the standard basis of operators. 
\end{itemize}
The last point will in principle benefit a lot of studies about effective field theory. For example, in matching between the UV new physics and the SMEFT operators, an independent and complete basis of operators is necessary for an unambiguous result. Therefore we need to identify the operator generated after integrating out heavy particles as a unique coordinate with respect to an independent and complete operator basis. Note that in reducing such an operator to our $y$-basis, terms eliminated by the EOM or the $[D,D]$ identity in this paper should be kept in the form of other types of operators. We will leave it for our future work.

\comments{
In principle, the last point will guarantee that any SMEFT operators generated by the matching in a study of the UV new physics with the top-down approach where heavy degrees of freedoms are integrated out, either with direct calculation of the relevant Feynman diagrams or with the covariant derivative expansion technique, can be converted to our standard basis of operators. 
However, unlike our treatment in finding a complete and independent basis of operators where we simply neglected EOMs and the anti-commutators of covariant derivatives, the full forms of EOMs are needed up to dimension 6 level during each step of the conversion processes and the anti-commutators of covariant derivatives should be replaced by the corresponding gauge boson field strength tensors, as the relations between the Wilson coefficients of different types become important.
Therefore, the automation of such processes remains a challenging programming problem that we leave for our future work.
}



\section*{Acknowledgements}

J.H.Y. would like to thank Yi Liao for correspondence on the same topic. H.L.L, Z.R. and J.H.Y. are supported by the National Science Foundation of China (NSFC) under Grants No. 11875003.  
J.H.Y. is also supported by the National Science Foundation of China (NSFC) under Grants No. 11947302. 
M.L.X. is supported by the National Natural Science Foundation of China (NSFC) under grant No.2019M650856 and the 2019 International Postdoctoral Exchange Fellowship Program.

\section*{Note added}

To our knowledge, Ref.~\cite{Liao:2020} also presents a list of the dimension nine operators in the standard model effective field theory which use the traditional method, imposing the constraints of EOM and IBP and finding the flavor relations after obtaining an over complete set of operators to eliminate the redundancies. Our method compared to theirs guarantees the independence and completeness with mathematical principles and separates different flavor permutation symmetries of the operators or the Wilson coefficients as independent irreducible flavor tensors enabling readers to directly written down the independent flavor components via enumerating certain semi-standard Young Tableau.

\appendix

\section{Notations and Conversions among Bases}
\label{app:A}
In this appendix, we present conversion relations of Lorentz structures between different notations for users' convenience. Relevant notations are Weyl spinors v.s. Dirac spinors and $SL(2,\mathbb{C})$ spinor indices v.s. $SO(3,1)$ Lorentz indices.\\

\noindent\underline{1. Converting four-component to two-component spinor} \\

In this part, $\Psi$ and $\bar{\Psi}$ denote 4-component spinors, $\xi$ and $\chi$ denote 2-component left-handed spinors, and their Hermitian conjugates $\xi^\dagger,\chi^\dagger$ denote 2-component right-handed spinors. 
Generally, a 4-component spinor consists of a 2-component left-handed spinor $\xi_{\alpha}$ and a 2-component right-handed spinor $\chi^{\dagger\dot\alpha}$
\begin{align}
\Psi=\left(\begin{array}{c} \xi_{\alpha}\\\chi^{\dagger\dot{\alpha}} \end{array} \right),\quad \bar{\Psi}=\Psi^{\dagger}\gamma^0=\left(\chi^{\alpha},\;\xi^{\dagger}_{\dot{\alpha}} \right)\;.
\end{align}
Here we provide some conversion relations for the following spinor bilinears:
\eq{\label{eq:bilinear}
	\bar{\Psi}_1\Psi_2=&\chi_1^{\alpha}\xi_{2\alpha}+\xi^{\dagger}_{1\dot{\alpha}}\chi^{\dagger\dot{\alpha}}_2\;,\\
	\bar{\Psi}_1\gamma^{\mu}\Psi_2=&\chi_1^{\alpha}\sigma^{\mu}_{\alpha\dot{\alpha}}\chi^{\dagger\dot{\alpha}}_2+\xi^{\dagger}_{1\dot{\alpha}}\bar{\sigma}^{\mu\dot{\alpha}\alpha}\xi_{2\alpha}\;,\\
	\bar{\Psi}_1\sigma^{\mu\nu}\Psi_2=&\chi_1^{\alpha}\left(\sigma^{\mu\nu}\right)_\alpha{}^\beta\xi_{2\beta}+\xi^{\dagger}_{1\dot{\alpha}}\left(\bar{\sigma}^{\mu\nu}\right)^{\dot{\alpha}}{}_{\dot{\beta}}\chi^{\dot{\beta}}_{2}\;,\\
	\Psi^{\rm{T}}_1C\Psi_2=&\xi_1^{\alpha}\xi_{2\alpha}+\chi^{\dagger}_{1\dot{\alpha}}\chi^{\dagger\dot{\alpha}}_2\;,\\
	\Psi^{\rm{T}}_1C\gamma^{\mu}\Psi_2=&\xi_1^{\alpha}\sigma^{\mu}_{\alpha\dot{\alpha}}\chi^{\dagger\dot{\alpha}}_2+\chi^{\dagger}_{1\dot{\alpha}}\bar{\sigma}^{\mu\dot{\alpha}\alpha}\xi_{2\alpha}\;,\\
	\Psi^{\rm{T}}_1C\sigma^{\mu\nu}\Psi_2=&\xi_1^{\alpha}\left(\sigma^{\mu\nu}\right)_\alpha{}^\beta\xi_{2\beta}+\chi^{\dagger}_{1\dot{\alpha}}\left(\bar{\sigma}^{\mu\nu}\right)^{\dot{\alpha}}{}_{\dot{\beta}}\chi^{\dagger\dot{\beta}}_2\;,\\
	\bar{\Psi}_1C\bar{\Psi}_2^{\rm{T}}=&\xi^{\dagger}_{1\dot{\alpha}}\xi^{\dagger\dot{\alpha}}_2+\chi^{\alpha}_1\chi_{2\alpha}\;,\\
	\bar{\Psi}_1\gamma^{\mu}C\bar{\Psi}_2^{\rm{T}}=&\chi_1^{\alpha}\sigma^{\mu}_{\alpha\dot{\alpha}}\xi^{\dagger\dot{\alpha}}_2+\xi^{\dagger}_{1\dot{\alpha}}\bar{\sigma}^{\mu\dot{\alpha}\alpha}\chi_{2\alpha}\;,\\
	\bar{\Psi}_1\sigma^{\mu\nu}C\bar{\Psi}_2^{\rm{T}}=&\xi^{\dagger}_{1\dot{\alpha}}\left(\bar{\sigma}^{\mu\nu}\right)^{\dot{\alpha}}{}_{\dot{\beta}}\xi^{\dagger\dot{\beta}}_2+\chi^{\alpha}_1\left(\sigma^{\mu\nu}\right)_\alpha{}^\beta\chi_{2\beta}\;.
}
where in the chiral representation $C=i\gamma^0\gamma^2=\begin{pmatrix} \epsilon_{\alpha\beta}&0\\0&\epsilon^{\dot{\alpha}\dot{\beta}}\end{pmatrix}=\begin{pmatrix} -\epsilon^{\alpha\beta}&0\\0&-\epsilon_{\dot{\alpha}\dot{\beta}}\end{pmatrix}$, $\gamma^{\mu}=\begin{pmatrix}
0&\sigma^{\mu}_{\alpha\dot{\beta}}\\\bar{\sigma}^{\mu\dot{\alpha}\beta}&0
\end{pmatrix}$ and $\sigma^{\mu\nu}=\dfrac{i}{2}[\gamma^\mu,\gamma^\nu]=\begin{pmatrix}
\left(\sigma^{\mu\nu}\right)_\alpha{}^\beta&0\\0&\left(\bar{\sigma}^{\mu\nu}\right)^{\dot{\alpha}}{}_{\dot{\beta}}
\end{pmatrix}$.

The Hermitian conjugates of 2-component spinor bilinears mentioned above are given by
\eq{
	\left(\chi\xi\right)^{\dagger}&
	=-\xi^{\dagger}\chi^{\dagger},\\
	\left(\chi\sigma^{\mu}\xi^{\dagger}\right)^{\dagger}&
	=-\xi\sigma^{\mu}\chi^{\dagger},\\
	\left(\chi^{\dagger}\bar{\sigma}^{\mu}\xi\right)^{\dagger}&
	=-\xi^{\dagger}\bar{\sigma}^{\mu}\chi,\\
	\left(\chi\sigma^{\mu\nu}\xi\right)^{\dagger}&
	=-\xi^{\dagger}\bar{\sigma}^{\mu\nu}\chi^{\dagger}\;,
}
and the Hermitian conjugates of 4-component spinor bilinears mentioned above are given by
\eq{
	\left(\bar{\Psi}_1\Psi_2\right)^\dagger &=-\bar{\Psi}_2\Psi_1,\\
	\left(\bar{\Psi}_1\gamma^{\mu}\Psi_2\right)^\dagger &=-\bar{\Psi}_2\gamma^{\mu}\Psi_1,\\
	\left(\bar{\Psi}_1\sigma^{\mu\nu}\Psi_2\right)^\dagger &=-\bar{\Psi}_2\sigma^{\mu\nu}\Psi_1,\\
	\left(\Psi^{\rm{T}}_1C\Psi_2\right)^\dagger &=-\bar{\Psi}_2C\bar{\Psi}^{\rm{T}}_1,\\
	\left(\Psi^{\rm{T}}_1C\gamma^{\mu}\Psi_2\right)^\dagger &=-\bar{\Psi}_2\gamma^{\mu}C\bar{\Psi}_1^{\rm{T}},\\
	\left(\Psi^{\rm{T}}_1C\sigma^{\mu\nu}\Psi_2\right)^\dagger &=-\bar{\Psi}_2\sigma^{\mu\nu}C\bar{\Psi}_1^{\rm{T}}.
}

\noindent\underline{2. $\sigma$ techniques}\\

The key of conversions between spinor indices and Lorentz indices is at the reduction of $\sigma$ products. We employ the following definitions: the metric $g_{\mu\nu}={\rm diag}(+1,-1,-1,-1)$; the Levi-Civita tensors $\epsilon^{0123}=-\epsilon_{0123}=+1$ and $\epsilon^{12}=\epsilon_{21}=+1$; the sigma matrices $\sigma^{\mu}_{\alpha\dot{\alpha}}=(\mathbbm{1}_{\alpha\dot{\alpha}},\tau^i_{\alpha\dot{\alpha}})^\mu$, $\bar{\sigma}^{\mu\dot{\alpha}\alpha}=(\mathbbm{1} ^{\dot\alpha\alpha},-\tau^{i\dot{\alpha}\alpha})^\mu$, with identity $\mathbbm{1}$ and Pauli matrices $\tau^i,\ i=1,2,3$.
$\sigma^{\mu}_{\alpha\dot{\alpha}}$ and $\bar{\sigma}^{\mu\dot{\alpha}\alpha}$ are related by raising and lowering indices with the $\epsilon$ tensor
$$\bar\sigma^{\mu\dot\alpha\alpha}=\epsilon^{\alpha\beta}\epsilon^{\dot{\alpha}\dot{\beta}}\sigma^{\mu}_{\beta\dot{\beta}}$$
We also define
\begin{align}
\left(\sigma^{\mu\nu}\right)_{\alpha}{}^{\beta}=&\frac{i}{2}\left(\sigma^{\mu}\bar{\sigma}^{\nu}-\sigma^{\nu}\bar{\sigma}^{\mu}\right)_{\alpha}{}^{\beta}\;,\label{eq:simunu}\\
\left(\bar{\sigma}^{\mu\nu}\right)^{\dot{\alpha}}{}_{\dot{\beta}}=&\frac{i}{2}\left(\bar{\sigma}^{\mu}\sigma^{\nu}-\bar{\sigma}^{\nu}\sigma^{\mu}\right)^{\dot{\alpha}}{}_{\dot{\beta}}\;,\label{eq:sibarmunu}
\end{align}
from which we can directly obtain the decomposition of two $\sigma$ products:
\begin{align}
\left(\sigma^{\mu}\bar{\sigma}^{\nu}\right)_{\alpha}{}^{\beta}=&g^{\mu\nu}\delta^{\beta}_{\alpha}-i\left(\sigma^{\mu\nu}\right)_{\alpha}{}^{\beta},\label{eq:2si}\\
\left(\bar{\sigma}^{\mu}\sigma^{\nu}\right)^{\dot{\alpha}}{}_{\dot{\beta}}=&g^{\mu\nu}\delta^{\dot{\alpha}}_{\dot{\beta}}-i\left(\bar{\sigma}^{\mu\nu}\right)^{\dot{\alpha}}{}_{\dot{\beta}},\label{eq:2sibar}
\end{align}
For a $\sigma$ chain consists of three or more $\sigma$'s, we may use the following three $\sigma$ decomposition
\begin{align}
\left(\sigma^{\mu}\bar{\sigma}^{\nu}\sigma^{\rho}\right)_{\alpha\dot{\beta}}=&g^{\mu\nu}\sigma^{\rho}_{\alpha\dot{\beta}}-g^{\mu\rho}\sigma^{\nu}_{\alpha\dot{\beta}}+g^{\nu\rho}\sigma^{\mu}_{\alpha\dot{\beta}}+i\epsilon^{\mu\nu\rho\lambda}\sigma_{\lambda\alpha\dot{\beta}},\label{eq:3si}\\
\left(\bar{\sigma}^{\mu}\sigma^{\nu}\bar{\sigma}^{\rho}\right)^{\dot{\alpha}\beta}=&g^{\mu\nu}\bar{\sigma}^{\nu\dot{\alpha}\beta}-g^{\mu\rho}\bar{\sigma}^{\nu\dot{\alpha}\beta}+g^{\nu\rho}\bar{\sigma}^{\mu\dot{\alpha}\beta}-i\epsilon^{\mu\nu\rho\lambda}\bar{\sigma}_{\lambda}^{\dot{\alpha}\beta},\label{eq:3sibar}
\end{align}
to recursively reduce it towards a linear combination of $\mathbbm{1},\sigma^{\mu},\bar{\sigma}^{\mu},\sigma^{\mu\nu}$, and $\bar{\sigma}^{\mu\nu}$.

To compute the trace of a $\sigma$ chain, one can simply reduce the chain to the basic forms above, and use the following equations
\eq{\label{eq:tr1}
	{\rm Tr}\;\mathbbm{1}=2,\quad {\rm Tr}\;\sigma^{\mu}=&{\rm Tr}\;\bar{\sigma}^{\mu}={\rm Tr}\;\sigma^{\mu\nu}={\rm Tr}\;\bar{\sigma}^{\mu\nu}=0\;.
}
At last we give a frequently used example of a four $\sigma$'s chain and its trace
\eq{
	&\sigma^{\mu}\bar{\sigma}^{\nu}\sigma^{\rho}\bar{\sigma}^{\kappa}=
	(g^{\mu\nu}g^{\rho\kappa}-g^{\mu\rho}g^{\nu\kappa}+g^{\nu\rho}g^{\mu\kappa}+i\epsilon^{\mu\nu\rho\kappa})\mathbbm{1} -i\left(g^{\mu\nu}\sigma^{\rho\kappa}-g^{\mu\rho}\sigma^{\nu\kappa}+g^{\nu\rho}\sigma^{\mu\kappa}+i\epsilon^{\mu\nu\rho\lambda}\sigma_{\lambda}{}^{\kappa}\right),\\
	&{\rm Tr}\; 
	\left(\sigma^{\mu}\bar{\sigma}^{\nu}\sigma^{\rho}\bar{\sigma}^{\kappa}\right)=2g^{\mu\nu}g^{\rho\kappa}-2g^{\mu\rho}g^{\nu\kappa}+2g^{\nu\rho}g^{\mu\kappa}+2i\epsilon^{\mu\nu\rho\kappa}, \\
	&{\rm Tr}\; \left(\bar\sigma^{\mu}{\sigma}^{\nu}\bar\sigma^{\rho}{\sigma}^{\kappa}\right)=2g^{\mu\nu}g^{\rho\kappa}-2g^{\mu\rho}g^{\nu\kappa}+2g^{\nu\rho}g^{\mu\kappa}-2i\epsilon^{\mu\nu\rho\kappa} .
}

\section{List of Classes up to Dimension 9}
\label{app:B}
We list all the classes of Lorentz structures from dimension 5 to dimension 9, 
where $\psi$ and $\psi^\dagger$ represent particle with helicity $-1/2$ and $1/2$ respectively, 
$F_L$ and $F_R$ represent gauge bosons with helicity $\mp 1$, $\phi$ represents scalar fields. 
The gray operators in each class are those not possible to form by SM $U(1)_Y$ singlets. 

\begin{table}
    \begin{align*}
        \begin{array}{cc|ll}
            \hline
            N & (n,\tilde{n}) & \multicolumn{2}{c}{\text{classes}} \\
            \hline
            3 & (2,0) & \color{gray}{F_{\rm L}\psi^2+h.c.} & \color{gray}{F^2_{\rm L}\phi+h.c.}\\
            \hline
            4 & (1,0) & \psi^2\phi^2+h.c. \\
            \hline
            5 & (0,0) & \color{gray}{\phi^5}\\
            \hline 
        \end{array}
    \end{align*}
    \caption{All the sub-classes of Lorentz structures at dimension 5}
\end{table}

\begin{table}
    \begin{align*}
        \begin{array}{cc|llll}
            \hline
            N & (n,\tilde{n}) & & \text{classes} & & \\
            \hline
            3 & (3,0) & F_{\rm L}^3+h.c. & & & \\
            \hline
            4 & (2,0) & \psi^4+h.c. & F^2_{\rm L}\psi^2\phi+h.c. & F^2_{\rm L}\phi^2+h.c. & \\
             & (1,1) & \psi^2\psi^{\dagger 2} & \psi\psi^{\dagger}\phi^2D & \phi^4D^2 & \\
            \hline
            5 & (1,0) & \psi^2\phi^3+h.c. & & & \\
            \hline
            6 & (0,0) & \phi^6 & & & \\
            \hline
        \end{array}
    \end{align*}
    \caption{All the sub-classes of Lorentz structures at dimension 6}
\end{table}

\begin{table}
    \begin{align*}
        \begin{array}{cc|lll}
            \hline
            N & (n,\tilde{n}) & & \text{classes} & \\
            \hline
            4 & (3,0) & \color{gray}{F^2_{\rm L}\phi^2+h.c.} & \color{gray}{F^3_{\rm L}\phi+h.c.} \\ 
             & (2,1) & \psi^3\psi^{\dagger} D+h.c. & \psi^2\phi^2D^2+h.c.  & \\
             & & \color{gray}{F^2_{\rm L}\psi^{\dagger 2}+h.c.} & \color{gray}{F_{\rm L}\psi\psi^{\dagger}\phi D+h.c.} & \color{gray}{F_{\rm L}\phi^3D^2+h.c.} \\
            \hline
            5 & (2,0) & \psi^4\phi+h.c. & F_{\rm L}\psi^2\phi^2+h.c. & \color{gray}{F^2_{\rm L}\phi^3+h.c.}\\
             & (1,1) & \psi^2\psi^{\dagger 2}\phi & \psi\psi^{\dagger}\phi^3D & \color{gray}{\phi^5D^2} \\
             \hline
            6 & (1,0) & \psi^2\phi^4 & \\
             & (0,0) & \color{gray}{\phi^7} \\
            \hline
        \end{array}
    \end{align*}
    \caption{All the sub-classes of Lorentz structures at dimension 7}
\end{table}

\begin{table}
    \begin{align*}
        \begin{array}{cc|llll}
            \hline
            N & (n,\tilde{n}) & & \text{classes} & & \\
            \hline
            4 & (4,0) & F^4_{\rm L}+h.c. & & & \\
             & (3,1) & F^2_{\rm L}\psi\psi^{\dagger}D+h.c. & \psi^4D^2+h.c. & F_{\rm L}\psi^2\phi D^2+h.c. & F^2_{\rm L}\phi^2D^2+h.c. \\
             & (2,2) & F^2_{\rm L}F^2_{\rm R} & F_{\rm L}F_{\rm R}\psi\psi^{\dagger} D & \psi^2\psi^{\dagger 2}D^2 & F_{\rm R}\psi^2\phi D+h.c. \\
             & & F_{\rm L}F_{\rm R}\phi^2D^2 & \psi\psi^{\dagger}\phi^2D^3 & \phi^4D^4 & \\
            \hline
            5 & (3,0) & F_{\rm L}\psi^4+h.c. & F^2_{\rm L}\psi^2\phi+h.c. & F^3_{\rm L}\phi^2+h.c. & \\
             & (2,1) & F_{\rm L}\psi^2\psi^{\dagger 2}+h.c. & F^2_{\rm L}\psi^{\dagger 2}\phi+h.c. & \psi^3\psi^{\dagger}\phi D+h.c. & F_{\rm L}\psi\psi^{\dagger}\phi^2D+h.c. \\
             & & \psi^2\phi^3D^2+h.c. & F_{\rm L}\phi^4D^2+h.c. & & \\
            \hline
            6 & (2,0) & \psi^4\phi^2+h.c. & F_{\rm L}\psi^2\phi^3+h.c. & F^2_{\rm L}\phi^4+h.c. & \\
             & (1,1) & \psi^2\psi^{\dagger 2}\phi^2 & \psi\psi^{\dagger}\phi^4D & \phi^6D^2 & \\
            \hline
            7 & (1,0) & \psi^2\phi^5+h.c. & & & \\
            \hline
            8 & (0,0) & \phi^8 \\
            \hline
        \end{array}
    \end{align*}
    \caption{All the sub-classes of Lorentz structures at dimension 8}
\end{table}

\begin{table}
    \begin{align*}
        \begin{array}{cc|llll}
            \hline
            N & (n,\tilde{n}) & & \text{classes} & & \\
            \hline
            4 & (4,1) & \color{gray}{F^2_{\rm L}\psi^2D^2+h.c.} & \color{gray}{F^3_{\rm L}\phi D^2+h.c.} & \\ 
             & (3,2) & \psi^3\psi^{\dagger}D^3+h.c. & \psi^2\phi^2D^4+h.c. & \color{gray}{F_{\rm L}F_{\rm R}\phi^2D^2+h.c.} & \\
             & & \color{gray}{F^2_{\rm L}\psi^{\dagger 2}D^2+h.c.} & \color{gray}{F^2_{\rm L}F_{\rm R}\phi D^2+h.c.} & \color{gray}{F_{\rm L}\psi\psi^{\dagger}\phi D^3+h.c.} & \color{gray}{F_{\rm L}\phi^3D^4+h.c.} \\
            \hline
            5 & (4,0) & \color{gray}{F^3_{\rm L}\psi^2+h.c.} & \color{gray}{F^4_{\rm L}\phi+h.c.} \\ 
             & (3,1) & F_{\rm L}\psi^3\psi^{\dagger}D+h.c. & \psi^4\phi D^2+h.c. & F_{\rm L}\psi^2\phi^2D^2+h.c. & \\
             & & \color{gray}{F^3_{\rm L}\psi^{\dagger 2}+h.c.} & \color{gray}{F^2_{\rm L}\psi\psi^{\dagger}\phi D+h.c.} & \color{gray}{F^2_{\rm L}\phi^3D^2+h.c.} \\
             & (2,2) & F_{\rm R}\psi^3\psi^{\dagger}D+h.c. & \psi^2\psi^{\dagger 2}\phi D^2 & F_{\rm R}\psi^2\phi^2D^2+h.c. & \psi\psi^{\dagger}\phi^3D^3 \\
             & & \color{gray}{F_{\rm L}F_{\rm R}^2\psi^2+h.c.} & \color{gray}{F^2_{\rm L}F^2_{\rm R}\phi} & \color{gray}{F_{\rm L}F_{\rm R}\psi\psi^{\dagger}\phi D} & \color{gray}{F_{\rm L}F_{\rm R}\phi^3D^2} \\
             & & \color{gray}{\phi^5D^4} \\
            \hline
            6 & (3,0) & \psi^6+h.c. & F_{\rm L}\psi^4\phi+h.c. & F^2_{\rm L}\psi^2\phi^2+h.c. & \color{gray}{F^3_{\rm L}\phi^3+h.c.} \\
             & (2,1) & \psi^4\psi^{\dagger 2}+h.c. & F_{\rm L}\psi^2\psi^{\dagger 2}\phi+h.c. & F^2_{\rm L}\psi^{\dagger 2}\phi^2+h.c. & \psi^3\psi^{\dagger}\phi^2D+h.c. \\
             & & F_{\rm L}\psi\psi^{\dagger}\phi^3D+h.c. & \psi^2\phi^4D^2+h.c. & \color{gray}{F_{\rm L}\phi^5D^2+h.c.} & \\
            \hline
            7 & (2,0) & \psi^4\phi^3+h.c. & F_{\rm L}\psi^2\phi^4+h.c. & \color{gray}{F^2_{\rm L}\phi^5+h.c.} & \\
             & (1,1) & \psi^2\psi^{\dagger 2}\phi^3 & \psi\psi^{\dagger }\phi^5D & \color{gray}{\phi^7D^5} & \\
            \hline
            8 & (1,0) & \psi^2\phi^6+h.c. \\
            \hline
            9 & (0,0) & \color{gray}{\phi^9} \\
            \hline
        \end{array}
    \end{align*}
    \caption{All the sub-classes of Lorentz structures at dimension 9}
\end{table}


\phantomsection
\addcontentsline{toc}{section}{\refname}

\bibliographystyle{JHEP}
\bibliography{Dim9EFTref}

\end{document}